\documentclass{aa} 
\usepackage{natbib, graphicx} 
\bibstyle{aa}
\def \etal   {{~et~al.~}}
\usepackage{txfonts}
\begin{document}
\title{Flux density measurements of a complete sample of faint blazars}
 
\author{ F. Mantovani \inst{ } \and
         M. Bondi     \inst{ } \and
	 K.-H. Mack   \inst{ }
         }

\offprints{Franco Mantovani\\
  \email{fmantovani@ira.inaf.it}}

\institute{Istituto di Radioastronomia -- INAF, via Gobetti 101,
 I--40129 Bologna, Italy 
}

\date{Received \today; accepted ???}

\abstract
{}
{We performed observations with the Effelsberg 100-m radio telescope to
measure flux densities and polarised emission of sources selected from 
the ``Deep X-ray Radio Blazar Survey" (DXRBS) to better define their 
spectral index behaviour in the radio band, with the aim to construct 
a homogeneous sample of blazars. }
{Sources were observed at four different frequencies with the Effelsberg 
100-m telescope.
We complemented these measurements with flux density data at 1.4\,GHz
derived from the NRAO VLA Sky Survey (NVSS).
}
{ The spectral indices of a sample of faint blazars were computed making use 
of almost simultaneous measurements. Sixty-six percent of the sources can be 
classified 
as ``bona fide'' blazars. Seven objects show a clearly inverted spectral index. 
Seventeen sources previously classified as flat spectrum radio quasars (FSRQ) 
are 
actually steep spectrum radio quasars (SSRQ). The flux densities obtained 
with the Effelsberg 100-m telescope at 5\,GHz are compared with the flux 
densities listed in the Green Bank (GB6) survey and in the Parkes-MIT-NRAO
(PMN) catalogue. About 43\% of 
the sources in our sample exhibit flux density variations on temporal scales 
of 19 or 22 years.
}
{We confirm that 75 out of 103 sources of the DXRBS are indeed FSRQs.
Twenty-seven sources show a spectral index steeper than $-0.5$ and should 
be classified
as SSRQs. Polarised emission was detected for 36 sources at 4.85\,GHz. The
median value of the percentage of polarised emission is ($5.8\pm0.9$)\%. Five 
sources show rotation measure (RM) values $>$ 200 rad m$^{-2}$.  }


\titlerunning{Flux density measurements of a sample of faint blazars}
\maketitle

\maketitle 
\section{Introduction} 

Blazars are an extreme class of active galactic nuclei (AGN)
characterized by high luminosity, rapid variability, and high
polarisation. In the radio band, blazars are core-dominated objects with
apparent superluminal speeds along relativistic jets pointed close to the
observer's line of sight. Blazars have flat spectral indices and include 
FSRQs and BL Lacertae objects, 
the counterparts of high- and low-luminosity radio galaxies. 
Owing to their orientation with respect to the line of sight,
blazars represent less than 5\% of all AGN, a quite rare class of sources.

The first blazar samples were selected at relatively high limiting
flux densities in the radio or X-ray band, $\sim$1 Jy or a few
times 10$^{-13}$ ergs cm$^{-2}$ s$^{-1}$, respectively 
\citep{stickel91,wood84,kollgaard96,gioia90,stocke91,perlman96}.

In the last decade of the past century, many efforts have been made to select
samples of blazars that could be representative of the whole population 
\citep{marcha94,laurent99,caccianiga99,bondi01}.
A deeper, more comprehensive sample of blazars has been constructed by 
\citet{perlman} and by \citet{landt}:  the 
``Deep X-ray Radio Blazar Survey" (DXRBS).
The DXRBS has been constructed by cross-correlating all ROSAT
sources of the WGCAT
catalogue \citep{white95} and radio sources with
flat radio spectra. The radio flux densities 
have been taken from available catalogues,
like the Green Bank GB6 \citep{gregory}, NORTH20CM \citep{white92}, and
Parkes-MIT-NRAO PMN \citep{griffith}.

The DXRBS sample is currently the faintest,
down to $\sim$50\,mJy at 5\,GHz and power $\sim$10$^{24}$ W\,Hz$^{-1}$,
and most extensive blazar sample with nearly complete optical spectroscopic
identifications.  The radio spectral information of the 
DXRBS sources is based on non-simultaneous measurements at just two
frequencies. Simultaneous flux density measurements at several frequencies   
are needed to obtain more reliable spectral indices.

In this paper we present the results from the simultaneous
multi-frequency campaign made with the Effelsberg 100-m telescope. 
In Section \ref{sec:observation} we summarise the observations and data
processing. In
Section \ref{sec:results} we present results from the observations. 
Conclusions are presented in sections \ref{sec:comm-sources} and
\ref{sec:conclusions}.
\section{Observations and data reduction} 
\label{sec:observation}

The sample of faint blazars, subject of the present investigation,
was constructed from the DXRBS catalogue. We selected all sources with 
Dec $\ge-20^\circ$ and obtained a sample of 103 blazars, 74 of which were
originally classified as FSRQs, 17 as SSRQs, and 12 as BL Lac obiects. 
They represent a complete
sample of objects with known optical spectroscopic identifications. 
Redshifts are available for 99 out of 103 sources.

We have made simultaneous flux density measurements  with the Effelsberg 
100-m telescope at 2.64\,GHz, 4.85\,GHz, 
8.35\,GHz, and/or 10.45\,GHz of the full sample of 103 sources.

The observations were carried out in the period 1--6 July 2009. 
All sources in the sample are point-like to the Effelsberg
telescope beams. They are also bright enough to be observed
by cross-scanning along the azimuth and elevation axes, with four to eight
subscans to determine the total intensity and polarisation characteristics.

Details about the observation mode, calibration and evaluation of flux 
density and polarised emission errors can be found in \citet{mantovani2009}. 
\subsection{Source coordinates and flux density measurements}
The positions for the sources listed in the DXRBS catalogue are taken
from the GB6 survey for 0$^\circ <$ Dec. $ < 70^\circ$
or PMN surveys for $-20^\circ <$ Dec. $ < 0^\circ$. 
To improve the positional accuracy even more, we have derived the positions 
of the DXRBS objects from 
the FIRST survey \citep{becker} and from the NVSS \citep{condon} 
at 1.4\,GHz. Images of a $15^\prime\times15^\prime$ field centred
at the GB6 or PMN positions were extracted from the two surveys, and 
new coordinates were obtained using the AIPS task JMFIT. 
Positions for 63 DXRBS objects were obtained from the FIRST catalogue. 
For the remaining sources the coordinates were from the NVSS.
The new coordinates are reported in 
Table~\ref{tab:fluxes} together with flux density measurements at 1.4\,GHz
from the NVSS and our own flux density measurements
with the Effelsberg 100-m telescope.
The source 3C\,286 was observed as flux density and polarisation
calibrator. 
The flux density measurements at all available frequencies are
on the \citet{baars} scale.  
\section{Results}
\label{sec:results}
Originally the spectral indices in DXRBS were defined from flux density 
measurements at 1.4\,GHz and 5\,GHz from the existing catalogues. 
This approach, however, causes a major problem. Blazars are variable sources at
all bands of the electromagnetic spectrum. 
Measurements at different frequencies obtained at different epochs may cause 
a misleading classification of the spectral indices.  
Consequently, we performed accurate flux density measurements 
of the DXRBS sample at several of the
frequencies available at the Effelsberg 100-m telescope.
\subsection{Spectral index behaviour}
Single-epoch spectral indices are determined making use of flux density
measurements taken at 2.64\,GHz, 4.85\,GHz, 8.35\,GHz, and/or 10.45\,GHz. 
Flux densities at 1.4\,GHz were taken from the NVSS. We make the assumption 
that sources alter very little in flux density with time 
at frequencies lower than about 2\,GHz (Padrielli et al. 1987). 
%
%
The spectral indices ($\alpha$ averaged over the whole spectra), have been 
classified as follows:

a) ``steep'' spectral index when $\alpha\le-0.5$; 
   ``steep $+$'' for $\alpha <-0.7$;

b) ``flat'' spectral index when $\alpha > -0.5$; 

c) ``steep--flat'' spectral index when $\alpha$ is steep at 
   lower frequencies and flattens at higher frequencies;

d) ``flat-steep'' spectral index when $\alpha$ is flat at lower frequencies and 
   steep at higher frequencies; 

e) a giga-Hertz peaked (``GPS'') spectral index when the shape is convex with
a peak at about the intermediate observing frequencies; 

f) ``inverted'' spectrum when the spectral index shape is mainly straight 
and the flux density clearly increases with frequency.

Examples of spectral index plots are shown in Fig.~\ref{fig:spixplot}.
The full set of spectral index plots is presented in Appendix A.
A crude classification of the spectral index behaviour is summarized
in Table~\ref{tab:spixtype}.

The spectral index type is listed in Table~\ref{tab:spix}. The $\alpha$
values between subsequent frequencies can be found in the spectral
index plots in Appendix A. For the source 
J0513.8$+$0156 only one flux density measurement is available due to 
technical problems during the observations. 
\subsection{Flux density variability}
Observations made at 4.85\,GHz have also allowed us to look for any flux 
density variability through a comparison with the measurements available in the
GB6 and PMN catalogues.

The GB6 survey
has been built from observations performed in 1986-87. The catalogue contains
sources with $S_{4.85\,GHz}>$18\,mJy. A 20\,mJy source has a typical 
error of 4-5\,mJy. The PMN survey was conducted in 1990 using the Parkes 64-m
radiotelescope and the NRAO seven-beam receiver at 4.85 GHz covering the
whole sky in the declination range $-87^\circ < \delta < 10^\circ$.
The PMN survey in its equatorial region, which is the one relevant for the
comparison with our southern sources, has a flux density limit of 40 mJy.
The sources in our sample observed by Effelsberg at 4.85 GHz and listed in the
GB6 and PMN catalogues are 67 and 35, respectively.
Effelsberg measurements are reported in Table~\ref{tab:fluxes}, 
while GB6 and PMN measurements are listed in Table~\ref{tab:spix}.

\begin{figure*}
   \centering
   \includegraphics[width=16cm]{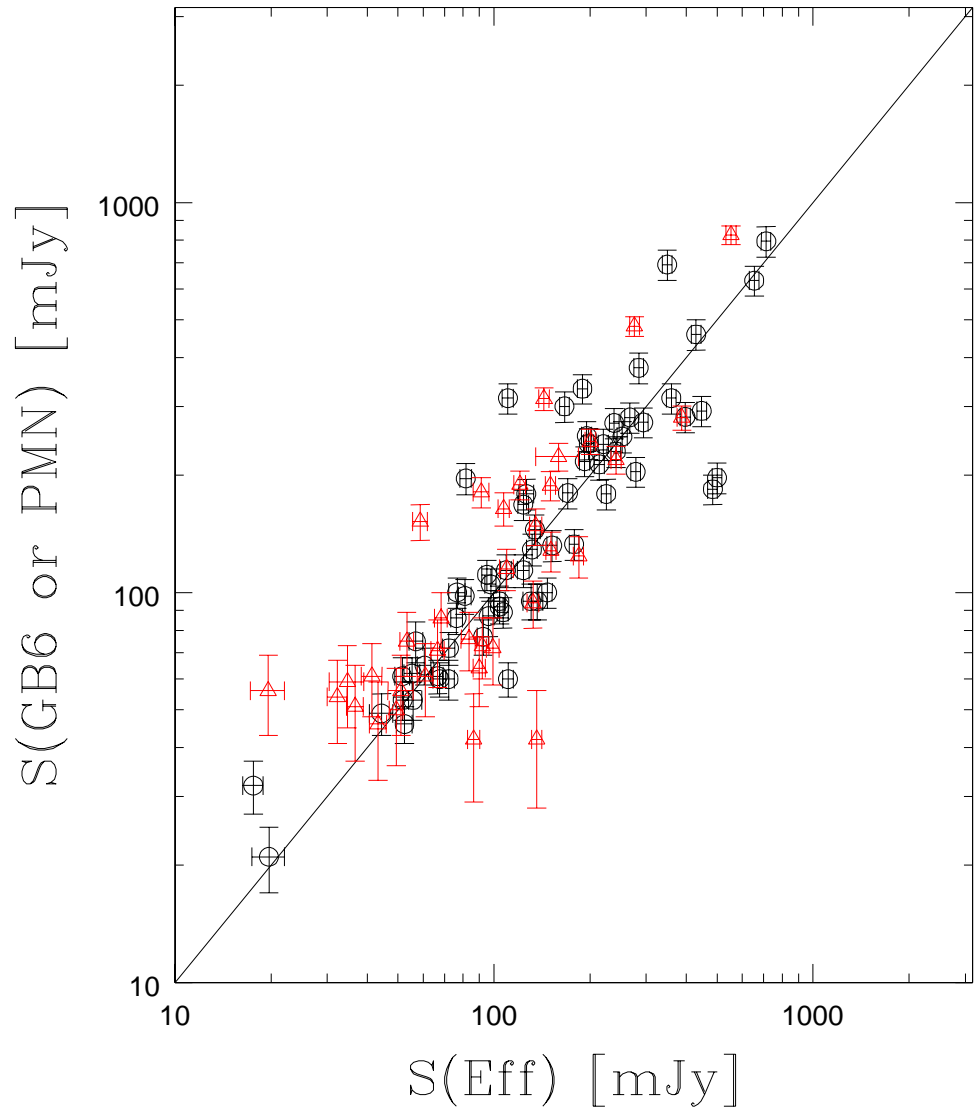}
      \caption{Effelsberg flux densities versus GB6 (points) or 
PMN  (triangles) flux densities at 5\,GHz.
The straight line means a ratio of 1. 
            \label{fig:fluxratio}
              }
   \end{figure*}

Firstly, we checked for any possible systematic difference between our
measurements and those from the GB6 and PMN catalogues.
In Fig.~\ref{fig:fluxratio} we plot the Effelsberg flux densities {\it versus} 
the GB6 (points) or PMN (triangle) flux densities at 5\,GHz.   
The flux density measurements are well-placed
around the equal flux density line, with a scatter given by the variability. 
More quantitatively, the mean value and dispersion for the ratio between the
Effelsberg and GB6 flux densities are 1.00 and 0.07. From the comparison
between the Effelsberg and the PMN flux density we obtain a slightly lower
value with a much higher dispersion 0.97 and 0.54, respectively.
This is consistent with the original finding from the comparison between the
PMN and GB6 surveys. 

For the sources in the GB6 catalogue we can compare observations
made at three epochs: our data taken in 2009 and the two measurements 
made in 1986
and 1987, which were used for the GB6 catalogue. We used the JAVA applet
available at the GB6 catalogue web page (http://pulsar.phas.ubc.ca/) 
to retrieve the flux
densities at epochs 1986 and 1987 for all our sources with declination 
$> 0^\circ$.

\begin{figure*}
   \centering
   \includegraphics[width=16cm]{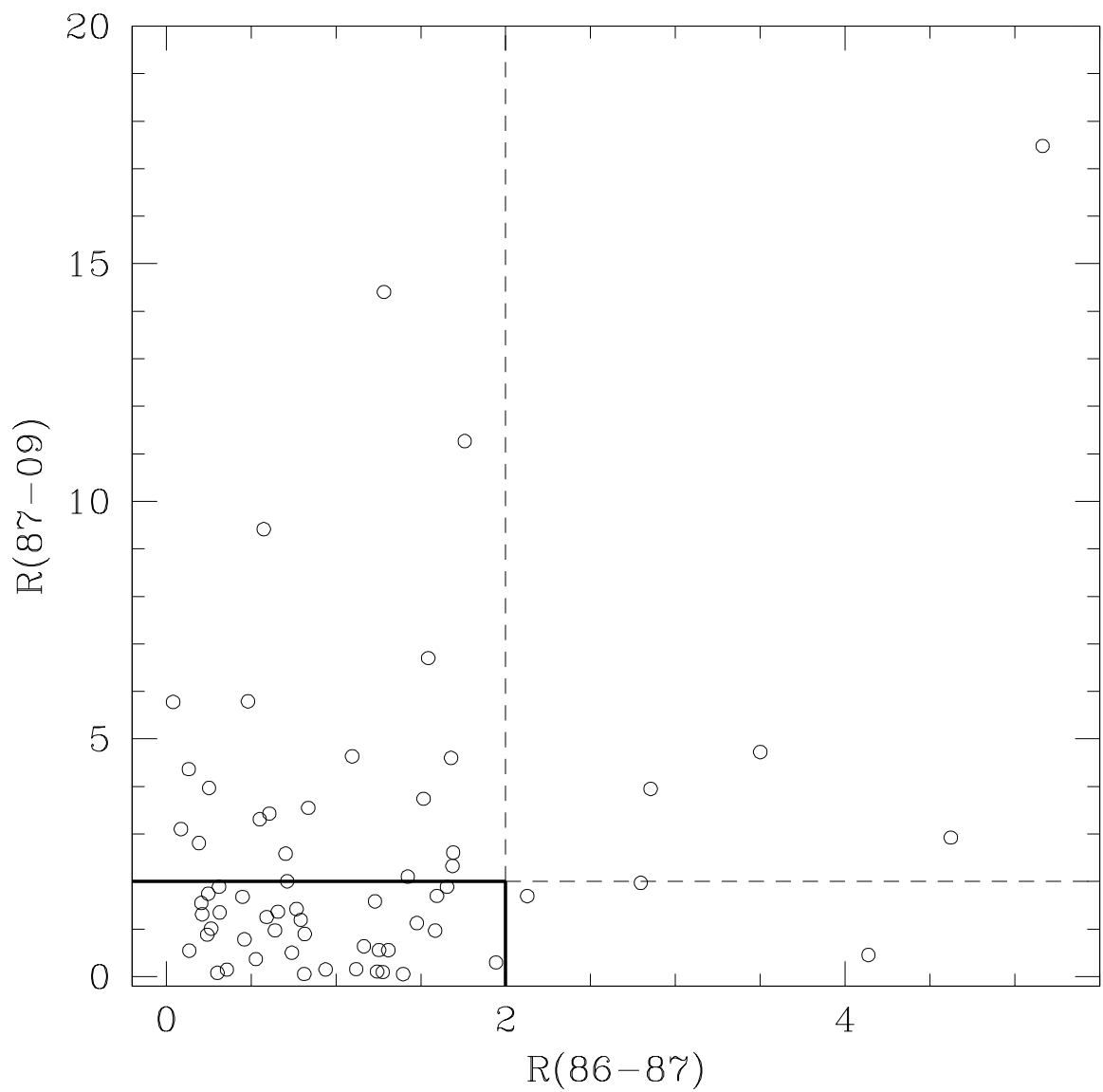}
      \caption{The $R$ values for both short and long-term variability.
            \label{fig:longterm}
              }
   \end{figure*}

The long-term variability in the GB6 catalogue is parametrized using the
 following quantity:
 
$$
R = \mid S_{87} - S_{86} \mid / \sigma~~,
$$

where $\sigma$ is the combined error of the two flux density measurements.
Out of the 67 sources in common, four do not possess a calculated $R$ value 
because
they lack either the 1986 or 1987 flux density measurement.
We calculated the same quantity $R$ using our 2009 Effelsberg observations and
the 1987 measurements from the GB6 survey and we plot the outcome in
Fig.~\ref{fig:longterm}. We consider variables the sources with 
$R > 2 $. We find that 55\% of the sources (35 objects) do not show significant 
variations both in the time intervals 1986 to 1987 and 1987 to 2009, 
11\% of the sources (seven objects) were
found variable between 1986-1987 compared to 40\% (twentyfive objects) of 
variable 
sources between
1987 and 2009; finally 5\% of the sources (three objects) showed significant 
variability between
1986 and 1987 but were found non-variable on the longer time 1987-2009.
Considering also the four objects for which we do not have the 
$R_{86-87}$ values,
we have a total of 28 radio sources (42\%) with $R_{87-09} > 2$ and therefore,
according to the threshold adopted, variable in a time interval of 22 years.

We calculated $R$ also for the 35 sources with $\delta\le 0$ that have the 
PMN measurements and we find that 45\% of the sources (16
objects) have $R_{90-09} > 2$. This fraction is consistent with that of
the northern sources and we can conclude that 43\% of the DXBRS
sample presented in this paper show significant variability on a time
interval of about 20 years. The quantity $R$ is reported in 
Table~\ref{tab:spix}.  
\subsection{Polarised emission}
The Effelsberg observations were carried out in full polarisation.
We consider a source to be polarised when the intensity of the polarised
emission is three times the rms error estimated for the polarised emission 
source by source . Results are presented in Table~\ref{tab:polar}.

At 1.4\,GHz, 27 sources show a polarised emission below the detection limits
according to the data extracted from the NVSS. One source in our
list was not found by the extraction process.
The remaining sources have a median value of the fractional polarisation m 
of $3.0^{+0.4}_{-0.8}$\%.

At 2.64\,GHz we found 25 sources polarised with a median value of m
of 4.9$^{+1.1}_ {-0.4}$\,\%. At 4.85\,GHz we 
found 36 sources polarised with a median m of  
($5.8\pm0.9$)\%. These integrated values are considered typical of AGN. They
can be compared with those achieved by 
\citet{mantovani2009} on a sample of compact steep-spectrum sources
and by \citet{klein2003} on a sample of steep spectrum extended
radio sources selected from the B3-VLA sample, as shown in
Table~\ref{tab:m-values}. The trend is a quick decrease in the fractional 
polarisation with increasing wavelength. However, cases of repolarisation, 
i.e. an
increase of fractional polarisation towards longer wavelengths, were reported, 
like that for the source 3C455 (see \citet{mantovani2009}, 
worth of further investigation. Repolarisation is seen for thirteen 
faint blazars in our sample, which show a peak of m at about 2.6\,GHz. 
Systematic instrumental effects to possibly explain this behaviour can be 
excluded. More accurate observations are required 
to confirm this trend of the fractional polarisation.

Rotation measure could be calculated for 27 sources, of which 
seventeen are classified as $bona fide$ blazars (FSRQs), nine as  
SSRQs, and one as a GPS object, according to our flux density measurements.
The RM values are in the range from 0 to 1950 rad m$^{-2}$
in the source rest frame (i.e. RM redshift corrected). 
The majority of the sources show a quite low
RM. No clear correlation of m with the spectral index type (SSRQs or FSRQs) 
could be found.
An example of RM and m plot is presented in Fig.~\ref{fig:RM-m-plot}.
The full set of plots can be found in Appendix B.
%
%
\section{Comments on individual sources}
\label{sec:comm-sources}
There are a few sources worth of brief comments.

{\bf J0204.8$+$1514 --}   
Adding 180$^\circ$ to the EVPA at 1.4\,GHz produces just a slightly worse fit. 
In this case the  RM is $-22.5$ rad m$^{-2}$.

\noindent
{\bf J1025.9$+$1253 --} 
A fit of equal goodness and a RM of $-$85.5 rad m$^{-2}$ results from rotation 
of the EVPA at 2.64\,GHz by 180$^\circ$ and the EVPA at 1.4\,GHz by 360$^\circ$.

\noindent
{\bf J0937.1$+$5008 --} 
The source J0937.1$+$5008 exhibits a high degree of variability at 5\,GHz,
of the order of 60\% to 70\% on a time scale of about 22 years comparing the
Effelsberg flux density and the GB6 flux density. Moreover, it shows an 
increase in flux density by a factor of 2.3 comparing the Effelsberg
flux density with the flux density measured few months later (22 Oct
2009) with the European VLBI Network observations at the same frequency
(Mantovani et al. in prep.).
Short-term variability was not found comparing
the two measurements made in 1986 and 1987 by the GB6. 
A preliminary milli-arcsecond
resolution image of J0937.1$+$5008 (resolution better than 5 mas; rms noise
1.5 mJy/beam) shows that the source is point-like \citep{mantovani2010}. 
J0937.1$+$5008 was detected by the Fermi Gamma-Ray Space Telescope \citep{abdo}.
This source presents a peculiar ``concave'' spectral index as shown
in Fig.~\ref{fig:j0937}.  A similar spectral index shape is also shown by 
the source J1028.6--0336.

\noindent
{\bf J1626.6$+$5809 --} 
Removing the EVPA at 1.4\,GHz, a good fit is
achieved using the three Effelsberg measurements and a rotation by 180$^\circ$
at 2.64\,GHz. The RM in this case is $-16.4$ rad m$^{-2}$.

\noindent
{\bf J1648.4$+$4104 --} 
A better fit to the data can be achieved by rotating
the EVPA at 1.4\,GHz by 360$^\circ$. The RM is now $170$ rad m$^{-2}$.  
\section{Conclusions}
\label{sec:conclusions}
Our investigation aims at veryfing the spectral index classification 
of the DXRBS sample. We define a spectral
index $\alpha\le-0.5$ as ``steep''. If $\alpha>-0.5$ it is considered
``flat''. However, the spectra often show a complex shape, i.e. not a steep or 
flat straight line. This makes the classification more difficult. 
We have attempted a ``crude'' classification source by source in
Table~\ref{tab:spix}, while spectral indices are available in the
plots we produced for each source (see Appendix A).

The majority of sources, 66 out of 102 (for one source we do not have
enough measurements) can be considered $bona fide$ blazars, namely
those with ``flat'', ``steep-flat'', ``inverted'', and possibly also 
``flat-steep'' spectra in the present classification. 

Several sources changed their
spectral index classification with respect to that reported in the original
list by \citet{perlman} and by \citet{landt}. 
Among them we have nine sources with a ``convex'' spectral index
peaking in the GHz regime, which should be more properly defined as
giga-Hertz peaked sources (GPS). These sources cannot be considered 
blazars.
We found that 17 sources previously classified as FSRQs are actually SSRQs. 
Moreover, we found six sources with spectral indices even steeper than $-0.7$, 
failing one of the original selection criteria. In total, we found that 
27 sources can be classified as SSRQs showing a spectral index $<-0.5$. 
This result suggests that a statistically meaningful comparison between
FSRQs and SSRQs in the same flux density limited complete sample 
is possible.

We can anticipate that all 42 sources belonging to our sample observed
so far with the EVN at 5\,GHz have been detected (Mantovani et al., in prep.).
Six sources, previously classified SSRQs (and found with $\alpha<-0.7$), 
host AGN. Two of them show a core-jet
structure, while the remaining four are point-like at a resolution better
than 5 milli-arcseconds. The EVN to Effelsberg flux density ratio ranges
from 0.05 to 0.8. Most of their radio emission is therefore
not coming from the cores of these sources.

Fourty-three percent of the DXRBS sample show significant variability on a 
time scale of 19--22 years; 11\% show short-term variability comparing Green
Bank flux density measurements taken one year apart in 1986 and 1987.

About 25\% of the sources show polarised emission above the detection
limits of our observations. We also extracted polarimetric information from the
NVSS to collect at least the three measurements needed to compute the 
RM diminishing the $n\pi$ ambiguity. The majority of the sources
show
$\mid$RM$\mid$ $<200$ rad m$^{-2}$ in the source rest frame,
possibly produced by the Faraday depth of the Galactic foreground. 
In nine cases, namely J0029.0$+$0509, 
J0322.6$-$1335, J0434.3$-$1443, J0510.0$+$1800, J0518.2$+$0624, 
J0744.8$+$2920, J1400.7$+$0425, J1419.1$+$0603, and J1648.4$+$4104
the rest frame RM are in the range of 200$<\mid$RM$\mid<$1940 rad m$^{-2}$.
The highest value found is 1938.8 rad m$^{-2}$ 
for the source J0434.3$-$1443. 
\begin{acknowledgements}
We thank an anonymous referee for his/her very helpful comments
and suggestions, and for a careful reading of the manuscript of this paper.
This work is based on observations with the 100-m telescope of the MPIfR
(Max-Planck-Institut f\"ur Radioastronomie) at Effelsberg. It has benefited
from research funding from the European Community's Framework Programme
under RadioNet R113CT 2003 5058187.
FM likes to thank Anton Zensus for the
kind hospitality at the Max-Planck-Institut f\"{u}r Radioastronomie, Bonn,
for a period during which part of this work has been done.
The authors like to thank Heinz Andernach for suggesting to check
for short-term flux density variability making use of Green Bank measurements.
This research has made use of the NASA/IPAC Extragalactic Database (NED), 
which is operated by the Jet Propulsion Laboratory, California Institute 
of Technology, under contract with the National Aeronautics and Space 
Administration. Part of this work was supported by the: COST Action MP0905
'Black Holes in a Violent Universe'.
\end{acknowledgements} 
\newpage
%
\begin{table*}
\caption{Flux densities at the available frequencies. 
}
\label{tab:fluxes}
\tabcolsep0.1cm
\begin{tabular}{lrrrrrrrrrrrr}
\hline
   name        & RA(J2000)      & Dec(J2000)    &S$_{1.4}$&  rms &S$_{2.64}$& rms &S$_{4.85}$ & rms &S$_{8.35}$& rms &S$_{10.45}$&  rms  \\
               &               &               &   mJy   &  mJy &   mJy    & mJy &   mJy    & mJy &   mJy   & mJy &   mJy    &   mJy\\
\hline
 J0012.5--1629 & 00 12 33.83  & --16 28 06.8 &   94.3  &  2.9 &   55.1 &  7.0 &   49.4  &   2.5  &       &     & 30.0 & 6.9      \\   
 J0015.5+3052  & 00 15 36.13  &  +30 52 24.0 &  234.0  &  7.0 &  145.6 &  3.2 &  106.7  &   4.0  &       &     & 51.9 & 7.9      \\  
 J0029.0+0509  & 00 29 03.60  &  +05 01 34.1 &  439.8  & 13.2 &  362.3 &  6.2 &  284.2  &   10.3 &       &     &147.9 & 15.7      \\  
 J0106.7--1034 & 01 06 44.09  & --10 34 10.0 &  276.0  &  9.7 &  186.4 &  3.9 &  150.3  &   5.6  &       &     & 93.7 & 10.1      \\  
 J0110.5--1647 & 01 10 35.20  & --16 48 31.0 &  105.6  &  3.7 &  100.1 &  3.7 &   99.1  &   4.5  &       &     & 95.6 & 10.8      \\  
 J0125.0+0146  & 01 25 05.48  &  +01 46 26.3 &  186.6  &  5.6 &  152.0 &  2.8 &  136.8  &   5.4  &       &     & 93.1 & 10.0      \\  
 J0126.2--0500 & 01 26 15.36  & --05 01 18.5 &   58.4  &  2.2 &   41.1 &  2.8 &   32.3  &   2.3  &       &     & 19.9 & 5.1      \\  
 J0204.8+1514  & 02 04 50.40  &  +15 14 11.2 & 4067.7  &122.0 & 2598.1 & 34.1 & 2132.2  &   73.7 &       &    &1239.9 & 117.8      \\  
 J0210.0--1004 & 02 10 00.14  & --10 03 52.8 &  265.5  &  8.0 &  263.8 &  5.0 &  200.4  &   7.6  &       &     & 90.4 & 11.1      \\  
 J0227.5--0847 & 02 27 32.08  & --08 48 13.1 &   65.3  &  2.0 &  118.9 &  2.9 &  109.5  &   5.4  &       &     & 66.7 & 15.4      \\    
 J0245.2+1047  & 02 45 15.05  &  +10 46 53.0 &  417.4  & 13.6 &  289.2 &  4.4 &  192.3  &   6.9  &       &     & 76.0 & 8.7      \\  
 J0304.9+0002  & 03 04 59.23  &  +00 02 33.4 &  123.8  &  4.3 &   77.6 &  4.3 &   53.3  &   2.6  & 32.8 & 5.1 & 29.0 & 6.0      \\   
 J0322.6--1335 & 03 22 38.45  & --13 35 17.8 &  266.6  &  8.0 &         &      &  107.3  &   4.2  & 87.2 & 1.8 &       &          \\  
 J0340.8--1814 & 03 40 48.05  & --18 14 00.0 &  299.6  & 10.3 &  197.6 & 11.7 &  135.2  &   5.9  & 79.8 & 2.1 &       &          \\  
 J0411.0--1637 & 04 11 00.60  & --16 36 07.9 &  183.5  &  6.0 &  152.5 &  8.7 &  184.2  &   6.7  &133.8 & 3.5 &       &          \\  
 J0414.0--1224 & 04 14 05.98  & --12 24 17.2 &   93.0  &  3.3 &   62.2 &  6.1 &   58.7  &   3.2  & 47.2 & 1.9 &       &          \\  
 J0414.0--1307 & 04 14 03.14  & --13 06 38.9 &  312.4  &  9.4 &  199.9 &  4.4 &  136.1  &   5.0  & 83.6 & 1.9 &       &          \\  
 J0421.5+1433  & 04 21 33.11  &  +14 34 03.0 &  273.2  &  9.3 &  163.5 &  4.4 &  109.1  &   4.1  & 56.5 & 1.5 &       &          \\  
 J0427.2--0756 & 04 27 14.21  & --07 56 24.8 &   74.1  &  2.6 &   57.9 &  2.7 &   51.1  &   4.6  & 45.6 & 1.3 & 37.1 & 7.0      \\  
 J0434.3--1443 & 04 34 19.05  & --14 42 55.9 &  205.2  &  6.2 &  384.3 &  5.8 &  387.9  &   13.6 &297.5 & 5.2 &       &          \\  
 J0435.1--0811 & 04 35 08.38  & --08 11 03.2 &   50.1  &  1.9 &   62.7 &  2.3 &   91.8  &   4.5  &       &     & 83.0 & 9.3      \\  
 J0447.9--0322 & 04 47 54.75  & --03 22 43.2 &   86.7  &  2.6 &   35.6 &  3.3 &   19.6  &   2.4  &       &     &       &          \\  
 J0502.5+1338  & 05 02 33.22  &  +13 38 11.3 &  544.8  & 16.3 &  344.7 &  5.2 &  430.0  &   15.0 &       &     &327.7 & 31.6      \\  
 J0510.0+1800  & 05 10 02.39  &  +18 00 41.8 &  704.3  & 21.1 &  533.0 &  8.1 &  712.8  &   24.7 &826.8 & 14.6&       &          \\  
 J0513.8+0156  & 05 13 51.94  &  +01 56 56.1 &  348.4  & 11.7 &  231.5 &  5.2 &          &        &       &     &       &          \\  
 J0518.2+0624  & 05 18 15.99  &  +06 24 22.5 &  571.0  & 17.1 &  375.0 &  5.7 &  241.2  &   8.5  &151.6 & 3.3 &       &          \\  
 J0535.1--0239 & 05 35 12.27  & --02 39 07.0 &   46.9  &  1.9 &   44.0 & 15.1 &   86.4  &   3.7  & 72.2 & 1.7 &       &          \\  
 J0646.8+6807  & 06 46 41.50  &  +68 07 42.9 &  116.0  &  4.0 &   76.9 &  3.0 &   72.2  &   3.1  &       &     & 74.5 & 8.3      \\  
 J0651.9+6955  & 06 51 54.56  &  +69 55 26.3 &  275.7  &  8.3 &  183.9 &  3.1 &  152.1  &   5.5  &       &     &109.1 & 11.1      \\  
 J0724.3--0715 & 07 24 17.32  & --07 15 19.7 &  330.4  &  9.9 &  255.4 &  4.3 &  275.6  &   9.7  &292.4 & 5.3 &       &          \\  
 J0744.8+2920  & 07 44 51.20  &  +29 20 08.0 &  337.8  & 11.8 &  272.6 &  4.7 &  224.9  &   7.9  &168.5 & 3.2 &       &          \\  
 J0816.0--0736 & 08 16 08.48  & --07 37 12.3 &   93.6  &  2.8 &   91.7 &  4.4 &   41.4  &   3.1  & 34.0 & 6.1 &       &          \\  
 J0829.5+0858  & 08 29 30.27  &  +08 58 21.3 &  333.9  & 10.0 &  234.3 &  3.6 &  170.3  &   6.2  &139.7 & 2.6 &       &          \\  
 J0847.2+1133  & 08 47 12.94  &  +11 33 50.2 &   32.8  &  1.1 &   21.0 &  2.4 &   17.6  &   1.3  & 17.6 & 3.6 &       &          \\  
 J0853.0+2004  & 08 53 02.75  &  +20 04 21.5 &  128.6  &  3.9 &   89.4 &  2.3 &   67.5  &   3.1  & 61.0 & 2.0 &       &          \\  
 J0908.2+5031  & 09 08 16.55  &  +50 31 05.4 &  171.6  &  5.2 &  102.9 &  2.7 &   76.3  &   3.2  &       &     & 59.7 &  8.0     \\  
 J0927.7--0900 & 09 27 46.93  & --09 00 22.2 &  148.3  &  4.5 &  130.8 &  3.8 &   91.2  &   5.2  &104.9 & 2.5 &       &          \\  
 J0931.9+5533  & 09 31 58.09  &  +55 33 13.1 &   95.7  &  3.5 &  101.8 &  2.8 &   55.7  &   3.7  &       &     & 42.7 & 12.1      \\  
 J0937.1+5008  & 09 37 12.32  &  +50 08 52.2 &  166.6  &  5.0 &   69.1 &  3.1 &  110.5  &   4.3  &       &     &269.6 & 26.4      \\  
 J0940.2+2603  & 09 40 14.73  &  +26 03 29.3 &  462.2  & 13.9 &  441.2 &  6.3 &  447.6  &   15.5 &       &     &378.7 & 36.4      \\  
J1006.1+3236   & 10 06 07.73  &  +32 36 27.7 &  473.8  & 16.7 &  282.5  &   4.9 &  194.9  &   7.0  &       &     &102.4 & 11.4      \\  
 J1006.5+0509  & 10 06 37.64  &  +05 09 53.5 &  172.4  &  5.2 &  108.2 &  4.1 &  126.3  &   6.6  &       &     &109.8 & 12.1      \\  
 J1010.8--0201 & 10 10 51.78  & --02 00 19.0 &  532.4  & 18.4 &  689.1 &  9.7 &  553.0  &   20.1 &       &     &499.5 & 48.1      \\  
 J1011.5--0423 & 10 11 30.16  & --04 23 29.6 &  158.4  &  4.8 &  151.7 &  3.9 &  120.5  &   5.0  &       &     &186.9 & 20.3      \\  
 J1025.9+1253  & 10 25 56.34  &  +12 53 48.8 &  539.0  & 16.2 &  508.6 &  7.5 &  654.4  &   22.7 &       &     &665.7 & 63.4      \\  
 J1026.4+6746  & 10 26 33.11  &  +67 46 12.5 &  234.0  &  7.9 &  167.7 &  3.5 &  131.6  &   5.0  &       &     & 92.6 & 10.5      \\  
 J1028.5--0236 & 10 28 34.05  & --02 36 59.7 &   86.6  &  2.6 &   98.1 &  3.2 &  132.4  &   5.5  &       &     &203.0 & 20.5      \\  
 J1028.6--0336 & 10 28 40.37  & --03 36 19.4 &  155.9  &  4.7 &   96.7 &  2.6 &   60.9  &   3.5  &       &     & 92.4 & 14.8      \\  
 J1032.1--1400 & 10 32 06.28  & --14 00 20.0 &  195.6  &  5.9 &  139.8 &  4.1 &  159.3  &  24.4 &       &     &134.5 & 13.9      \\  
 J1101.8+6241  & 11 01 53.80  &  +62 41 56.0 &  726.5  & 24.2 &  462.2 &  7.1 &  348.9  &   12.2 &       &     &186.5 & 18.7      \\  
 J1105.3--1813 & 11 05 19.19  & --18 13 14.0 &   21.6  &  0.8 &   26.5 &  2.9 &   34.7  &   4.3  &       &     &       &          \\  
 J1116.1+0828  & 11 16 09.96  &  +08 29 22.1 &  244.7  &  7.4 &  443.7 & 16.2 &  397.8  &   14.8 &       &     &310.7 & 30.8      \\  
 J1120.4+5855  & 11 20 27.27  &  +58 56 13.0 &   90.7  &  2.7 &   73.5 &  2.6 &   52.4  &   2.9  &       &     & 36.6 & 9.5      \\  
 J1150.4+0156  & 11 50 24.83  &  +01 56 19.0 &  153.2  &  5.5 &  120.2 &  6.7 &  130.6  &   7.4  &       &     & 91.7 & 9.4      \\  
 J1204.2--0710 & 12 04 16.70  & --07 10 09.6 &  167.7  &  6.0 &  173.5 &  4.9 &  150.8  &   5.6  &       &     &128.6 & 14.5      \\  
 J1206.2+2823  & 12 06 19.62  &  +28 22 54.4 &   37.8  &  1.2 &   27.0 &  2.8 &   19.7  &   2.3  & 18.6 & 1.2 &       &          \\  
 J1213.0+3248  & 12 13 03.81  &  +32 47 36.8 &  139.5  &  4.2 &   91.7 &  2.5 &   66.6  &   2.7  & 60.2 & 1.8 &       &          \\  
 J1213.2+1443  & 12 13 15.24  &  +14 44 02.5 &  113.5  &  3.8 &   77.9 &  2.6 &   51.6  &   2.8  & 31.5 & 1.7 &       &          \\  
 J1217.1+2925  & 12 17 08.31  &  +29 25 34.0 &   66.2  &  2.0 &   44.4 &  3.0 &   44.4  &   3.7  & 41.4 & 1.4 &       &          \\  
 J1222.6+2934  & 12 22 43.14  &  +29 34 40.4 &   95.0  &  2.9 &  107.0 &  3.4 &  110.6  &   4.3  & 89.9 & 3.4 &       &          \\  
 J1223.9+0650  & 12 23 54.66  &  +06 50 02.0 &  275.7  &  8.3 &  266.4 &  4.4 &  252.3  &   8.9  &       &     &210.1 & 20.9      \\  
 J1224.5+2613  & 12 24 33.39  &  +26 13 14.3 &  681.0  & 21.4 &  386.9 &  6.5 &  238.1  &   8.4  &135.9 & 2.7 &       &          \\  
 J1225.5+0715  & 12 25 31.22  &  +07 15 52.1 &  123.9  &  4.3 &   75.1 &  3.8 &   56.9  &   2.8  &       &     & 27.7 & 6.1      \\  
 J1229.5+2711  & 12 29 34.26  &  +27 11 56.8 &  160.2  &  4.8 &  107.8 &  3.3 &  123.5  &   4.8  &133.6 & 2.7 &       &          \\  
 J1231.7+2848  & 12 31 43.82  &  +28 47 50.9 &  140.5  &  5.5 &  147.5 &  3.4 &  123.5  &   5.3  & 99.2 & 2.1 &       &          \\   
 J1311.3--0521 & 13 11 17.78  & --05 21 21.7 &   69.8  &  2.5 &   51.0 &  6.6 &   43.3  &   2.7  &       &     & 37.0 & 7.8      \\  
 J1315.1+2841  & 13 15 13.65  &  +28 40 52.9 &  110.3  &  3.3 &  118.7 &  3.5 &  103.9  &   4.4  & 71.1 & 1.9 &       &          \\  
 J1320.4+0140  & 13 20 26.78  &  +01 40 36.5 &  670.7  & 20.1 &  534.3 &  7.3 &  486.0  &   16.9 &491.1 & 8.8 &       &          \\  
 J1329.0+5009  & 13 29 05.87  &  +50 09 27.0 &  246.3  &  7.4 &  173.8 &  4.1 &  178.3  &   6.4  &156.2 & 3.0 &       &          \\  
 J1332.7+4722  & 13 32 45.27  &  +47 22 22.3 &  233.2  &  7.0 &  222.2 &  3.7 &  189.3  &   6.9  &183.9 & 3.3 &       &          \\  
 J1337.2--1319 & 13 37 14.84  & --13 19 17.0 &  113.8  &  4.1 &   88.8 &  2.9 &   66.6  &   3.1  & 60.8 & 1.5 &       &          \\
\hline
\end{tabular}
\end{table*}
\newpage
\begin{table*}
\tabcolsep0.1cm
\begin{tabular}{lrrrrrrrrrrrr}
\hline
   name        & RA(J2000)      & Dec(J2000)    &S$_{1.4}$&  rms &S$_{2.64}$& rms &S$_{4.85}$ & rms &S$_{8.35}$& rms &S$_{10.45}$&  rms  \\
               &               &               &   mJy   &  mJy &   mJy    & mJy &   mJy    & mJy &   mJy   & mJy &   mJy    &   mJy\\
\hline
 J1359.6+4010  & 13 59 38.104  &  +40 11 37.98 &  162.8  &  4.9 &  199.4 &  3.1 &  266.5  &   9.4  &265.3 & 4.7 &       &          \\  
 J1400.7+0425  & 14 00 48.405  &  +04 25 30.50 &  322.5  &  9.7 &  197.8 &  3.7 &  153.9  &   5.6  &150.4 & 2.8 &       &          \\  
 J1404.2+3413  & 14 04 16.732  &  +34 13 16.10 &  141.0  &  4.9 &   92.3 &  3.5 &   55.3  &   3.1  &       &     & 25.5 & 3.7      \\  
 J1406.9+3433  & 14 06 53.860  &  +34 33 37.06 &  169.2  &  5.1 &  248.2 &  3.7 &  278.1  &   9.8  &       &     &189.0 & 18.8      \\  
 J1416.4+1242  & 14 16 28.639  &  +12 42 13.53 &  110.2  &  3.3 &   95.9 &  2.6 &   81.0  &   3.7  & 75.7 & 2.2 &       &          \\  
 J1417.5+2645  & 14 17 30.372  &  +26 44 56.81 &   75.6  &  2.3 &   86.3 &  2.6 &   92.4  &   3.7  &       &     & 82.0 & 9.2      \\  
 J1419.1+0603  & 14 19 09.312  &  +06 03 30.24 &  374.6  & 11.2 &  283.1 &  5.8 &  220.2  &   7.9  &177.7 & 3.3 &       &          \\  
 J1420.6+0650  & 14 20 40.958  &  +06 51 03.74 &  533.6  & 18.2 &  326.5 &  4.8 &  197.9  &   7.1  &129.9 & 3.0 &       &          \\   
 J1423.3+4830  & 14 23 17.955  &  +48 30 15.47 &  208.5  &  6.3 &         &      &  146.8  &   5.4  &       &     &103.9 & 11.1      \\  
 J1427.9+3247  & 14 27 58.325  &  +32 47 40.05 &   88.0  &  3.5 &   98.2  &  2.8 &   60.5  &   3.0  &       &     & 31.1 & 6.0      \\  
 J1442.3+5236  & 14 42 19.598  &  +52 36 21.09 &  187.2  &  5.6 &  132.2 &  3.0 &   95.0  &   3.6  &       &     & 51.3 & 6.3      \\  
 J1507.9+6214  & 15 07 57.346   &  +62 13 34.64 &  549.7  & 16.5 &  327.9 &  5.0 &  213.9  &  7.6  &       &     &103.4 & 11.3      \\  
 J1539.1--0658 & 15 39 09.627  & --06 58 43.58 &   46.8  &  1.5 &   71.1 &  3.3 &   83.5  &   4.6  & 77.0 & 2.1 &       &          \\  
 J1543.6+1847  & 15 43 43.810  &  +18 47 20.31 &  356.2  & 10.7 &  206.8 &  3.9 &  166.4  &   5.9  &       &     &127.3 & 12.9      \\  
 J1606.0+2031  & 16 06 05.922  &  +20 32 09.27 &  180.8  &  5.4 &  105.2 &  3.3 &   81.6  &   3.3  & 66.8 & 1.6 &       &          \\  
 J1626.6+5809  & 16 26 37.429  &  +58 09 11.48 &  532.9  & 18.3 &  400.6 &  5.8 &  359.7  &  12.5  &298.9 & 5.5 &       &          \\  
 J1629.7+2117  & 16 29 47.534  &  +21 17 17.51 &  264.4  &  7.9 &  151.9 &  3.4 &   97.6  &   3.9  & 55.6 & 1.4 &       &          \\  
 J1648.4+4104  & 16 48 29.314  &  +41 04 05.84 &  243.0  &  7.3 &  373.8 &  5.6 &  499.9  &  17.4  &534.5 & 9.4 &       &          \\  
 J1656.6+5321  & 16 56 39.687  &  +53 21 48.51 &   94.4  &  2.9 &  135.9 &  2.8 &  134.3  &   5.3  &112.6 & 3.2 &       &          \\   
 J1656.8+6012  & 16 56 48.308  &  +60 12 16.09 &  292.2  &  8.8 &  475.5 &  7.1 &  484.7  &   16.8  &435.0 & 7.8 &       &          \\  
 J1722.3+3103  & 17 22 18.547  &  +31 03 26.23 &  129.4  &  4.2 &   82.5 &  2.9 &   51.7  &   2.3  & 29.4 & 1.3 &       &          \\  
 J1804.7+1755  & 18 04 42.496  &  +17 55 58.76 &  218.0  &  7.2 &  152.5 &  3.2 &  104.1  &   4.0  & 59.8 & 1.4 &       &          \\  
 J1840.9+5452  & 18 40 57.444  &  +54 52 14.12 &  207.3  &  7.3 &  188.4 &  4.2 &  195.1  &   7.1  &191.6 & 3.4 &180.0 & 17.6      \\  
 J2109.7--1332 & 21 09 49.865  & --13 32 46.21 &   61.9  &  1.9 &   46.4 &  3.2 &   36.7  &   2.2  &       &     & 26.3 & 6.2      \\
 J2154.1--1501 & 21 54 07.563  & --15 01 23.14 &  311.5  &  9.9 &  235.0 &  5.9 &  241.2  &   8.8  &       &     &149.7 & 5.1     \\ 
 J2159.3--1500 & 21 59 20.202  & --15 00 35.52 &   82.5  &  2.5 &   90.1 &  3.2 &   68.2  &   3.1  &       &     & 25.9 & 14.9     \\ 
 J2239.7--0631 & 22 39 46.534  & --06 31 57.37 &  117.2  &  4.2 &   84.4 &  3.6 &   89.9  &   4.0  &       &     & 54.9 & 6.9      \\  
 J2320.6+0032  & 23 20 37.954  &  +00 31 40.10 &   82.0  &  2.5 &   77.9 &  3.3 &   96.1  &   3.8  & 85.1 & 1.9 &       &          \\  
 J2322.0+2114  & 23 22 01.946  &  +21 13 50.17 &  151.6  &  4.9 &   92.5 &  2.5 &   76.6  &   4.5  & 38.1 & 1.4 &       &          \\  
 J2329.0+0834  & 23 29 05.799  &  +08 34 15.22 &  172.9  &  5.2 &  222.5 &  4.2 &  293.6  &   10.3 &313.8 & 7.3 &       &          \\  
 J2333.2--0131 & 23 33 16.675  & --01 31 07.61 &  258.3  &  7.8 &  145.4 &  3.9 &  143.5  &   5.4  &146.4 & 2.7 &       &          \\  
 J2347.6+0852  & 23 47 37.444  &  +08 52 56.58 &  147.7  &  5.4 &   93.4 &  3.1 &   72.1  &   3.1  &       &     & 39.8 & 5.5      \\
\hline
\hline
\end{tabular}
\normalfont
\smallskip\noindent
\flushleft{\normalsize {
The table is organised as follows:
col. 1: source name; cols. 2 and 3: right ascension and 
declination in J2000 coordinates, taken from FIRST or NVSS images; 
cols. 4--13: flux density measurements and associated errors at the
NVSS and Effelsberg 100-m telescope observing frequencies. }}
\end{table*}
\newpage
\tabcolsep0.3cm
\begin{table*}
\caption{Spectral index classification.}
\label{tab:spixtype}
\begin{tabular}{lrrl}
\hline
Sp.In. type  & No.    & \%   &  notes   \\
\hline
                      &        &      &          \\
Steep $\alpha\le-0.5$ & 27     & 26.2   &  16 with $\alpha<-0.7$ \\
Steep--flat           & 18     & 17.5 &          \\
Flat--steep           &  4     &  3.9 &          \\
Flat                  & 37     & 35.9 &          \\
GPS                   &  9     &  8.7 &          \\
Inverted              &  7     &  6.8 &          \\
Undefined             &  1     &    1 &          \\
\hline
\end{tabular}
\end{table*}
\newpage
\tabcolsep0.1cm
\begin{table*}
\tiny
\caption{ Variability and spectral class.}
\label{tab:spix} 
\begin{tabular}{lllrrcllll}
\hline
   name         &  $z$  & O.I. &S$_{G-P}$ & rms & S$_{EF}$/S$_{G-P}$ & $R_{87/90-09}$ & $R_{86-87}$ & S.I. & comments     \\
                &       &      & mJy      & mJy &                   &                 &               &      &   \\
\hline
 J0012.5--1629  & 0.151 & FSRQ &  50 & 14  & 0.99 & 0.04  & & steep & SSRQ \\  
 J0015.5+3052   & 1.619 & FSRQ &  89 & 12  & 1.20 & 2.60  & & steep $+$ & SSRQ \\  
 J0029.0+0509   & 1.633 & FSRQ & 377 & 49  & 0.75 & 2.10  & & flat-steep & break at $\sim$5\,GHz  \\  
 J0106.7--1034  & 0.469 & FSRQ & 188 & 16  & 0.80 & 2.22  & & steep & SSRQ \\  
 J0110.5--1647  & 0.780 & FSRQ &  72 & 14  & 1.37 & 1.84  & & flat &  \\  
 J0125.0+0146   & 1.559 & FSRQ &  95 & 13  & 1.45 & 4.63  & & flat &  \\  
 J0126.2--0500  & 0.411 & FSRQ &  54 & 13  & 0.60 & 1.64  & & steep & SSRQ \\  
 J0204.8+1514   & 0.405 & FSRQ &3073 &273  & 0.69 & 2.57  & & steep & SSRQ \\  
 J0210.0--1004  & 1.976 & FSRQ & 244 & 18  & 0.82 & 2.23  & & flat-steep & break at $\sim$2.7\,GHz  \\  
 J0227.5--0847  & 2.228 & FSRQ & 115 & 14  & 0.95 & 0.36  & & GPS & peak at $\sim$3.2\,GHz \\    
 J0245.2+1047   & 0.070 & BLLac& 217 & 19  & 0.89 & 0.55  & & steep & SS-BLLac \\  
 J0304.9+0002   & 0.563 & FSRQ &  75 &  ?  & 0.71 & 1.52  & & steep & SSRQ \\   
 J0322.6--1335  & 1.468 & FSRQ & 164 & 16  & 0.65 & 3.42  & & steep & SSRQ \\  
 J0340.8--1814  & 0.195 & RG ? & 148 & 16  & 0.91 & 0.75  & & steep & SS-RG \\  
 J0411.0--1637  & 0.622 & FSRQ & 124 & 15  & 1.48 & 3.66  & & flat &  \\  
 J0414.0--1224  & 0.463 & FSRQ & 152 & 16  & 0.39 & 5.71  & & steep $+$ & SSRQ \\  
 J0414.0--1307  & 0.569 & FSRQ &  42 & 14  & 3.24 & 6.32  & & steep-flat &break at $\sim2.7$\,GHz  \\  
 J0421.5+1433   & ?     & BLLac& 114 & 11  & 0.96 & 0.50  & & steep $+$ & SS-BLLac \\  
 J0427.2--0756  & 1.375 & FSRQ &  56 & 13  & 0.91 & 0.35  & & flat &  \\  
 J0434.3--1443  & 1.899 & FSRQ & 281 & 20  & 1.38 & 4.41  & & GPS & break at $\sim$3.2\,GHz \\  
 J0435.1--0811  & 0.791 & FSRQ &  73 & 13  & 1.26 & 1.36  & & GPS & possibly inverted \\  
 J0447.9--0322  & 0.774 & FSRQ &  56 & 13  & 0.35 & 2.75  & & steep $+$ & SSRQ \\  
 J0502.5+1338   & ?     & BLLac& 459 & 41  & 0.94 & 1.69  & 2.12 & steep-flat & break at $\sim$2.7\,GHz \\  
 J0510.0+1800   & 0.416 & FSRQ & 796 & 71  & 0.90 & 0.89  & & flat &  \\  
 J0513.8+0156   & 0.084 & BLLac& 131 & 13  &      &       & & -- &  \\
 J0518.2+0624   & 0.891 & FSRQ & 230 & 21  & 1.05 & 0.10  & & steep $+$ & SSRQ \\  
 J0535.1--0239  & 1.033 & FSRQ &  42 & 13  & 2.06 & 3.28  & & inverted &  \\  
 J0646.8+6807   & 0.927 & FSRQ &  72 &  7  & 1.00 & 0.36  & & steep-flat &  \\  
 J0651.9+6955   & 1.367 & FSRQ & 132 & 12  & 1.15 & 2.58  & & flat &  \\  
 J0724.3--0715  & 0.270 & FSRQ & 482 & 28  & 0.57 & 6.96  & & flat &  \\  
 J0744.8+2920   & 1.168 & FSRQ & 179 & 16  & 1.26 & 3.74  & & flat &  \\  
 J0816.0--0736  & 0.040 & BLLac&  61 & 13  & 0.68 & 1.46  & & flat-steep  & break at $\sim$2.7\,GHz \\  
 J0829.5+0858   & 0.866 & FSRQ & 180 & 16  & 0.95 & 0.15  & & flat &  \\  
 J0847.2+1133   & 0.119 & BLLac&  32 &  5  & 0.55 & 1.74  & & steep-flat &break at $\sim$2.7\,GHz  \\  
 J0853.0+2004   & 1.930 & SSRQ &  60 &  6  & 1.13 & 0.05  & & steep-flat &break at $\sim$3.0\,GHz  \\  
 J0908.2+5031   & 0.917 & SSRQ &  86 &  8  & 0.89 & 1.00  & & steep-flat &break at $\sim$3.0\,GHz  \\  
 J0927.7--0900  & 0.254 & FSRQ & 181 & 16  & 0.50 & 5.33  & & flat &  \\  
 J0931.9+5533   & 0.266 & SSRQ &  53 &  6  & 1.05 & 0.97  & & GPS & peak at $\sim$2.7\,GHz \\  
 J0937.1+5008   & 0.275 & FSRQ & 315 & 28  & 0.35 & 11.2  & & inverted ? & peculiar spectral index \\  
 J0940.2+2603   & 0.498 & FSRQ & 292 & 26  & 1.53 & 4.60  & & flat &  \\  
 J1006.1+3236   & 1.020 & SSRQ & 231 & 21  & 0.84 & 2.77  & & steep $+$ &  \\  
 J1006.5+0509   & 1.216 & FSRQ & 179 & 16  & 0.71 & 3.10  & & steep-flat &break at $\sim2.7$\,GHz  \\  
 J1010.8--0201  & 0.896 & FSRQ & 826 & 45  & 0.67 & 5.53  & & flat &  \\  
 J1011.5--0423  & 1.588 & FSRQ & 189 & 16  & 0.64 & 4.08  & & flat &  \\  
 J1025.9+1253   & 0.663 & FSRQ & 631 & 56  & 1.04 & 0.14  & & flat &  \\  
 J1026.4+6746   & 1.181 & FSRQ & 129 & 12  & 1.02 & 0.15  & & flat &  \\  
 J1028.5--0236  & 0.476 & FSRQ &  94 & 13  & 1.41 & 2.72  & & inverted &  \\  
 J1028.6--0336  & 1.781 & SSRQ &  61 & 13  & 1.00 & 0.001 & & steep-flat  & possibly inverted at higher frequencies \\  
 J1032.1--1400  & 1.039 & FSRQ & 223 & 18  & 0.71 & 2.10  & & flat &  \\  
 J1101.8+6241   & 0.663 & FSRQ & 693 & 61  & 0.50 & 9.41  & & steep &  SSRQ\\  
 J1105.3--1813  & 0.578 & FSRQ &  59 & 14  & 0.59 & 1.65  & & inverted &  \\  
 J1116.1+0828   & 0.486 & FSRQ & 282 & 25  & 1.41 & 3.42  & & GPS & peak at $\sim$2.7\,GHz \\  
 J1120.4+5855   & 0.158 & NLRG &  46 &  5  & 1.14 & 0.05  & & flat &  \\  
 J1150.4+0156   & 1.502 & FSRQ &  95 & 10  & 1.38 & 2.80  & & flat &  \\  
 J1204.2--0710  & 0.185 & BLLac& 128 & 15  & 1.18 & 1.42  & & flat &  \\  
 J1206.2+2823   & 0.708 & FSRQ &  21 &  4  & 0.94 & 1.12  & & steep-flat &break at $\sim$5.0\,GHz  \\  
 J1213.0+3248   & 2.502 & FSRQ &  61 &  6  & 1.09 & 0.54  & & steep-flat &break at $\sim$5.0\,GHz  \\  
 J1213.2+1443   & 0.718 & SSRQ &  61 &  7  & 0.85 & 1.42  & & steep $+$ &  \\  
 J1217.1+2925   & 0.974 & FSRQ &  49 &  6  & 0.91 & 0.55  & & steep-flat &break at $\sim$2.7\,GHz  \\  
 J1222.6+2934   & 0.401 & SSRQ &  60 &  6  & 1.84 & 5.79  & & flat &  \\  
 J1223.9+0650   & 1.189 & FSRQ & 251 & 23  & 1.01 & 1.25  & & flat &  \\  
 J1224.5+2613   & 0.687 & FSRQ & 272 & 24  & 0.88 & 0.29  & & steep $+$ & SSRQ \\  
 J1225.5+0715   & 1.120 & SSRQ &  75 &  9  & 0.76 & 0.09  & & steep $+$ &  \\  
 J1229.5+2711   & 0.490 & NLRG & 165 & 15  & 0.75 & 1.97  &2.79 & steep-flat &break at $\sim$2.7\,GHz; possibly inverted  \\  
 J1231.7+2848   & ?     & BLLac& 114 & 11  & 1.08 & 1.19  & & flat &  \\   
 J1311.3--0521  & 0.160 & BLLac&  46 & 13  & 0.94 & 0.20  & & flat &  \\  
 J1315.1+2841   & 1.576 & FSRQ &  95 &  9  & 1.09 & 1.88  & & flat-steep &possibly GPS, break at $\sim$5.0\,GHz  \\  
 J1320.4+0140   & 1.235 & BLLac& 541 & 48  & 0.90 & 1.07  & & flat &  \\  
 J1329.0+5009   & 2.650 & SSRQ & 133 & 12  & 1.34 & 3.31  & & steep-flat &break at $\sim$2.7\,GHz  \\  
 J1332.7+4722   & 0.668 & SSRQ & 333 & 29  & 0.57 & 4.72  &3.50 & flat &  \\  
 J1337.2--1319  & 3.475 & FSRQ &  71 & 14  & 0.94 & 0.30  & & flat &  \\  
\hline
\end{tabular}
\end{table*}
\newpage
\tabcolsep0.1cm
\begin{table*}
\begin{tabular}{lllrrcllll}
\hline
   name         &  $z$  & O.I. &S$_{G-P}$ & rms & S$_{EF}$/S$_{G-P}$ & $R_{87/90-09}$ & $R_{86-87}$ & S.I. & comments     \\
                &       &      & mJy      & mJy &                   &                 &               &      &   \\
\hline
 J1359.6+4010   & 0.407 & FSRQ & 281 & 25  & 0.95 & 0.96   &    & inverted &  \\  
 J1400.7+0425   & 2.550 & FSRQ & 267 & 24  & 0.58 & 4.52   &    & steep-flat &break at $\sim$2.7\,GHz  \\  
 J1404.2+3413   & 0.937 & SSRQ &  62 &  6  & 0.89 & 1.36   &    & steep $+$ &  \\  
 J1406.9+3433   & 2.556 & FSRQ & 204 & 18  & 1.36 & 3.96   &    & GPS & peak at $\sim$5.0\,GHz \\  
 J1416.4+1242   & 0.335 & FSRQ &  98 & 10  & 0.83 & 2.32   &    & flat &  \\  
 J1417.5+2645   & 1.455 & FSRQ &  77 &  8  & 1.20 & 1.89   &    & flat &  \\  
 J1419.1+0603   & 2.389 & FSRQ & 240 & 22  & 0.92 & 3.94   & 2.85   & flat &  \\  
 J1420.6+0650   & 0.236 & FSRQ & 241 &  9  & 0.82 & 1.34   &    & steep $+$ & SSRQ \\   
 J1423.3+4830   & 0.569 & SSRQ & 100 &  9  & 1.47 & 4.36   &    & flat &  \\  
 J1427.9+3247   & 0.568 & SSRQ &  65 &  7  & 0.93 & 1.69   &    & GPS &  peak at $\sim$2.7\,GHz \\  
 J1442.3+5236   & 1.800 & FSRQ & 111 & 10  & 0.86 & 1.54   &    & steep & SSRQ \\  
 J1507.9+6214   & 1.478 & SSRQ & 213 & 19  & 1.00 & 0.07   &    & steep $+$ &  \\  
 J1539.1--0658  & ?     & BLLac&  76 & 13  & 1.10 & 0.54   &    & GPS & peak at $\sim$5.0\,GHz \\  
 J1543.6+1847   & 1.396 & FSRQ & 300 & 27  & 0.55 & 5.77   &    & steep-flat &break at $\sim$2.7\,GHz  \\  
 J1606.0+2031   & 0.383 & FSRQ & 196 & 18  & 0.42 & 6.70   &    & steep-flat &break at $\sim$2.7\,GHz  \\  
 J1626.6+5809   & 0.748 & SSRQ & 315 & 28  & 1.14 & 1.58   &    & flat &  \\  
 J1629.7+2117   & 0.833 & FSRQ & 105 & 10  & 0.93 & 0.78   &    & steep $+$ & SSRQ \\  
 J1648.4+4104   & 0.851 & SSRQ & 197 & 18  & 2.54 & 14.4   &    & inverted &  \\  
 J1656.6+5321   & 1.555 & FSRQ & 145 & 13  & 0.93 & 2.92   & 4.62   & flat & possibly GPS \\   
 J1656.8+6012   & 0.623 & FSRQ & 184 & 16  & 2.63 & 17.4   & 5.16   & flat & possibly GPS \\  
 J1722.3+3103   & 0.305 & SSRQ &  53 &  6  & 0.98 & 0.63   &    & steep $+$ &  \\  
 J1804.7+1755   & 0.435 & FSRQ &  92 &  9  & 1.13 & 1.30   &    & steep $+$ & SSRQ \\  
 J1840.9+5452   & 0.646 & BLLac& 252 & 22  & 0.77 & 3.54   &    & flat &  \\  
 J2109.7--1332  & 1.226 & FSRQ &  51 & 14  & 0.72 & 1.00   &    & flat &  \\
 J2154.1--1501  & 1.208 & FSRQ & 219 & 18  & 1.10 & 1.10   &    & flat & possibly flat-steep \\ 
 J2159.3--1500  & 2.270 & FSRQ &  86 & 14  & 0.79 & 1.24   &    & GPS & peak at $\sim$2.7\,GHz \\ 
 J2239.7--0631  & 0.264 & SSRQ &  64 & 13  & 1.40 & 1.90   &    & flat-steep &break at $\sim$3.5\,GHz  \\  
 J2320.6+0032   & 1.894 & FSRQ &  87 & 10  & 1.10 & 0.88   &    & flat &  \\  
 J2322.0+2114   & 0.707 & FSRQ & 100 &  9  & 0.77 & 1.67   &    & steep $+$ & SSRQ \\  
 J2329.0+0834   & 0.948 & FSRQ & 273 & 24  & 1.08 & 0.45   & 4.13   & inverted &  \\  
 J2333.2--0131  & 1.062 & FSRQ & 314 & 21  & 0.46 & 7.86   &    & steep-flat &break at $\sim$2.7\,GHz  \\  
 J2347.6+0852   & 0.292 & FSRQ & 154 &  7  & 0.47 & 2.00   &    & steep & SSRQ \\
\hline
\hline
\end{tabular}
\normalfont
\smallskip\noindent
\flushleft{\normalsize {
The table is organised as follows: 
col. 1: source name; 
cols. 2 and 3: redshift and spectral class respectively, taken from 
\citet{perlman} and \citet{landt}; 
col. 4: GB6 flux density at 4.85\,GHz for sources with 
$0^\circ <$ Dec.$ < 75^\circ$ 
  and PMN flux density at 4.85\,GHz for sources with
  $-20^\circ <$ Dec.$ < 0^\circ$ taken from \citet{perlman} and \citet{landt}; 
col. 5: rms error;
col. 6: Effelsberg to GB6 or PMN flux density ratio; 
col. 7: $R$ parameter for the long-term variability;
col. 8: $R$ parameter for the short-term variability; 
col. 9: spectral index type;
col. 10: new spectral index classification and comments. 
}}
\end{table*}
\newpage
\tabcolsep0.06cm
\begin{table*}
\tiny
\caption{ Percentage of polarised emission, Electric Vector Position Angles, 
and Rotation Measures}
\label{tab:polar}
\begin{tabular}{lrrrrrrrrrrrrrrrrrrrrrr}
\hline
name      &m$_{1.4}$  &$\Delta$m& $\chi_{1.4}$  &$\Delta\chi$&m$_{2.64}$ & $\Delta$m& $\chi_{2.64}$ &$\Delta\chi$ &m$_{4.85}$& $\Delta$m &  $\chi_{4.85}$  & $\Delta\chi$ &m$_{8.35}$&$\Delta$m & $\chi_{8.35}$ &$\Delta\chi$& m$_{10.45}$& $\Delta$m & $\chi_{10.45}$ & $\Delta\chi$& RM  \\
          & \%       & \%      & $^\circ$   & $^\circ$    &  \%      & \%       & $^\circ$   & $^\circ$     &   \%    &  \%       &  $^\circ$  &  $^\circ$   &  \%     &   \%     &  $^\circ$  & $^\circ$  &    \%        & \%         & $^\circ$  & $^\circ$ & rad m$^{-2}$ \\
\hline
 J0012.5--1629 & $<1.4$ &     &      &      &$<3$ &     &       &      &$<2$ &      &      &     &      &      &      &      & $<1$ &      &     &     &      \\
 J0015.5+3052  & 6.3    & 0.4 & 43.8 &  0.6 & 6.1 & 1.3 &--14.8 &  8.0 & 8.5 & 1.4  & 10.1 & 6.5 &      &      &      &      & $<1$ &      &     &     & 14.4 \\
 J0029.0+0509  & 2.6    & 0.2 &--45.0 &  0.8 & 4.9 & 0.6 &  76.2 &  4.5 &4.8  & 0.5  &--115.2& 5.1 &      &      &      &      & $<1$ &      &     &     &29.8 \\
 J0106.7--1034 & 4.5    & 0.3 & 60.0 &  0.7 & 4.5 & 1.0 & 105.7 &  9.0 &3.9  & 1.2  & --61.4&10.4 &      &      &      &      & $<1$ &      &     &     &$-$24.3 \\
 J0110.5--1647 & 2.8    & 0.6 &--18.1 &  3.2 & $<3$&     &       &      &$<2$ &      &      &     &      &      &      &      & $<1$ &      &     &     &      \\
 J0125.0+0146  & 8.1    & 0.5 &--19.9 &  0.6 & $<3$&     &       &      &7.9  & 1.2  & 66.0 & 5.7 &      &      &      &      & $<1$ &      &     &     &      \\
 J0126.2--0500 & $<$1.7 &      &     &      & $<3$&     &       &      &$<2$ &      &      &     &      &      &      &      & $<1$ &      &     &     &      \\
 J0204.8+1514  & 0.04   & 0.01 & 20.7 &  4.9 & 1.5  & 0.1  &70.2 & 1.7 &0.5  & 0.1  & 58.0 & 5.0 &      &      &      &      & $<1$ &      &     &     & 62.9 \\
 J0210.0--1004 & 1.1    & 0.2  & 78.9 &  2.9 & 4.3  & 0.9  &74.8 & 7.9 &5.7  & 0.7  & 64.0 & 5.1 &      &      &      &      & $<1$ &      &     &     & 7.2  \\
 J0227.5--0847 & $<2.0$ &      &      &      & $<3$ &      &     &     &$<2$ &      &      &     &      &      &      &      & $<1$ &      &     &     &      \\
 J0245.2+1047  & 2.5    & 0.7  &--67.6 &  2.0 & 2.7  &  0.7 &62.9 & 8.8 &$<2$ &      &      &     &      &      &      &      & $<1$ &      &     &     &      \\
 J0304.9+0002  & 5.2    & 0.6  &--47.3 &  1.5 & 6.8  & 2.3  &71.1 &13.1 &$<2$ &      &      &     &$<1$  &      &      &      & $<1$ &      &     &     &      \\
 J0322.6--1335 & 3.8    & 0.3  & 49.2 &  0.8 &      &      &     &     &7.3  & 1.8  & 82.2 & 9.8 & 5.17 & 1.12 &  67.5&  8.8 &      &      &     &     & 62.4 \\
 J0340.8--1814 & 1.7    & 0.2  &--13.9 &  1.8 & $<3$ &      &     &     &4.9  & 1.3  & 75.1 &10.8 & $<1$ &      &      &      &      &      &     &     &      \\
 J0411.0--1637 & 3.6    & 0.6  & 37.2 &  1.7 & $<3$ &      &     &     &$<2$ &      &      &     & $<1$ &      &      &      &      &      &     &     &      \\
 J0414.0--1307 & 1.8    & 0.5  & 49.1 &  5.2 & $<3$ &      &     &     &14.6 & 3.7  &--63.5 & 9.8 & $<1$ &      &      &      &      &      &     &     &      \\
 J0414.0--1224 & 3.4    & 0.2  & --4.0 &  0.8 & $<3$ &      &     &     &$<2$ &      &      &     & $<1$ &      &      &      &      &      &     &     &      \\
 J0421.5+1433  & 3.3    & 0.7  & 77.8 &  1.8 & 10.8 &  1.1 &31.0 & 3.9 &13.5 & 1.5  & 60.6 & 4.2 & 10.8 & 1.6  & 74.1 & 5.4  &      &      &     &     &$-$68.6 \\
 J0427.2--0756 & $<2.7$ &      &      &      & $<3$ &      &     &     &$<2$ &      &      &     &$<1$  &      &      &      & $<1$ &      &     &     &      \\
 J0434.3--1443 & 1.0    & 0.2  & 43.6 &  4.1 & 2.8  & 0.6  &108.5& 7.7 &3.75 & 0.50 & 63.6 & 5.1 &  5.1 & 0.3  & 90.8 &  2.7 &      &      &     &     &$-$230.7\\
 J0435.1--0811 & $<1.1$ &      &      & $<3$ &      &      &     &15.4 & 3.0 &  5.7 & 9.0  &     &      &      &      &      & $<1$ &      &     &     &      \\
 J0447.9--0322 & $<1.6$ &      &      &      & $<3$ &      &     &     &$<2$ &      &      &     &      &      &      &      &      &      &     &     &      \\
 J0502.5+1338  &    0.6 &  0.1 &--14.1 &  2.7 & $<3$ &      &     &     &1.8  & 0.3  &--39.3 & 7.7 &      &      &      &      & $<1$ &      &     &     &      \\
 J0510.0+1800  & 1.8    & 0.1  &--58.3 &  0.8 & 1.9  & 0.3  &115.3& 6.4 &1.3  & 0.2  &  1.6 & 7.8 &  1.7 & 0.1  &--28.7 & 2.9  &      &      &     &     &$-$127.8  \\
 J0513.8+0156  & $<1.5$ &      &      & $<3$ &      &      &     &     &     &      &      &     &      &      &      &      &      &      &     &     &      \\
 J0518.2+0624  & 0.6    & 0.1  & 85.4 &  3.0 & 4.6  & 0.5  &82.0 & 4.4 &3.7  & 0.6  & 35.5 & 6.9 & 2.58 & 0.62 & 22.8 & 9.8  &      &      &     &     &95.4 \\
 J0535.1--0239 & $<4.0$ &      &      &      & $<3$ &      &     &     &$<2$ &      &      &     & $<1$ &      &      &      &      &      &     &     &      \\
 J0646.8+6807  & 6.5    & 0.7  &--78.4 &  1.3 & $<3$ &      &     &     &$<2$ &      &      &     &      &      &      &      & $<1$ &      &     &     &      \\
 J0651.9+6955  & 3.0    & 0.2  &--62.2 &  1.0 & 4.9  &  1.1 & 9.3 & 7.9 &7.1  & 1.3  & 15.4 & 6.6 &      &      &      &      & $<1$ &      &     &     &$-$34.4 \\
 J0724.3--0715 & 3.1    & 0.2  &  1.3 &  0.9 & $<3$ &      &     &     &5.4  & 0.6  &--81.8 & 4.8 &  3.1 &  0.4 &--96.9 &  4.7 &      &      &     &     &109.1 \\
 J0744.8+2920  & 4.5    &  0.3 & 65.2 &  0.6 & 3.6  & 0.8  &80.2 & 8.0 &4.1  & 1.0  & 43.4 & 9.0 & 2.6  &  0.5 & 47.5 &  8.1 &      &      &     &     & 81.5 \\
 J0816.0--0736 & $<1.4$ &      &      &      & $<3$ &      &     &     &$<2$ &      &      &     & $<1$ &      &      &      &      &      &     &     &      \\
 J0829.5+0858  & 3.8    & 0.2  & --5.6 &  0.7 & 5.5  & 0.9  &80.6 & 5.8 & 5.6 & 1.0  & 66.8 & 6.7 &  3.6 &  0.7 & 70.5 &  7.0 &      &      &     &     & 44.0 \\
 J0847.2+1133  & $<3.8$ &      &      &      & $<3$ &      &     &     &$<2$ &      &      &     &$<1$  &      &      &      &      &      &     &     &      \\
 J0853.0+2004  & 4.8    & 0.5  &  6.7 &  1.3 & $<3$ &      &     &     &$<2$ &      &      &     &$<1$  &      &      &      &      &      &     &     &      \\
 J0908.2+5031  & 5.7    & 0.4  &--51.7 &  0.8 & 5.0  & 1.5  &-32.3&12.1 &$<2$ &      &      &     &      &      &      &      & $<1$ &      &     &     &      \\
 J0927.7--0900 & $<0.9$ &      &      &      & $<3$ &      &     &     &$<2$ &      &      &     &$<1$  &      &      &      &      &      &     &     &      \\
 J0931.9+5533  & $<7.1$ &      &      &      & $<3$ &      &     &     &$<2$ &      &      &     &      &      &      &      & $<1$ &      &     &     &      \\
 J0937.1+5008  & 4.6    &  0.4 &--18.6 &  1.0 & $<3$ &      &     &     &$<2$ &      &      &     &      &      &      &      &  $<1$&      &     &     &      \\
 J0940.2+2603  & 1.4    & 0.1  &--34.0 &  1.3 &  3.8 &  0.5 &71.7 & 5.4 &4.8  & 0.4  &--103.3& 3.0 &      &      &      &      &  3.6 &  1.1 & 90.3& 10.0&$-$49.3 \\
 J1006.1+3236  &        &      &      &      & $<3$ &      &     &     &6.5  & 1.0  &--129.0& 6.5 &      &      &      &      &  $<1$&      &     &     &      \\
 J1006.5+0509  & 1.6    & 0.3  &--20.9 &  3.0 & $<3$ &      &     &     &$<2$ &      &      &     &      &      &      &      &  $<1$&      &     &     &      \\
 J1010.8--0201 & 3.0    & 0.2  & 29.4 &  0.6 &  2.3 &  0.3 &59.7 & 4.6 &$<2$ &      &      &     &      &      &      &      &  $<1$&      &     &     &      \\
 J1011.5--0423 & 6.9    & 0.5  &--67.0 &  0.8 & $<3$ &      &     &     &7.5  & 1.7  &  9.6 & 9.1 &      &      &      &      &  $<1$&      &     &     &      \\
 J1025.9+1253  & 3.0    & 0.2  & 43.9 &  0.5 &  2.6 &  0.5 &41.1 & 6.3 &1.4  & 0.3  &--122.3& 7.0 &      &      &      &      &  $<1$&      &     &     & $-$13.0 \\
 J1026.4+6746  & 4.6    & 0.4  & --72.5&  0.9 & 4.8  & 1.2  &146.1& 9.6 &$<2$ &      &      &     &      &      &      &      &  $<1$&      &     &     &      \\
 J1028.5--0236 & 2.7    & 0.6  & 85.9 &  3.9 & $<3$ &      &     &     &6.7  & 1.5  & 66.1 & 9.0 &      &      &      &      &  $<1$&      &     &     &      \\
 J1028.6--0336 & 8.8    & 0.6  &--35.1 &  0.6 & 11.2 & 2.0  &147.1& 7.3 &15.7 & 3.1  &--33.1 & 8.5 &      &      &      &      &  $<1$&      &     &    &$-$1.0 \\
 J1032.1--1400 & 3.2    & 0.3  & 77.5 &  1.4 & $<3$ &      &     &     &$<2$ &      &      &     &      &      &      &      &  $<1$&      &     &     &      \\
 J1101.8+6241  & 3.0    &  0.2 & 67.8 &  0.4 & 3.7  & 0.5  &85.7 & 5.1 &2.0  & 0.5  & 81.4 &10.0 &      &      &      &      &  $<1$&      &     &    &$-$8.1 \\
 J1105.3--1813 & $<6.7$ &      &      &      & $<3$ &      &     &     &$<2$ &      &      &     &      &      &      &      &      &      &     &     &     \\
 J1116.1+0828  & 1.5    & 0.2  &--84.6 &  2.3 & $<3$ &      &     &     &5.8  & 1.0  & 41.1 & 7.7 &      &      &      &      &  $<1$&      &     &     &     \\
 J1120.4+5855  & $<1.4$ &      &      &      & $<3$ &      &     &     &$<2$ &      &      &     &      &      &      &      &  $<1$&      &     &     &      \\
 J1150.4+0156  & 7.5    &  0.6 &--33.4 & 0.8  & $<3$ &      &     &     & 5.8 & 1.6  &--59.1 &10.3 &      &      &      &      &  $<1$&      &     &     &      \\
 J1204.2--0710  & 3.0    &  0.4 &--66.0 & 1.9  & $<3$ &      &     &     & 4.7 & 1.1  &--104.3&11.4 &      &      &      &      &  $<1$&      &     &     &      \\
 J1206.2+2823  & $<3.3$ &      &      &      & $<3$ &      &     &     &$<2$ &      &      &     &$<1$  &      &      &      &      &      &     &     &      \\
 J1213.0+3248  & 1.4    &  0.4 &--26.9 &  4.2 & $<3$ &      &     &     &$<2$ &      &      &     &$<1$  &      &      &      &      &      &     &     &      \\
 J1213.2+1443  & 4.0    & 0.1  & 20.3 &  2.8 & $<3$ &      &     &     &$<2$ &      &      &     &$<1$  &      &      &      &      &      &     &     &      \\
 J1217.1+2925  & 2.0    & 0.7  & 27.0 &  6.1 & $<3$ &      &     &     &$<2$ &      &      &     &$<1$  &      &      &      &      &      &     &     &      \\ J1222.6+2934  & $<0.4$ &      &      &      & $<3$ &      &     &     &$<2$ &      &      &     &$<1$  &      &      &      &      &      &     &     &      \\
 J1223.9+0650  & 4.6    & 0.3  & 26.9 &  0.7 & $<3$ &      &     &     &$<2$ &      &      &     &      &      &      &      &  $<1$&      &     &     &      \\
 J1224.5+2613  & 2.2    & 0.2  &--49.6 &  0.9 & 2.8  & 0.6  &117.4& 7.5 &$<2$ &      &      &     &$<1$  &      &      &      &      &      &     &     &      \\
 J1225.5+0715  & 2.0    & 0.5  &--10.9 &  4.0 & $<3$ &      &     &     &$<2$ &      &      &     &      &      &      &      &  $<1$&      &     &     &      \\
 J1229.5+2711  & $<1.3$ &      &      &      & $<3$ &      &     &     &$<2$ &      &      &     &$<1$  &      &      &      &      &      &     &     &      \\
 J1231.7+2848  & 3.3    & 0.5  &--56.5 &  2.0 & $<3$ &      &     &     &$<2$ &      &      &     & 2.4  &  0.8 & --8.4 & 15.4 &      &      &     &     &      \\
 J1311.3--0521 & 3.6    & 1.0  & 74.6 &  4.3 & $<3$ &      &     &     &$<2$ &      &      &     &      &      &      &      &  $<1$&      &     &     &      \\
 J1315.1+2841  & 3.6    & 0.5  & 81.6 &  2.2 & $<3$ &      &     &     &$<2$ &      &      &     &$<1$  &      &      &      &      &      &     &     &      \\
 J1320.4+0140  & 5.3    & 0.2  &--37.4 &  0.3 & 6.2  & 0.4  &152.4& 2.4 &6.4  & 0.4  &--19.4 & 2.2 & 3.6  & 0.2  &--0.2  &  1.8 &      &      &     &     &$-$16.4  \\
 J1329.0+5009  & 1.3    & 0.2  & 80.7 &  2.3 & $<3$ &      &     &     &$<2$ &      &      &     &  1.9 &  0.6 & 65.4 & 13.0 &      &      &     &     &      \\
 J1332.7+4722  & $<0.5$ &      &      &      & $<3$ &      &     &     &4.8  & 0.8  & --4.4 & 6.4 &  3.9 &  0.6 & 152.8&  5.7 &      &      &     &     &      \\
 J1337.2--1319 & 1.3    &  0.5 &--41.6 &  6.0 & $<3$ &      &     &     &$<2$ &      &      &     &$<1$  &      &      &      &      &      &     &     &      \\
\hline
\end{tabular}
\end{table*}
\newpage
\tabcolsep0.06cm
\begin{table*}[h]
\tiny
\begin{tabular}{lrrrrrrrrrrrrrrrrrrrrr}
\hline
name      &m$_{1.4}$  &$\Delta$m& $\chi_{1.4}$  &$\Delta\chi$&m$_{2.64}$ & $\Delta$m& $\chi_{2.64}$ &$\Delta\chi$ &m$_{4.85}$& $\Delta$m &  $\chi_{4.85}$  & $\Delta\chi$ &m$_{8.35}$&$\Delta$m & $\chi_{8.35}$ &$\Delta\chi$& m$_{10.45}$& $\Delta$m & $\chi_{10.45}$ & $\Delta\chi$& RM  \\
          & \%       & \%      & $^\circ$   & $^\circ$    &  \%      & \%       & $^\circ$   & $^\circ$     &   \%    &  \%       &  $^\circ$  &  $^\circ$   &  \%     &   \%     &  $^\circ$  & $^\circ$  &    \%        & \%         & $^\circ$  & $^\circ$ & rad m$^{-2}$ \\
\hline
 J1359.6+4010  & $<0.8$ &      &      &      & $<3$ &      &     &     &$<2$ &      &      &     & $<1$ &      &      &      &      &      &     &     &      \\
 J1400.7+0425  & 1.3    & 0.2  & --9.8 &  2.1 &  4.1 &  0.9 &118.7& 8.0 & 4.9 & 1.1  &--22.7 & 8.0 & 3.2  & 0.5  & --16.5& 6.8  &      &      &     &     &$-$60.5 
\\
 J1404.2+3413  & 5.8    & 0.6  & 72.6 &  1.1 & $<3$ &      &     &     &$<2$ &      &      &     &      &      &      &      & $<1$ &      &     &     &      \\
 J1406.9+3433  & 1.3    & 0.3  & 85.2 &  3.8 & $<3$ &      &     &     &1.7  & 0.5  &--69.3 &11.7 &      &      &      &      &  $<1$&      &     &     &      \\
 J1416.4+1242  & 4.0    & 0.5  & --5.8 &  2.0 & $<3$ &      &     &     &$<2$ &      &      &     & $<1$ &      &      &      &      &      &     &     &      \\
 J1417.5+2645  & 1.9    & 0.6  & --5.2 &  5.7 & $<3$ &      &     &     &$<2$ &      &      &     &      &      &      &      &  $<1$&      &     &     &      \\
 J1419.1+0603  & 0.4    & 0.1  &--49.1 &  5.5 & $<3$ &      &     &     &4.1  & 1.0  &--27.4 & 8.7 &  2.9 & 0.5  & --12.9& 7.9  &      &      &     &     &$-$84.3 \\
 J1420.6+0650  & 5.4    & 0.4  &--25.4 &  0.5 & 6.2  & 0.6  &150.3& 4.0 &7.2  & 1.1  &--27.1 & 5.6 & 4.5  & 0.7  & --23.3&  6.1 &      &      &     &     &$-$0.5 ?   \\
 J1423.3+4830  & 1.2    & 0.2  & --3.2 &  3.1 &      &      &     &     &$<2$  &     &      &     &      &      &      &      &  $<1$&      &     &     &      \\
 J1427.9+3247  & $<1.8$ &      &      &      & $<3$ &      &     &     &$<2$  &     &      &     &      &      &      &      &  $<1$&      &     &     &      \\
 J1442.3+5236  & 4.9    & 0.4  &--89.8 &  0.9 & 3.8  & 1.3  &208.1&13.1 &$<2$  &     &      &     &      &      &      &      &  $<1$&      &     &     &      \\
 J1507.9+6214  & 0.9    & 0.1  & 65.6 &  1.7 &  2.0 &  0.6 &187.2&10.8 &$<2$  &     &      &     &      &      &      &      &  $<1$&      &     &     &      \\
 J1539.1--0658 & $<3.0$ &      &      &      & $<3$ &      &     &     &$<2$  &     &      &     & $<1$ &      &      &      &      &      &     &     &   \\
 J1543.6+1847  & 3.5    & 0.2  &--59.1 &  0.7 &  6.3 &  0.9 &96.3 & 5.3 & 6.2  & 1.1 & 81.7 & 6.1 &      &      &      &      &  $<1$&      &     &     &--13.6   \\
 J1606.0+2031  & 5.4    & 0.4  & 44.9 &  0.9 &  5.4 &  1.8 &170.1&13.0 & 7.2  & 1.8 &--27.8 & 9.5 &  5.1 & 1.4  & --26.3& 10.7 &      &      &     &     & 29.0 \\
 J1626.6+5809  & 0.5    & 0.1  &--85.9 &  3.0 & 2.1  & 0.4  &203.9& 8.0 & 2.1  & 0.5 & 32.3 & 8.7 &  1.4 & 0.3  & 34.9 &  7.7 &      &      &     &     & --50.2 \\
 J1629.7+2117  & $<0.5$ &      &      &  3.8 &  1.1 & 97.4 & 11.1&$<2$ &      &     &      &$<1$ &      &      &      &      &      &      &     &     &      \\
 J1648.4+4104  & 2.2    & 0.2  & 69.9 &  1.6 & $<3$ &      &     &     & 1.7  & 0.4 & 18.2 & 9.0 &  3.4 & 0.2  &--2.2  & 2.0  &      &      &     &     &170.0 \\ J1656.6+5321  & 1.8    & 0.5  &--42.2 &  4.4 & $<3$ &      &     &     &$<2$  &     &      &     &  4.0 & 0.8  & 107.0& 8.3  &      &      &     &     &      \\ J1656.8+6012  & 0.7    & 0.2  &--25.2 &  3.9 &  3.9 &  0.3 &147.0& 3.5 & 4.9  & 0.3 &133.6 & 2.5 &  3.1 & 0.2  & 129.8&  2.3 &      &      &     &     & 10.9 \\  
 J1722.3+3103  & $<2.5$ &      &      &      & $<3$ &      &     &     &$<2$  &     &      &     &  $<1$&      &      &      &      &      &     &     &      \\ J1804.7+1755  & 3.8    & 0.5  & 54.2 &  1.5 & $<3$ &      &     &     & 6.7  & 1.3 &--51.8 & 7.8 &  $<1$&      &      &      &      &      &     &     &      \\ J1840.9+5452  & 3.1    & 0.3  & 84.6 &  1.3 &  5.1 &  1.5 &261.5&11.0 &$<2$  &     &      &     &      &      &      &      &  $<1$&      &     &     &      \\ J2109.7--1332 & $<2.2$ &      &      &      & $<3$ &      &     &     &$<2$  &     &      &     &  $<1$&      &      &      &      &      &     &     &      \\
 J2154.1--1500 & 2.3    & 0.4  &--54.5 &  2.0 &  4.6 &  1.2 &92.7 &10.3 & 3.1  & 0.6 &--91.7 & 8.7 &      &      &      &      &  $<1$&      &     &     & 15.4 \\ J2159.3--1501 & $<1.6$ &      &      &      & $<3$ &      &     &     &$<2$  &     &      &     &      &      &      &      &  $<1$&      &     &     &      \\ J2239.7--0631 & $<0.4$ &      &      &      & $<3$ &      &     &     &$<2$  &     &      &     &      &      &      &      &  $<1$&      &     &     &      \\ J2320.6+0032  & 2.9    &  0.7 &  2.8 &  3.8 & $<3$ &      &     &     &$<2$  &     &      &     &      &      &      &      & $<1$ &      &     &     &      \\ J2322.0+2114  & $<7.1$ &      &      &      & $<3$ &      &     &     & $<2$ &     &      &     & $<1$ &      &      &      &      &      &     &     &      \\ J2329.0+0834  & 1.9    &  0.3 & --8.1 &  2.6 & $<3$ &      &     &     & $<2$ &     &      &     & $<1$ &      &      &      &      &      &     &     &      \\ J2333.2--0131 & 1.7    &  
 0.2 & 64.2 &  2.1 & $<3$ &      &     &     & $<2$ &     &      &     &  4.6 &  0.7 &91.7  &  5.2 &      &      &     &     &      \\ J2347.6+0852  & 7.0    & 0.9  &  8.4 &  1.2 & $<3$ &      &     &     & $<2$ &     &      &     &  2.8 &  0.6 & --26.1&  9.2 &      &      &     &     &      \\
\hline
\hline
\end{tabular}
\normalfont
\smallskip\noindent
\flushleft{\normalsize {
The table is organised as follows: col. 1: source name; 
cols. 2 and 3: percentage of polarised flux density and associated error, 
cols. 4 and 5: EVPA $\chi$ and associated error at 1.4\,GHz extracted from 
  \citet{condon}; 
cols. 6--9:  percentage of polarised flux density and associated error, 
$\chi$ and associated error at 2.64\,GHz; 
cols. 10--13:  percentage of polarised flux density and associated error, $\chi$ and 
associated error at 4.85\,GHz; 
cols. 14--17: percentage of polarised flux density  and associated error, $\chi$ and associated 
  error at 8.35\,GHz;  
  10.45\,GHz respectively; 
cols. 18--21: percentage of polarised flux density and associated error, $\chi$ and associated 
  error at 10.45\,GHz;
col. 22: rotation measure (observer's frame)
}}
\end{table*}
\tabcolsep0.3cm
\begin{table*}
\caption{Comparison between percentage of polarised emission for
different samples of objects.}
\label{tab:m-values}
\begin{tabular}{clll}
\hline
Freq.(GHz)  &  m blazars           & m B3-VLA  &  m CSS                   \\
\hline
            &                      &           &                          \\
1.4         &  $3.0^{+0.4}_{-0.8}$ & 2.2       &  $1.6^{+2.2}_{-1.0}$     \\
            &                      &           &                          \\
2.64        &  $4.9^{+1.1}_{-0.4}$ & 3.7       &  $4.2^{+6.1}_{-0.6}$     \\
            &                      &           &                          \\
4.85        &  $5.8\pm0.9$         & 5.2       &  $3.7^{+5.7}_{-1.0}$     \\
            &                      &           &                          \\
\hline
\end{tabular}
\end{table*}
\newpage
%
\begin{figure*}[t]
\addtocounter{figure}{+0}
\centering
\includegraphics[width=8cm]{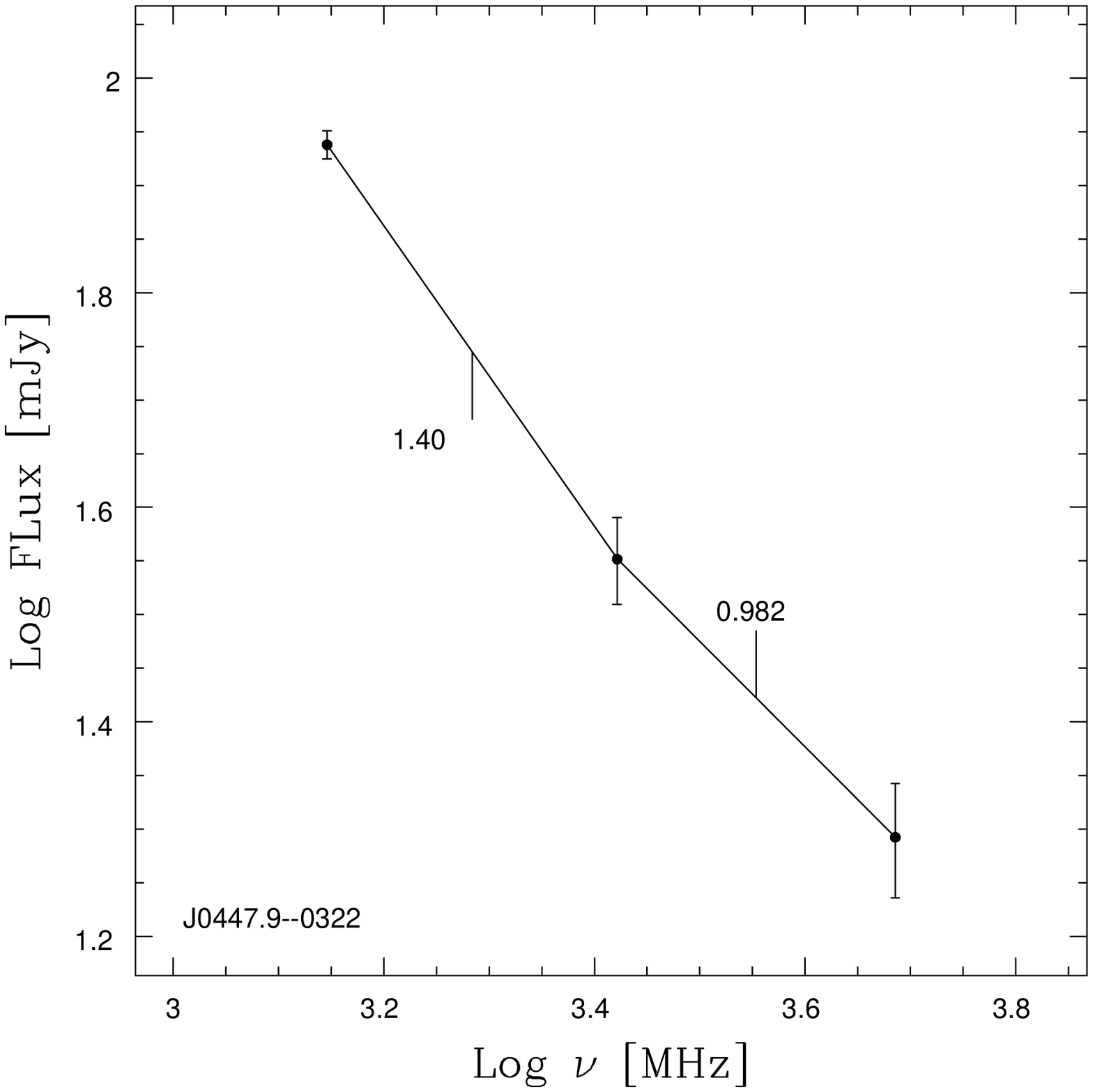}
\includegraphics[width=8cm]{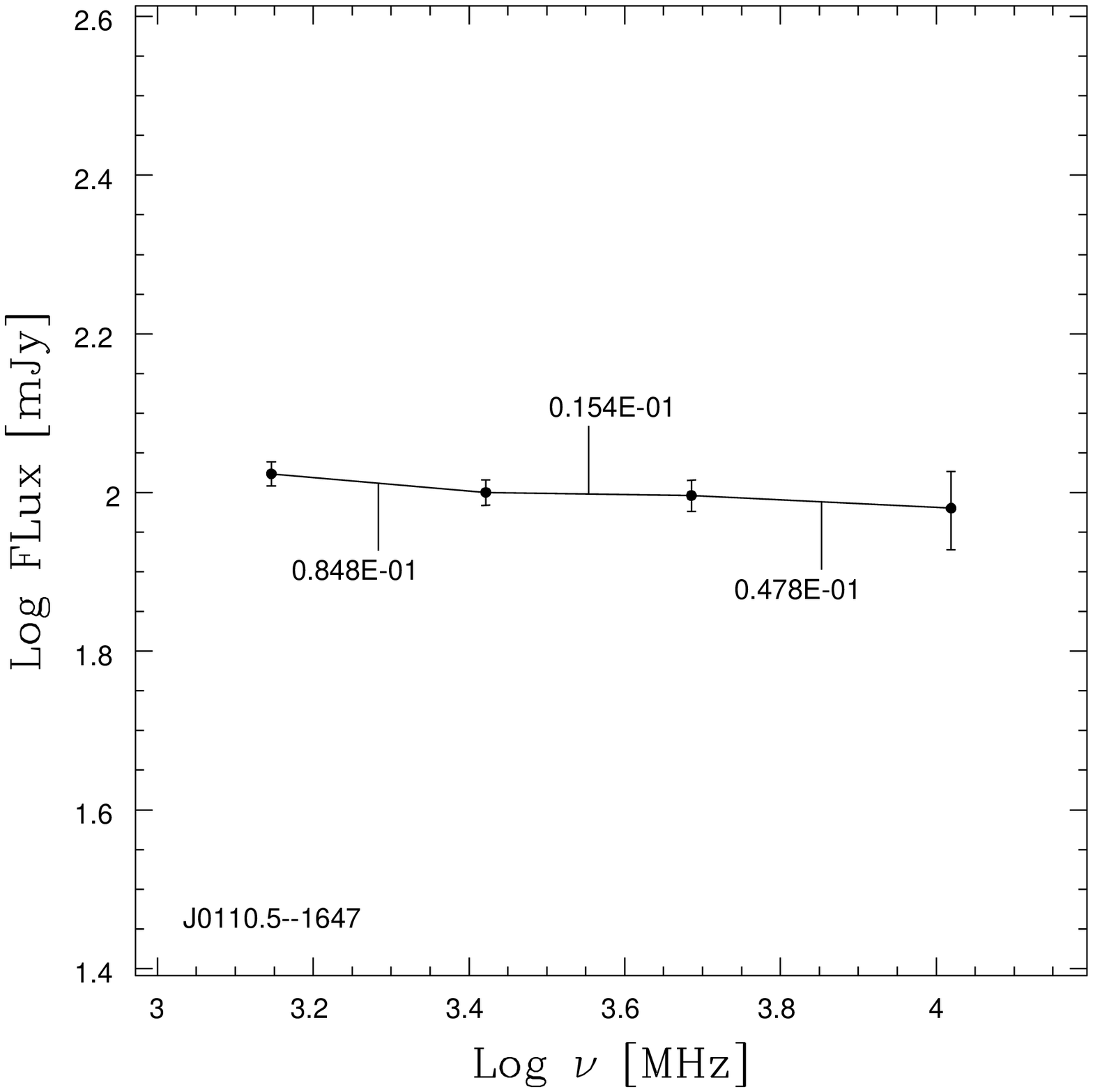}
\includegraphics[width=8cm]{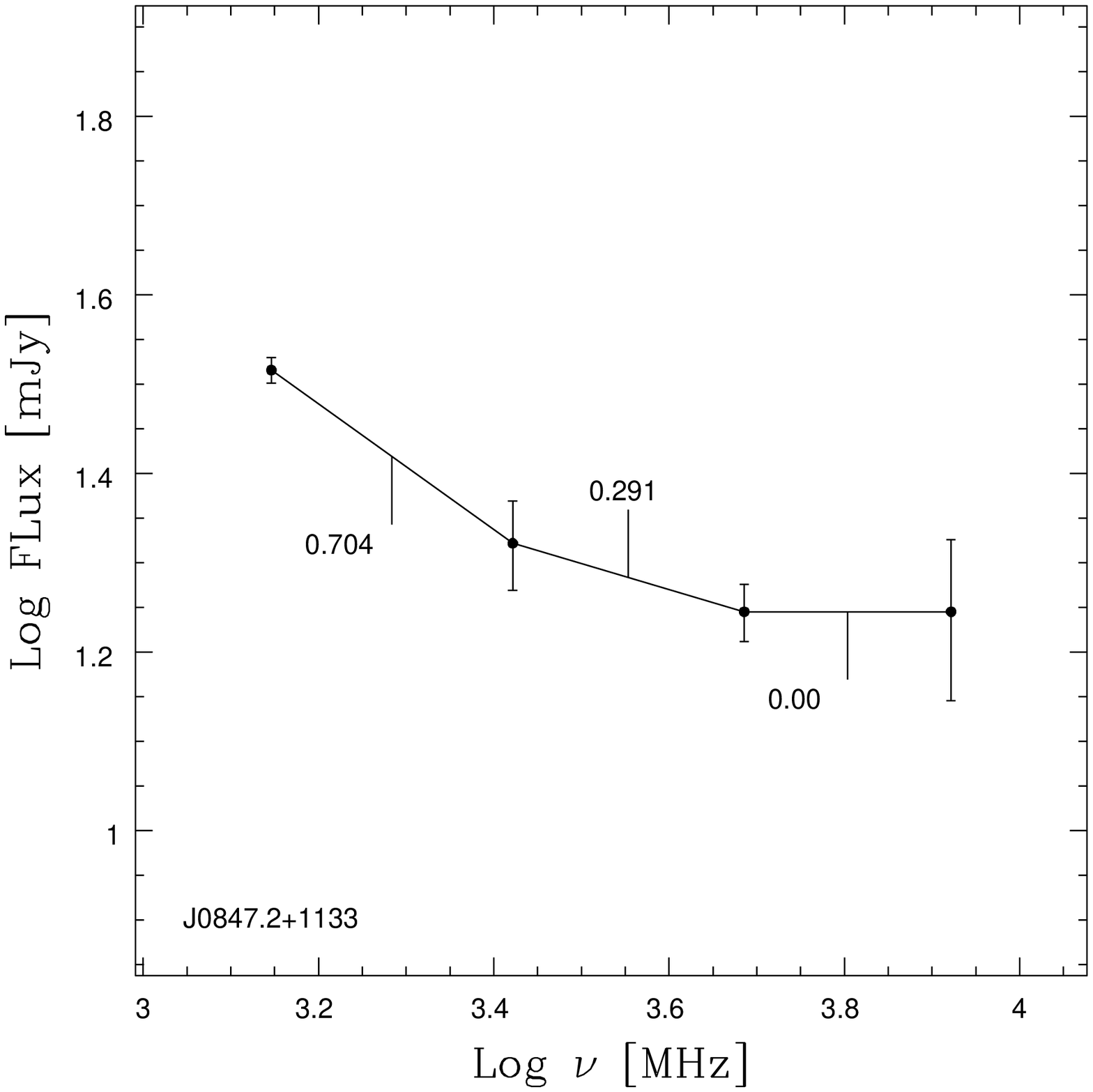}
\includegraphics[width=8cm]{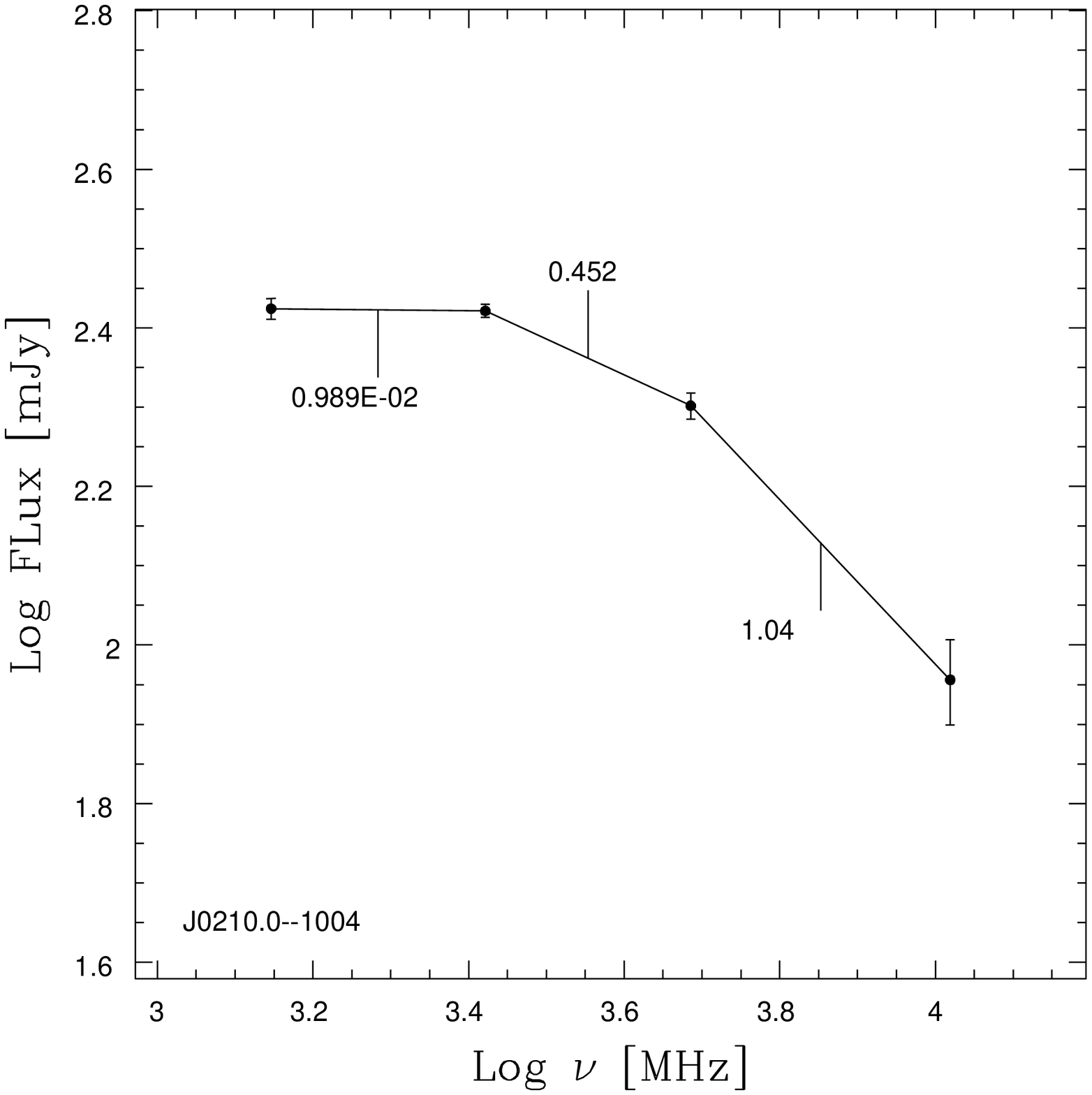}
\includegraphics[width=8cm]{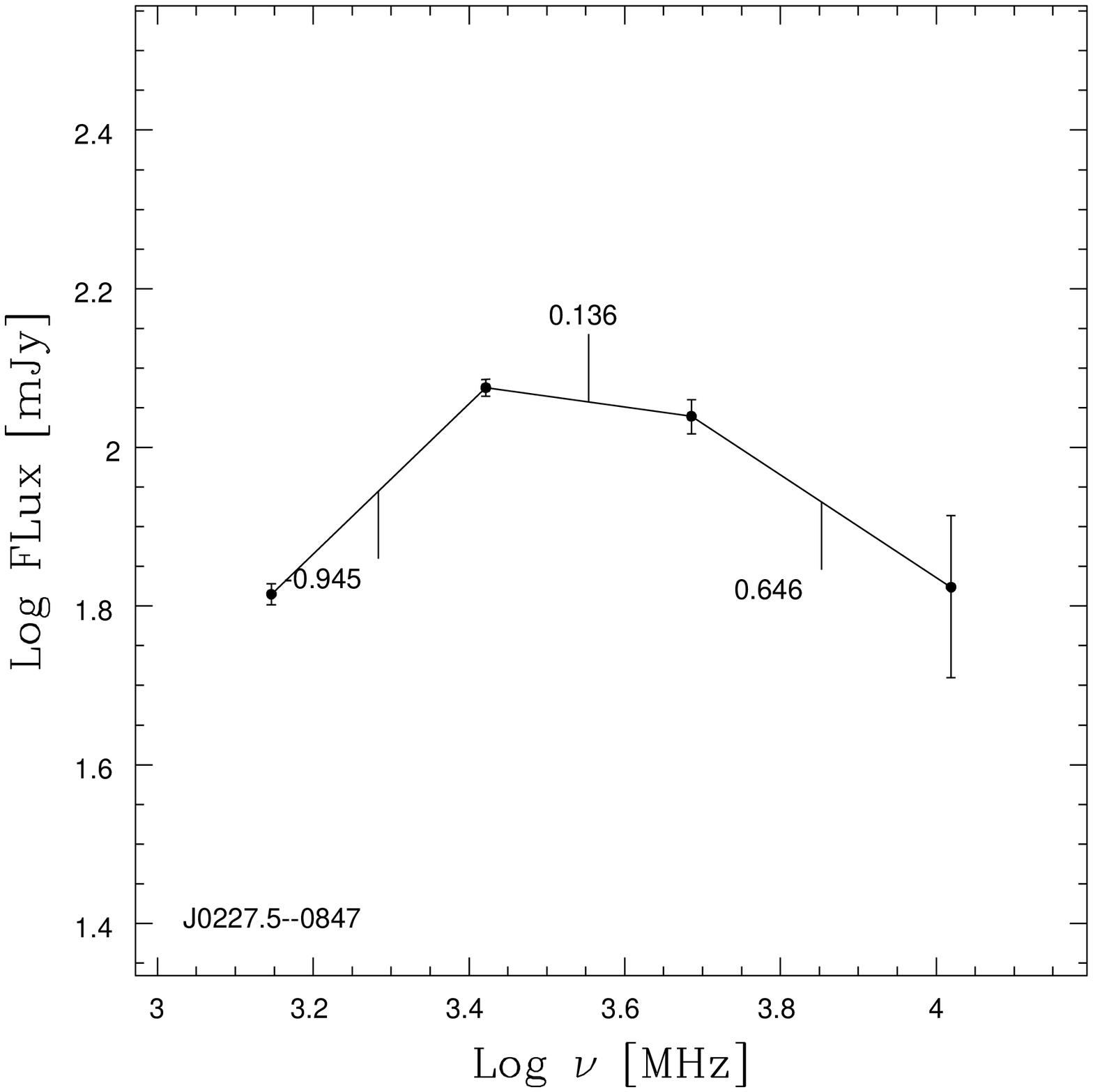}
\includegraphics[width=8cm]{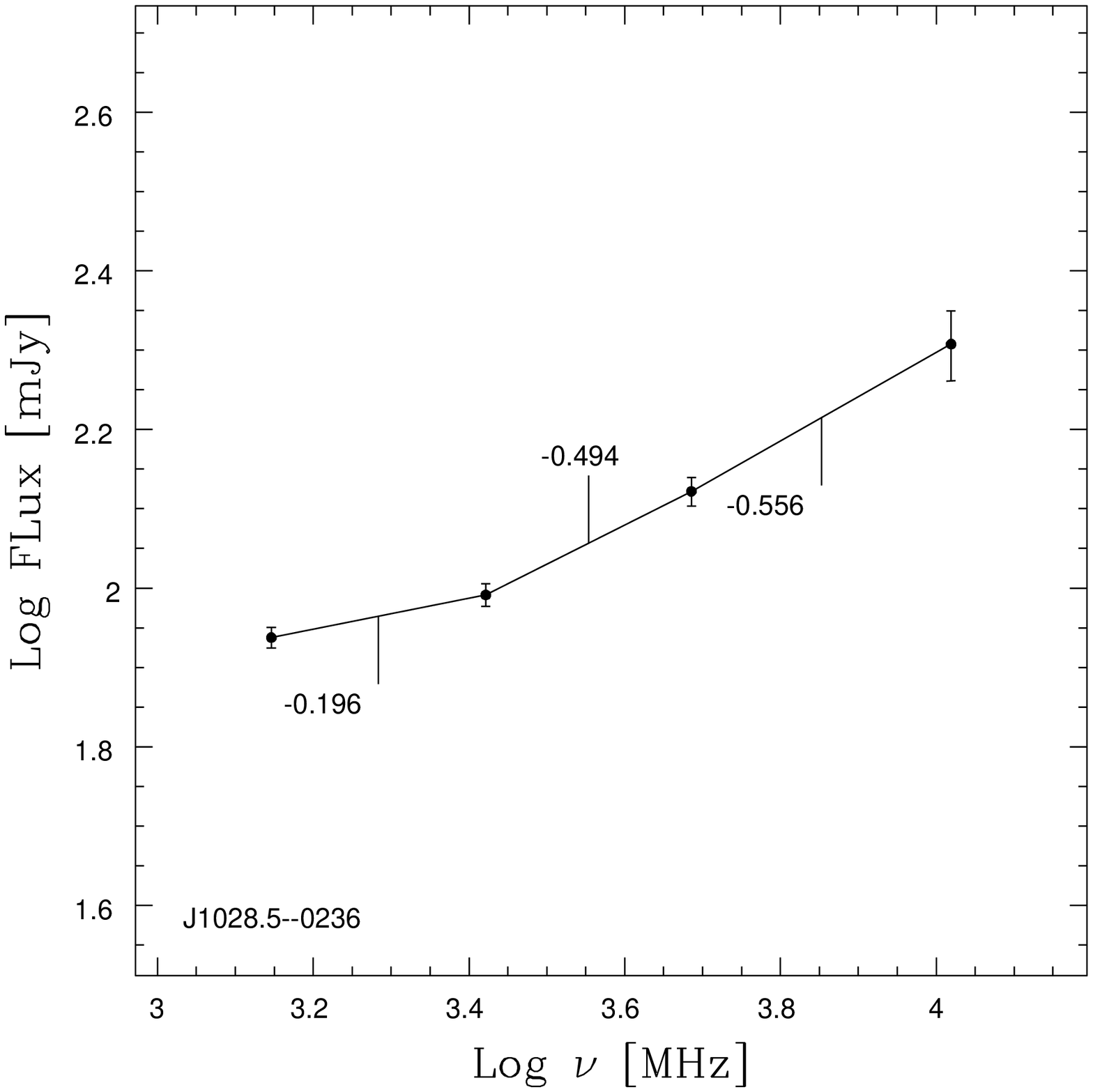}
\caption{Examples of spectral index behaviour, namely: ``steep'' 
(J0447.9$-$0322),
``flat'' (J0110.5$-$1647), ``steep--flat'' (J0847.2$+$1133), 
``flat--steep'' (J0210.0$-$1004), ``GPS'' (J0227.5$-$0847), ``inverted'' 
(J1028.5$-$0236).
\label{fig:spixplot}}
\end{figure*}
\newpage
%
\begin{figure*}[t]
\addtocounter{figure}{+0}
\centering
\includegraphics[width=8cm]{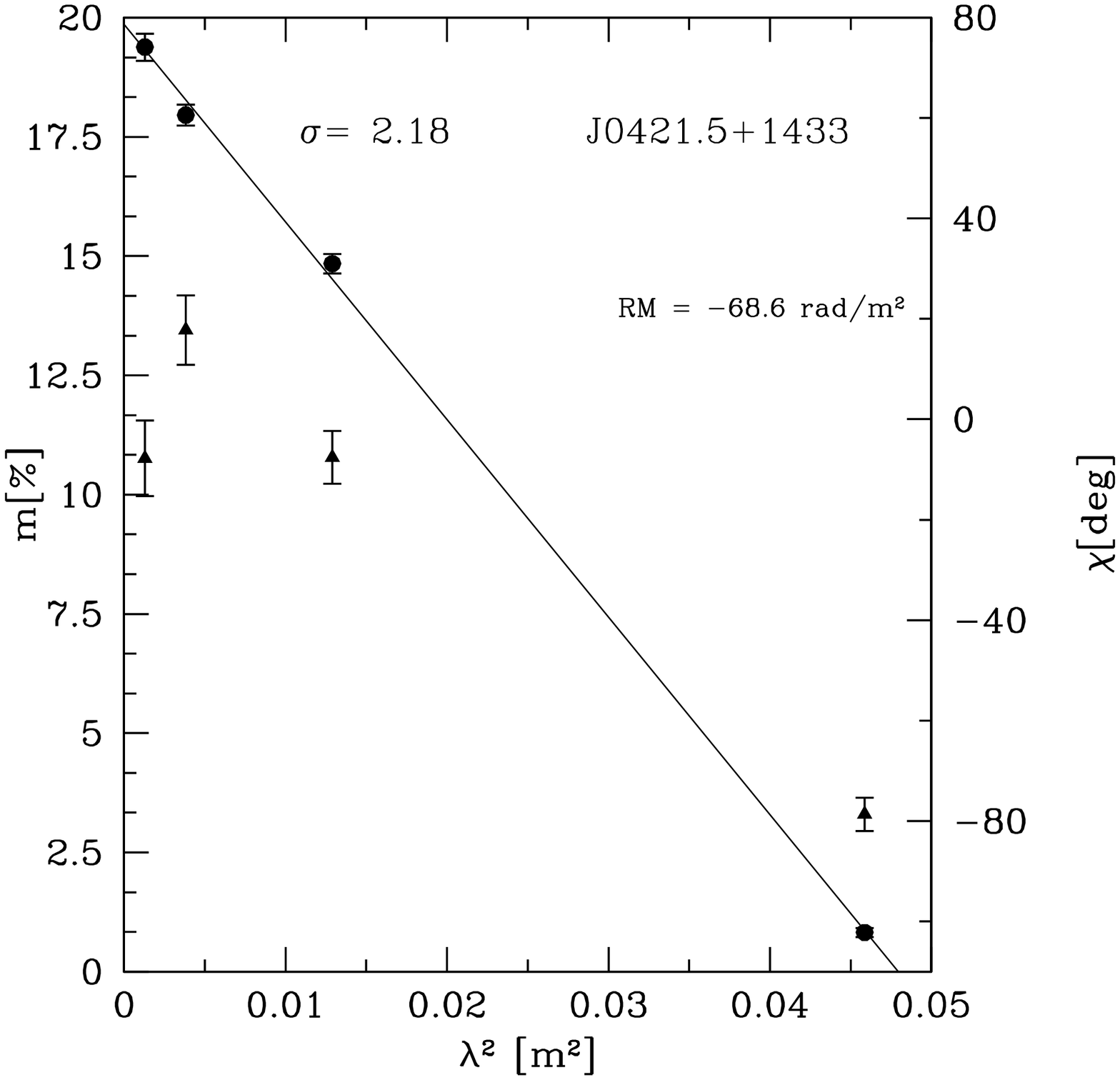}
\includegraphics[width=8cm]{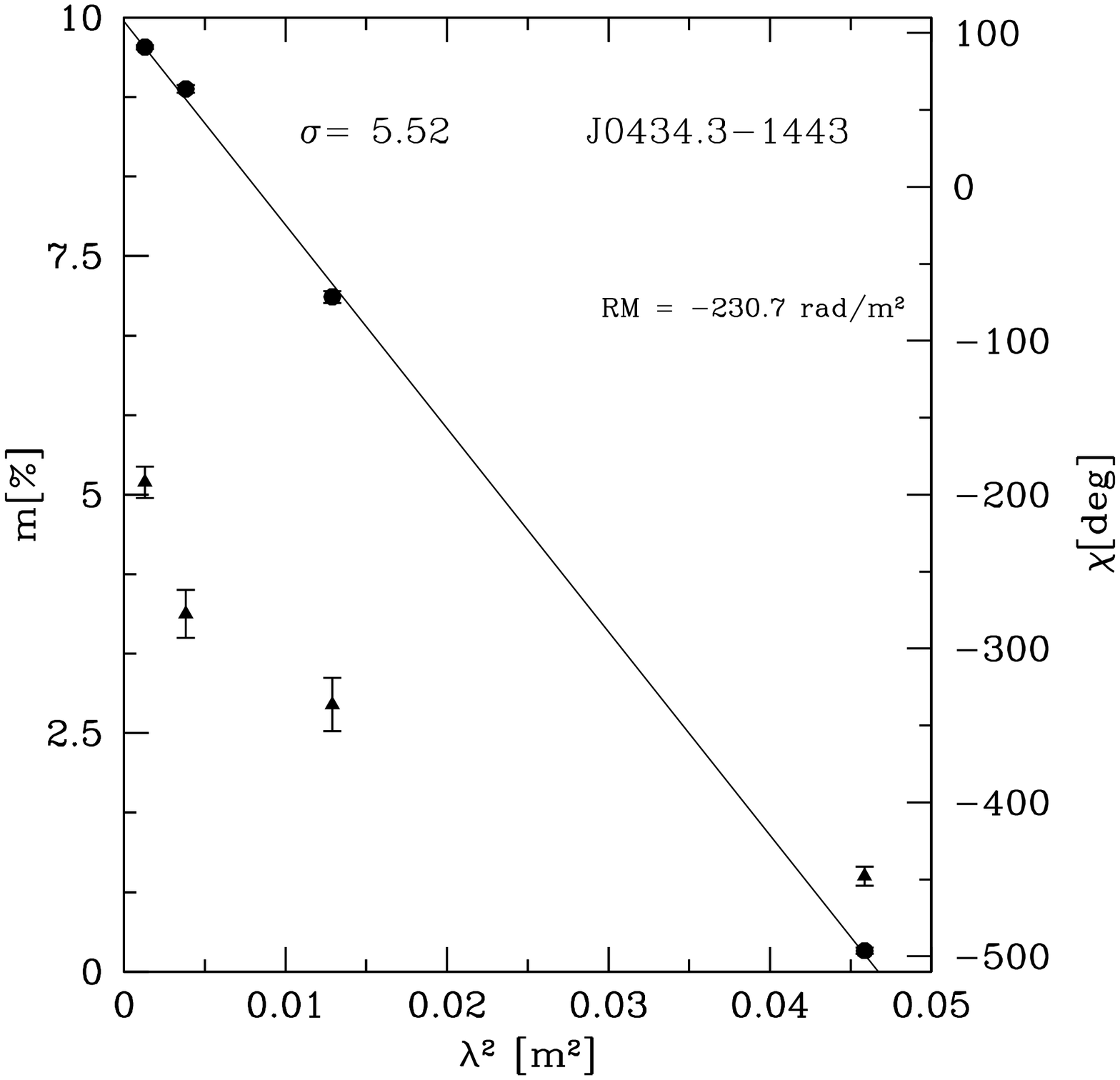}
\caption{Examples of position angles of the electric vector $\chi$ (dots)
and fractional polarisation $m$ (triangles) versus $\lambda^2$ plot. $\sigma$
values assess the quality of the best fit.
\label{fig:RM-m-plot}}
\end{figure*}
\newpage
%
%
%
%
%
%
\begin{figure*}
   \centering
   \includegraphics[width=16cm]{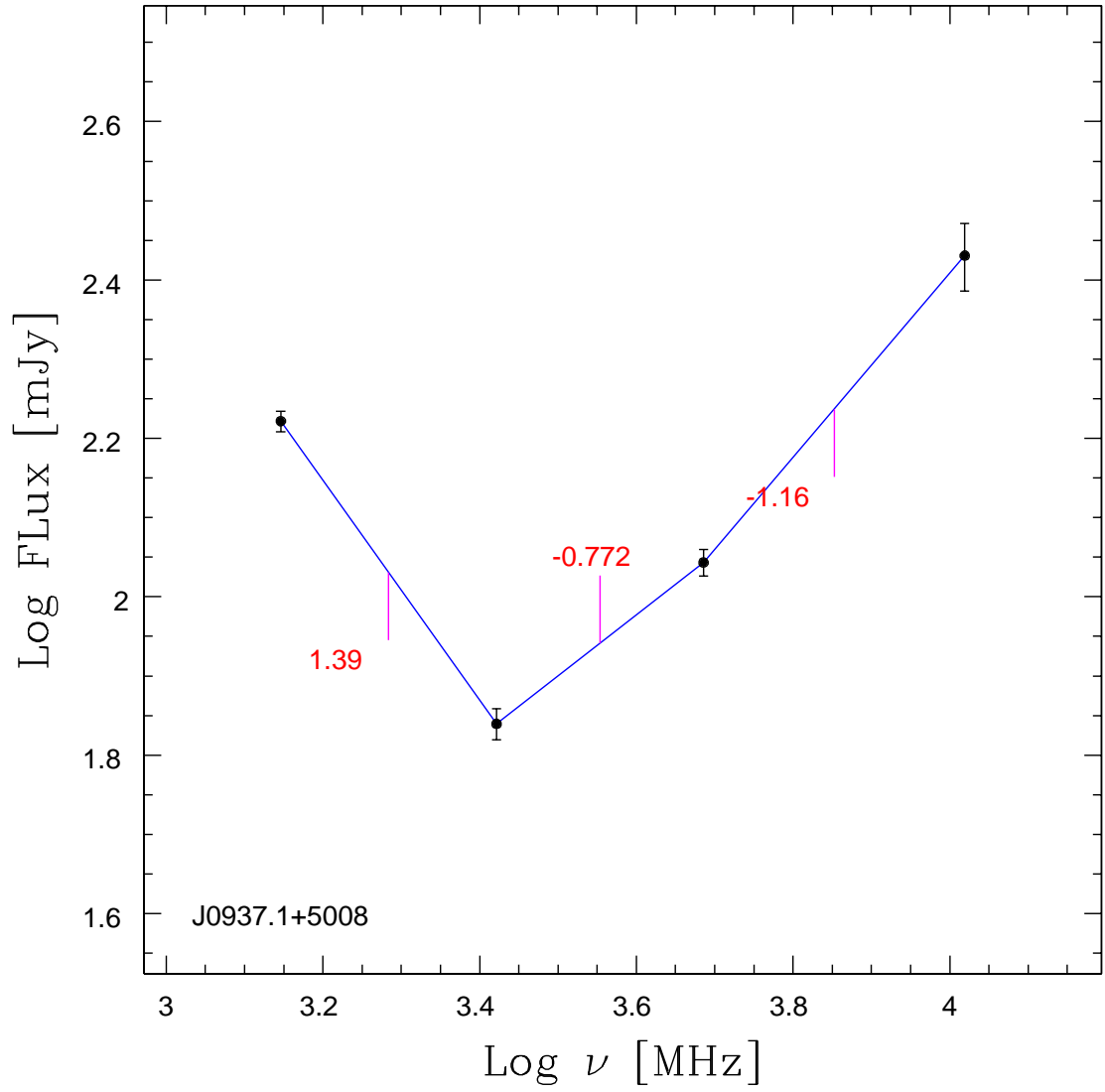}
      \caption{The ``concave'' spectral index of the source J0937.1$+$5008.
            \label{fig:j0937}
              }
   \end{figure*}
\newpage
\clearpage
%
%
\appendix
\section{Spectral index plots}
%
%
\begin{figure*}[t]
\addtocounter{figure}{+0}
\centering
\includegraphics[width=8cm]{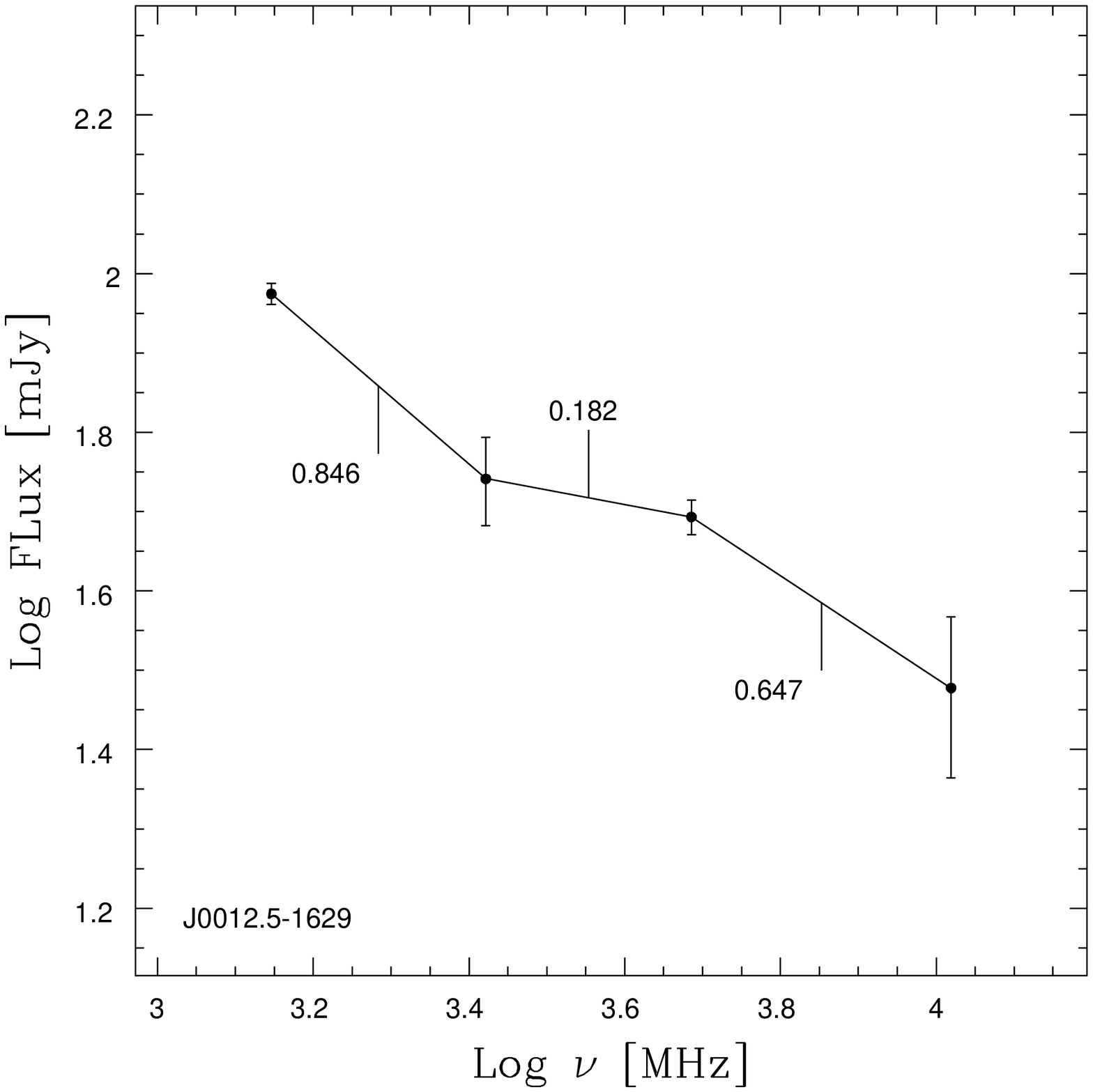}
\includegraphics[width=8cm]{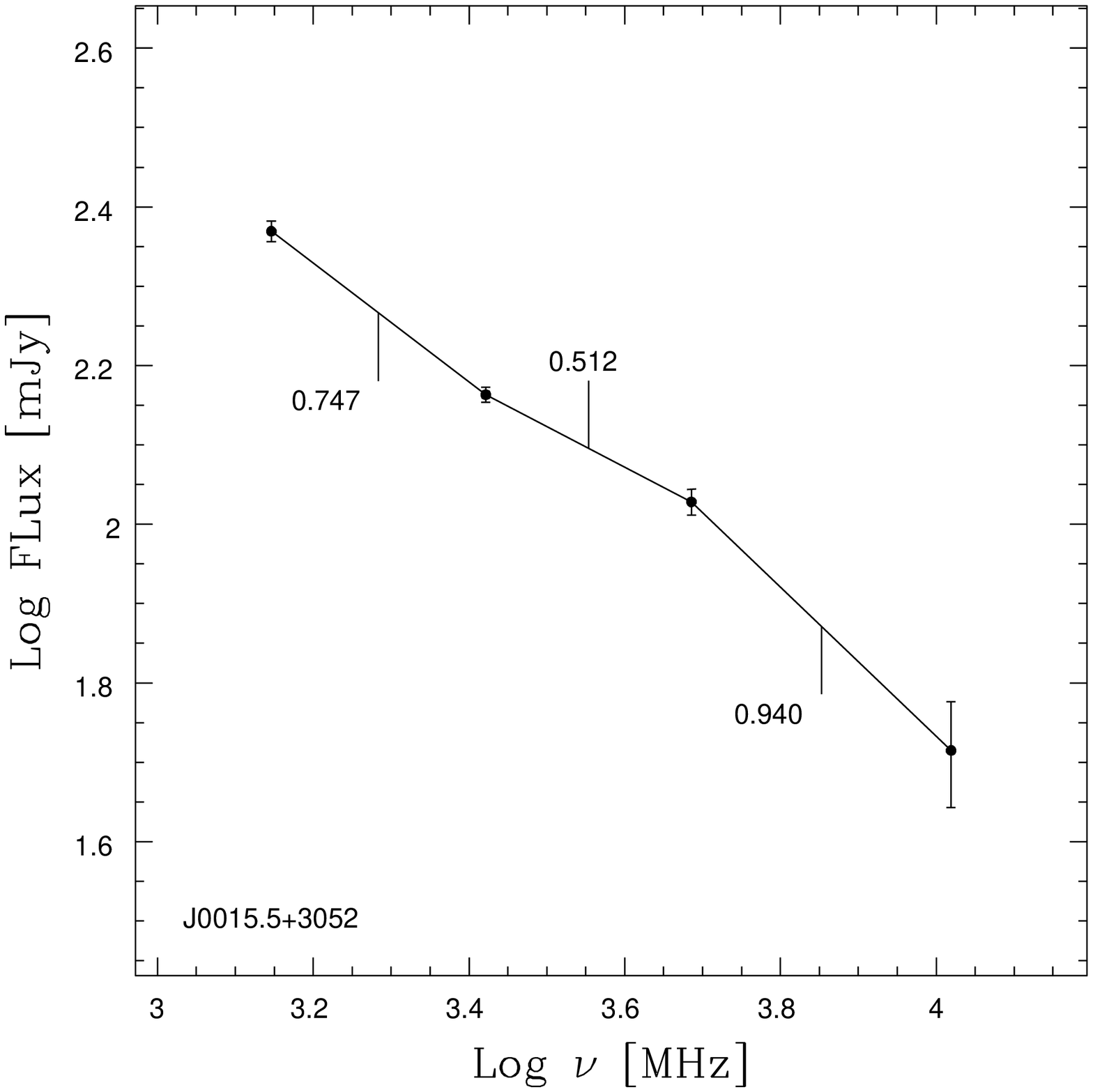}
\includegraphics[width=8cm]{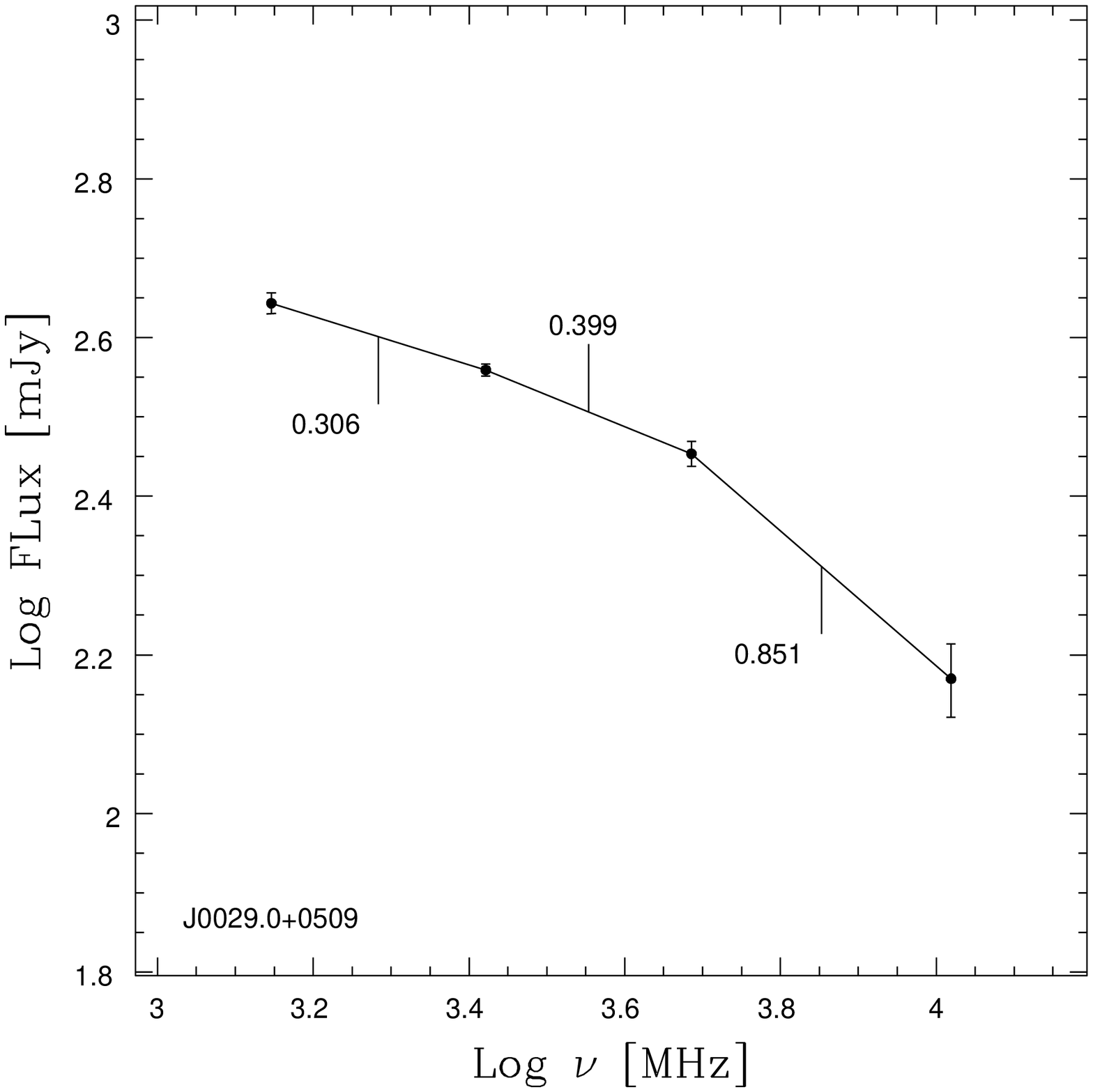}
\includegraphics[width=8cm]{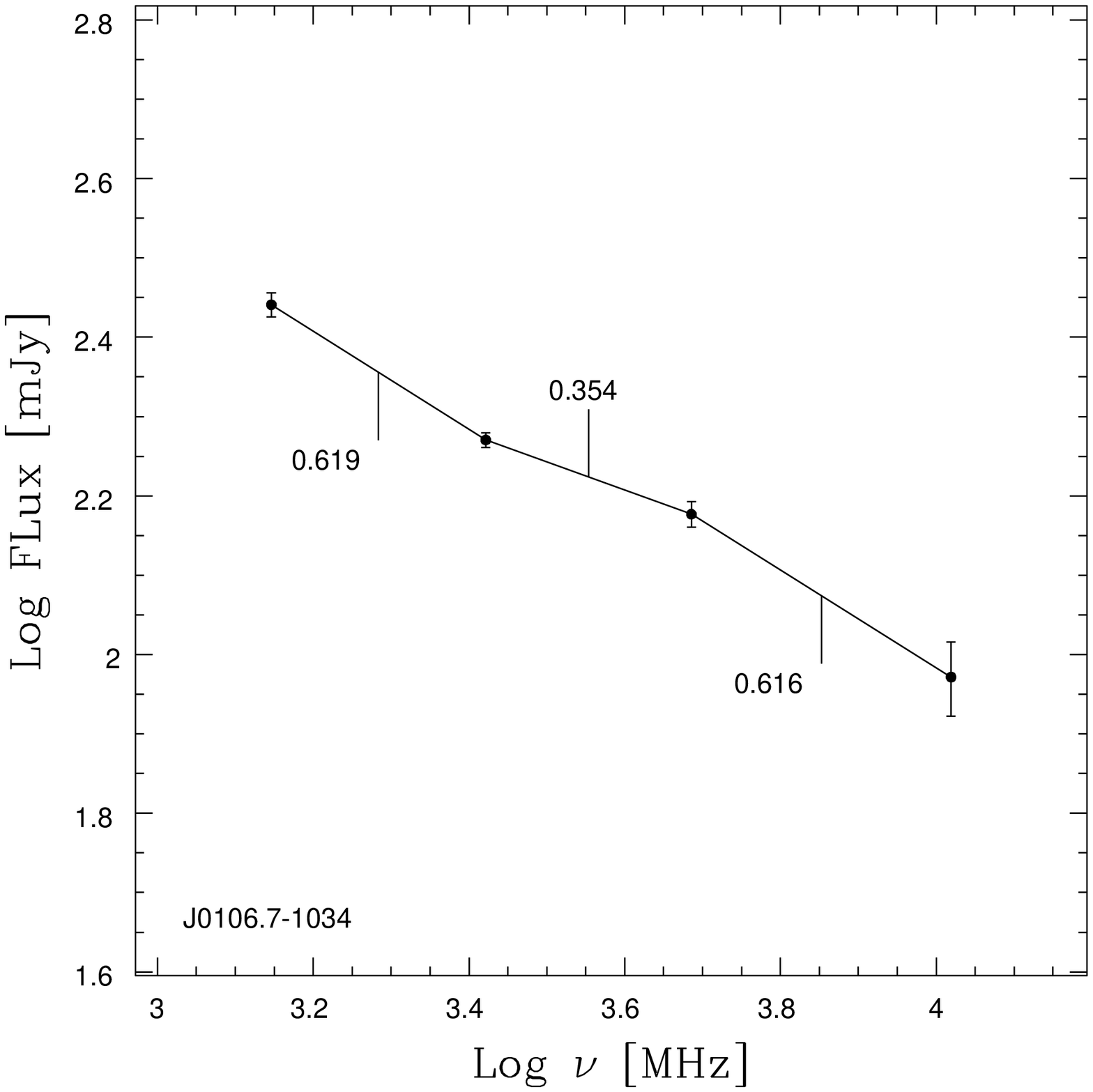}
\includegraphics[width=8cm]{j0110.5_e.ps}
\includegraphics[width=8cm]{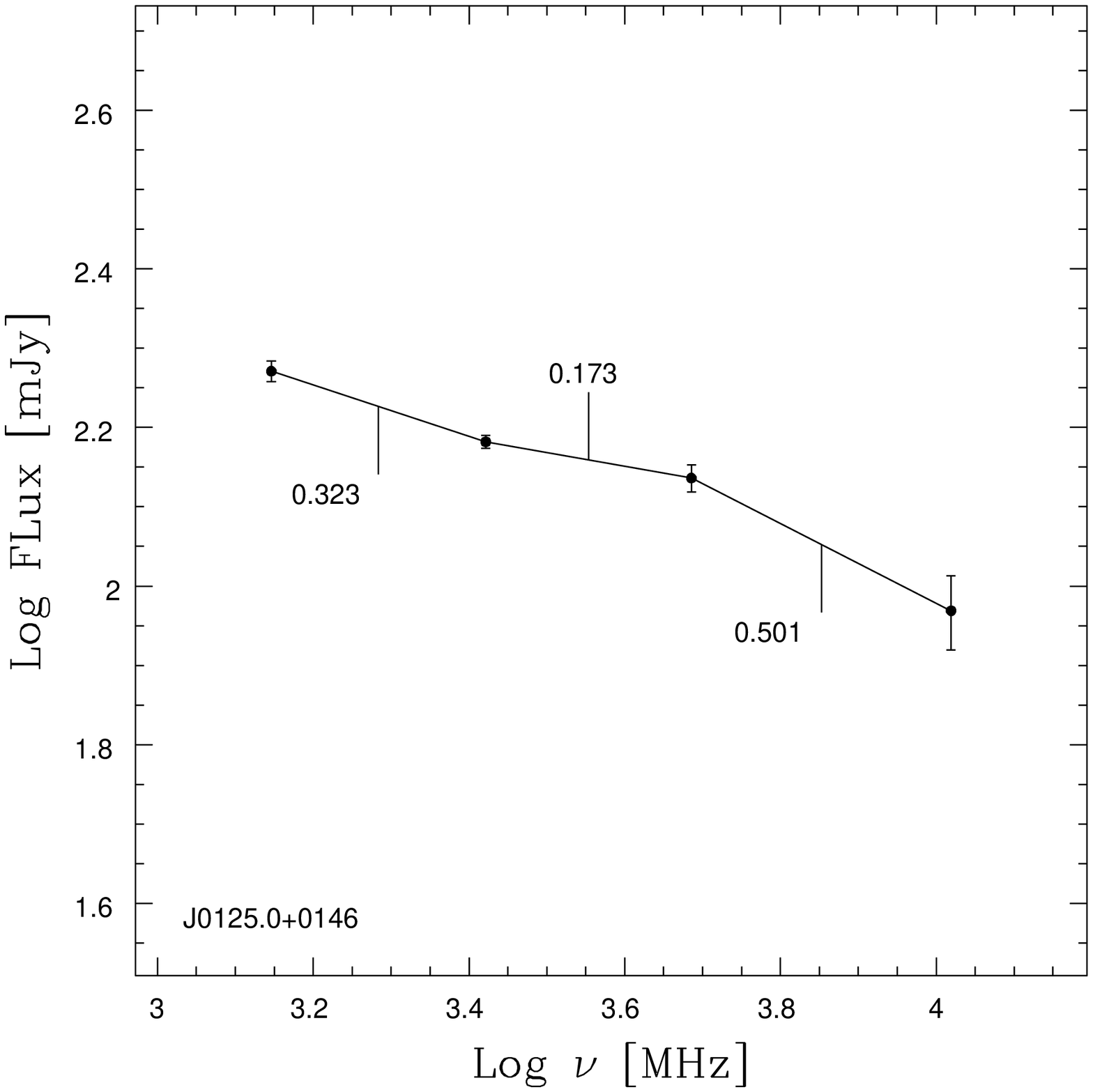}
\caption{Spectral index plots of sources in Table 3.}
\end{figure*}
\newpage
\begin{figure*}[t]
\addtocounter{figure}{+0}
\centering
\includegraphics[width=8cm]{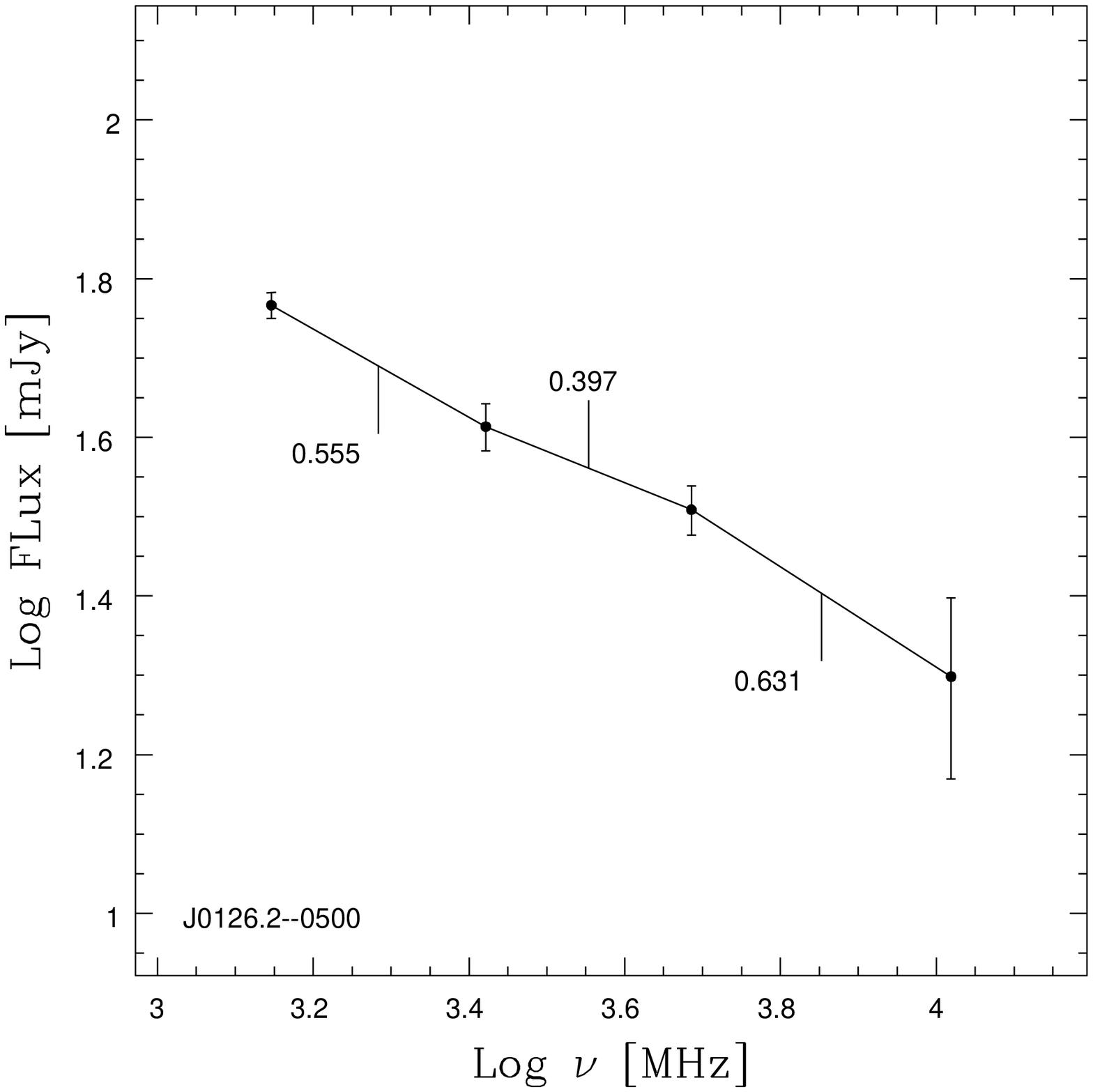}
\includegraphics[width=8cm]{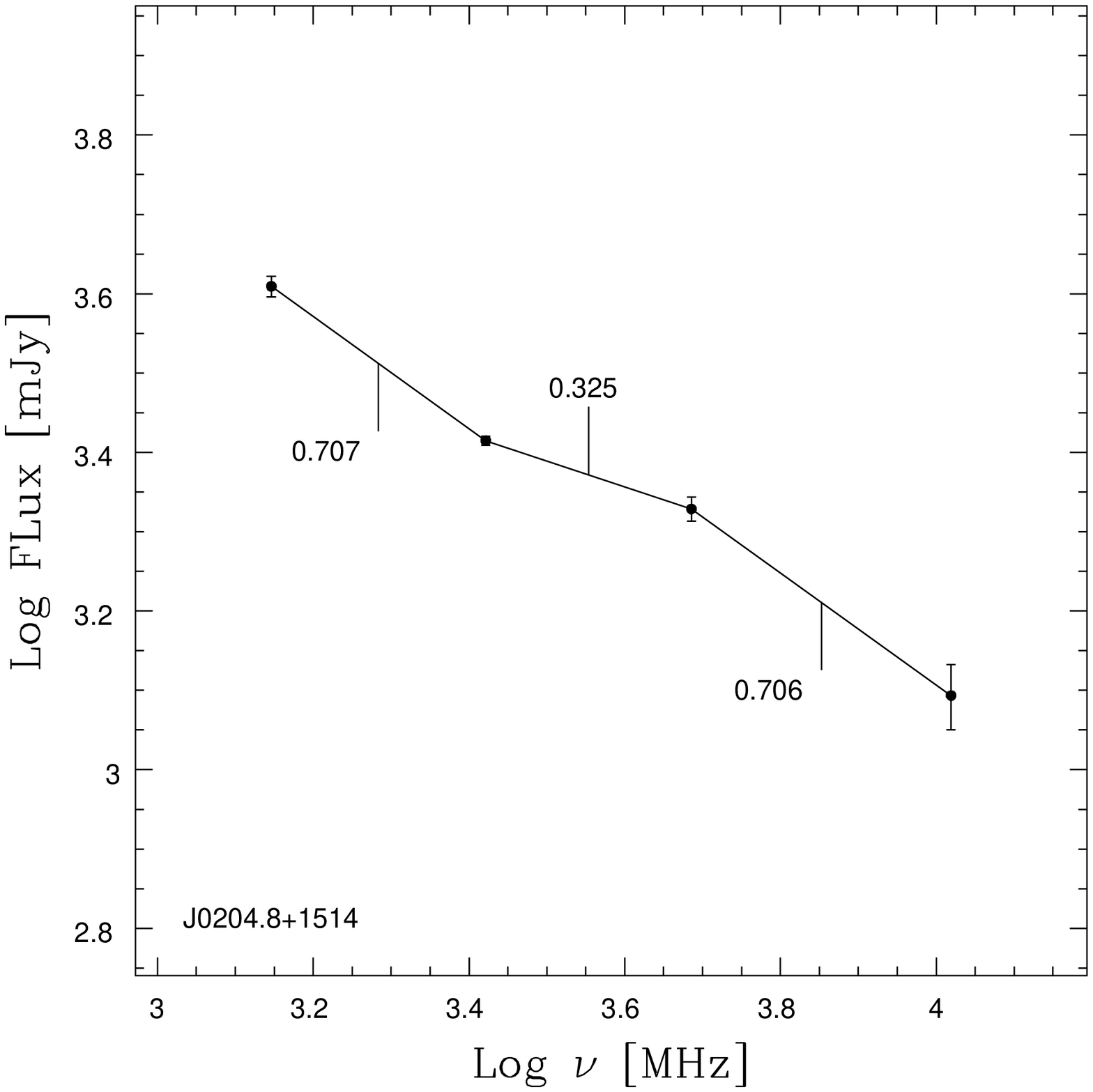}
\includegraphics[width=8cm]{j0210.0_e.ps}
\includegraphics[width=8cm]{j0227.5_e.ps}
\includegraphics[width=8cm]{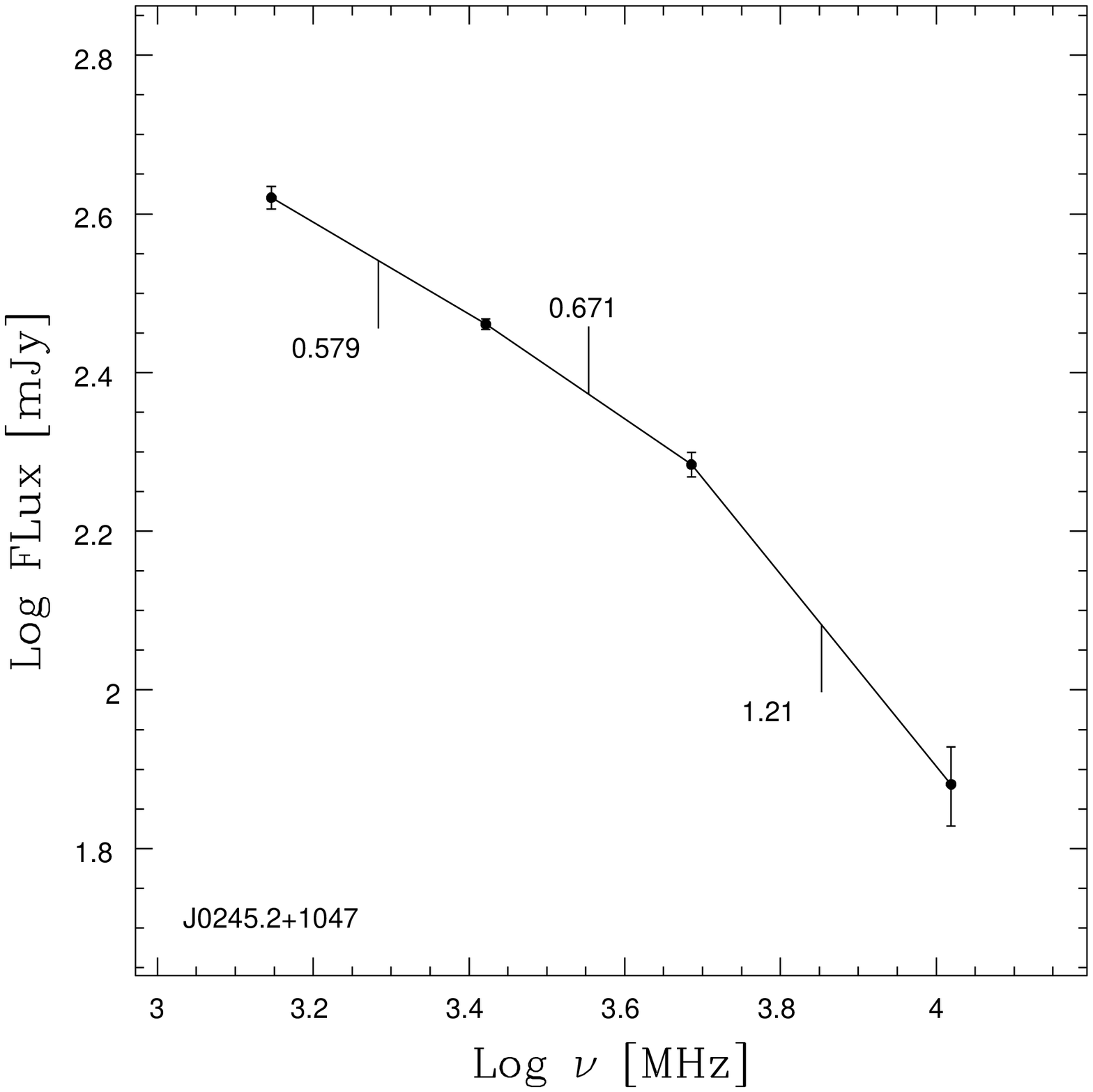}
\includegraphics[width=8cm]{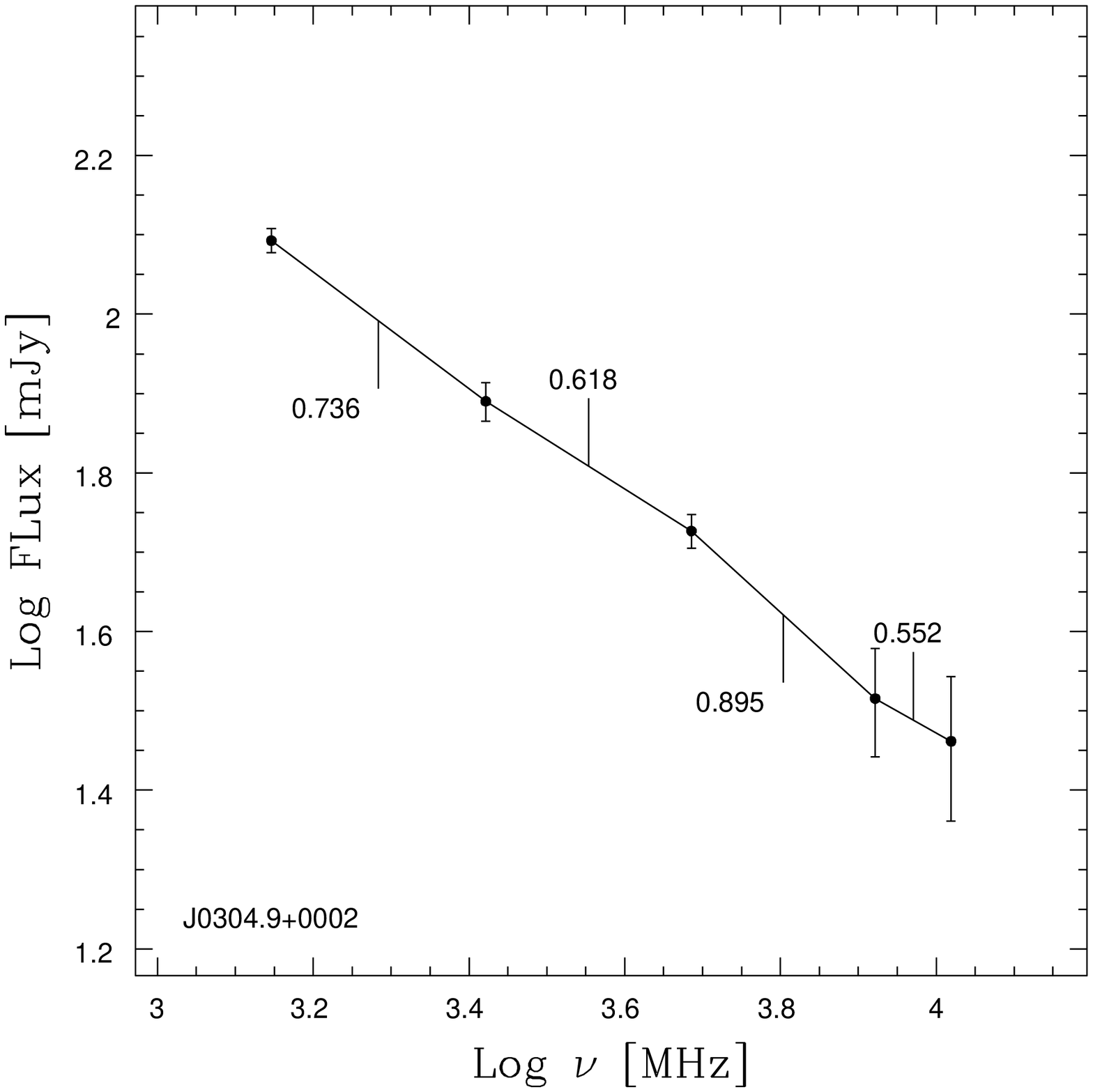}
\caption{Spectral index plots of sources in Table 3.}
\end{figure*}
\newpage
\begin{figure*}[t]
\addtocounter{figure}{+0}
\centering
\includegraphics[width=8cm]{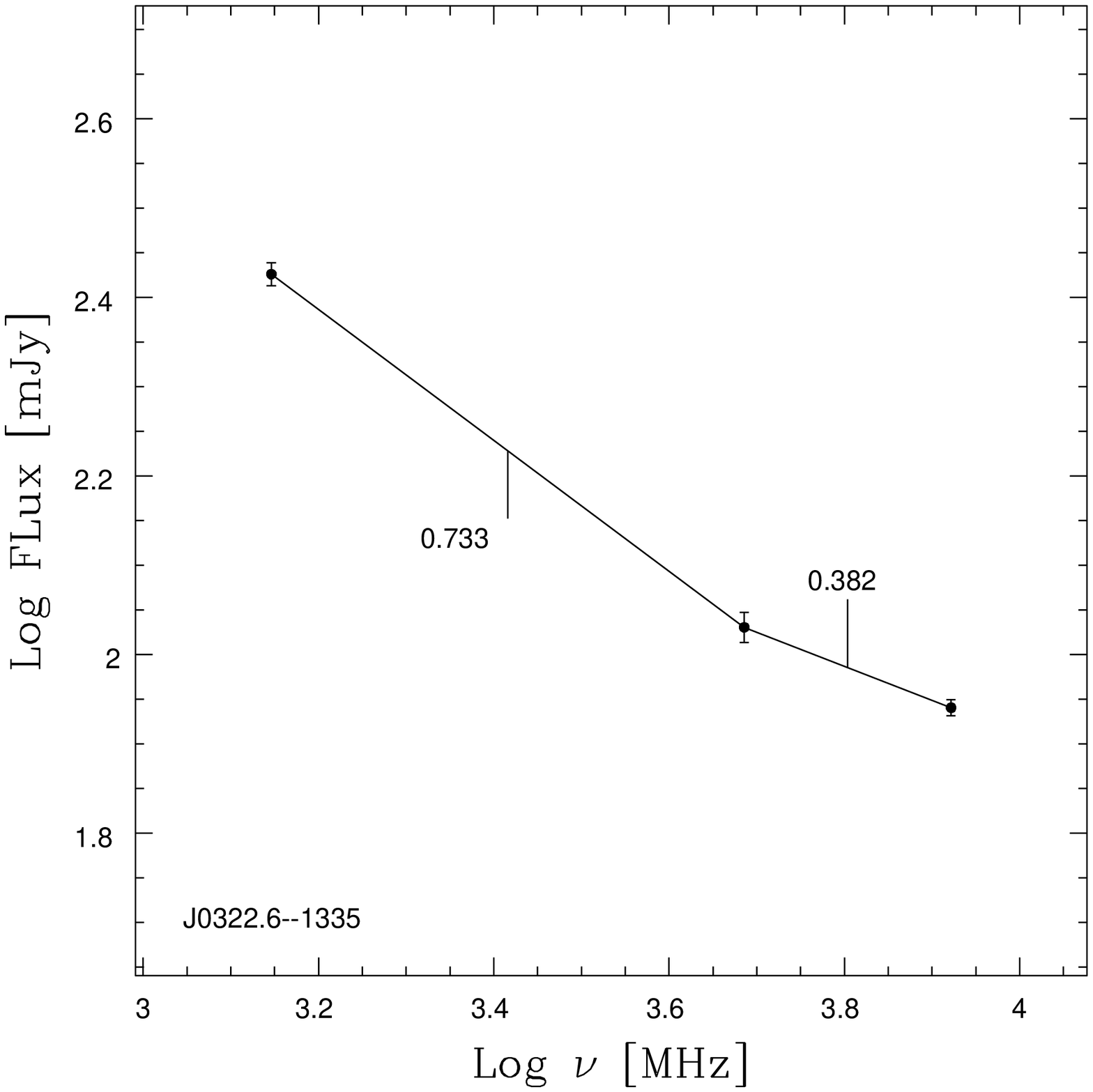}
\includegraphics[width=8cm]{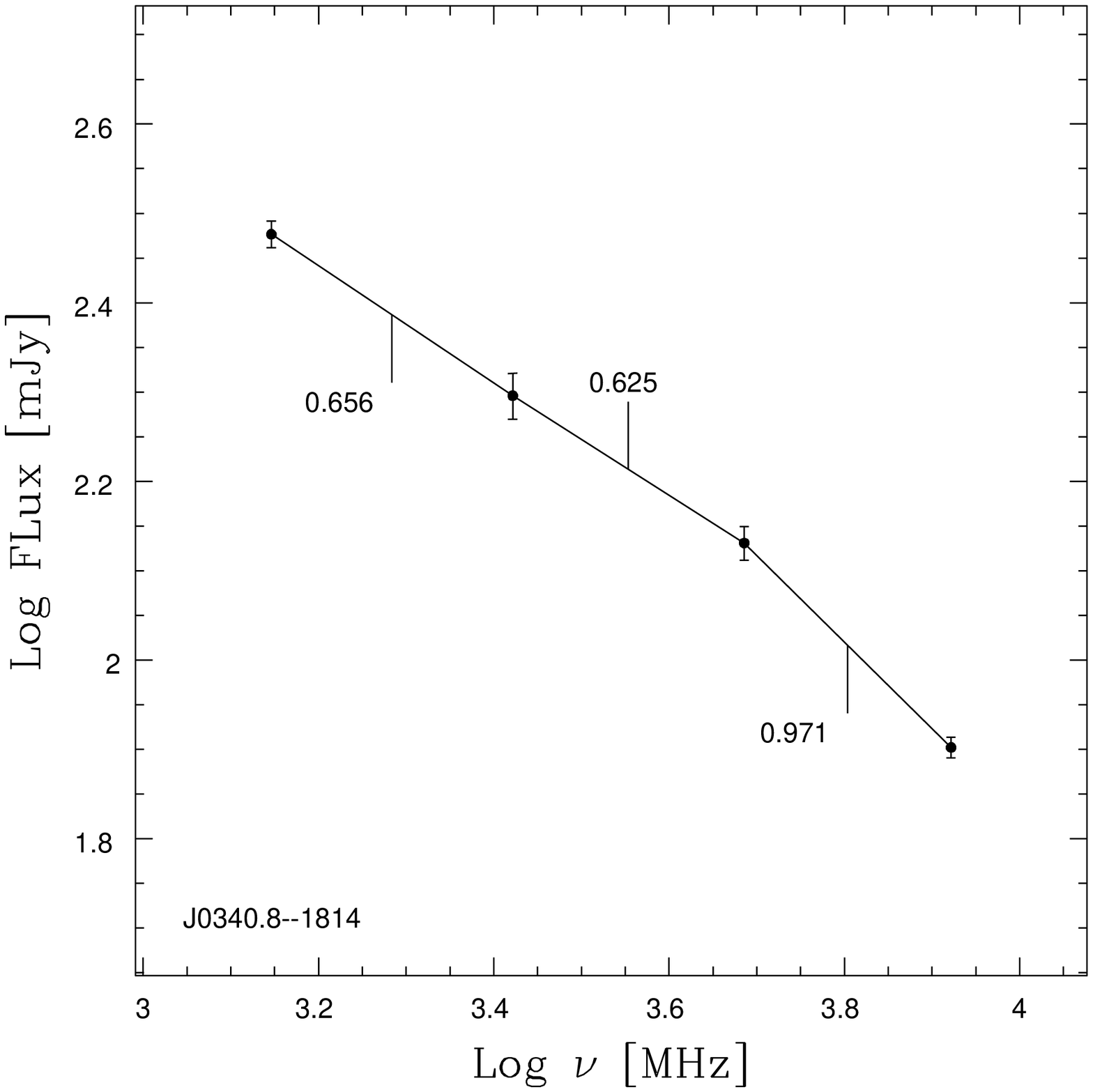}
\includegraphics[width=8cm]{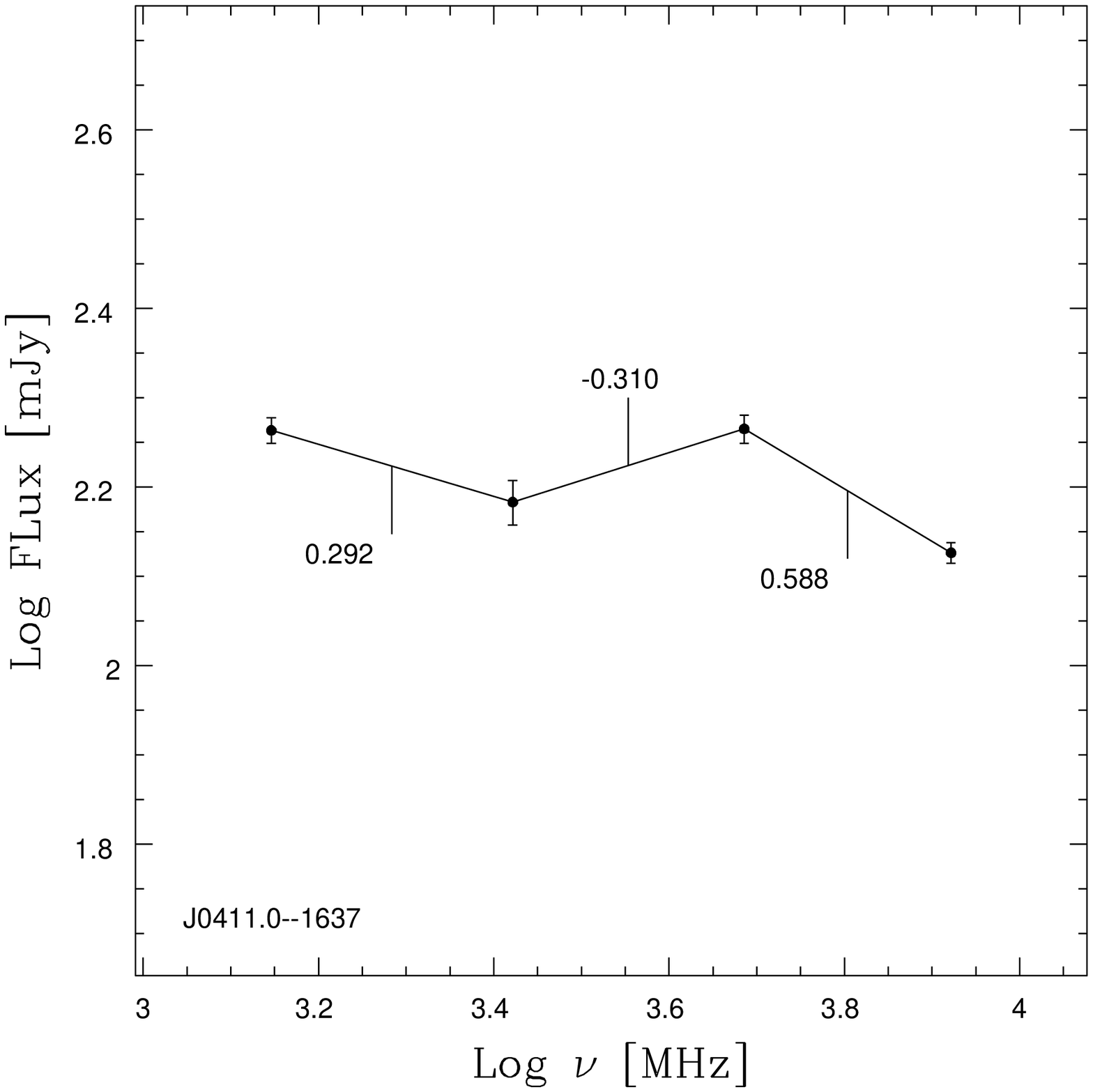}
\includegraphics[width=8cm]{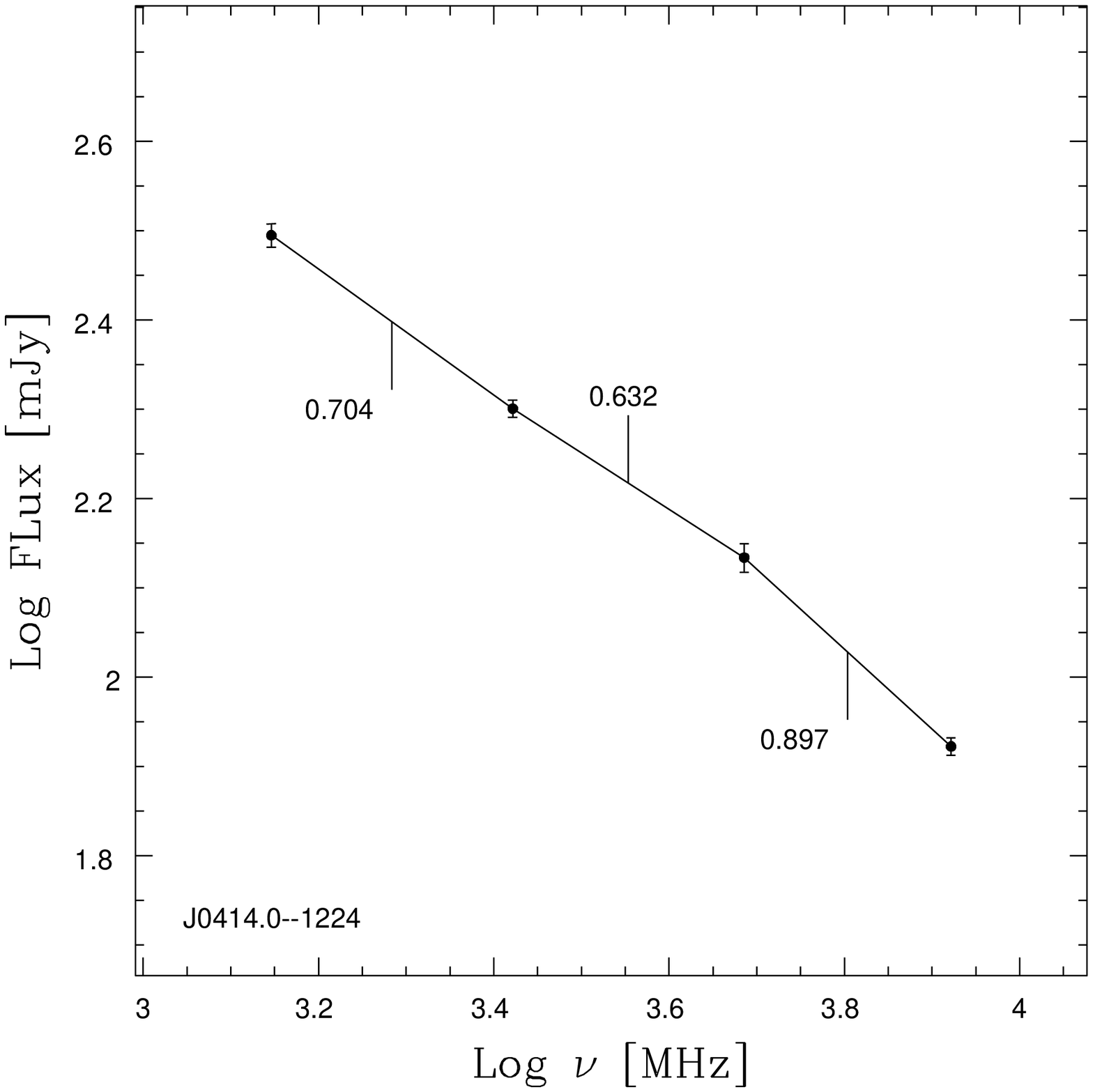}
\includegraphics[width=8cm]{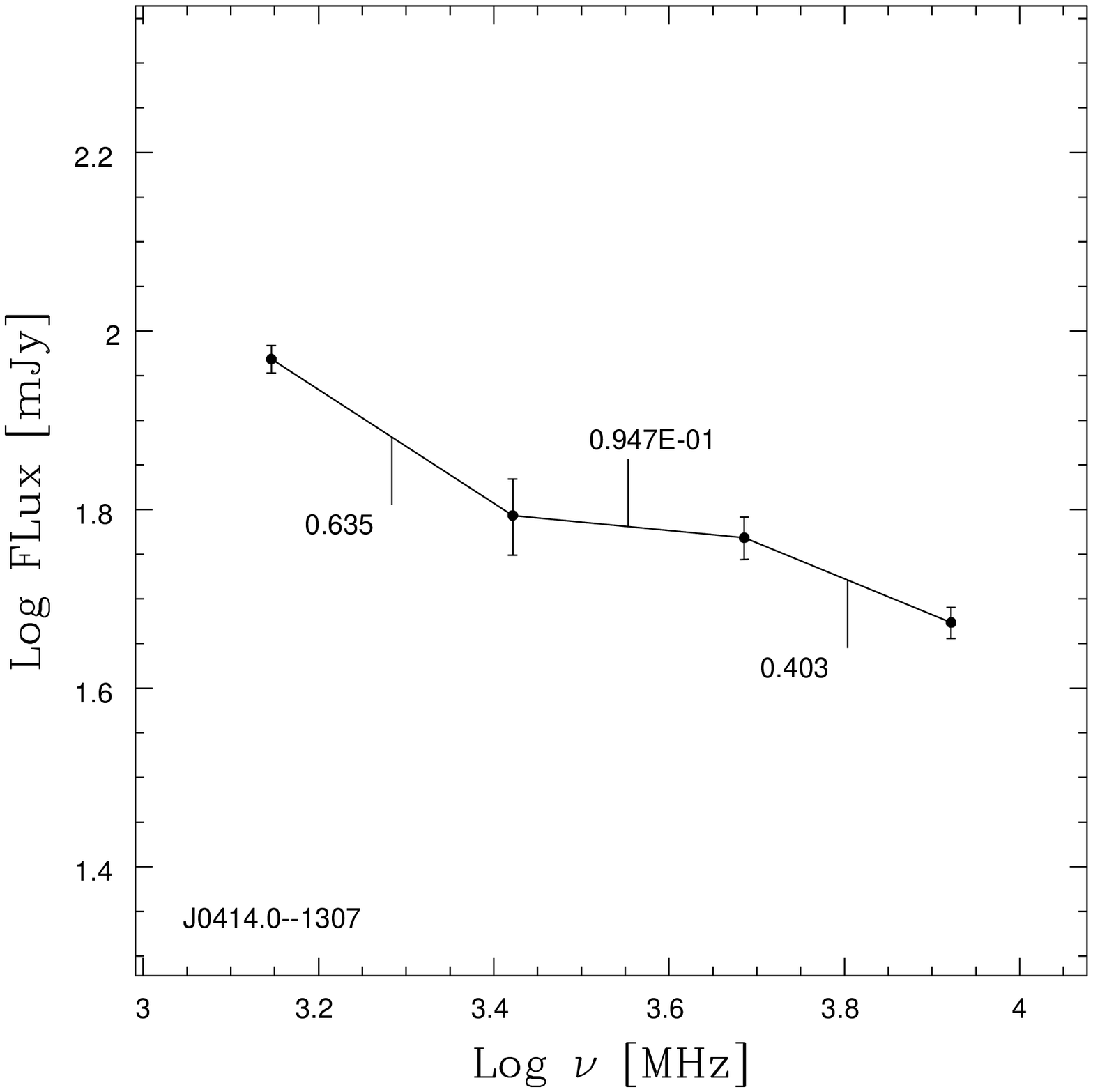}
\includegraphics[width=8cm]{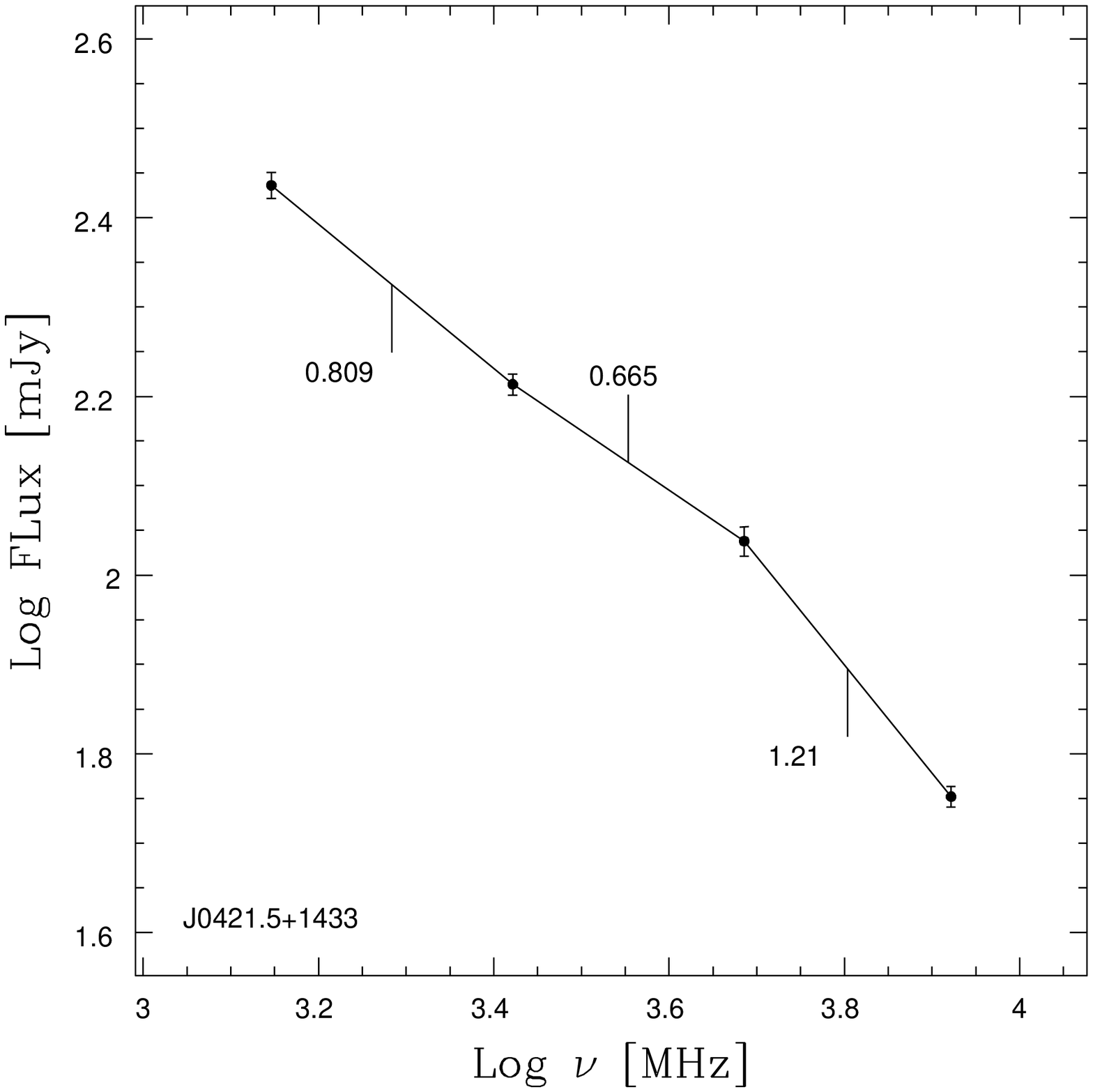}
\caption{Spectral index plots of sources in Table 3.}
\end{figure*}
\newpage
\begin{figure*}[t]
\addtocounter{figure}{+0}
\centering
\includegraphics[width=8cm]{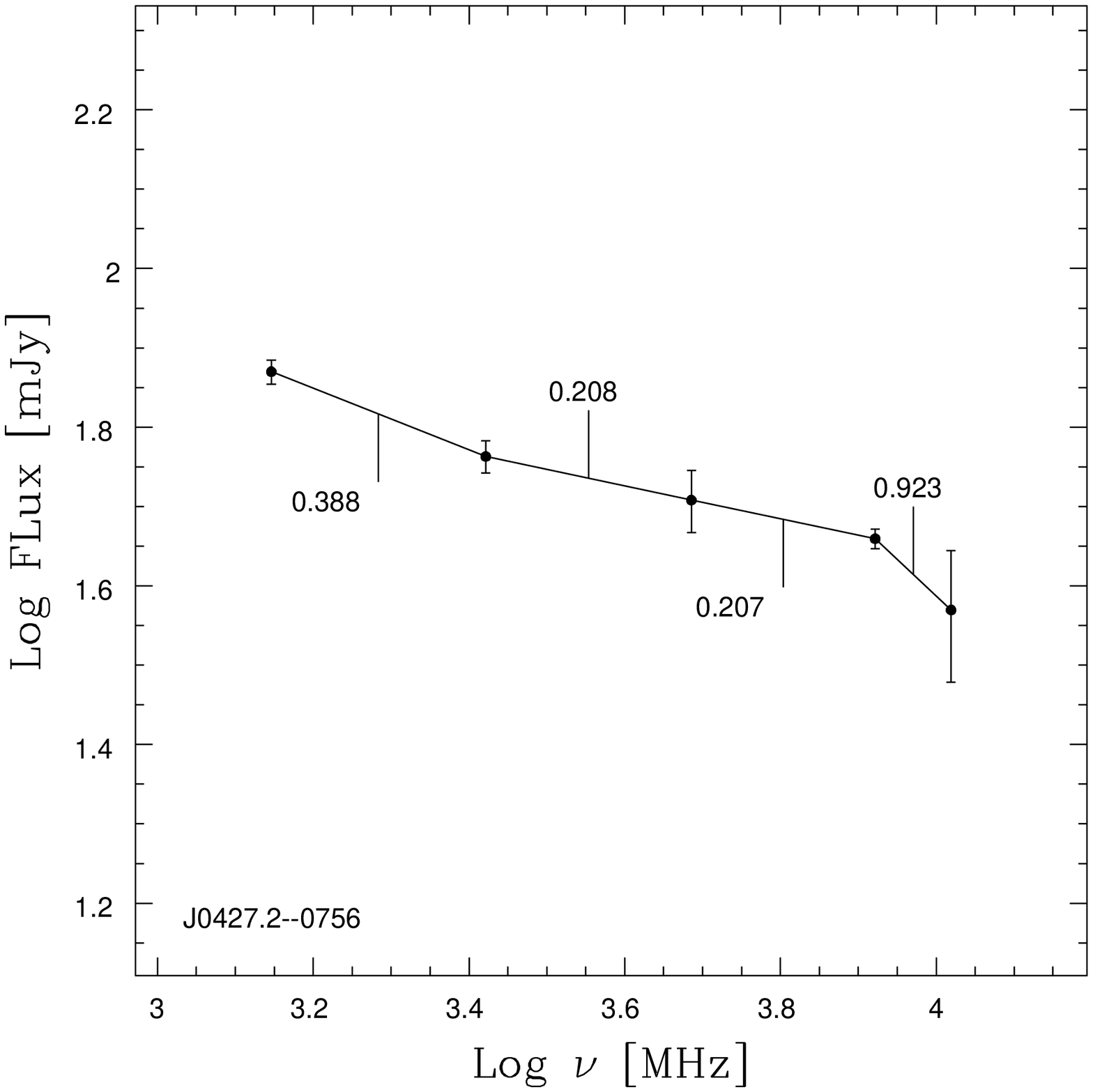}
\includegraphics[width=8cm]{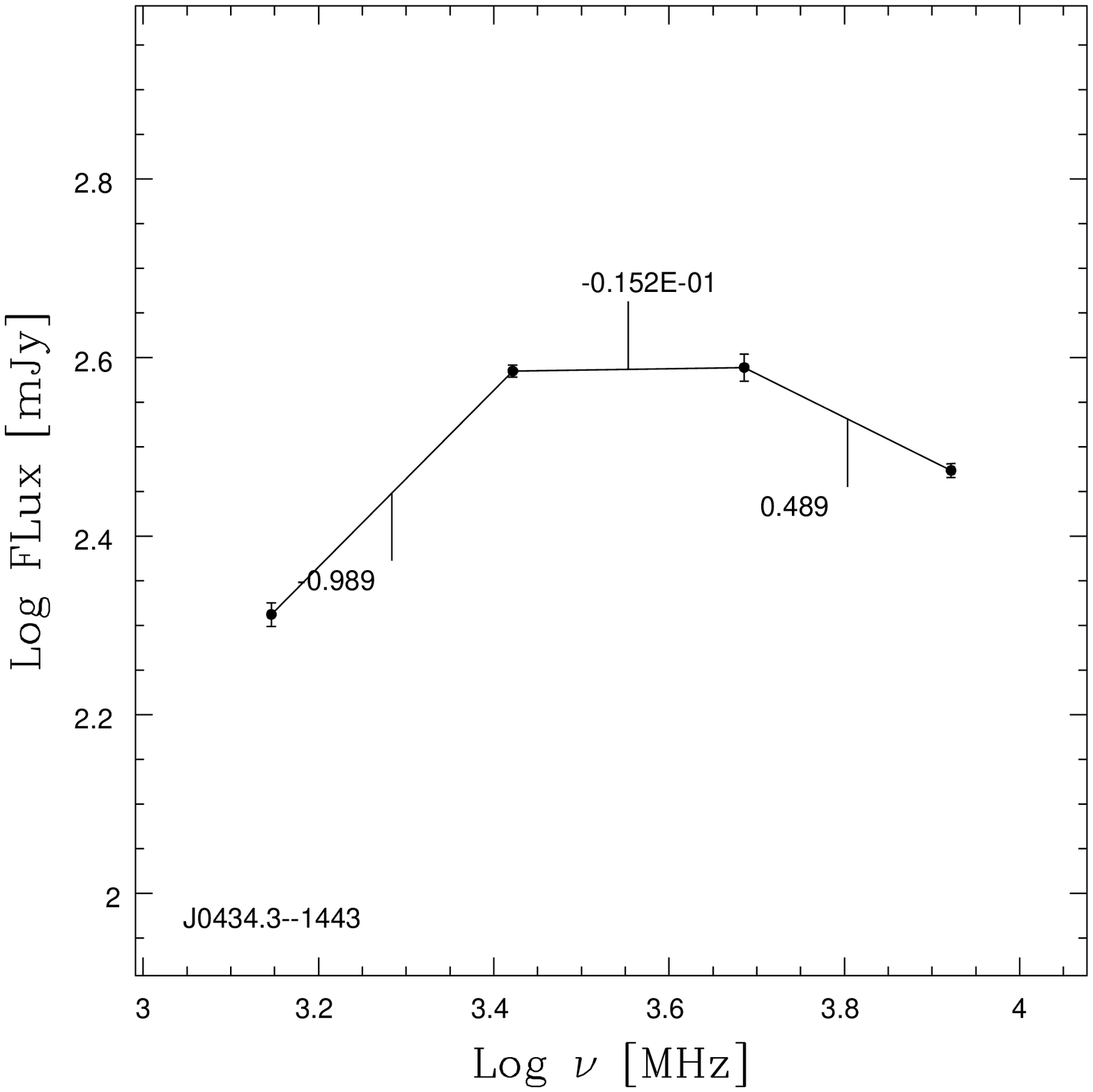}
\includegraphics[width=8cm]{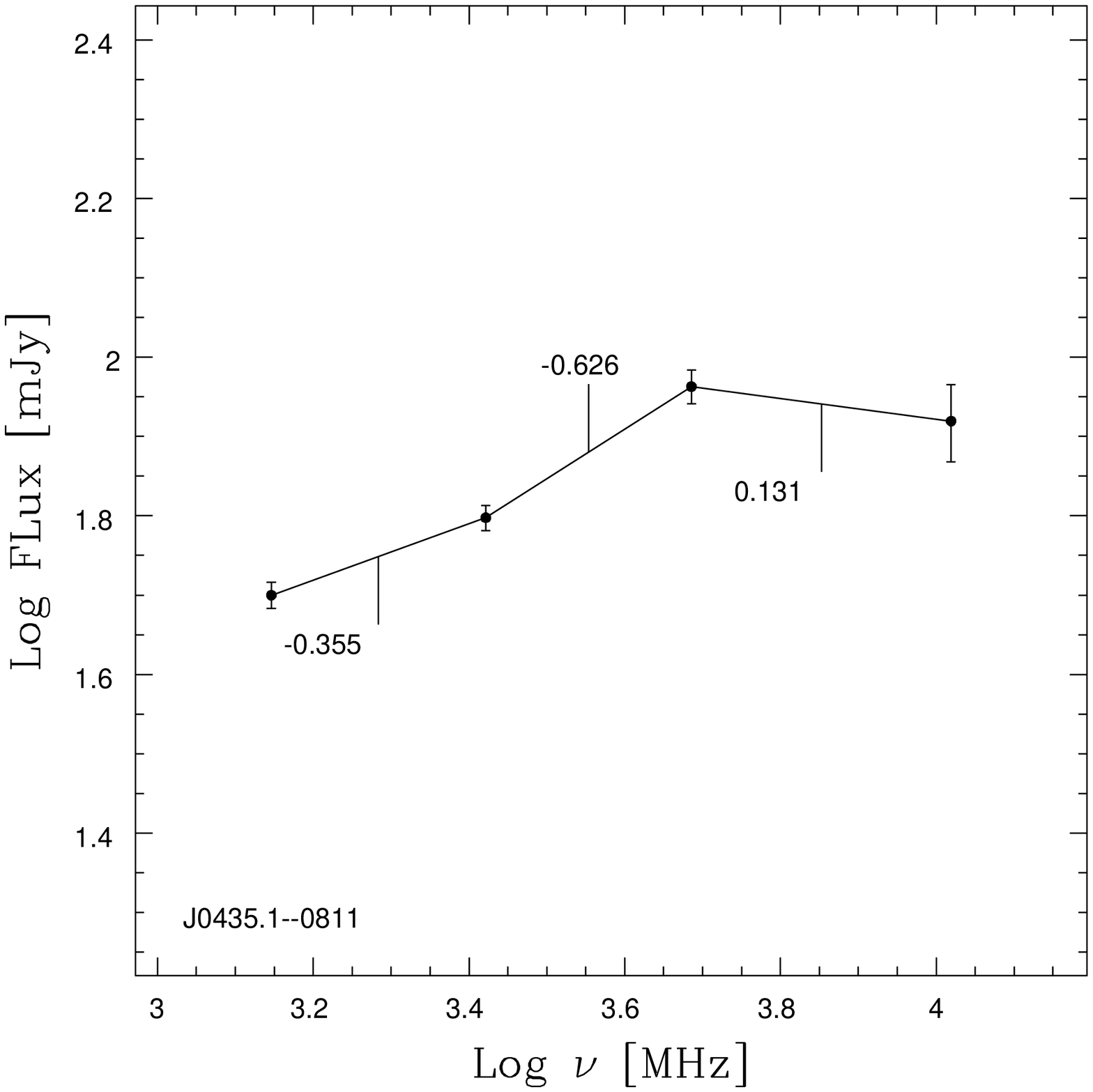}
\includegraphics[width=8cm]{j0447.9_e.ps}
\includegraphics[width=8cm]{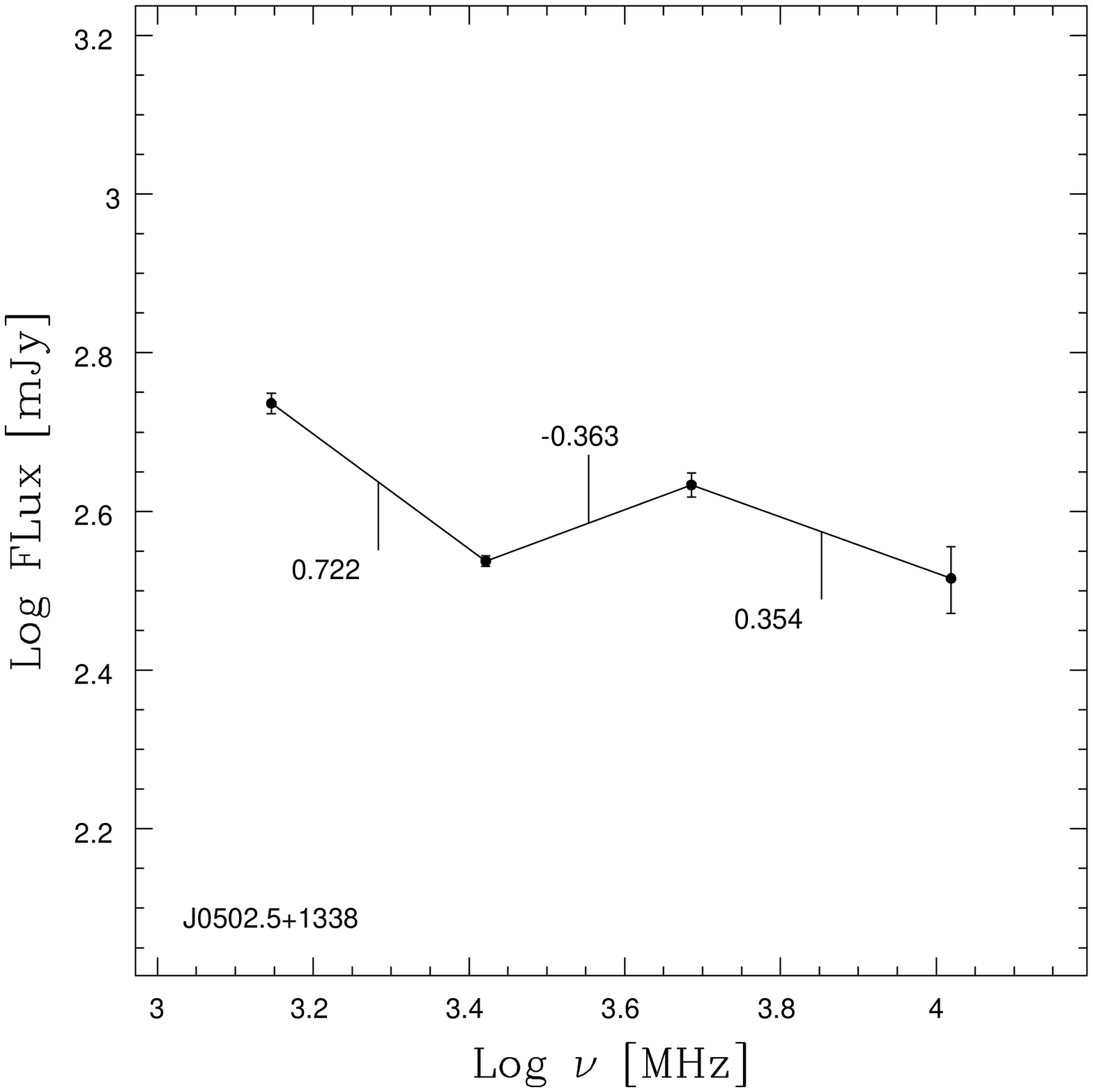}
\includegraphics[width=8cm]{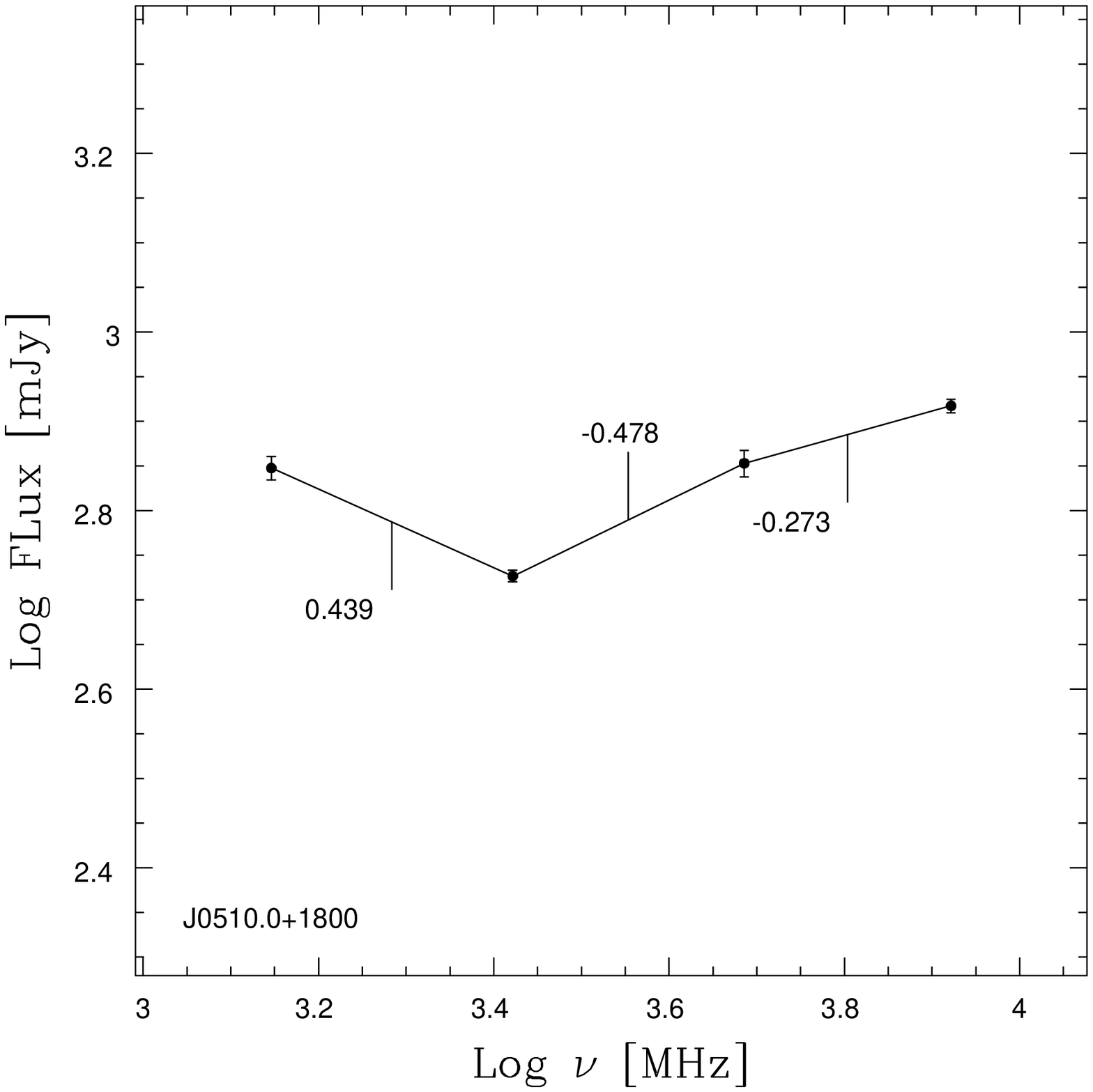}
\caption{Spectral index plots of sources in Table 3.}
\end{figure*}
\newpage
\begin{figure*}[t]
\addtocounter{figure}{+0}
\centering
\includegraphics[width=8cm]{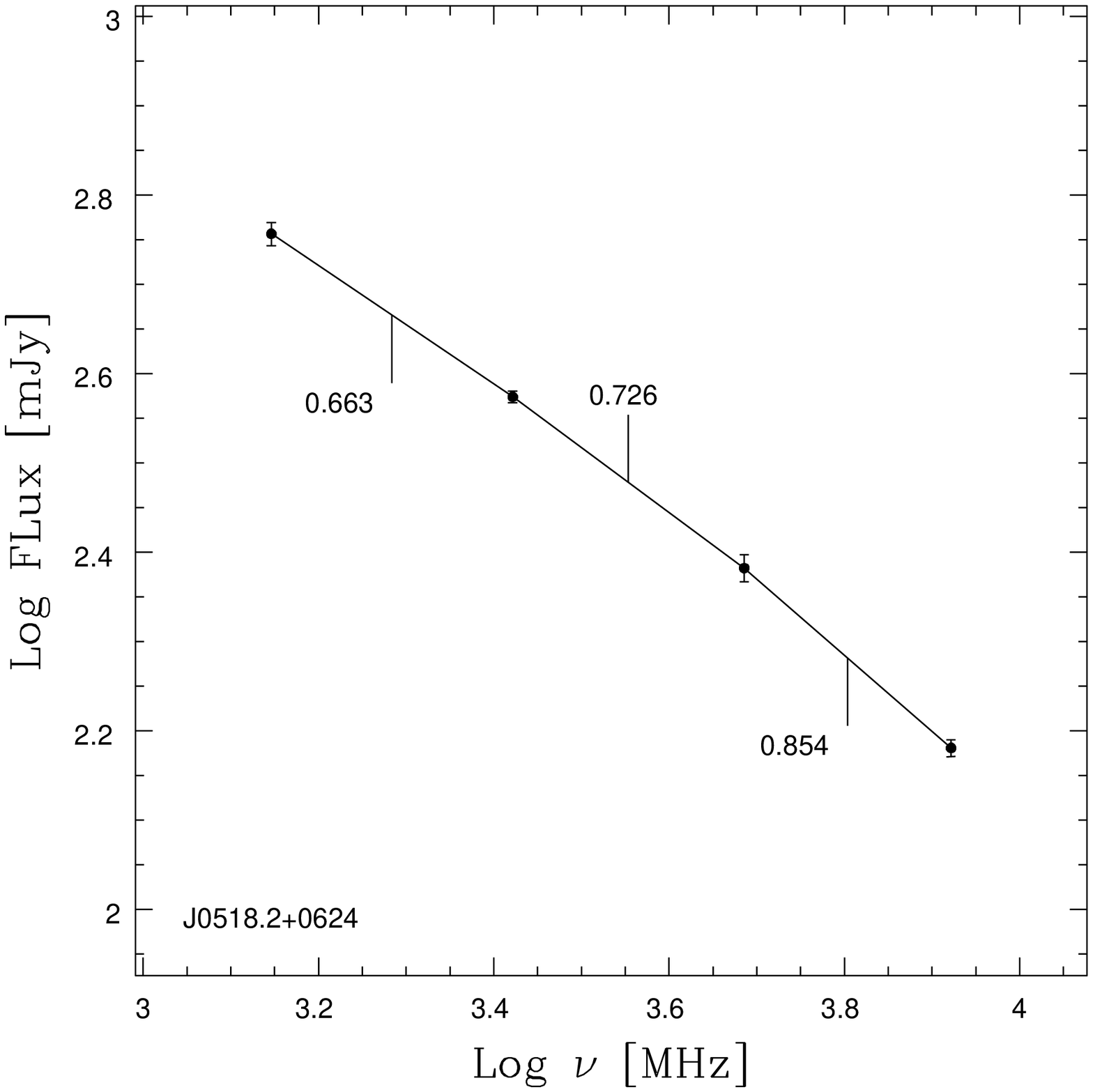}
\includegraphics[width=8cm]{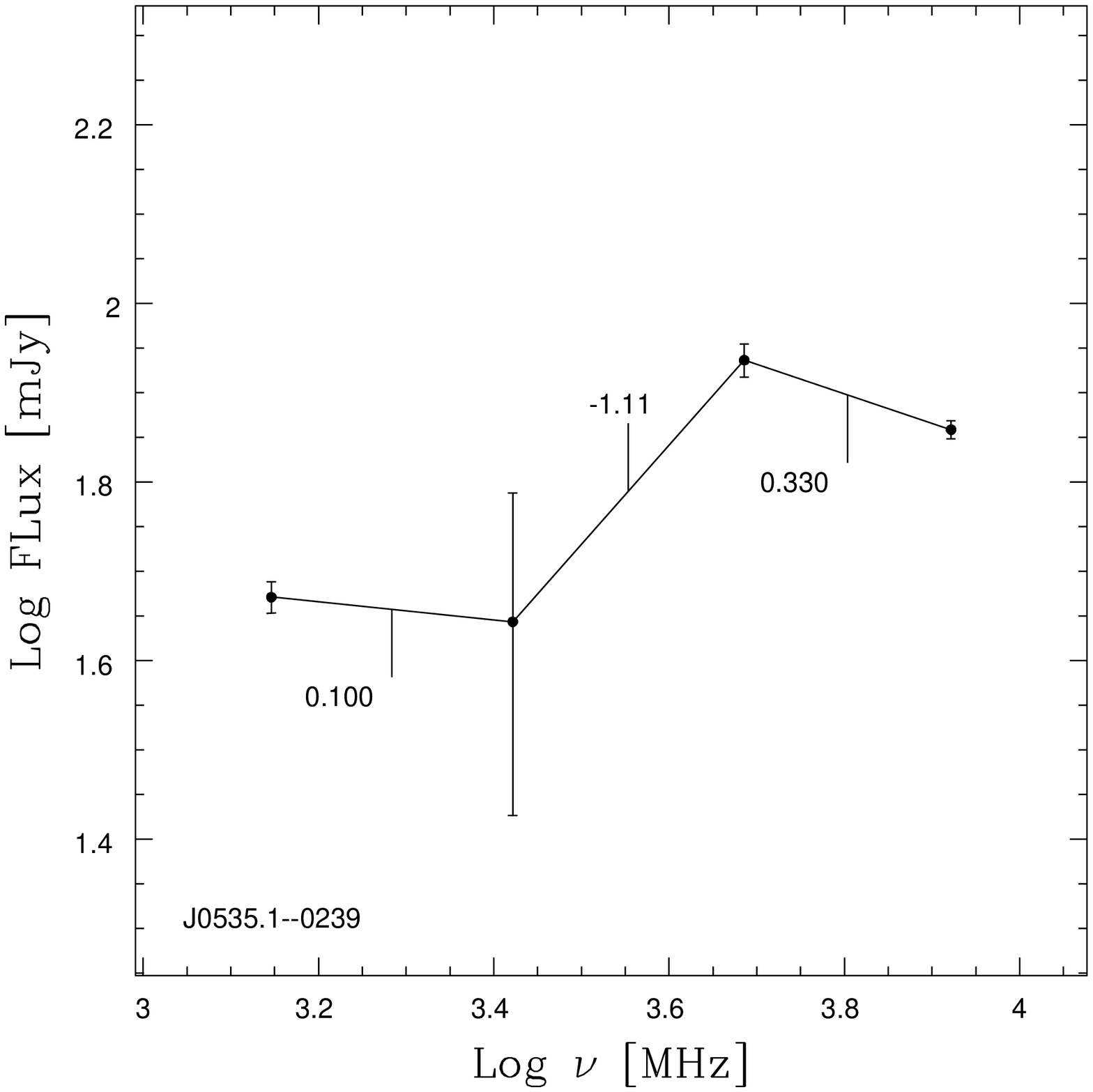}
\includegraphics[width=8cm]{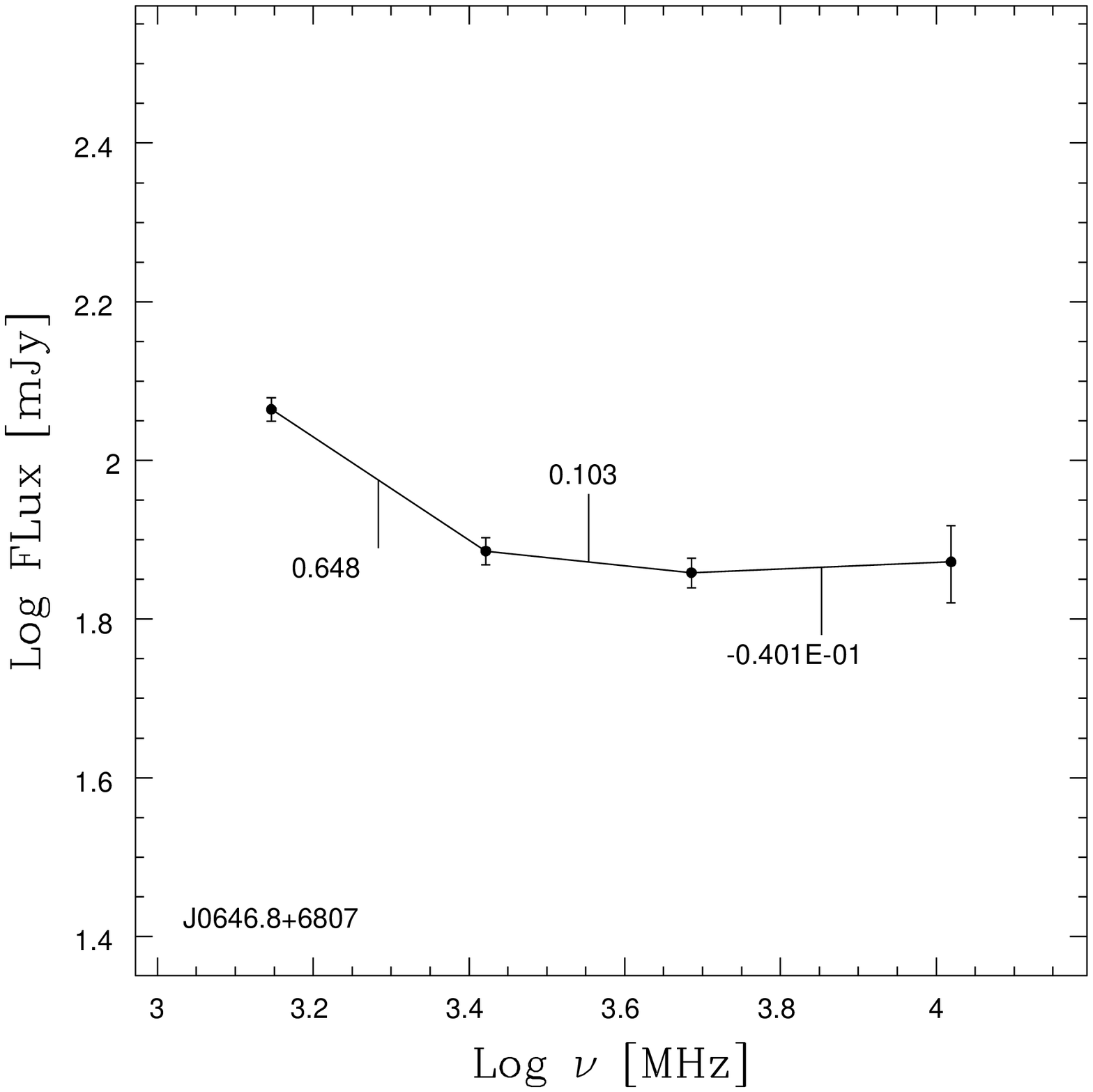}
\includegraphics[width=8cm]{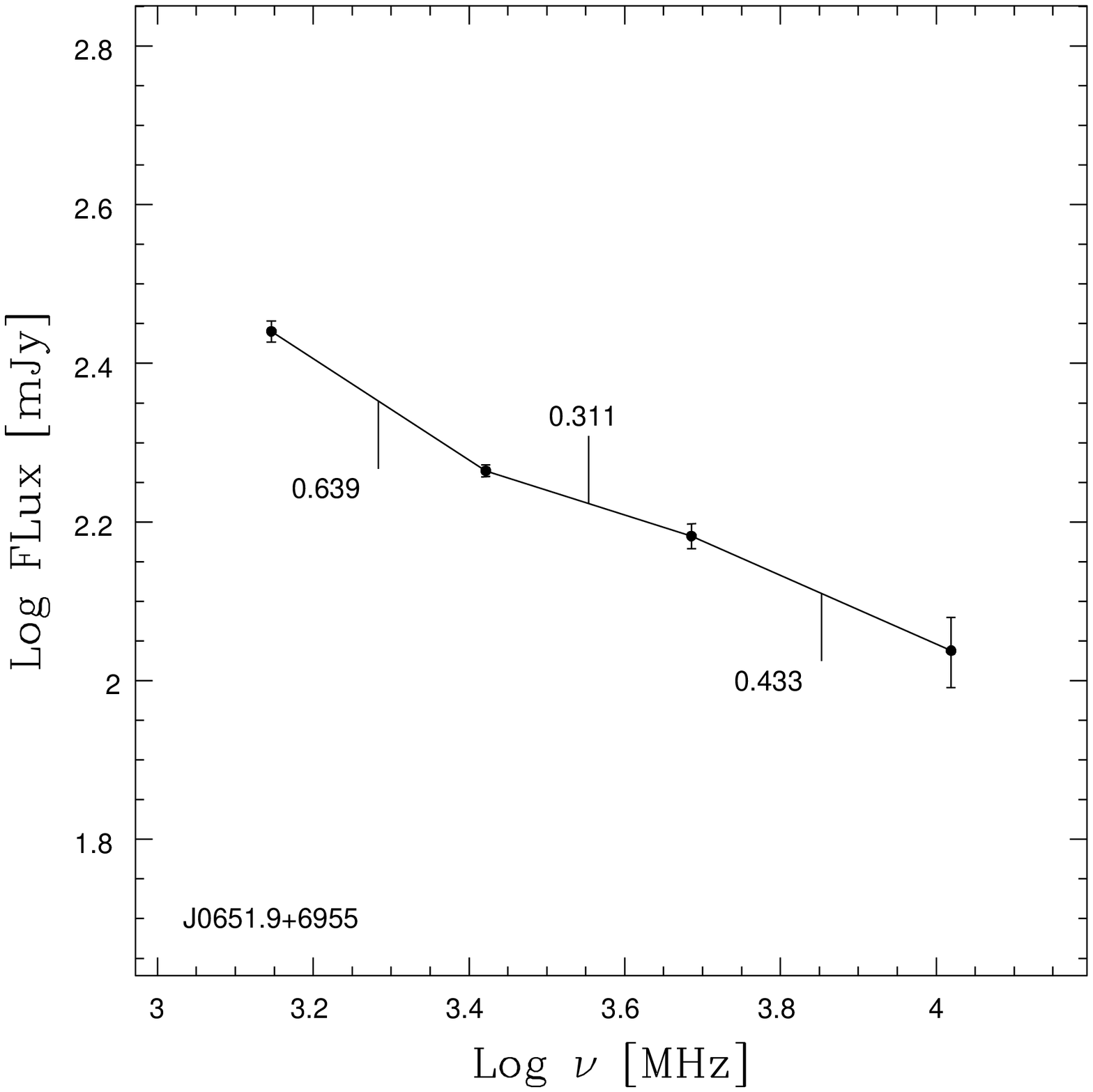}
\includegraphics[width=8cm]{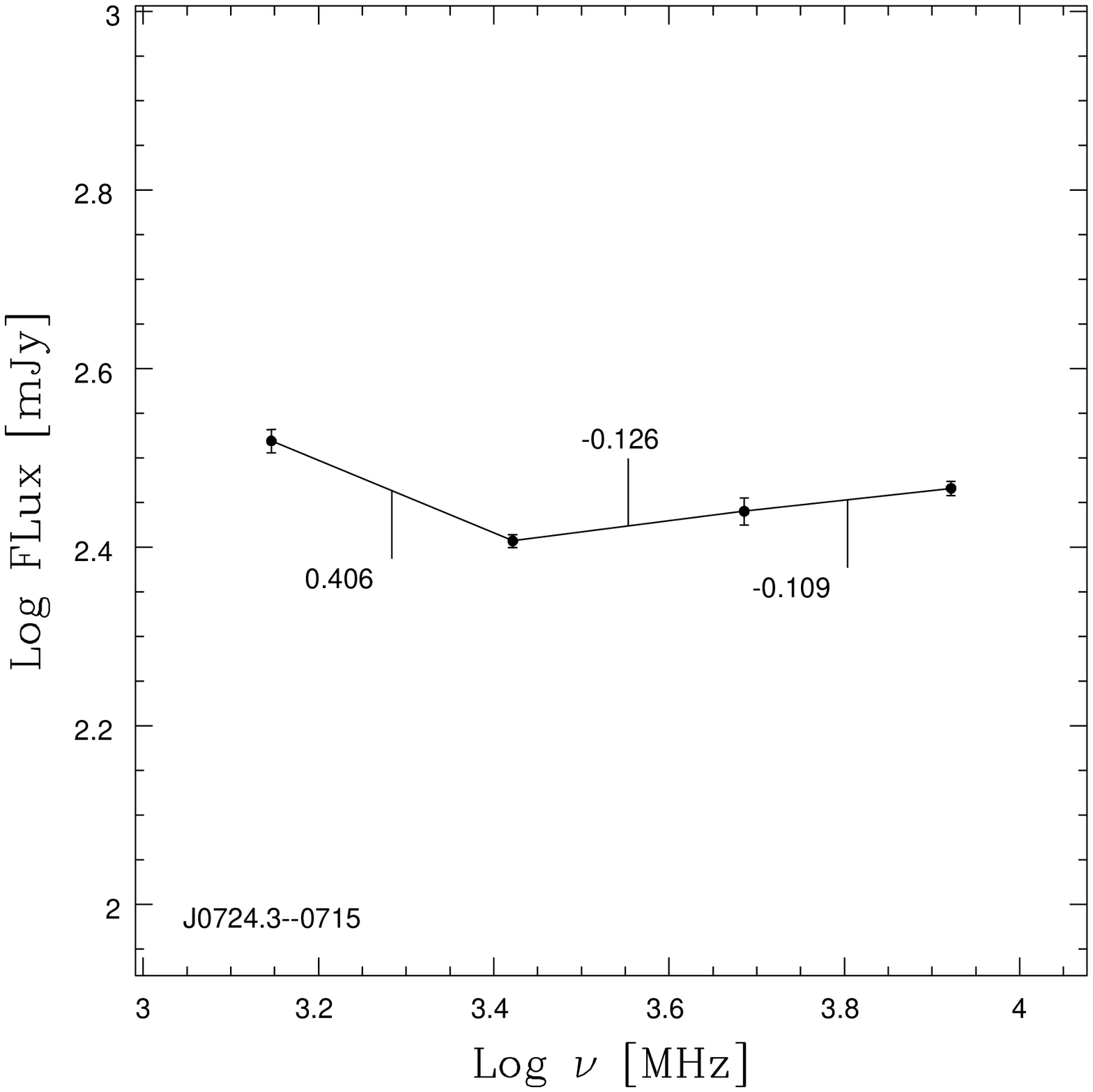}
\includegraphics[width=8cm]{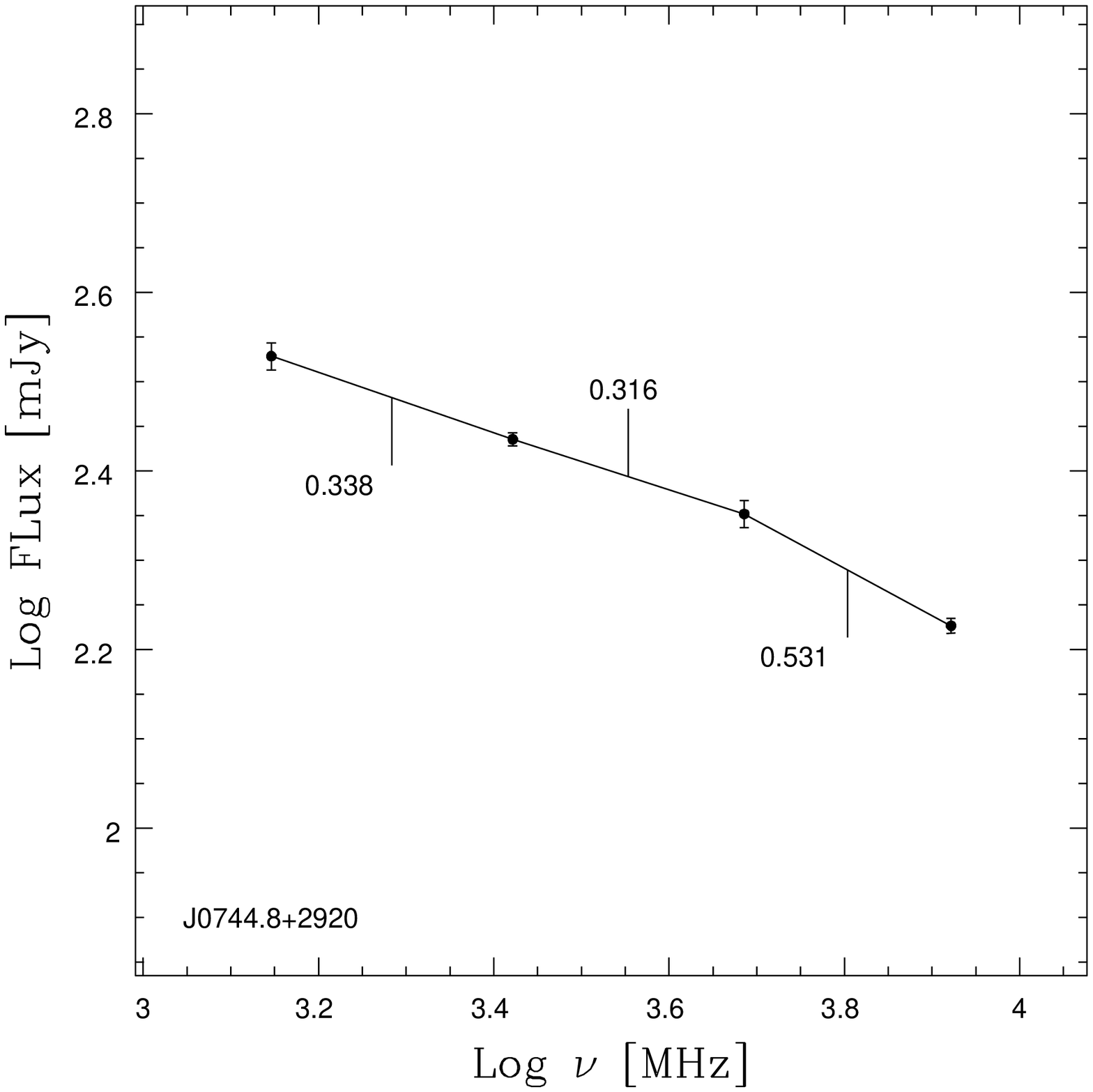}
\caption{Spectral index plots of sources in Table 3.}
\end{figure*}
\newpage
\begin{figure*}[t]
\addtocounter{figure}{+0}
\centering
\includegraphics[width=8cm]{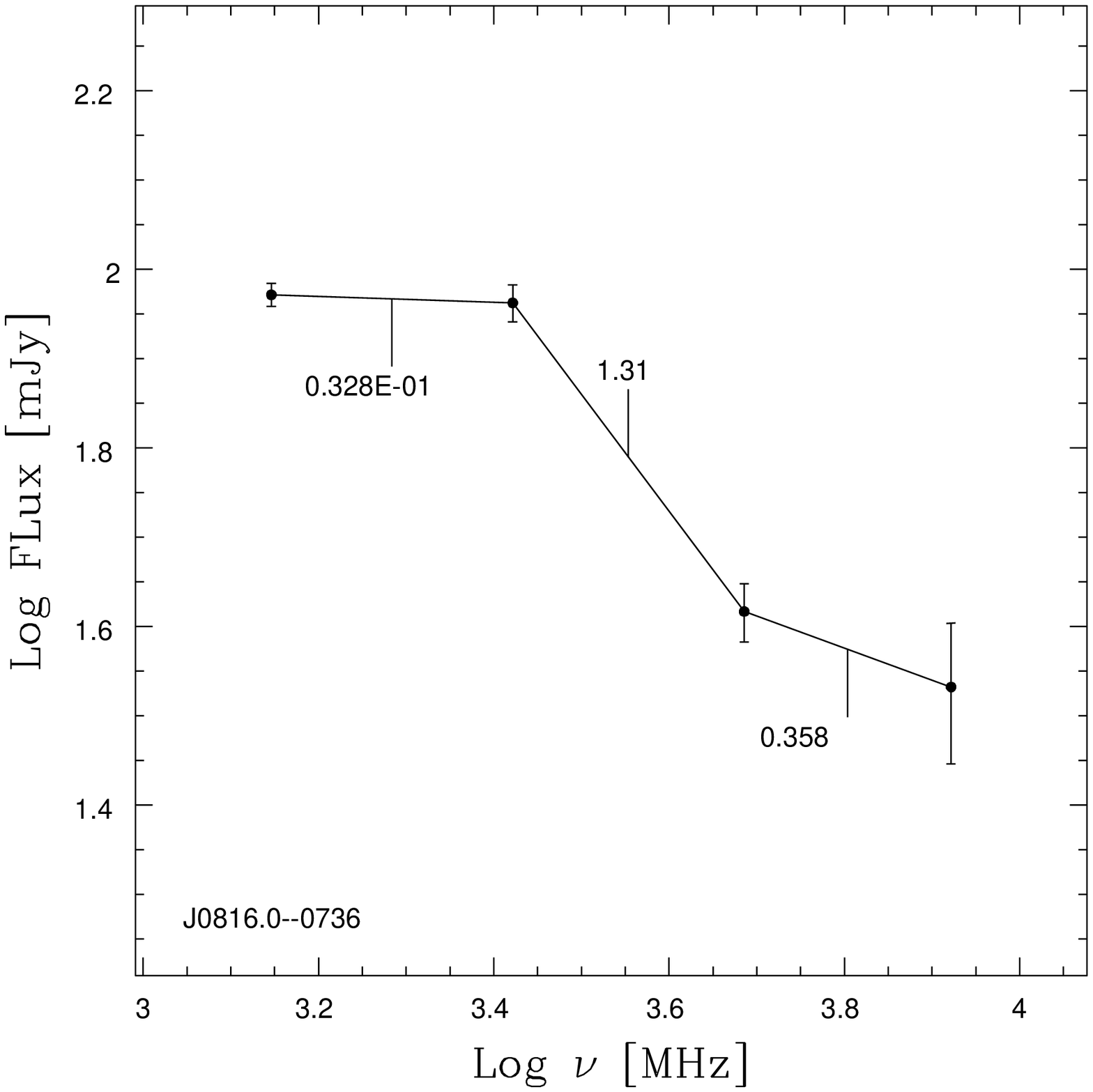}
\includegraphics[width=8cm]{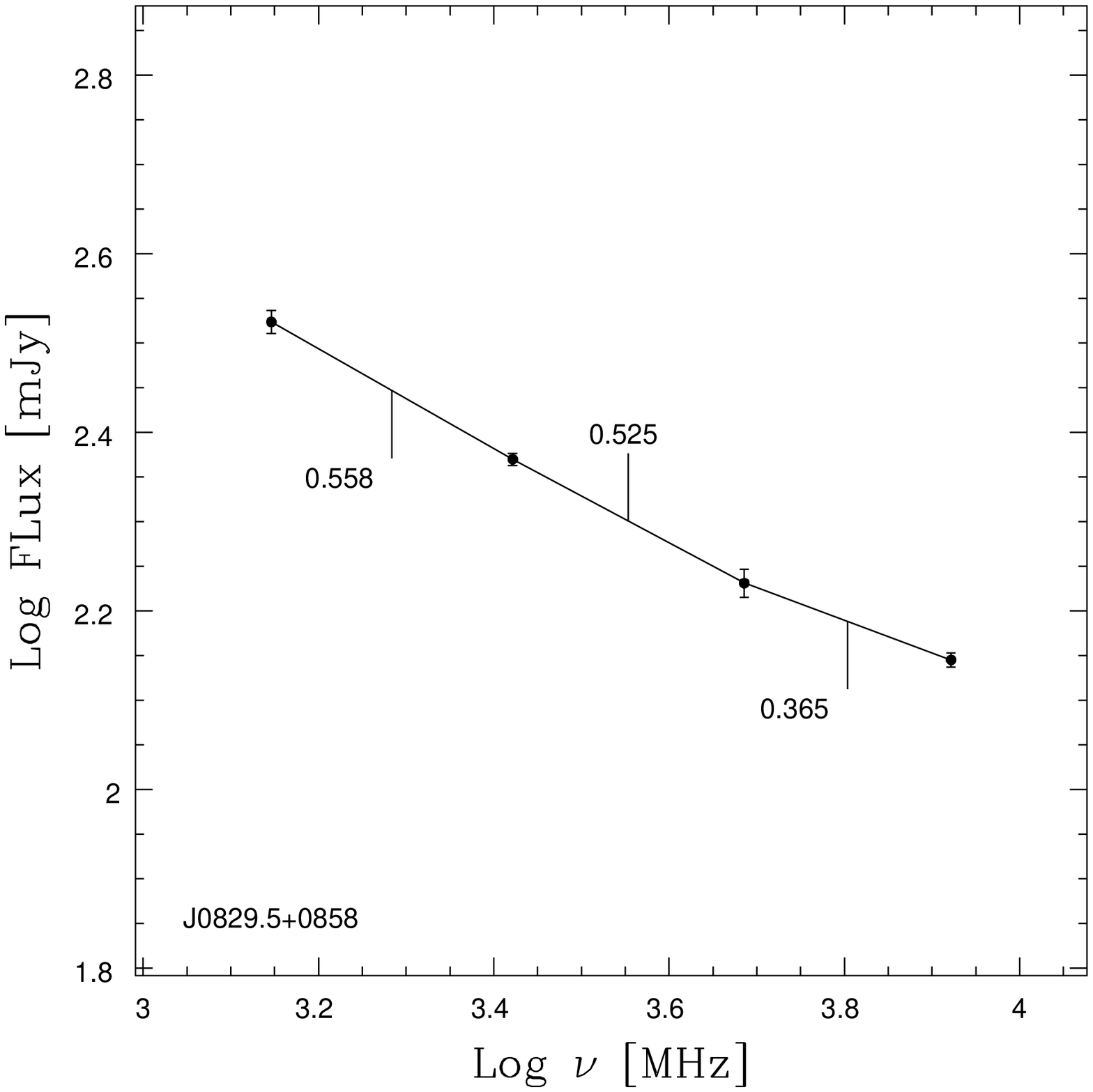}
\includegraphics[width=8cm]{j0847.2_e.ps}
\includegraphics[width=8cm]{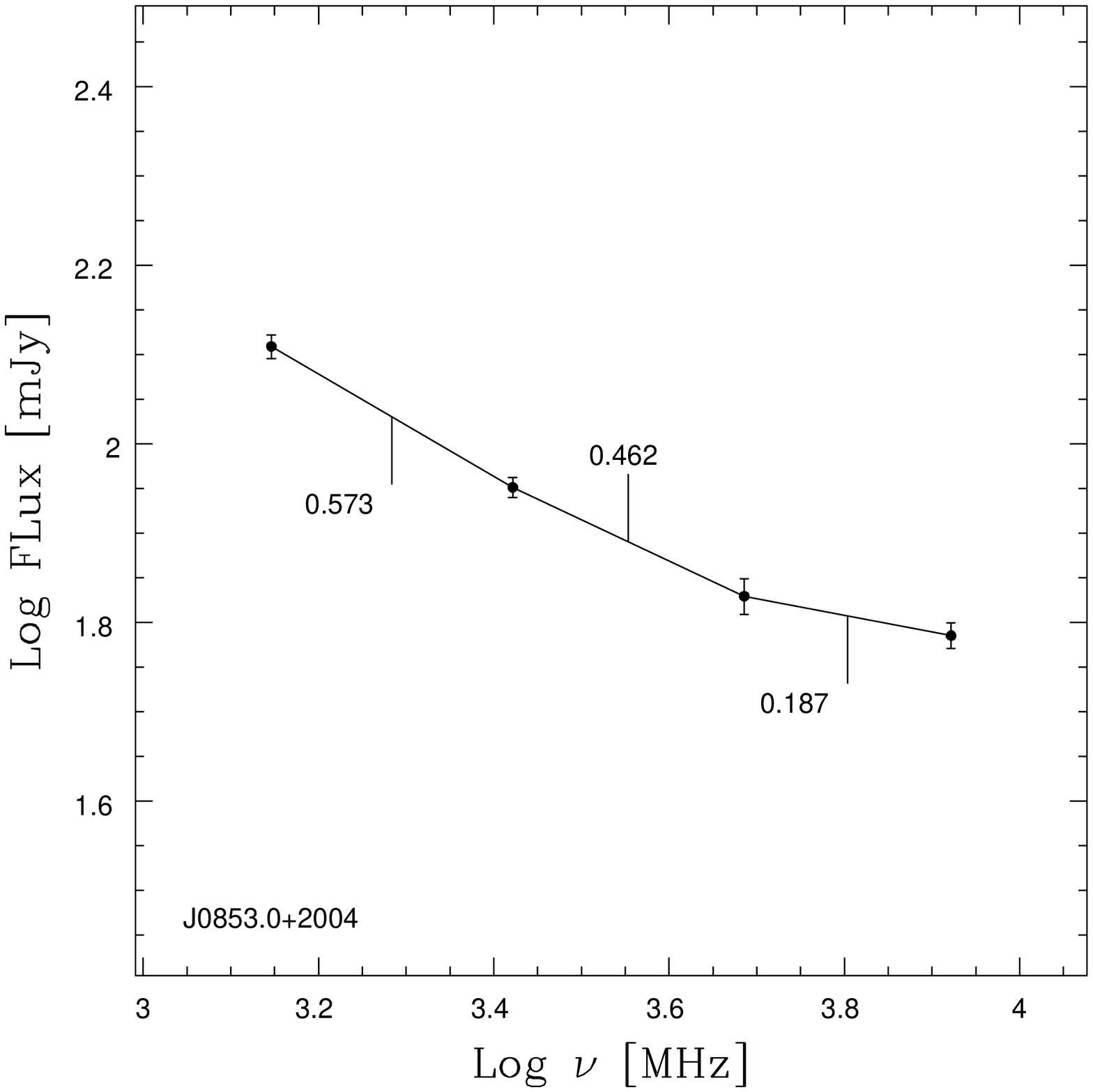}
\includegraphics[width=8cm]{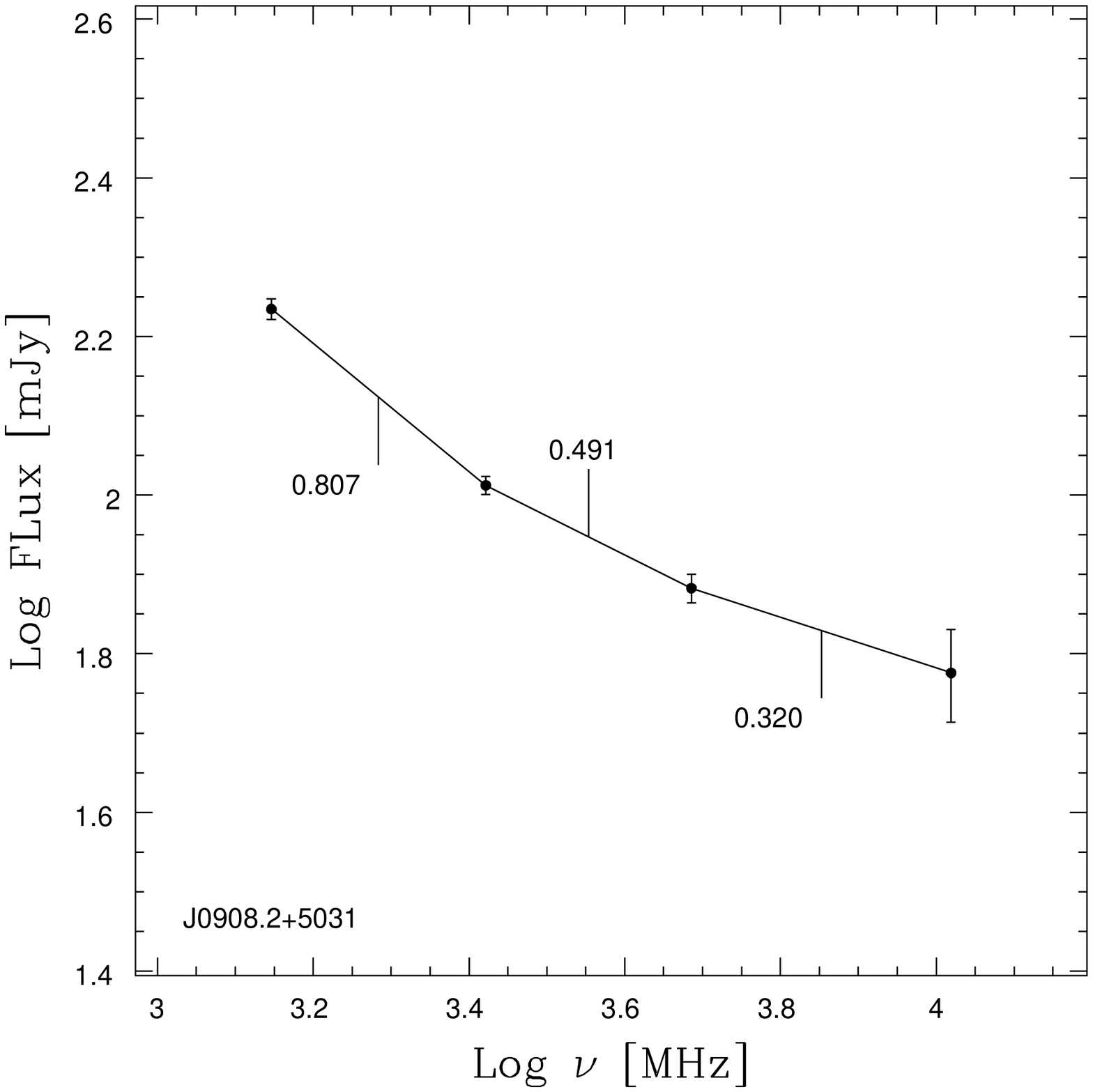}
\includegraphics[width=8cm]{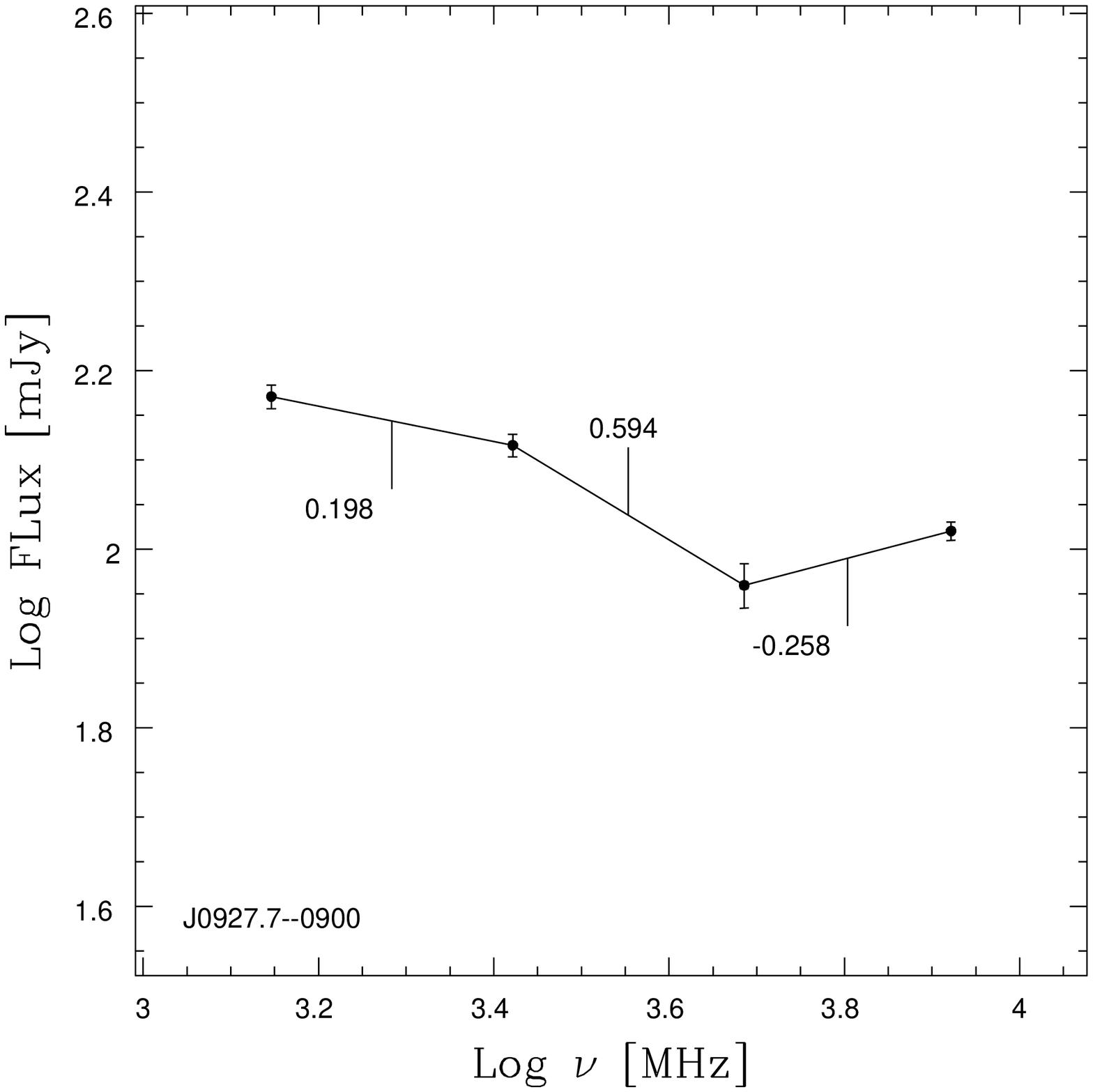}
\caption{Spectral index plots of sources in Table 3.}
\end{figure*}
\newpage
\begin{figure*}[t]
\addtocounter{figure}{+0}
\centering
\includegraphics[width=8cm]{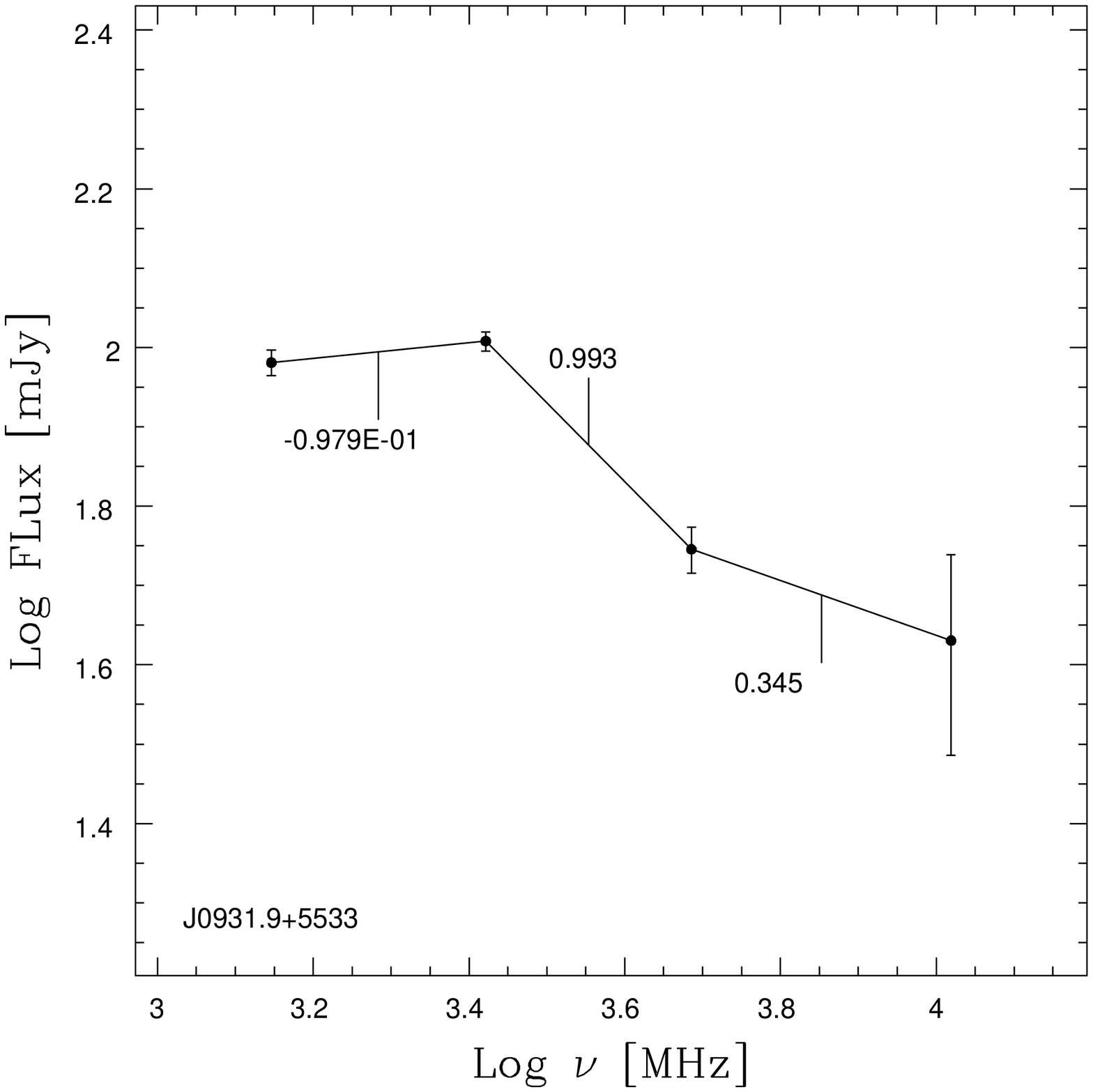}
\includegraphics[width=8cm]{j0937.1_e.ps}
\includegraphics[width=8cm]{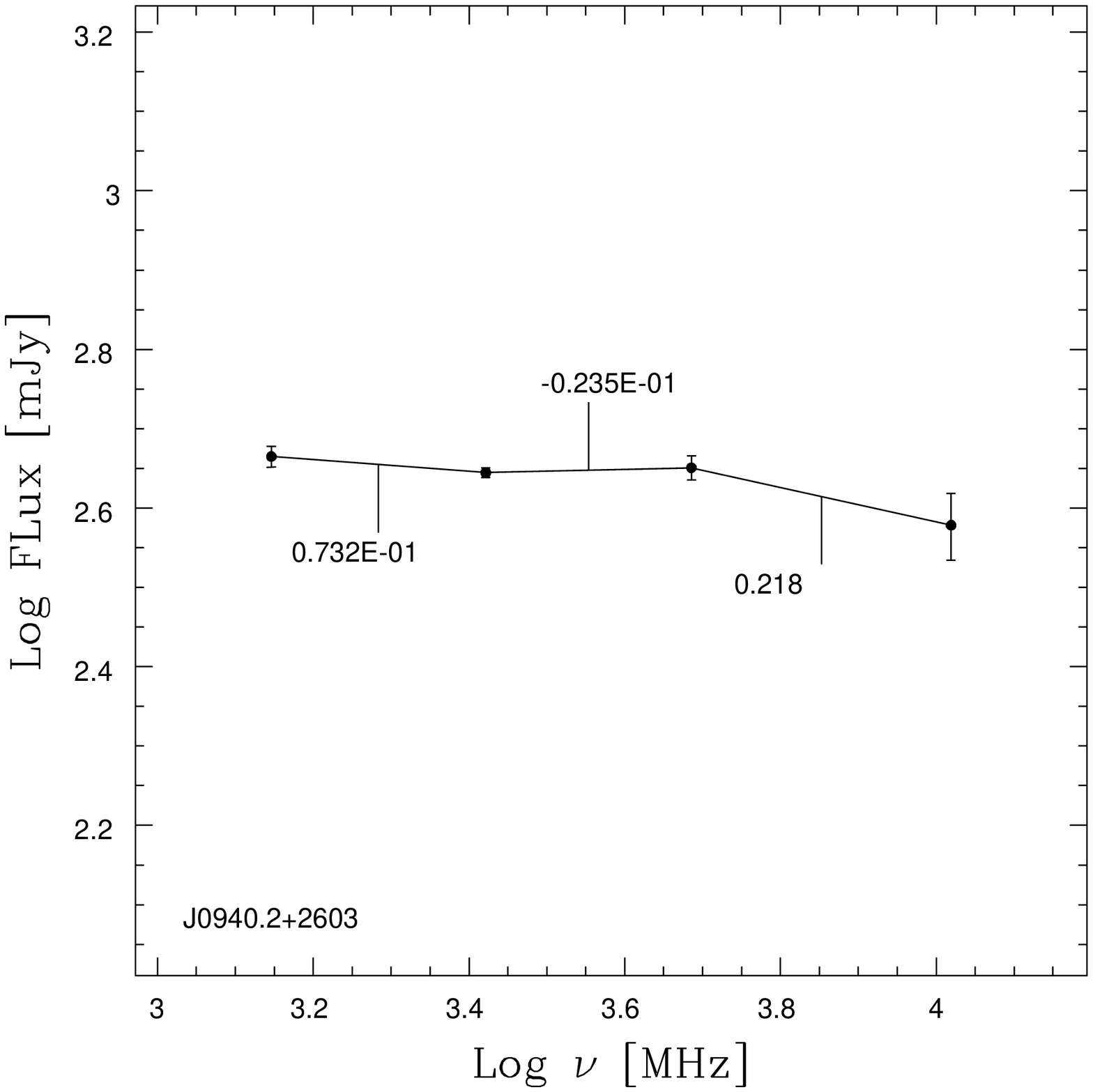}
\includegraphics[width=8cm]{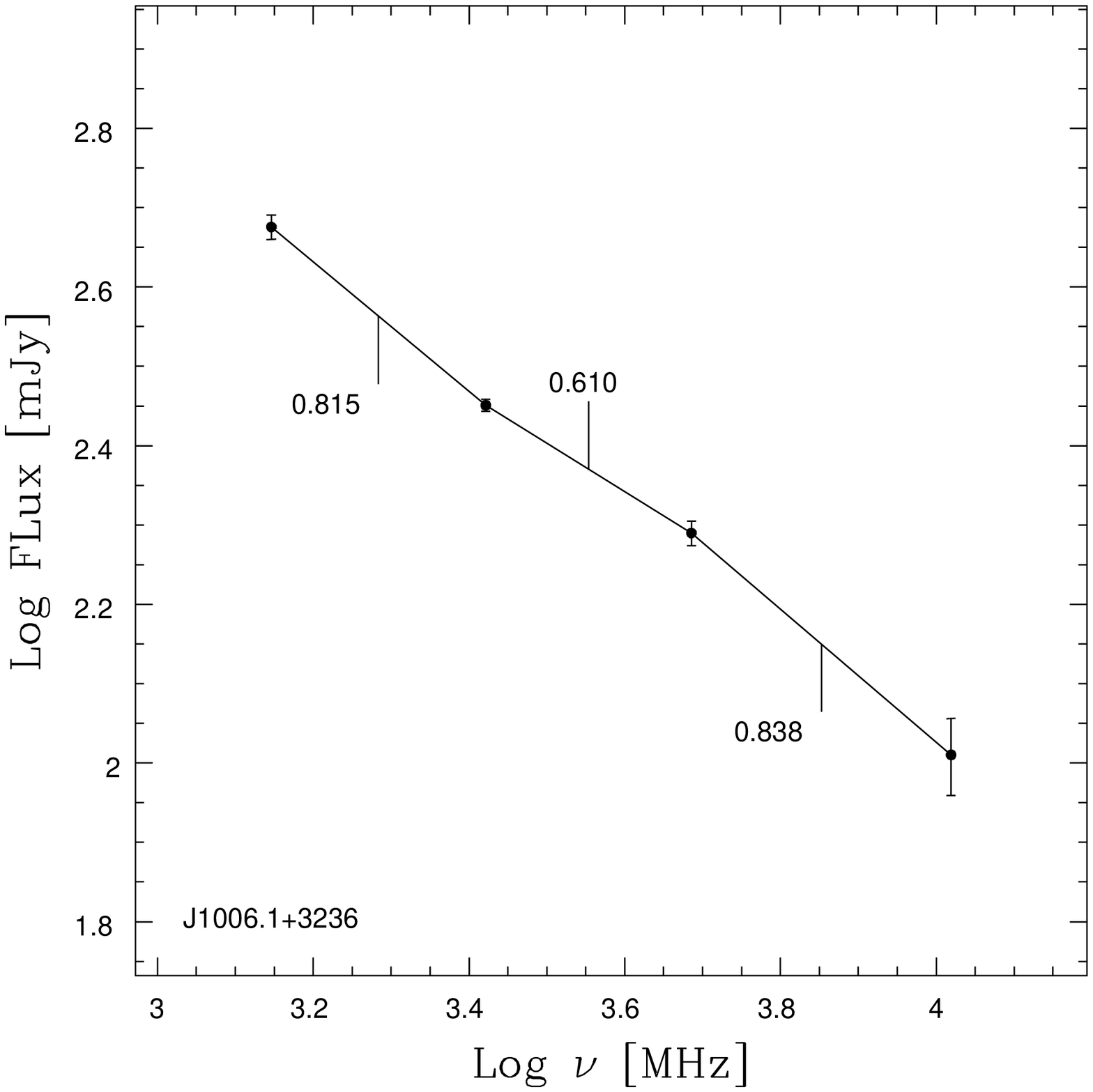}
\includegraphics[width=8cm]{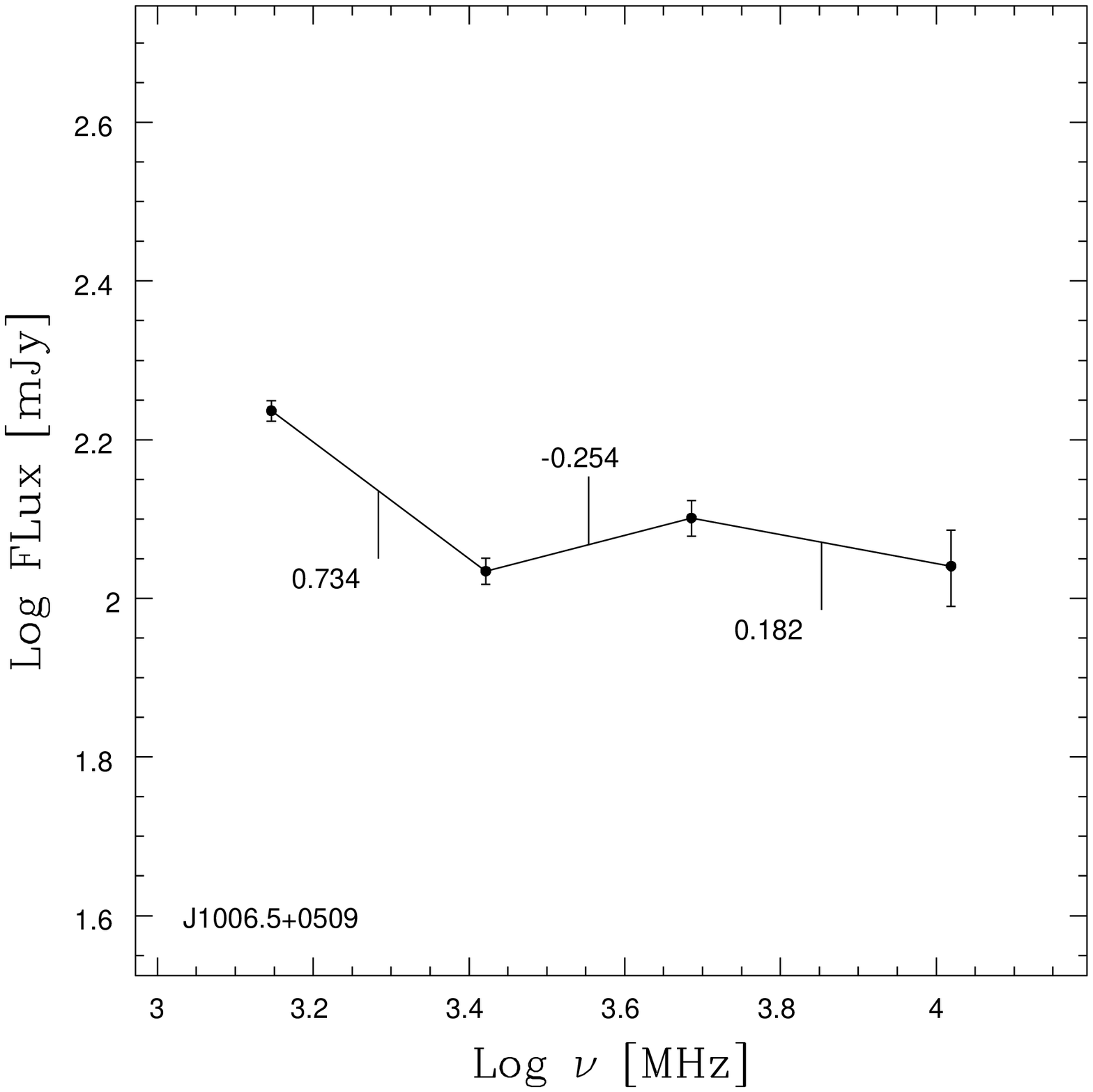}
\includegraphics[width=8cm]{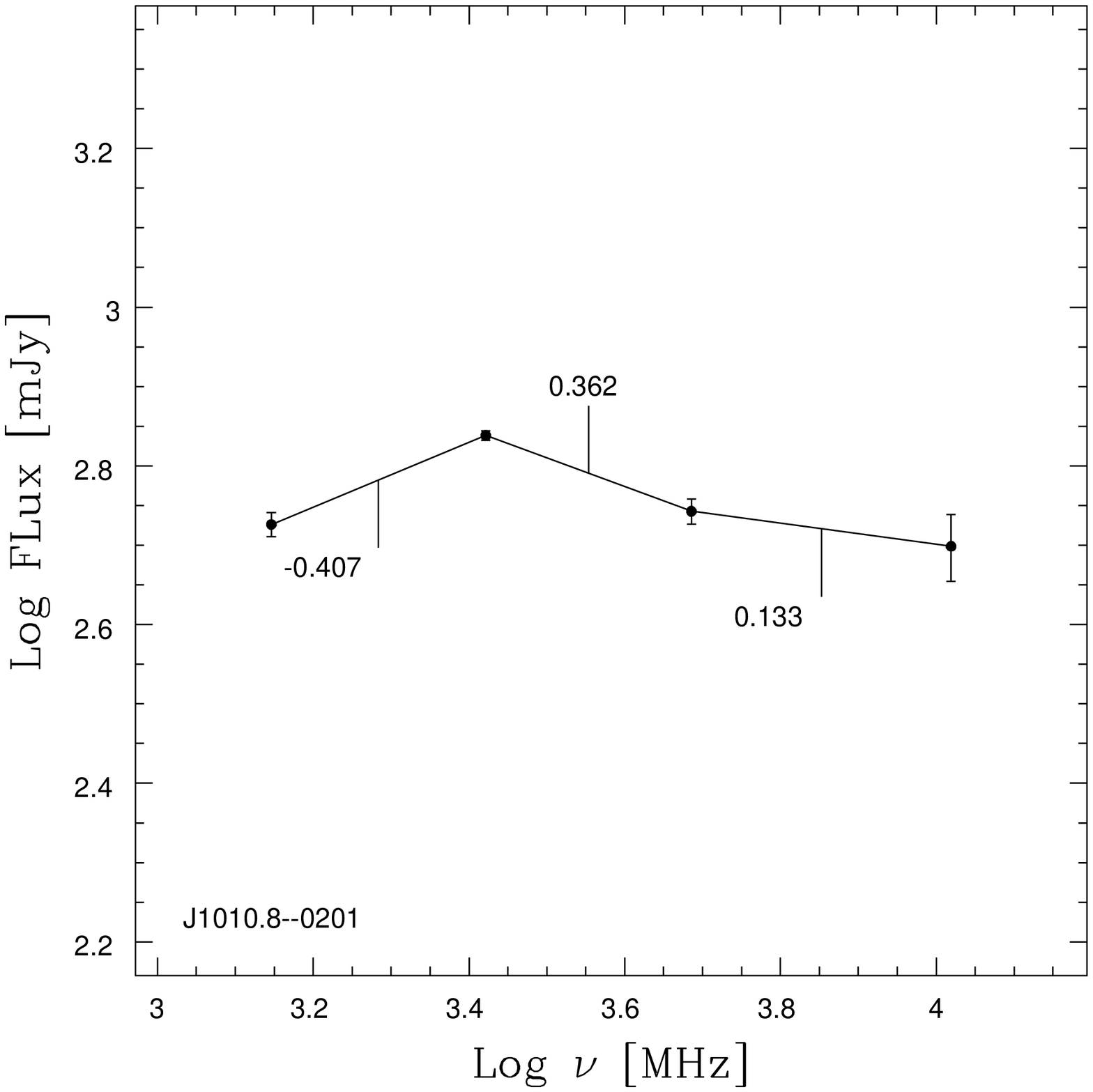}
\caption{Spectral index plots of sources in Table 3.}
\end{figure*}
\newpage
\begin{figure*}[t]
\addtocounter{figure}{+0}
\centering
\includegraphics[width=8cm]{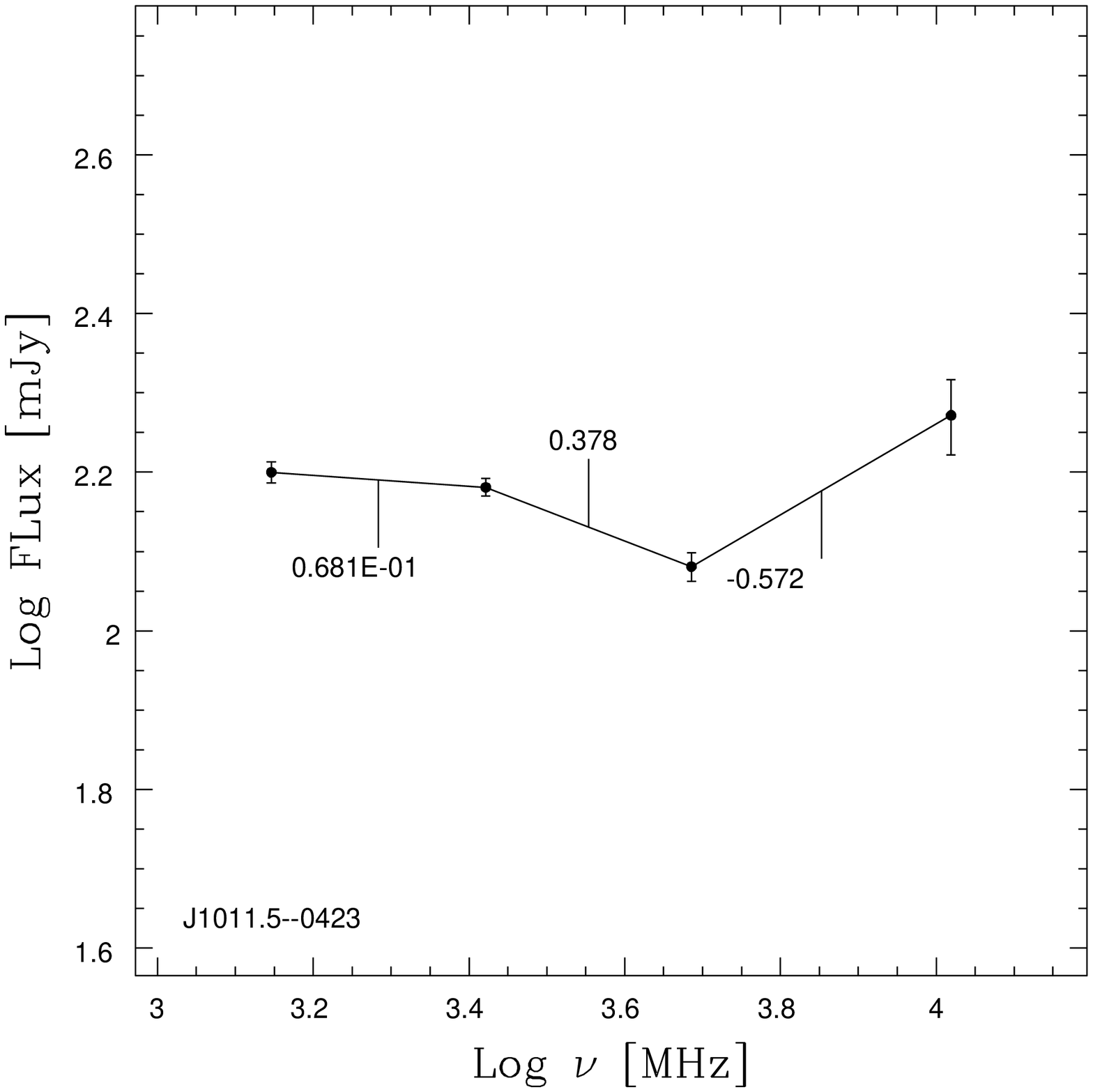}
\includegraphics[width=8cm]{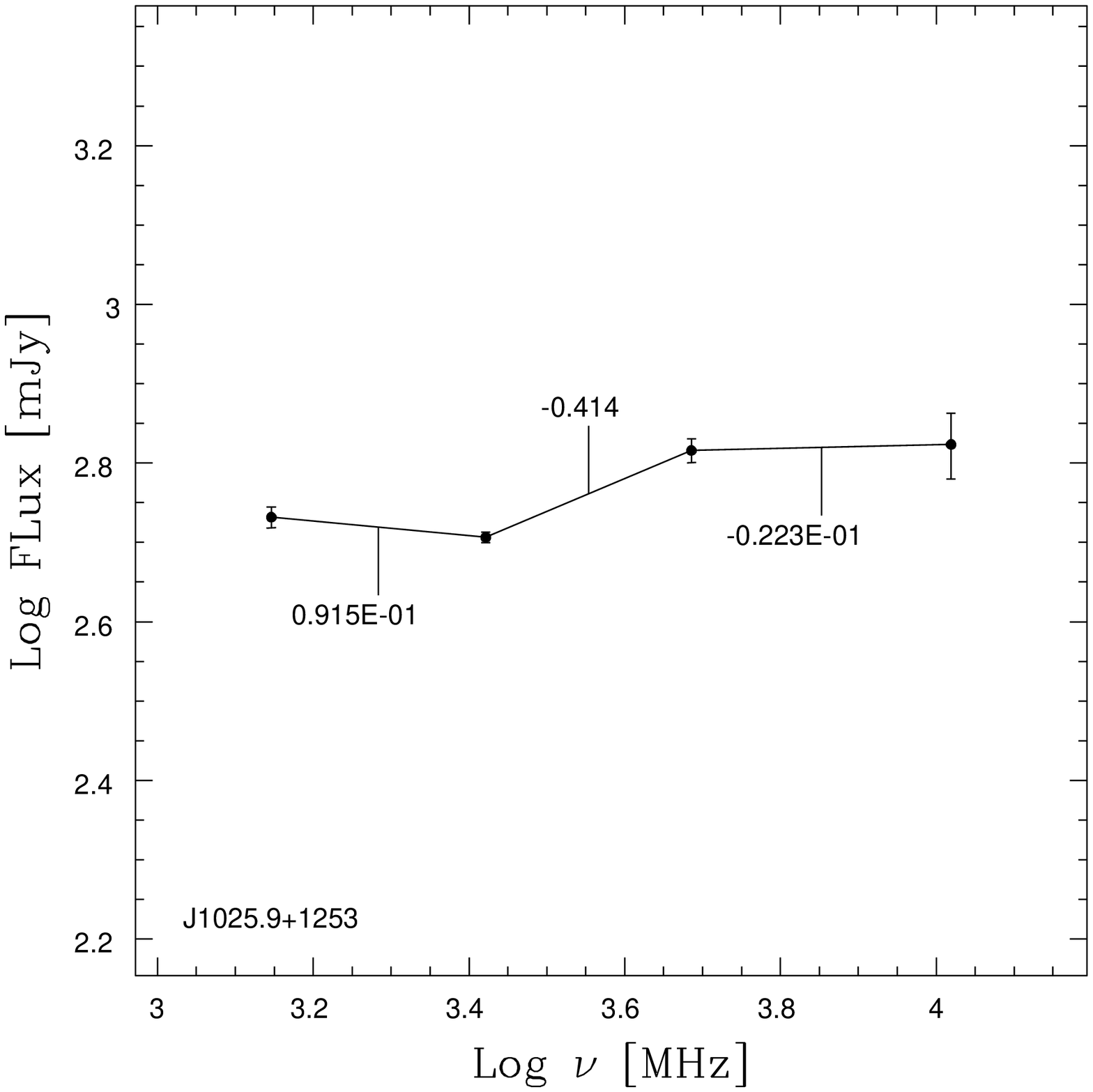}
\includegraphics[width=8cm]{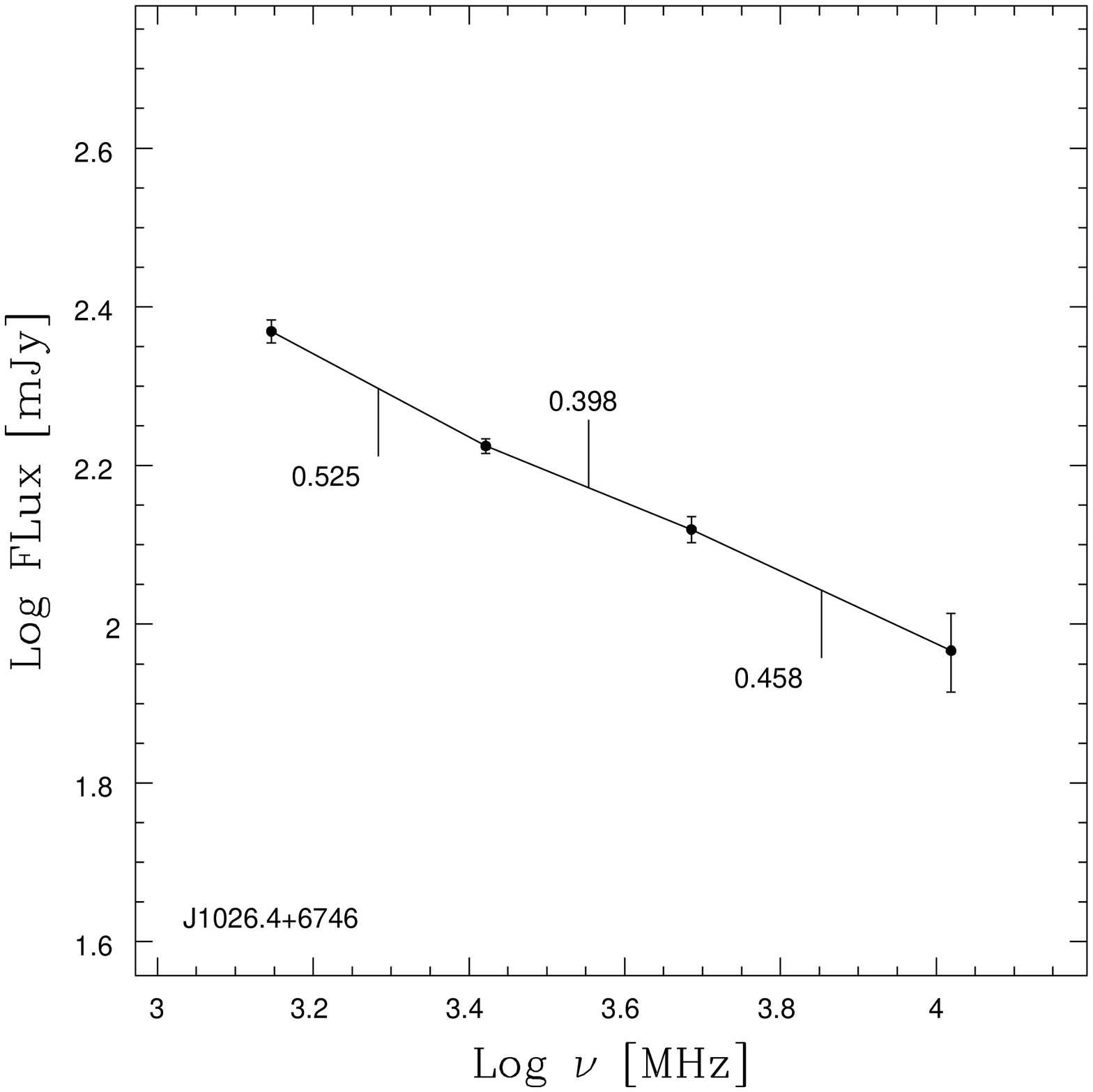}
\includegraphics[width=8cm]{j1028.5_e.ps}
\includegraphics[width=8cm]{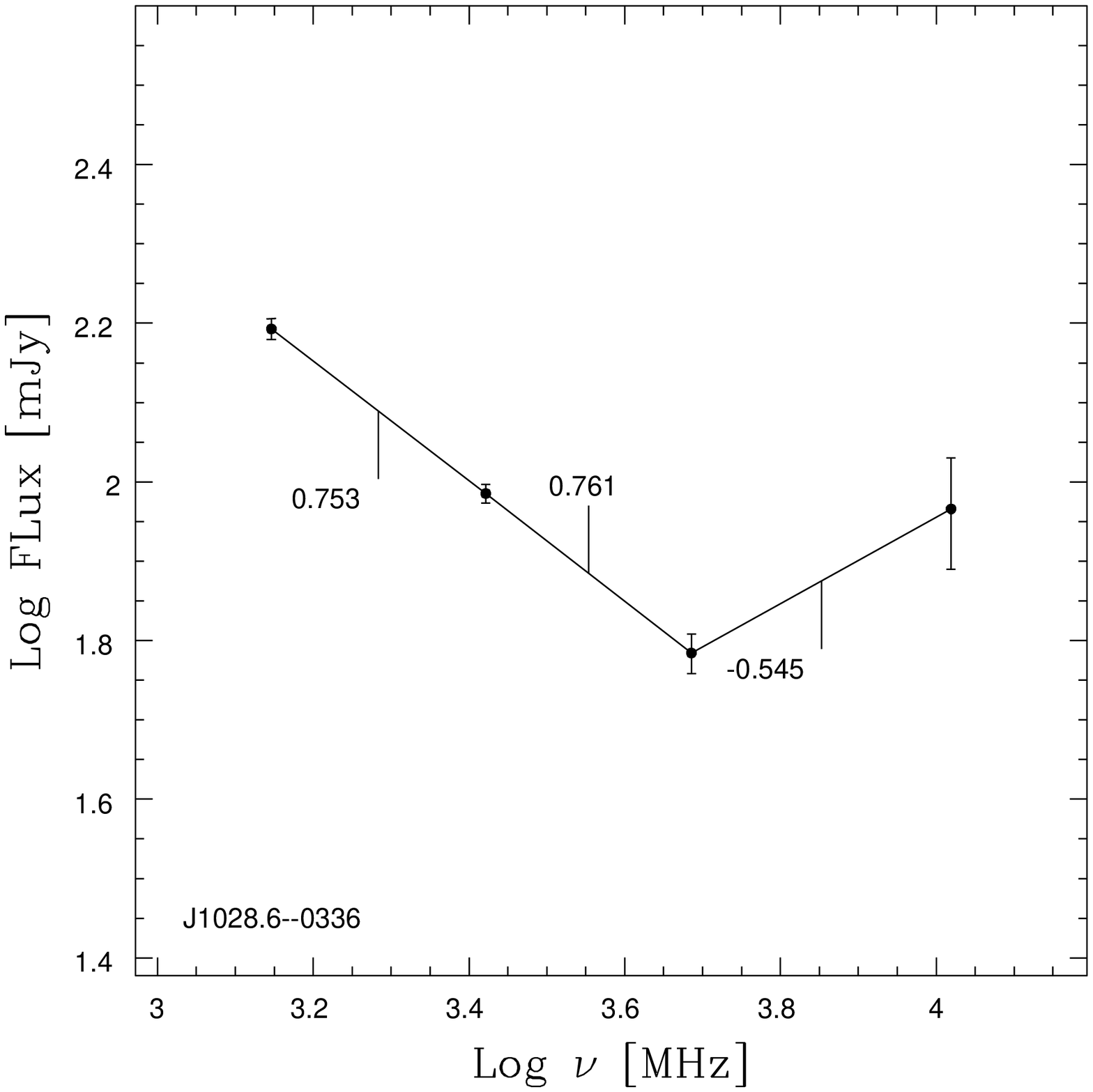}
\includegraphics[width=8cm]{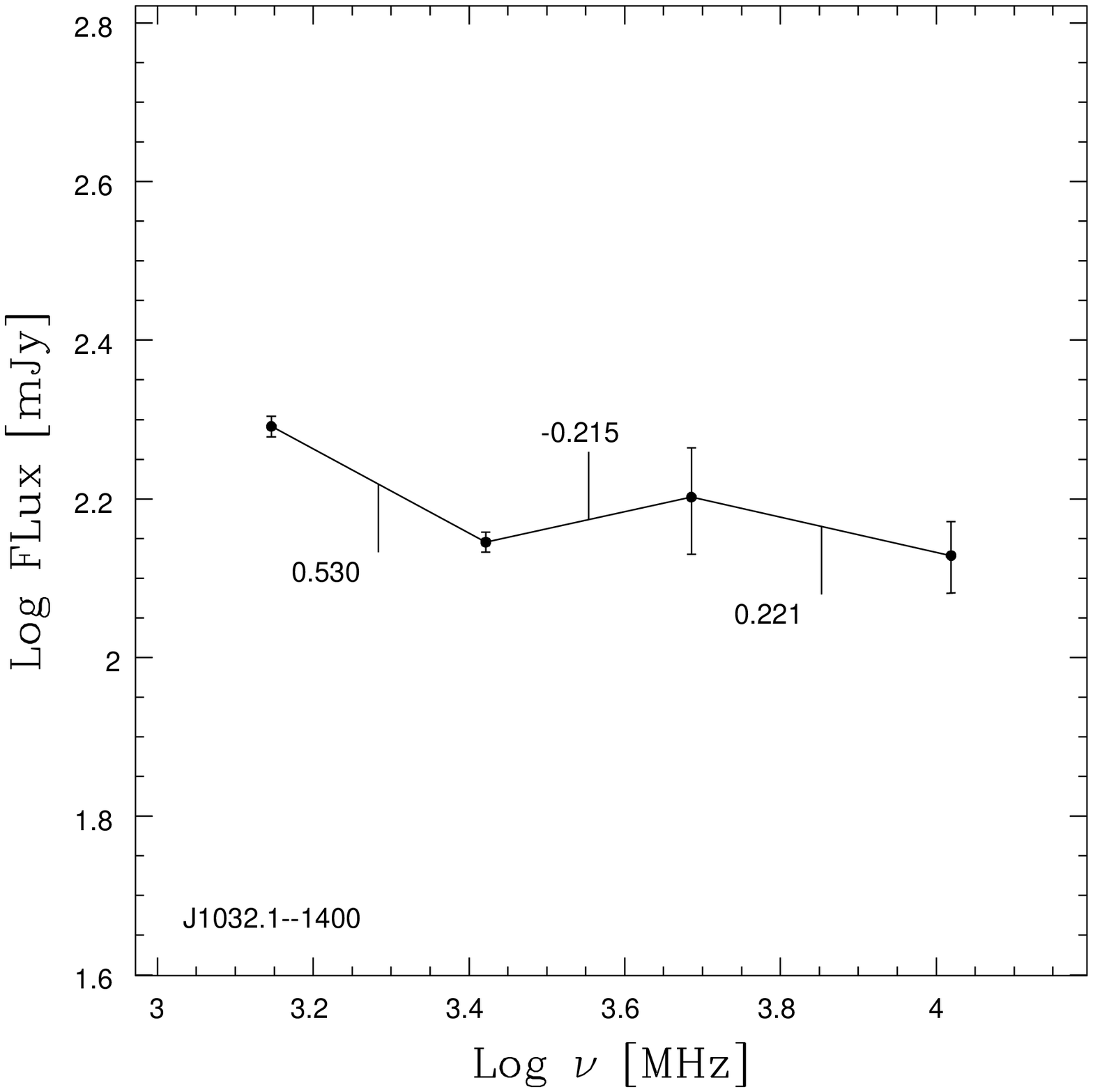}
\caption{Spectral index plots of sources in Table 3.}
\end{figure*}
\newpage
\begin{figure*}[t]
\addtocounter{figure}{+0}
\centering
\includegraphics[width=8cm]{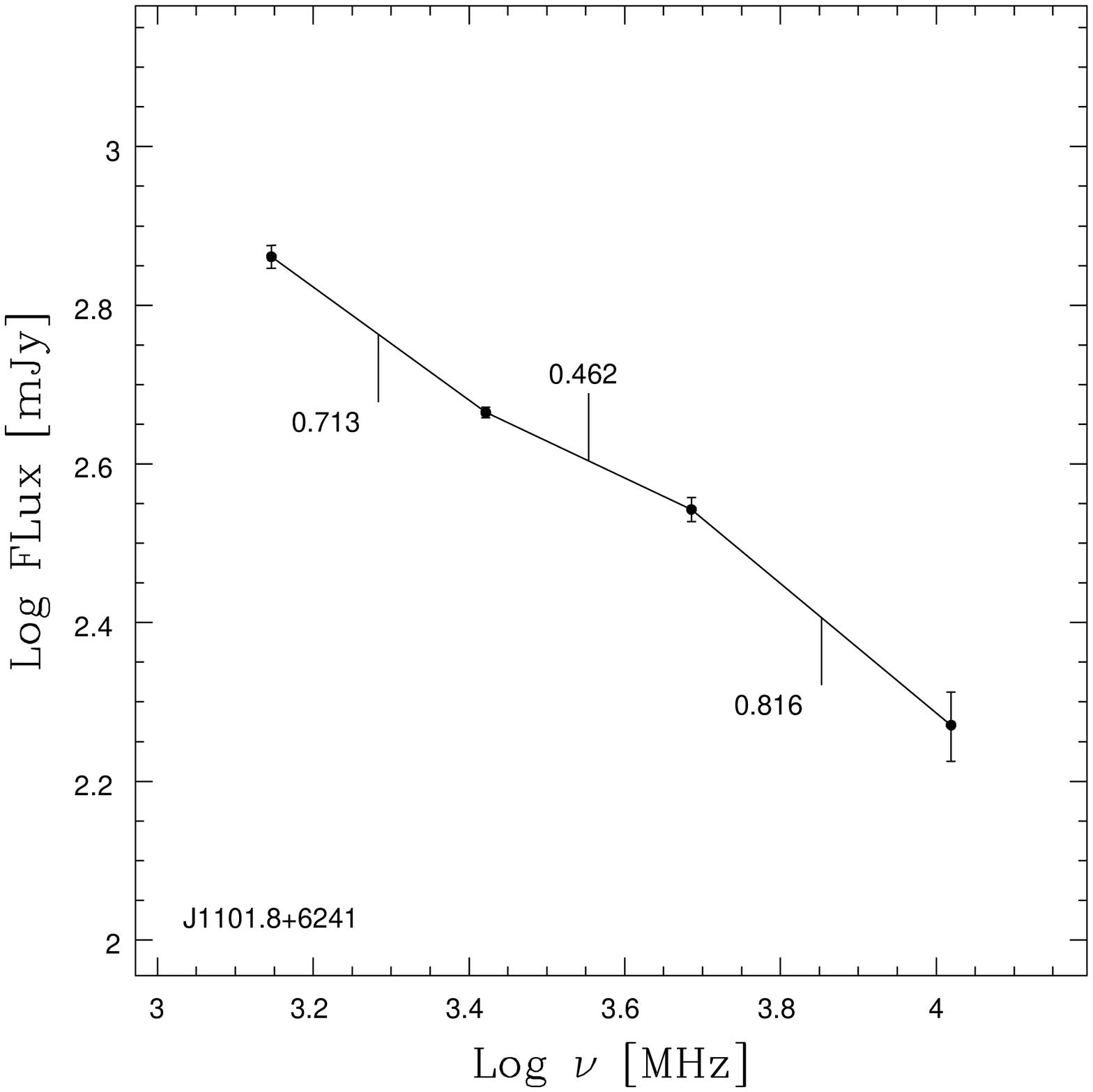}
\includegraphics[width=8cm]{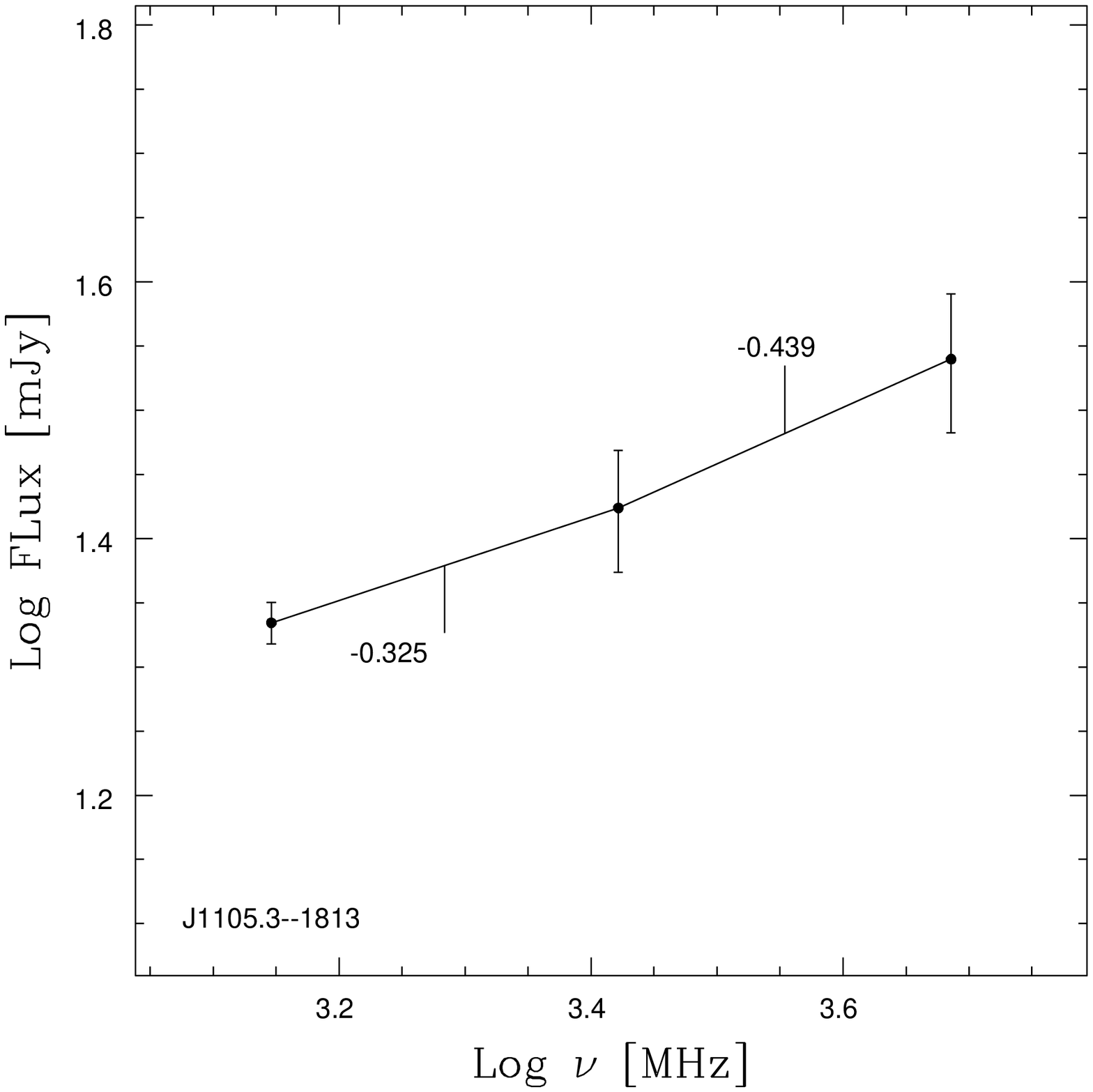}
\includegraphics[width=8cm]{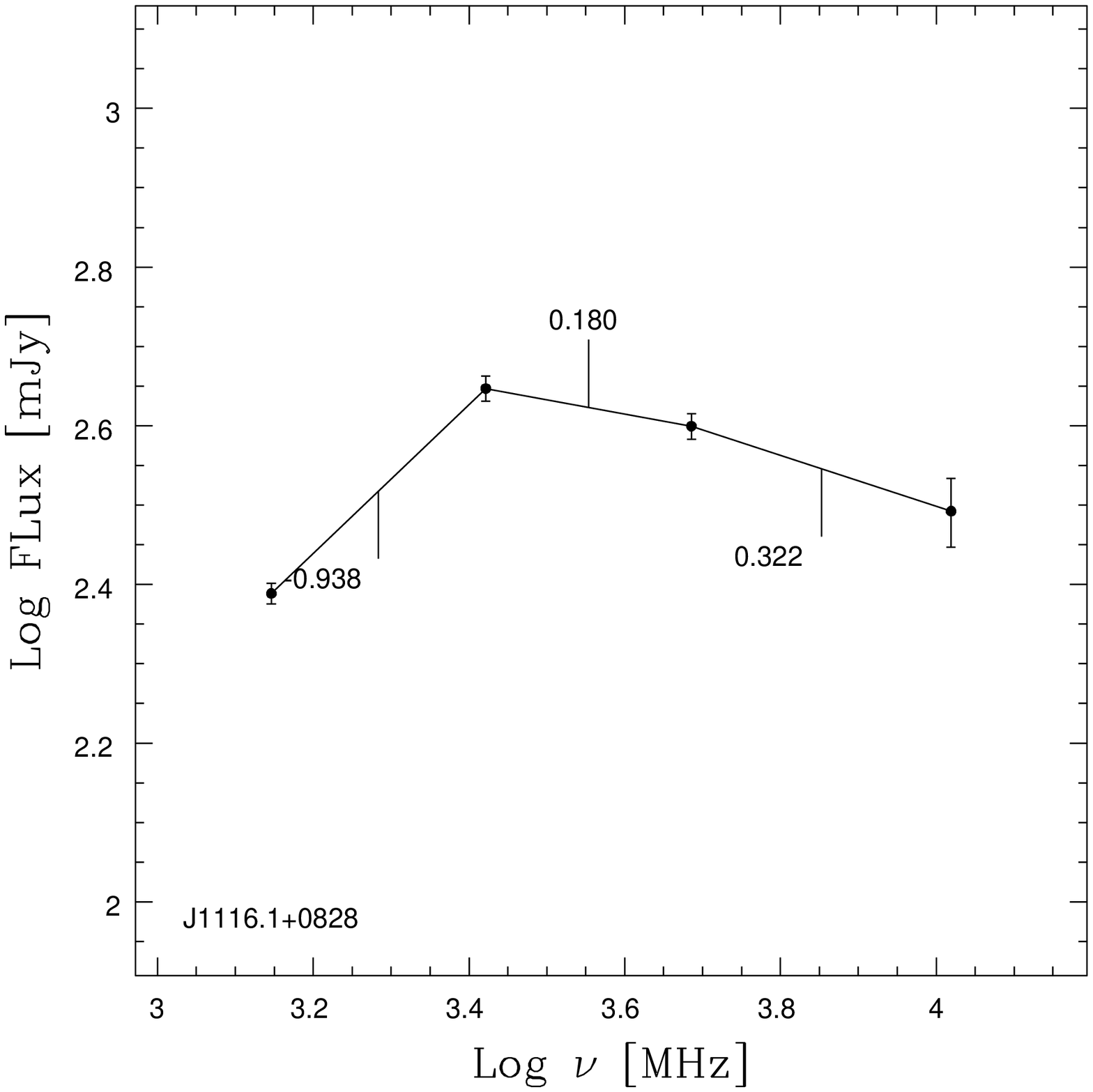}
\includegraphics[width=8cm]{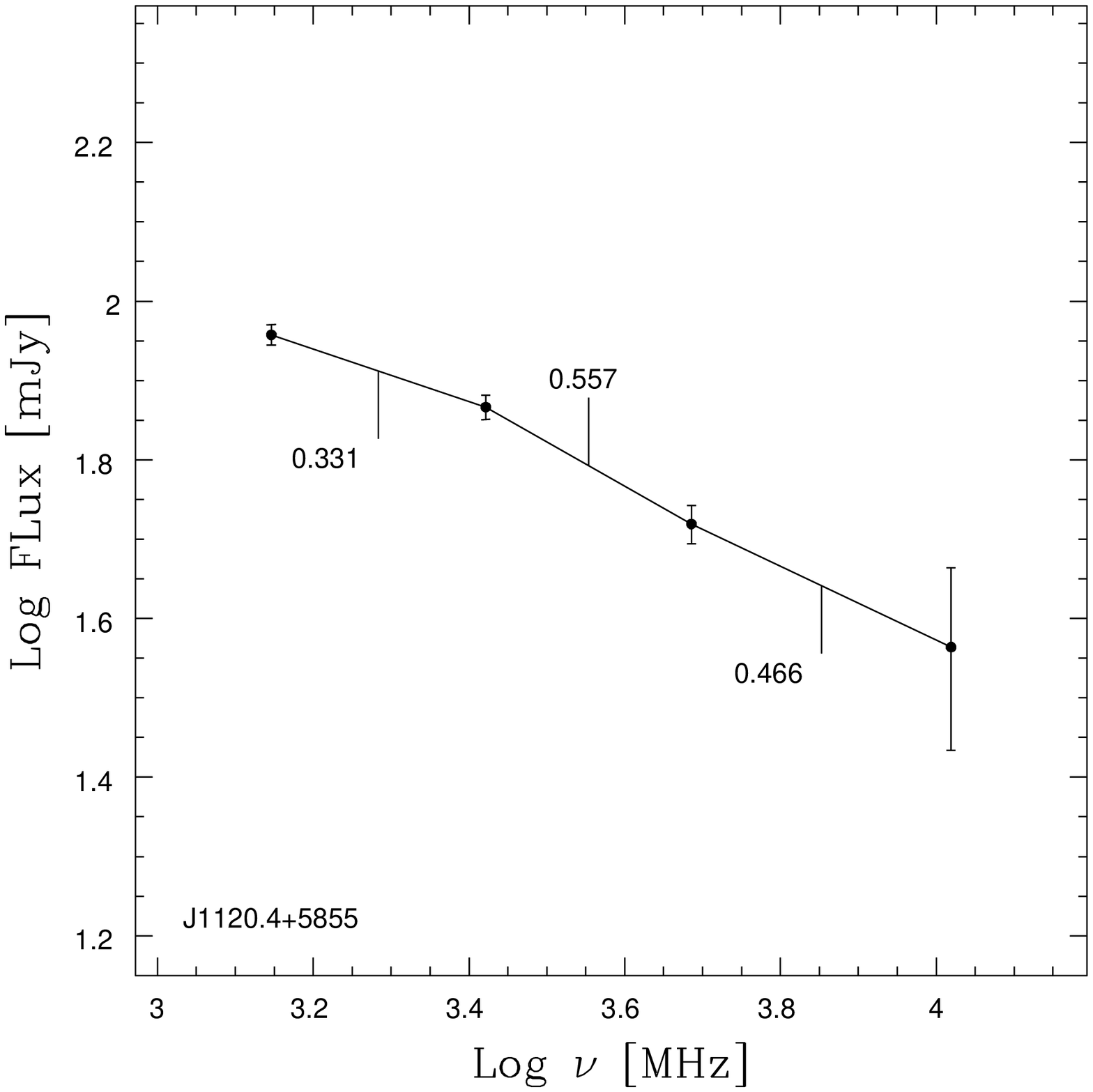}
\includegraphics[width=8cm]{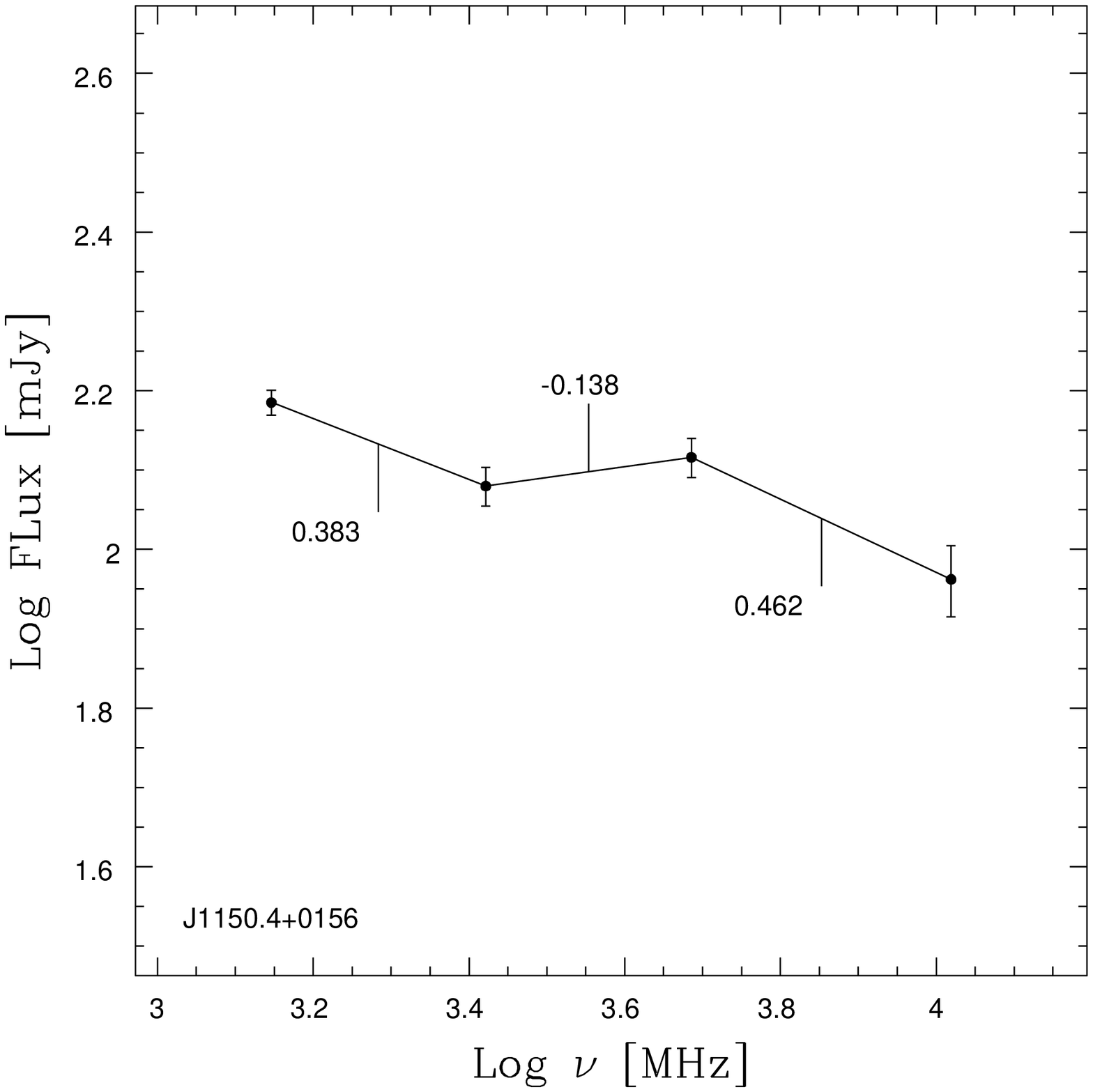}
\includegraphics[width=8cm]{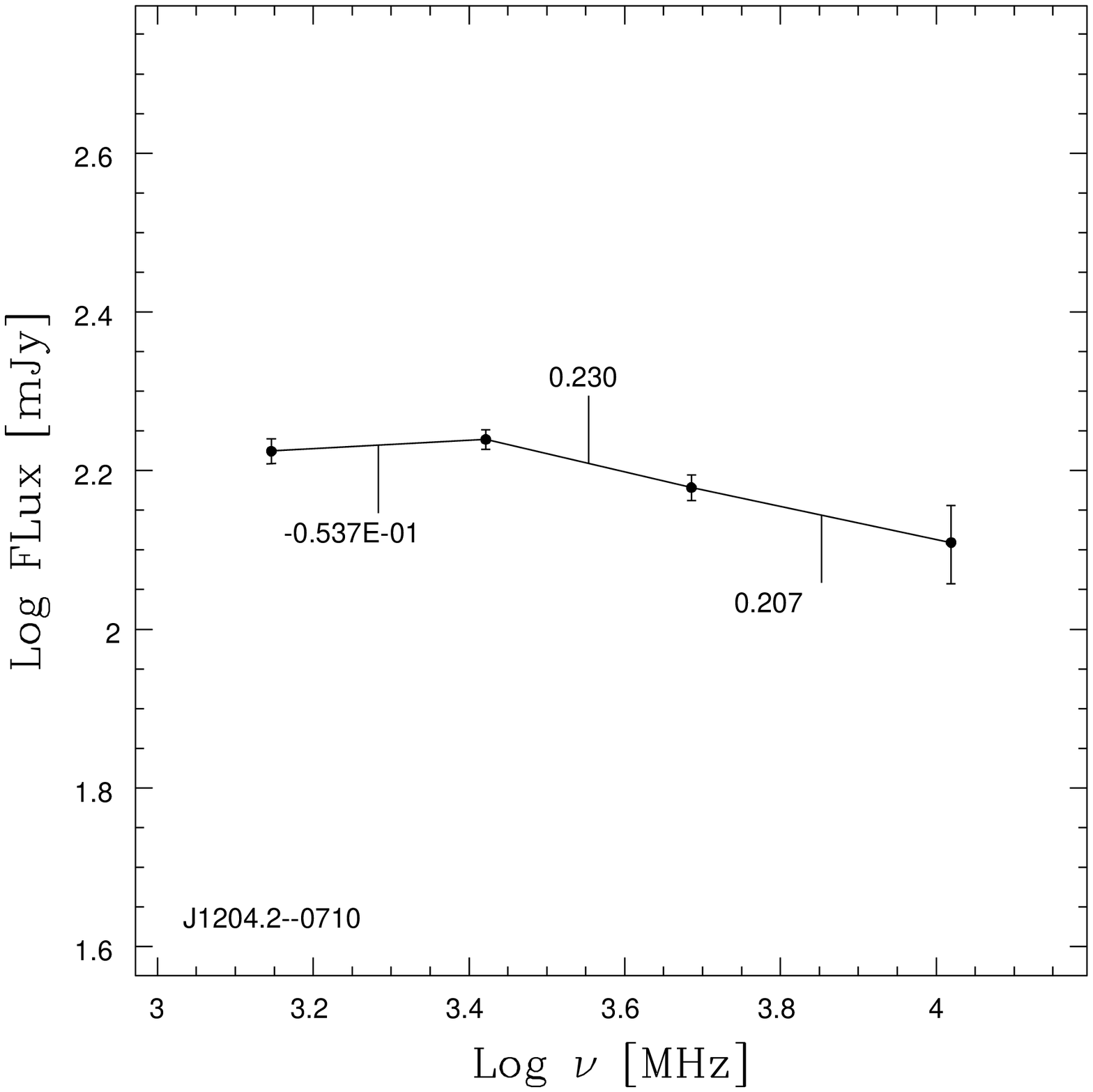}
\caption{Spectral index plots of sources in Table 3.}
\end{figure*}
\clearpage
\newpage
\begin{figure*}[t]
\addtocounter{figure}{+0}
\centering
\includegraphics[width=8cm]{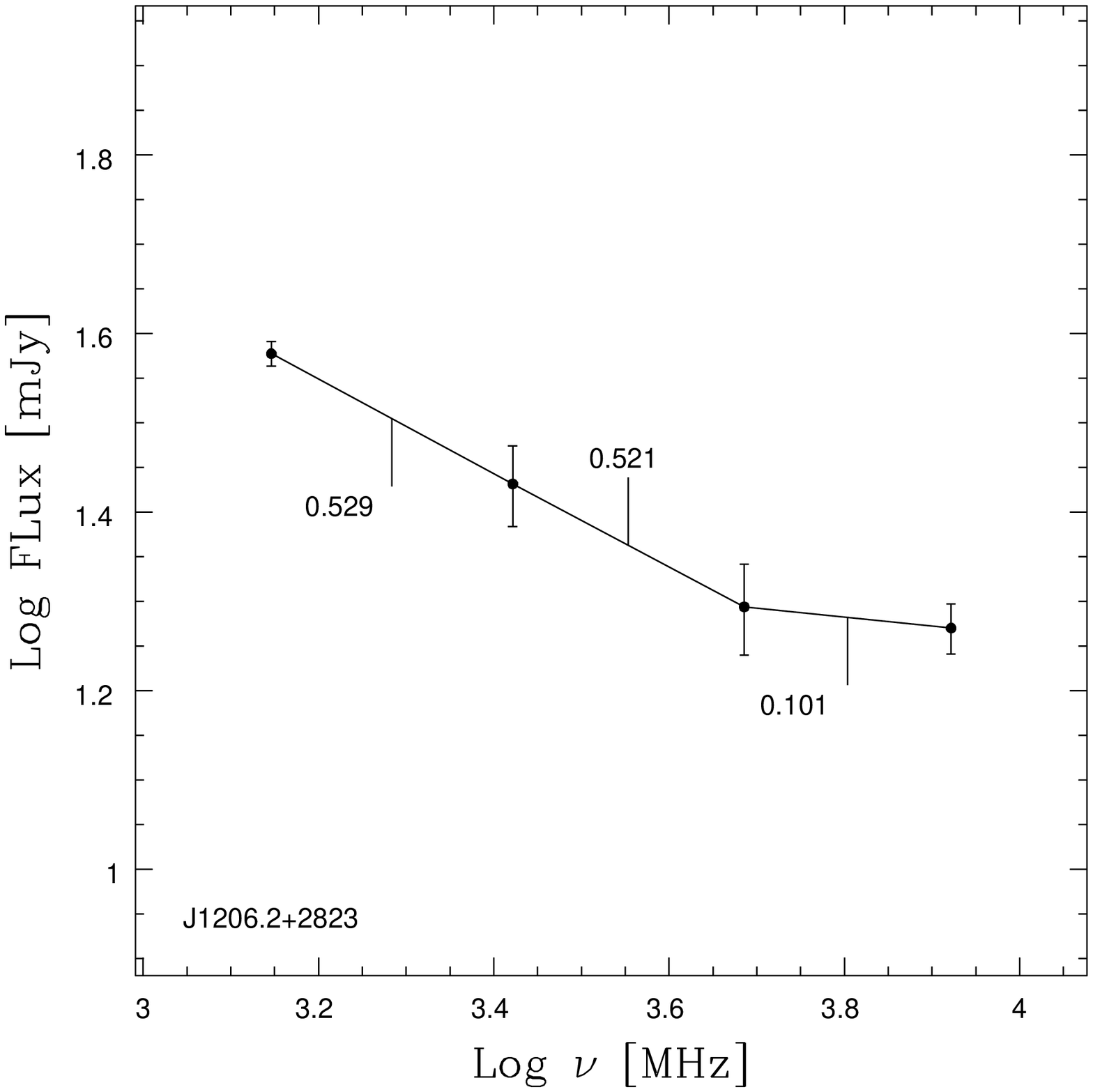}
\includegraphics[width=8cm]{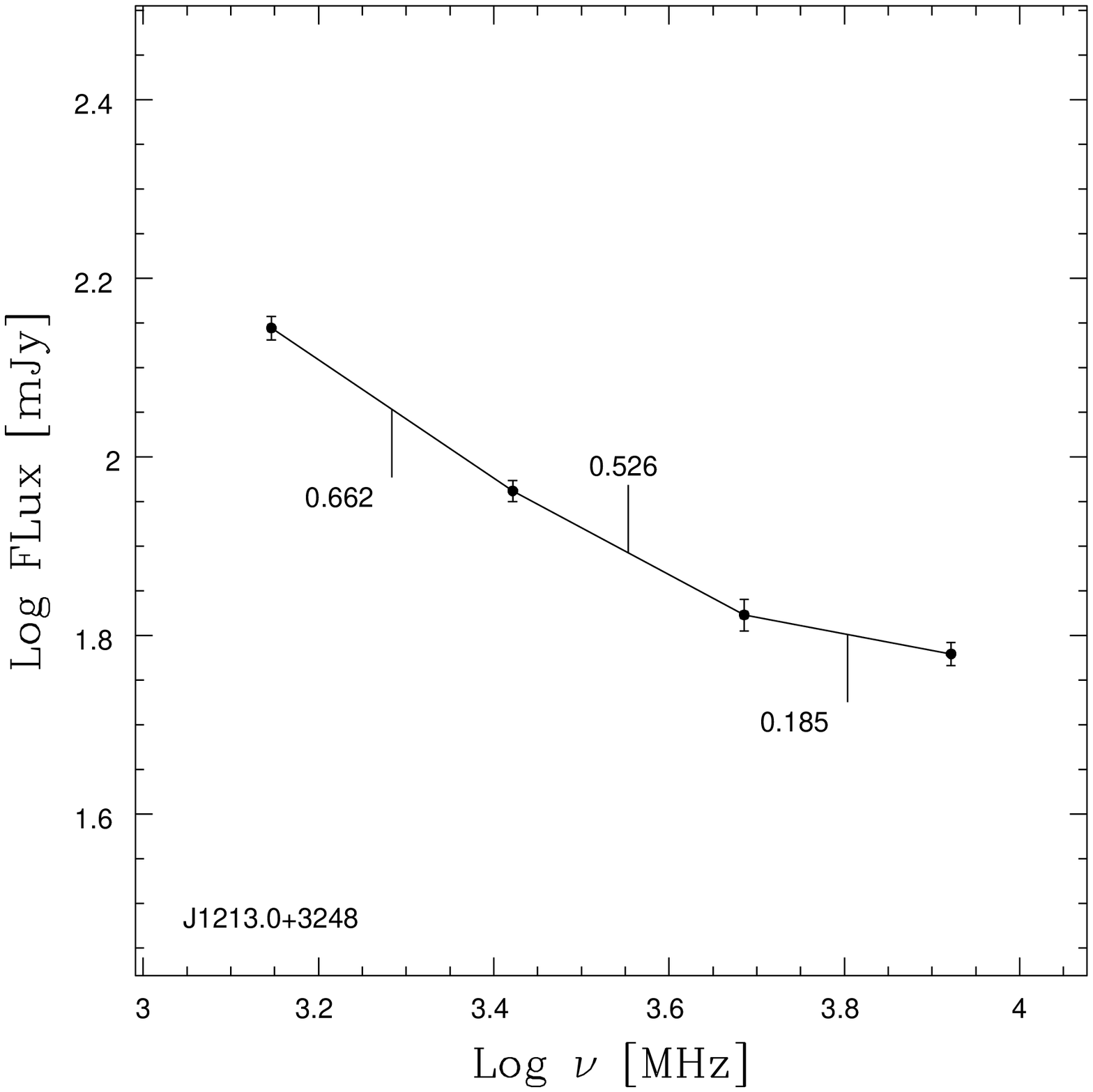}
\includegraphics[width=8cm]{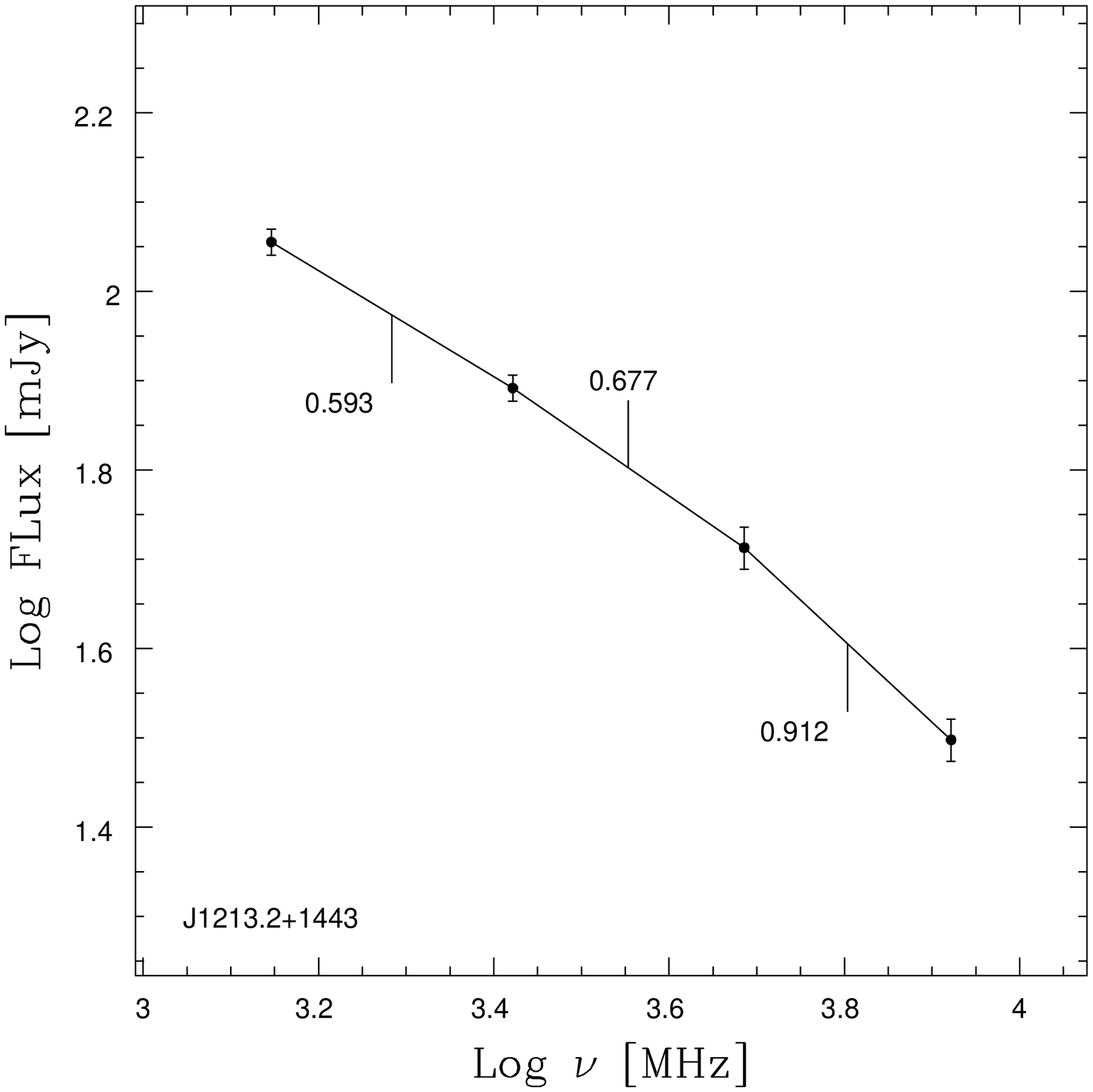}
\includegraphics[width=8cm]{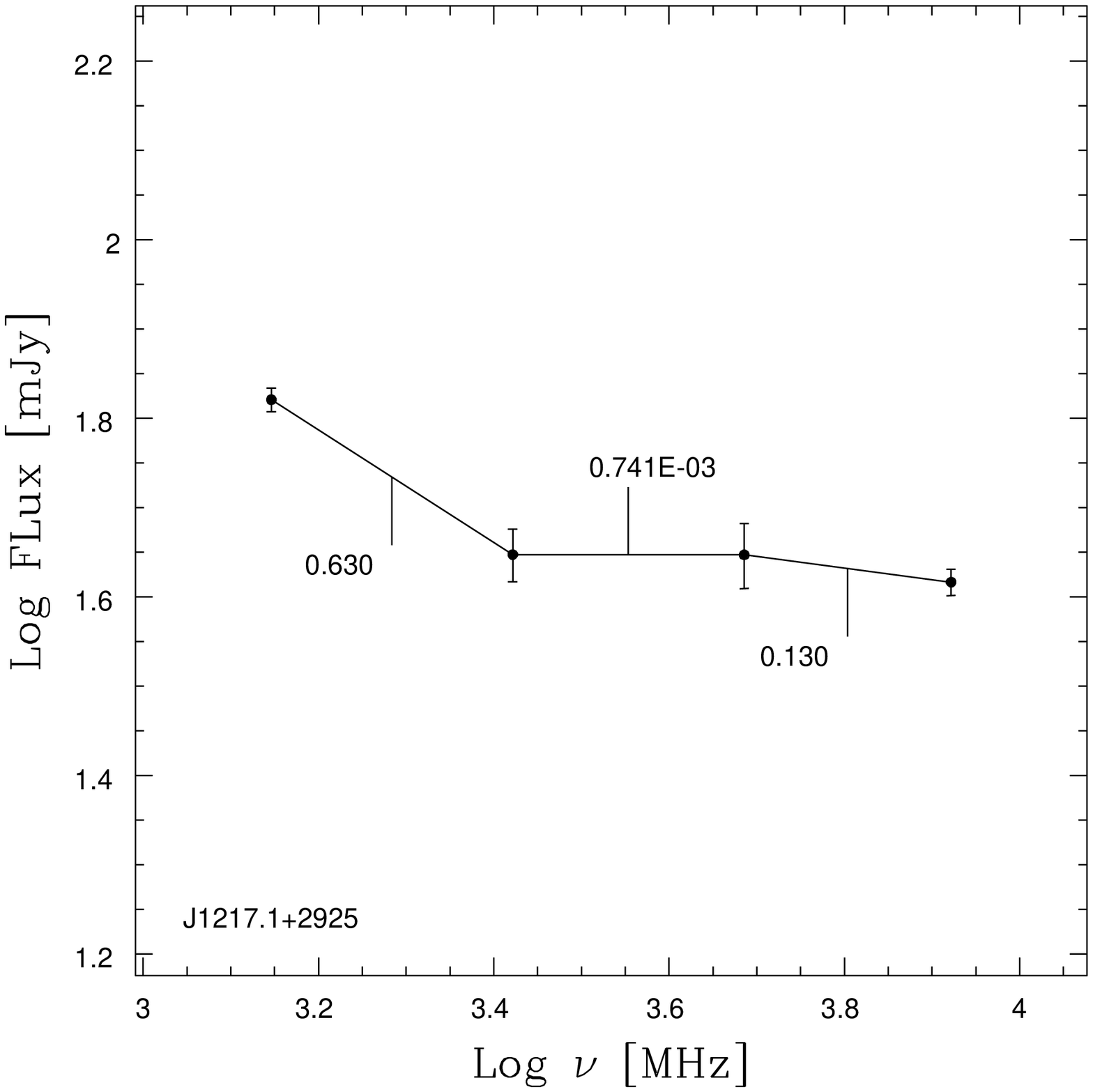}
\includegraphics[width=8cm]{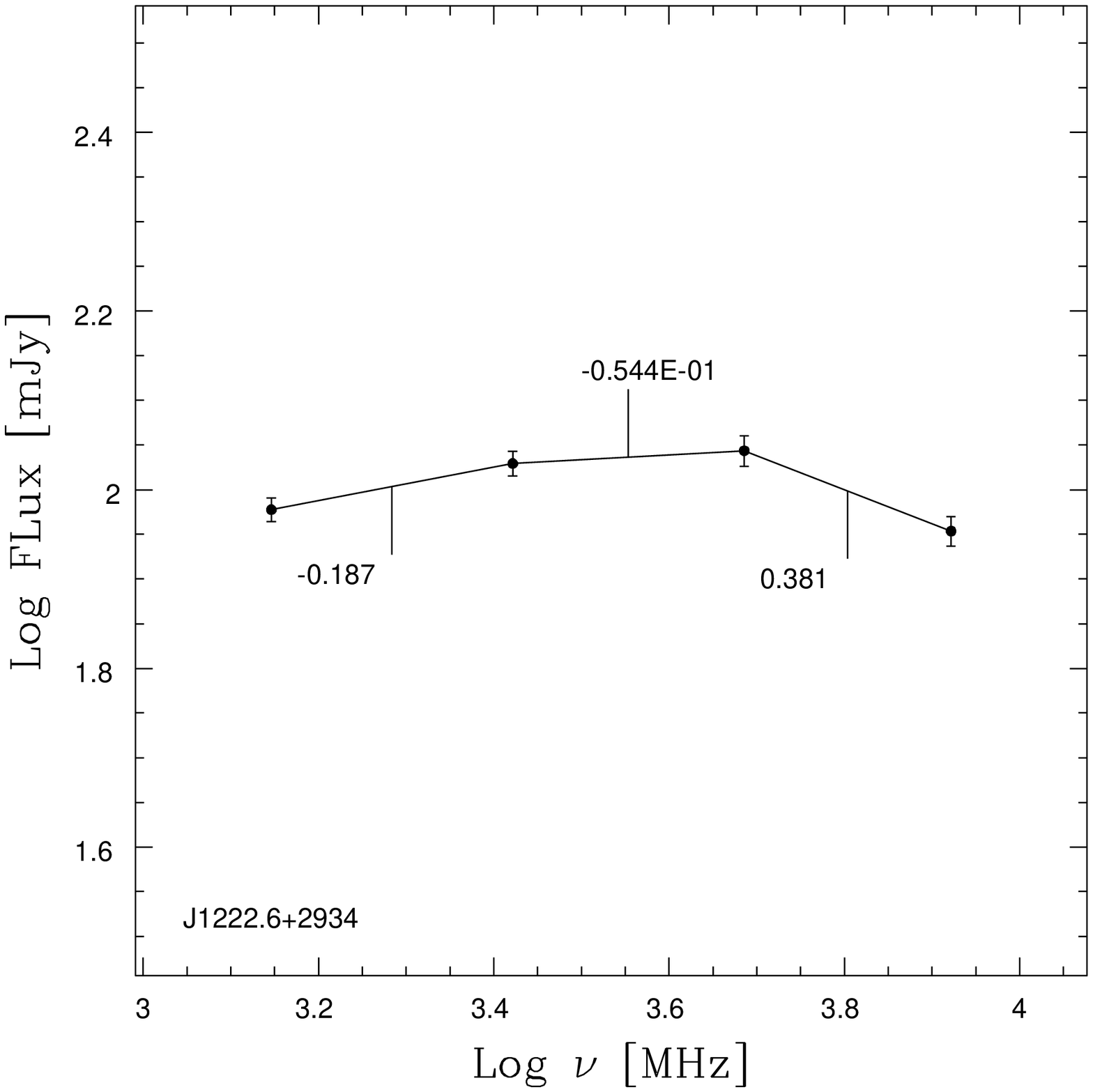}
\includegraphics[width=8cm]{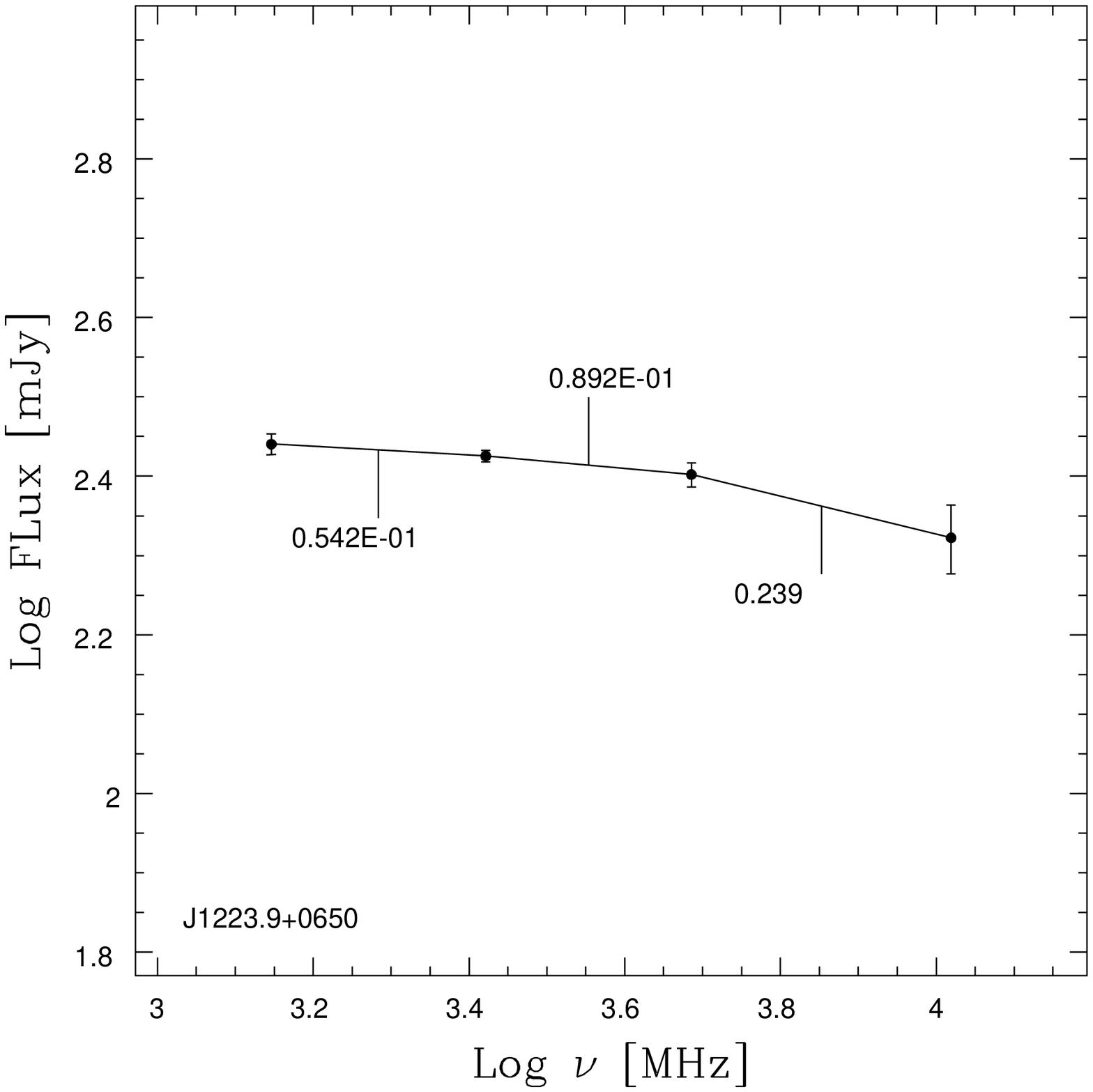}
\caption{Spectral index plots of sources in Table 3.}
\end{figure*}
\clearpage
%
\begin{figure*}[t]
\addtocounter{figure}{+0}
\centering
\includegraphics[width=8cm]{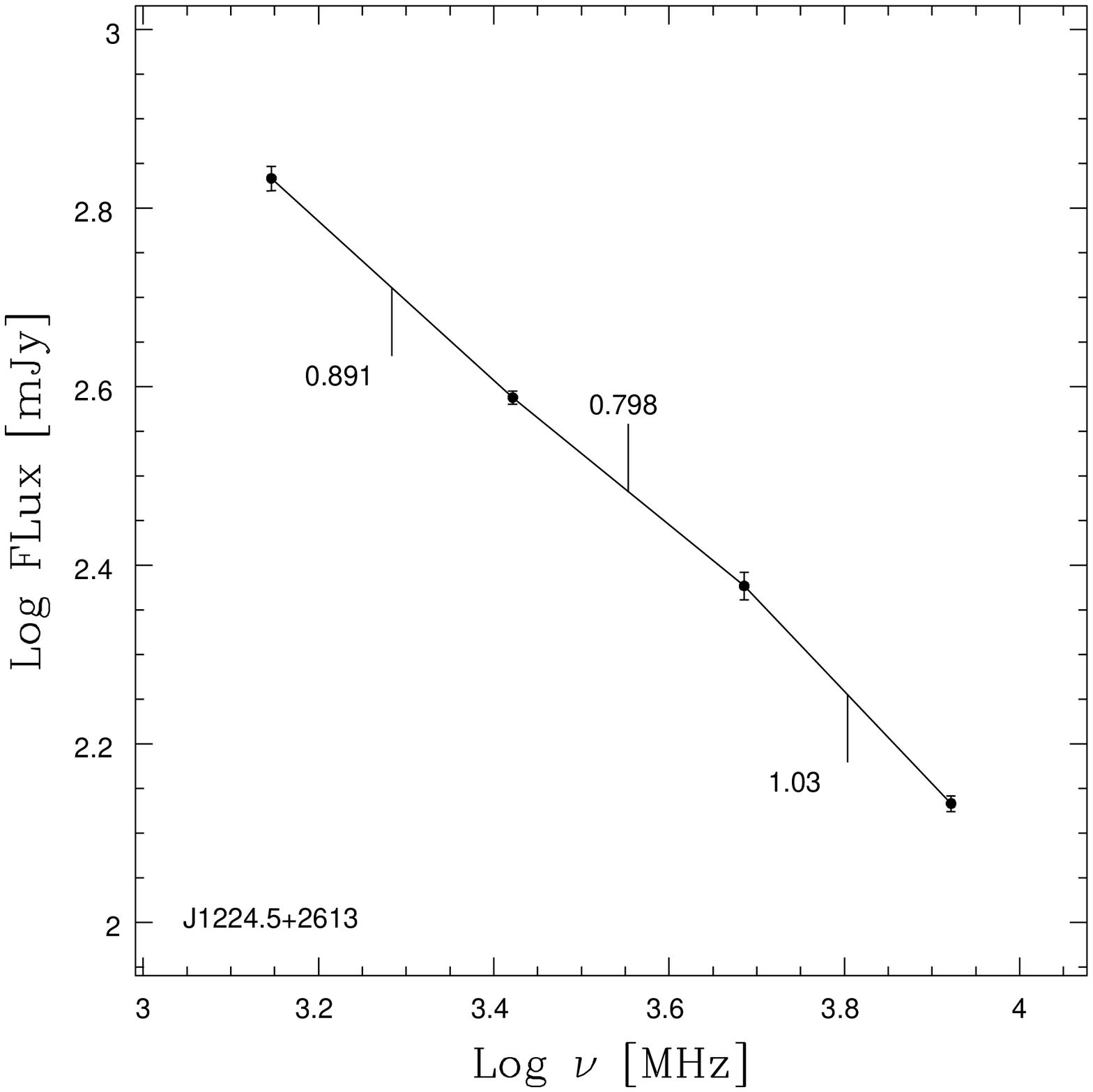}
\includegraphics[width=8cm]{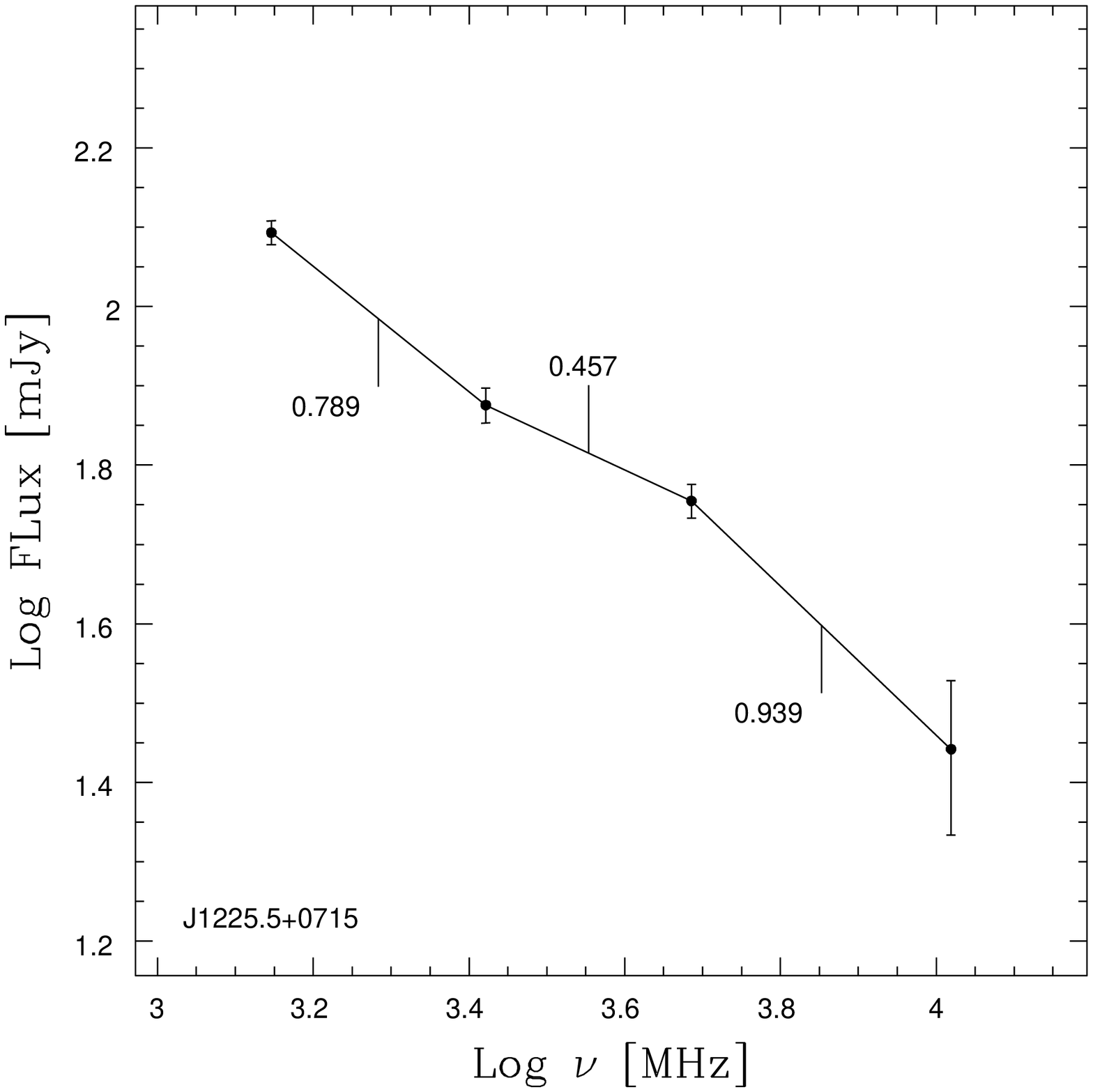}
\includegraphics[width=8cm]{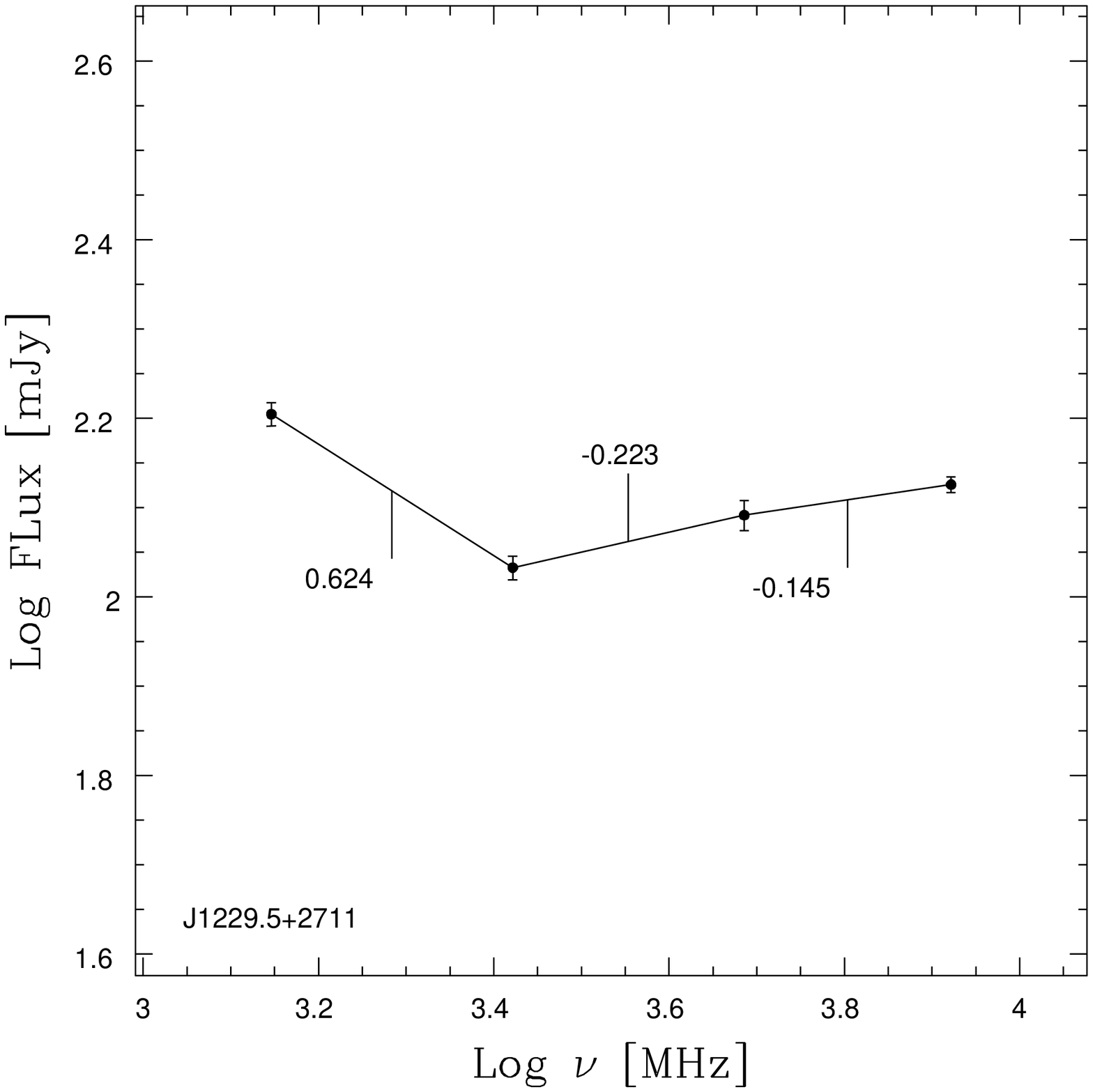}
\includegraphics[width=8cm]{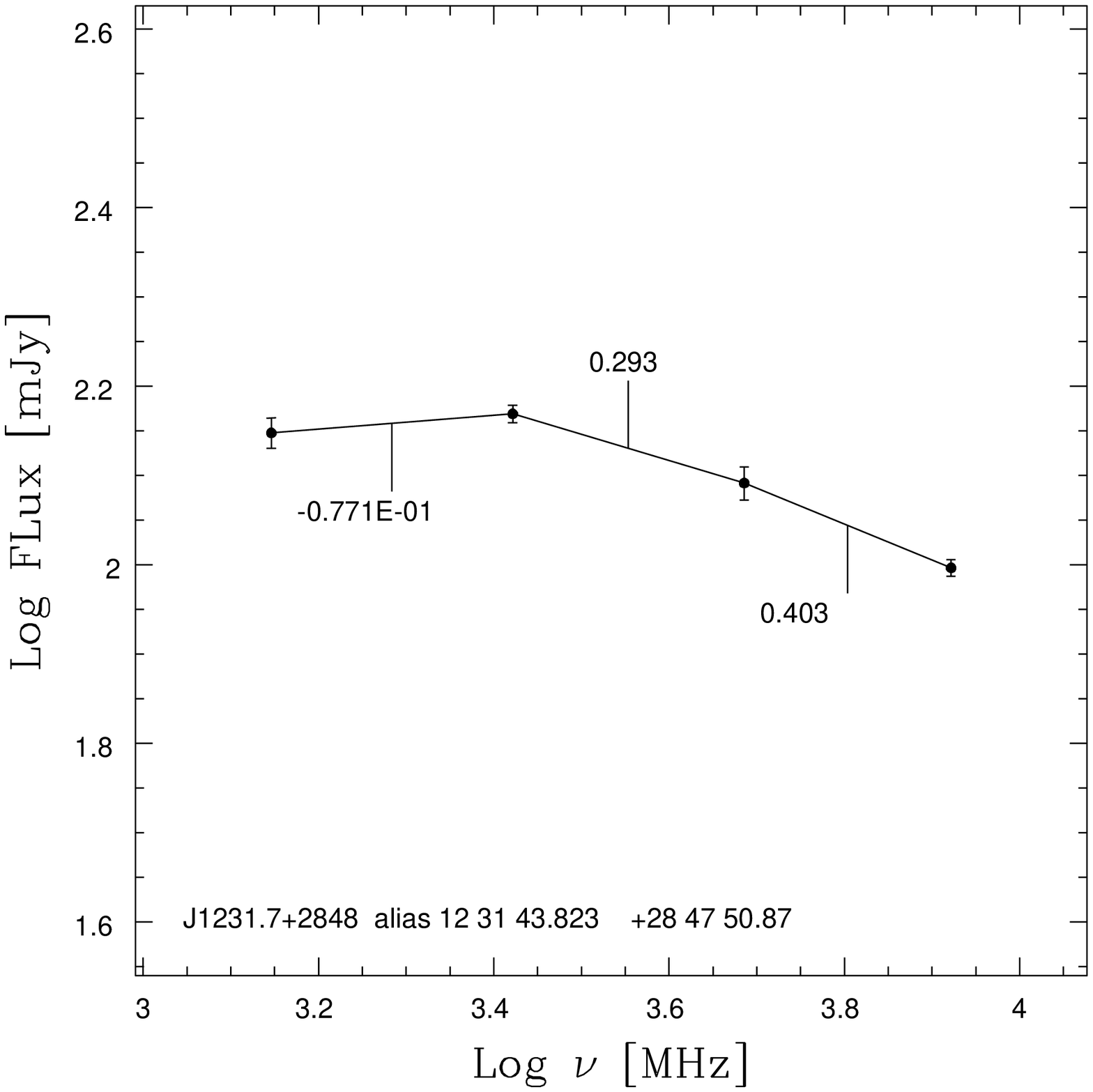}
\includegraphics[width=8cm]{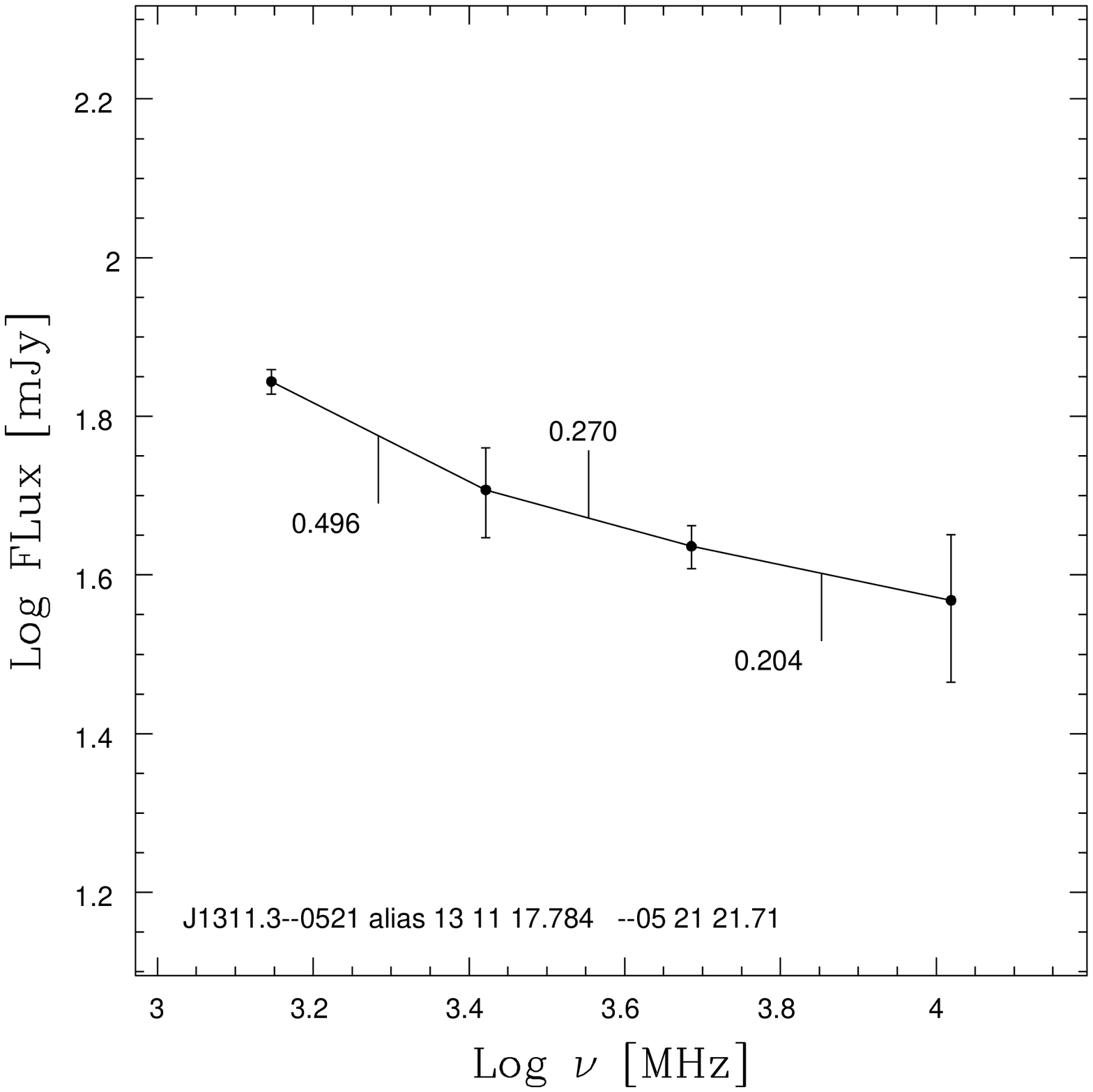}
\includegraphics[width=8cm]{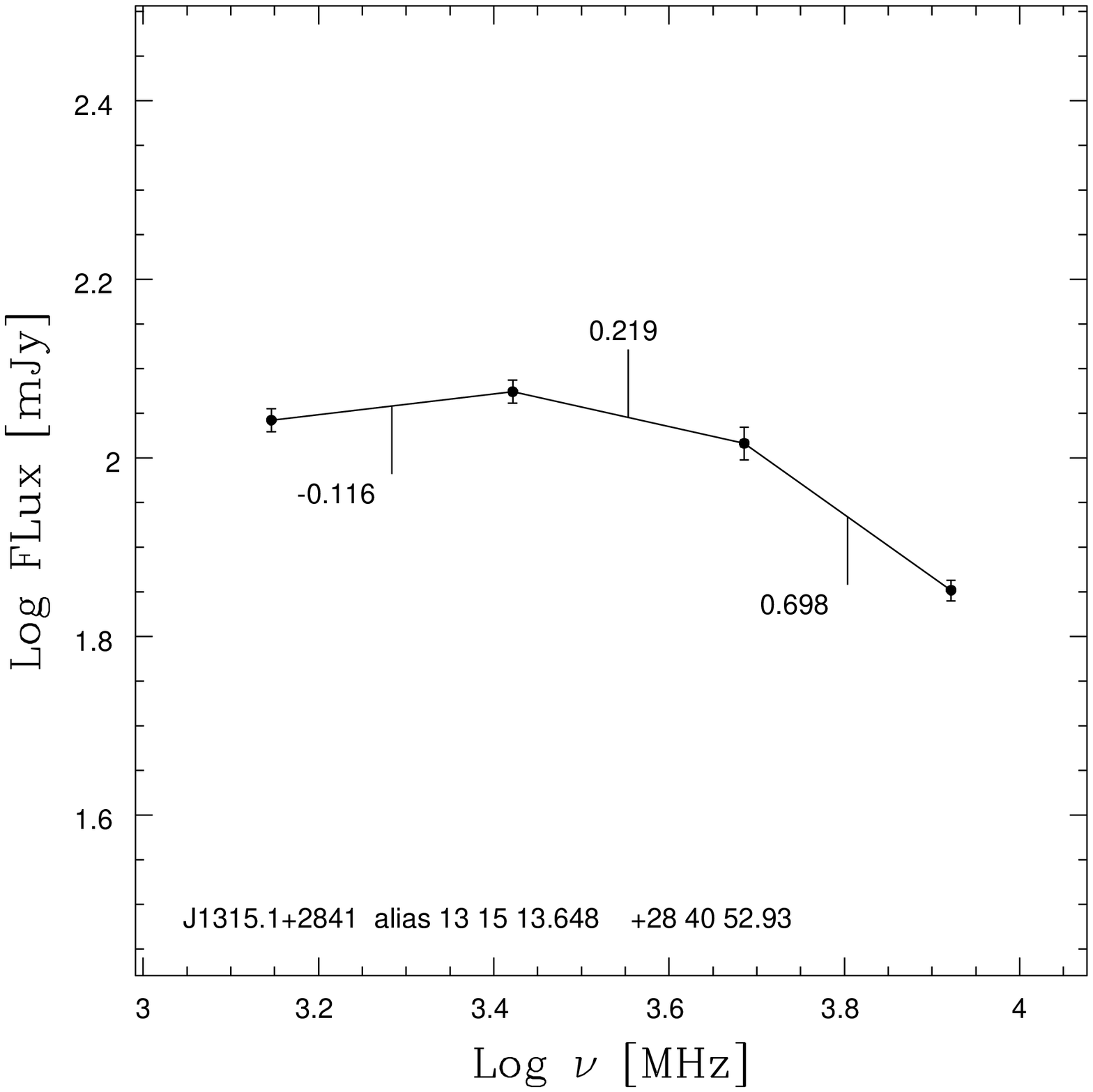}
\caption{Spectral index plots of sources in Table 3.}
\end{figure*}
\newpage
\begin{figure*}[t]
\addtocounter{figure}{+0}
\centering
\includegraphics[width=8cm]{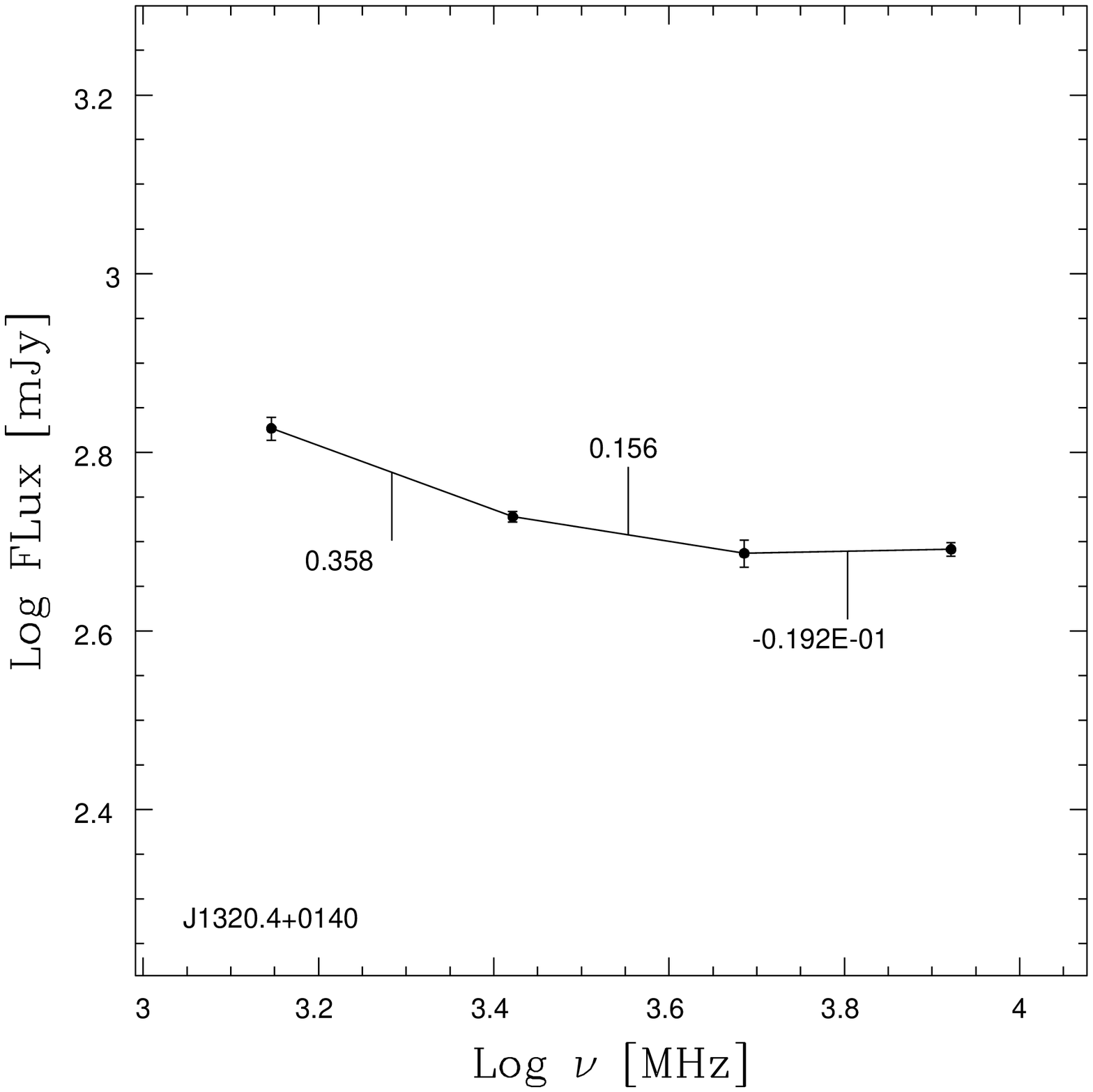}
\includegraphics[width=8cm]{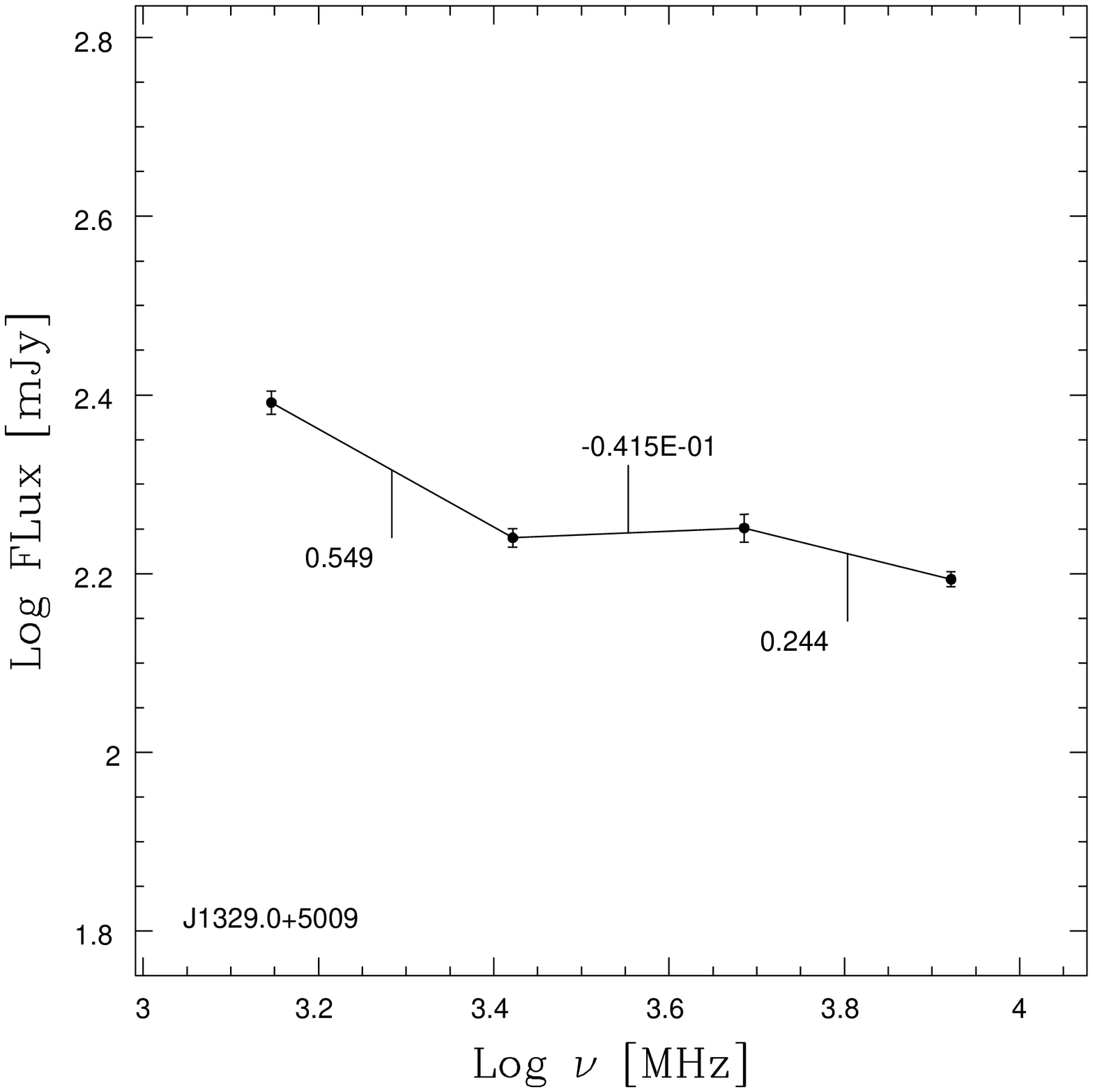}
\includegraphics[width=8cm]{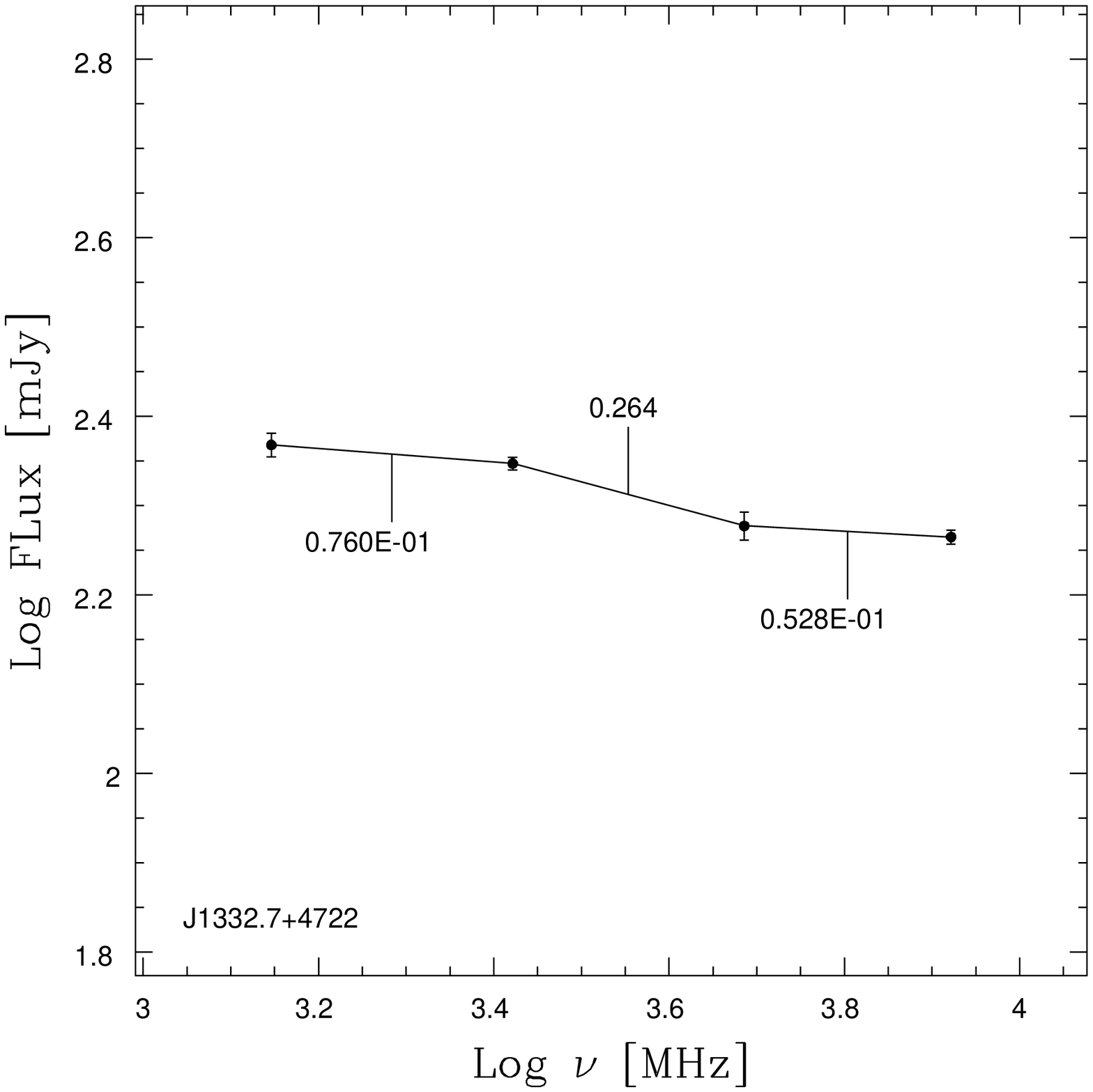}
\includegraphics[width=8cm]{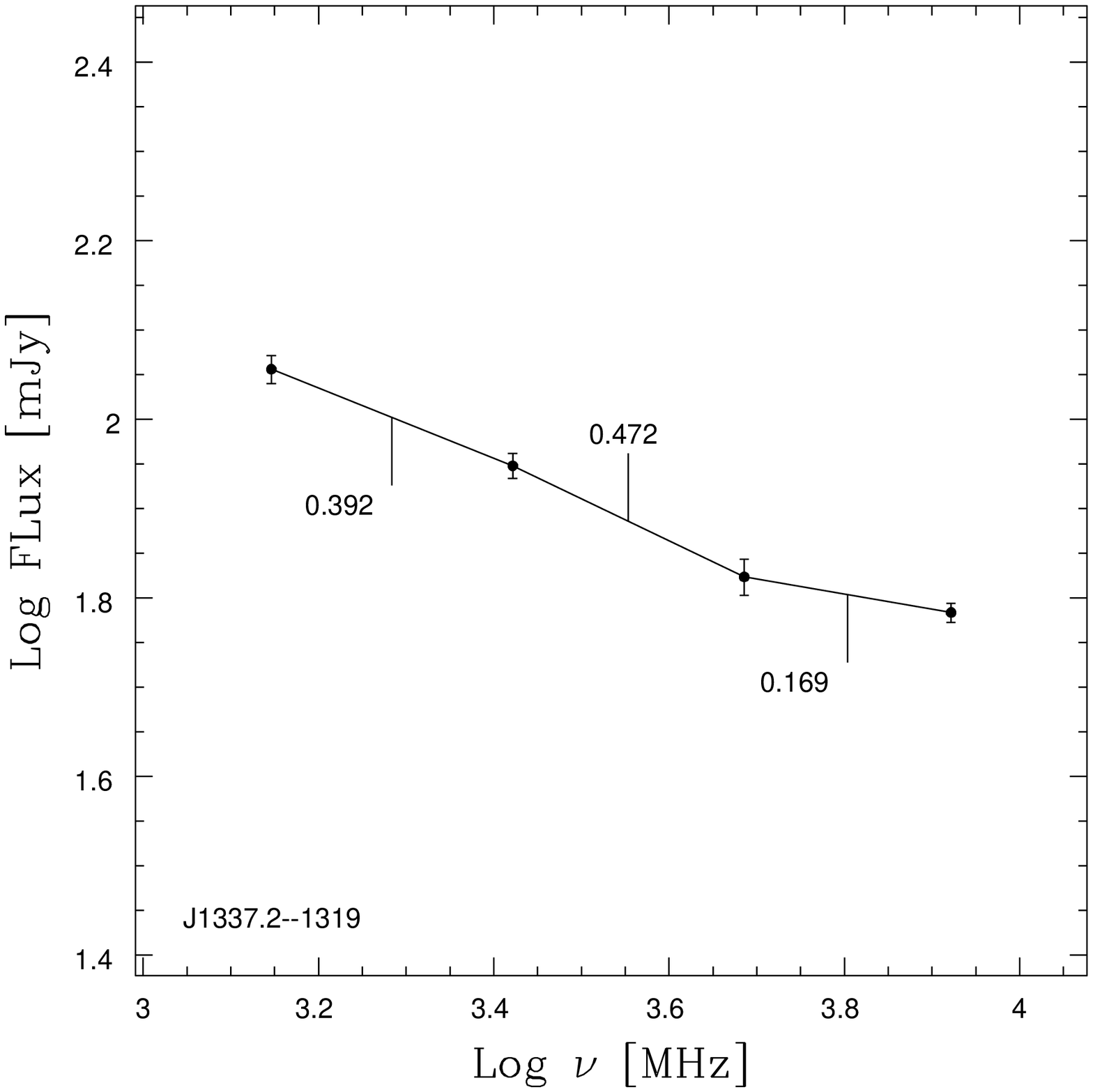}
\includegraphics[width=8cm]{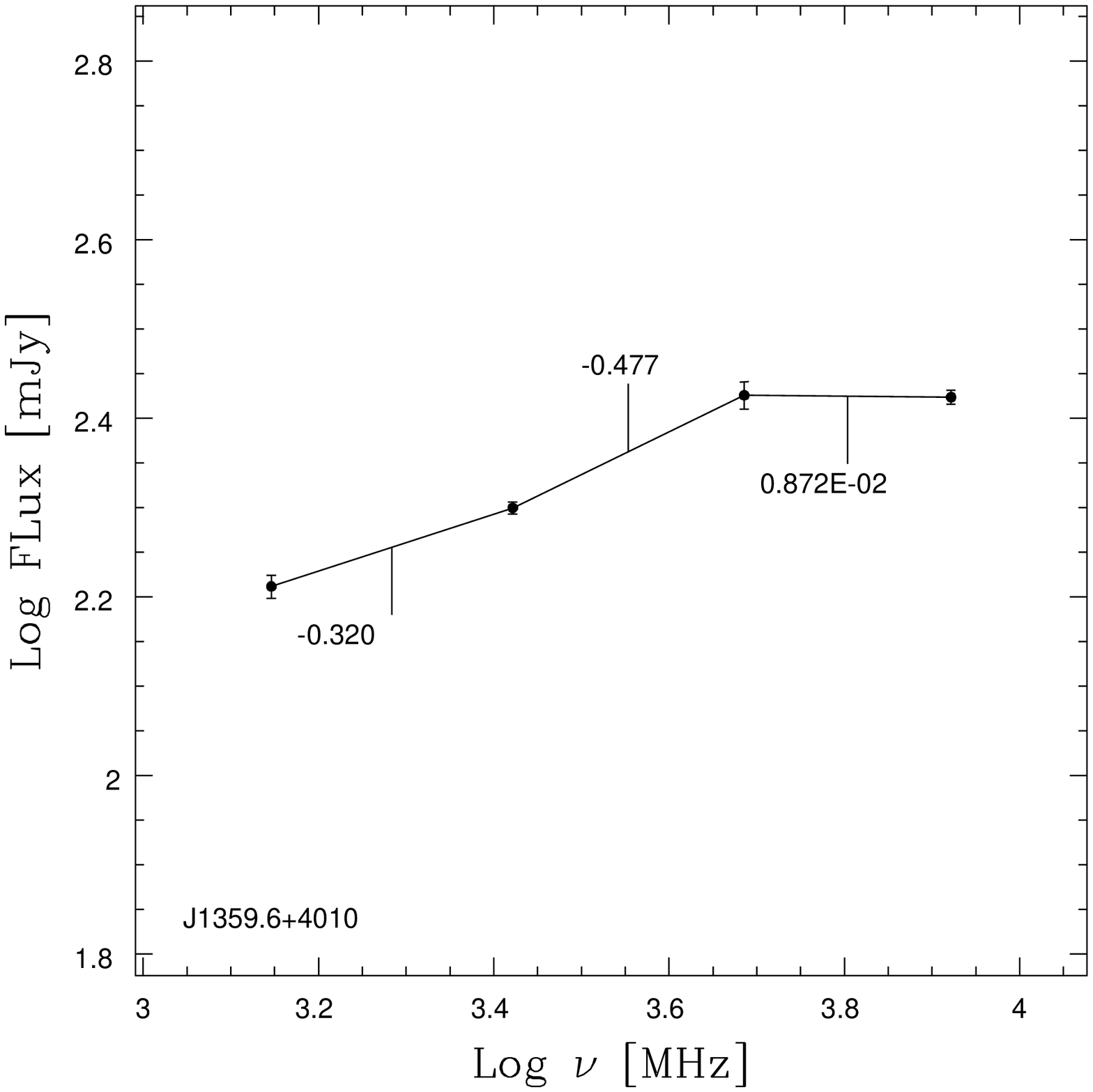}
\includegraphics[width=8cm]{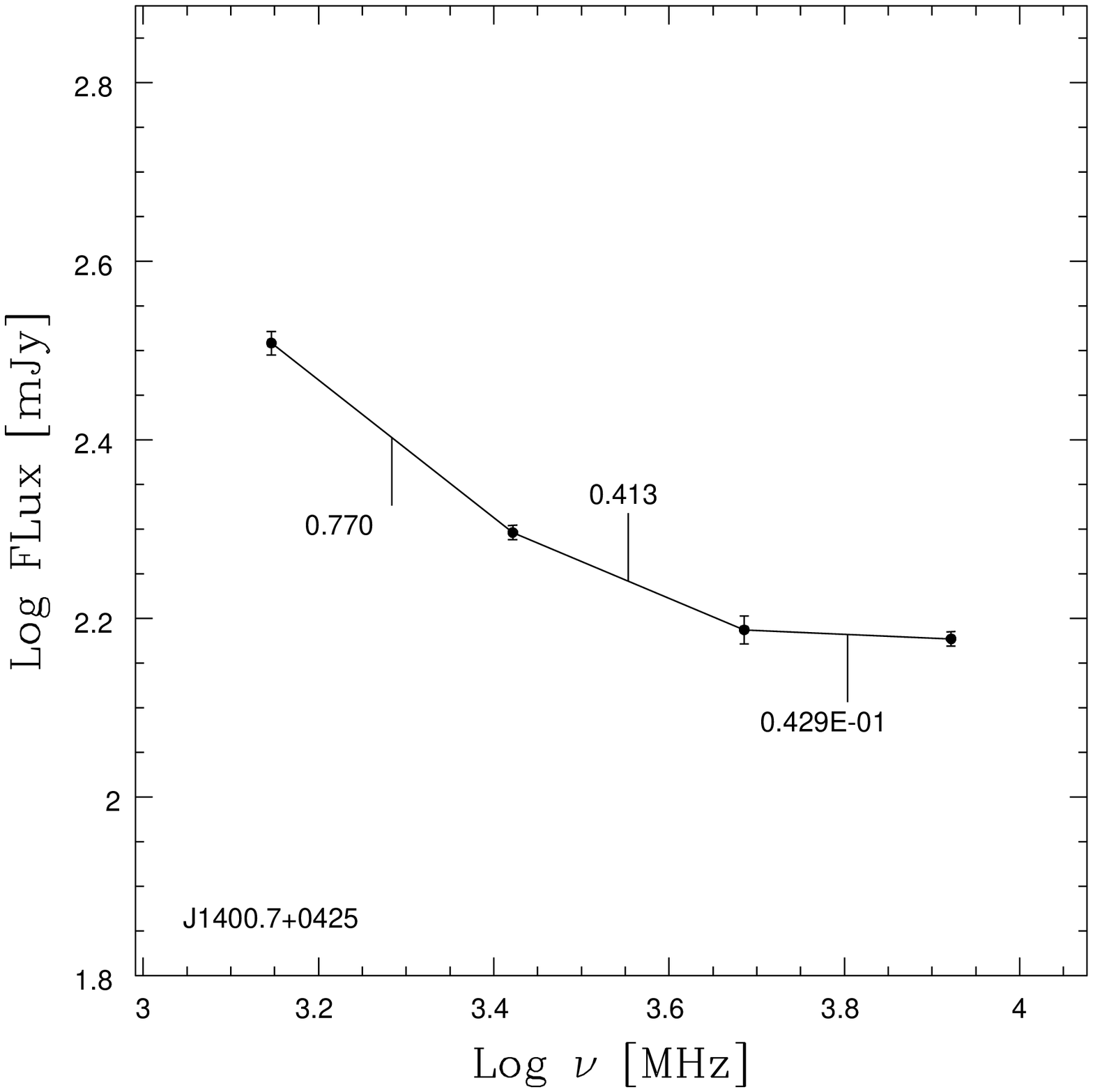}
\caption{Spectral index plots of sources in Table 3.}
\end{figure*}
%
%
%
\newpage
\begin{figure*}[t]
\addtocounter{figure}{+0}
\centering
\includegraphics[width=8cm]{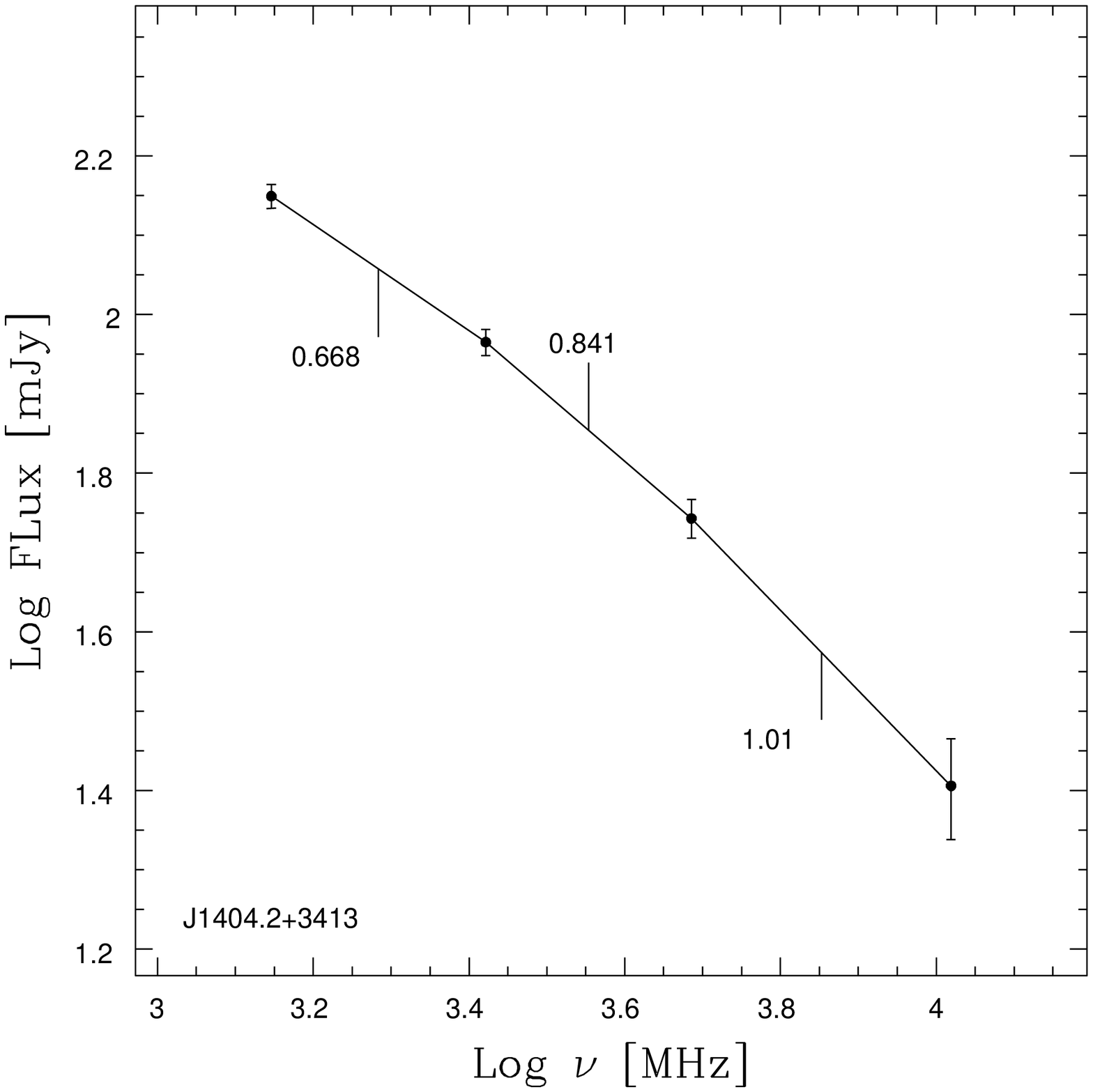}
\includegraphics[width=8cm]{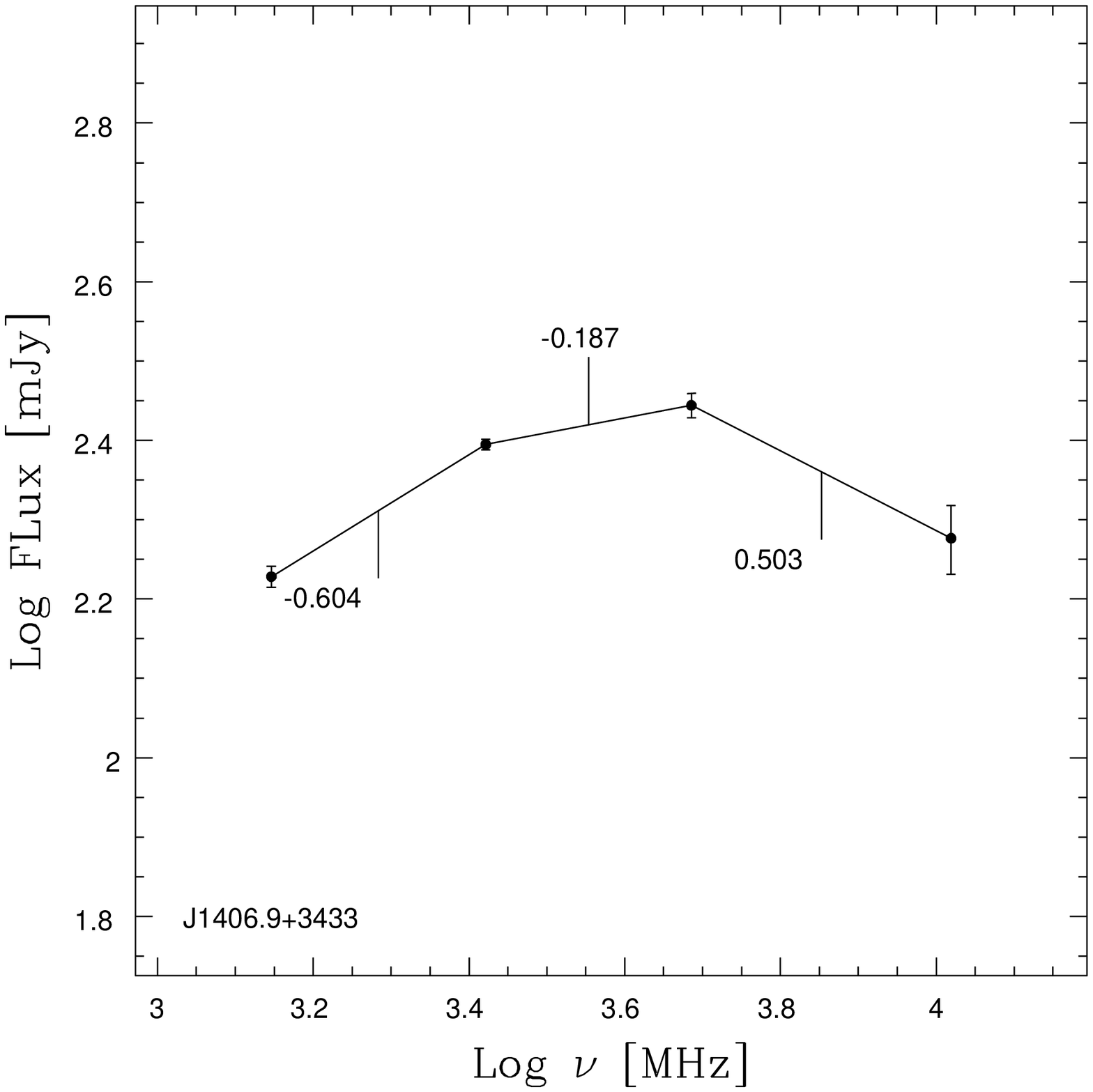}
\includegraphics[width=8cm]{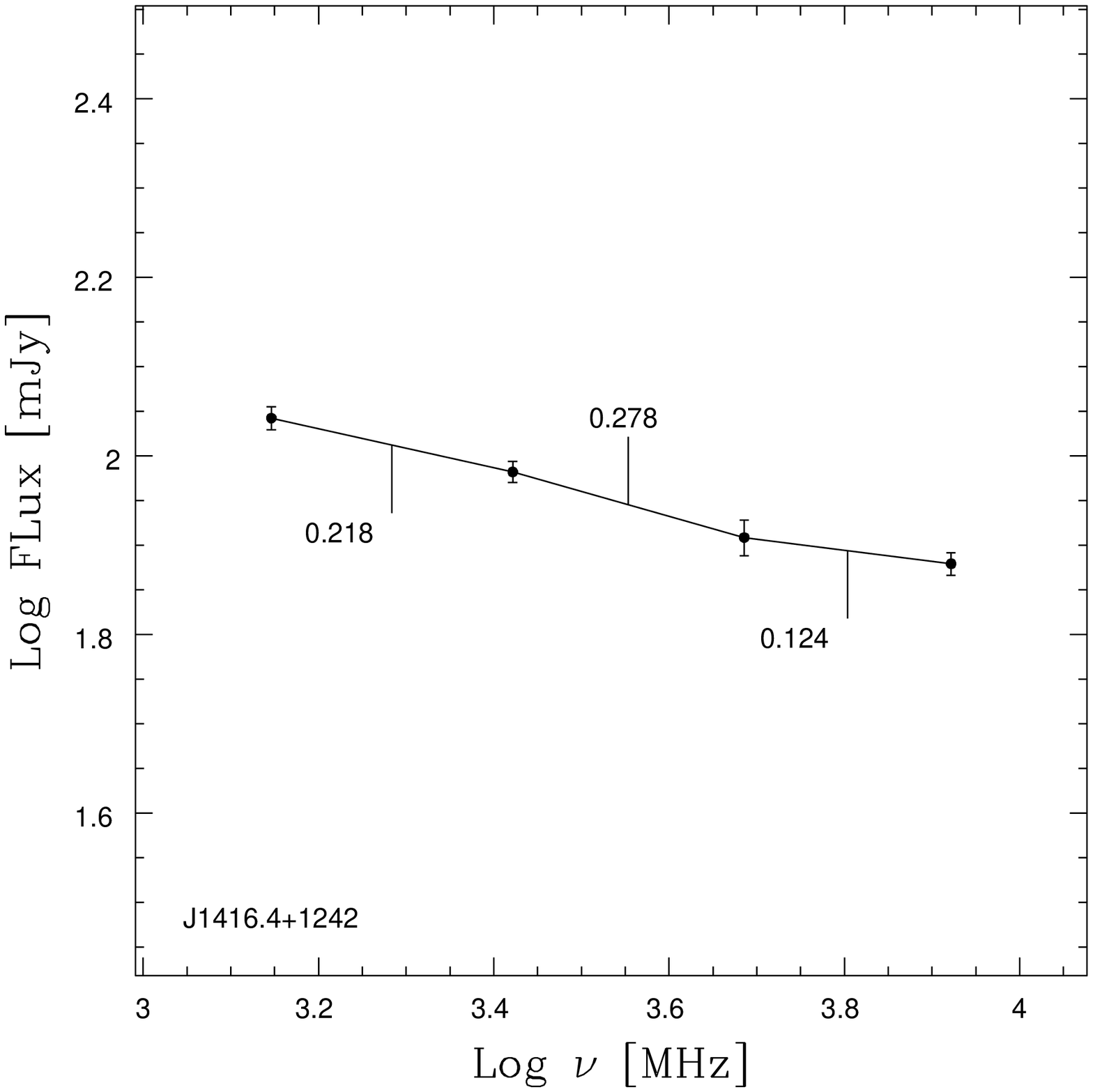}
\includegraphics[width=8cm]{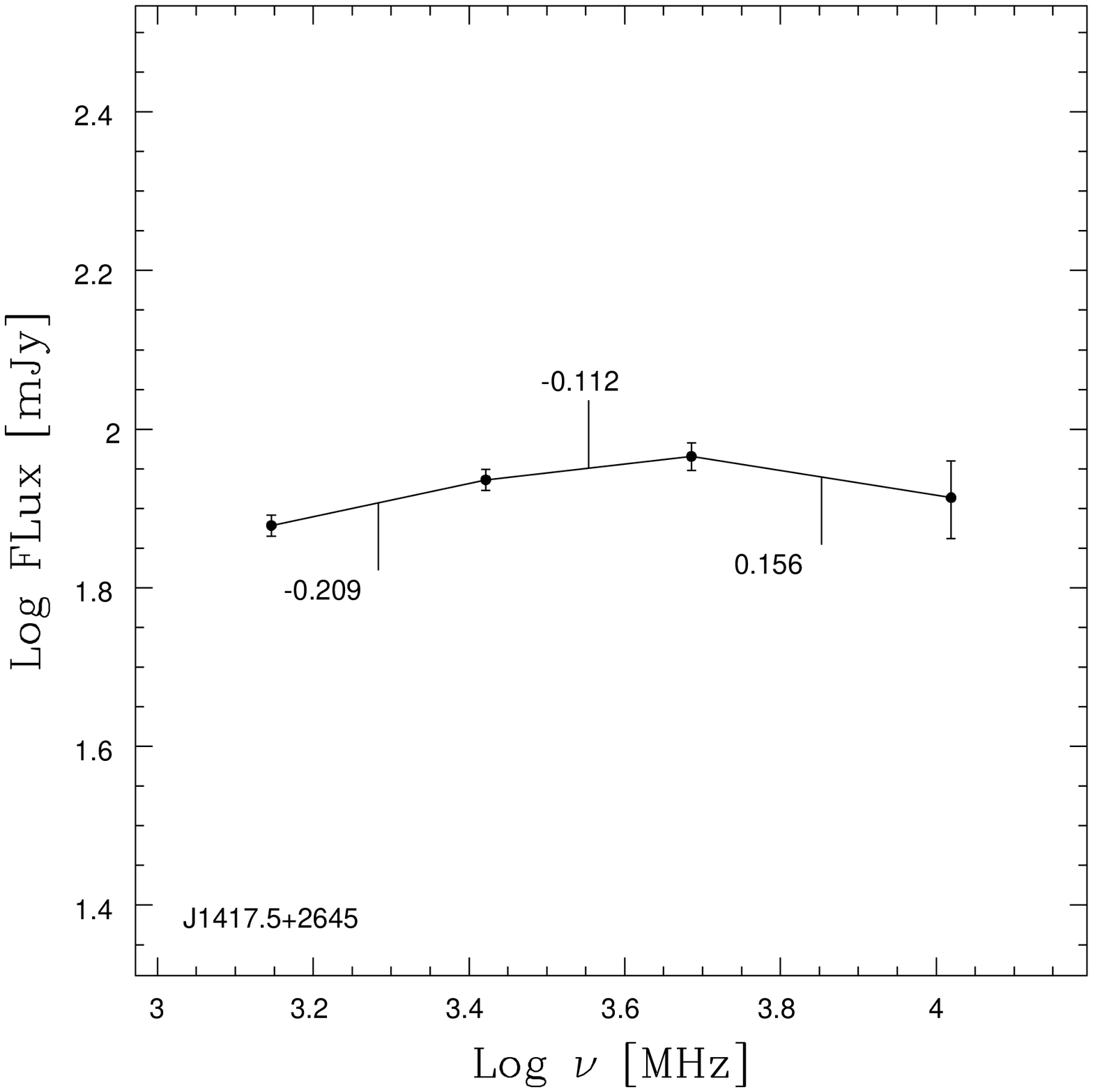}
\includegraphics[width=8cm]{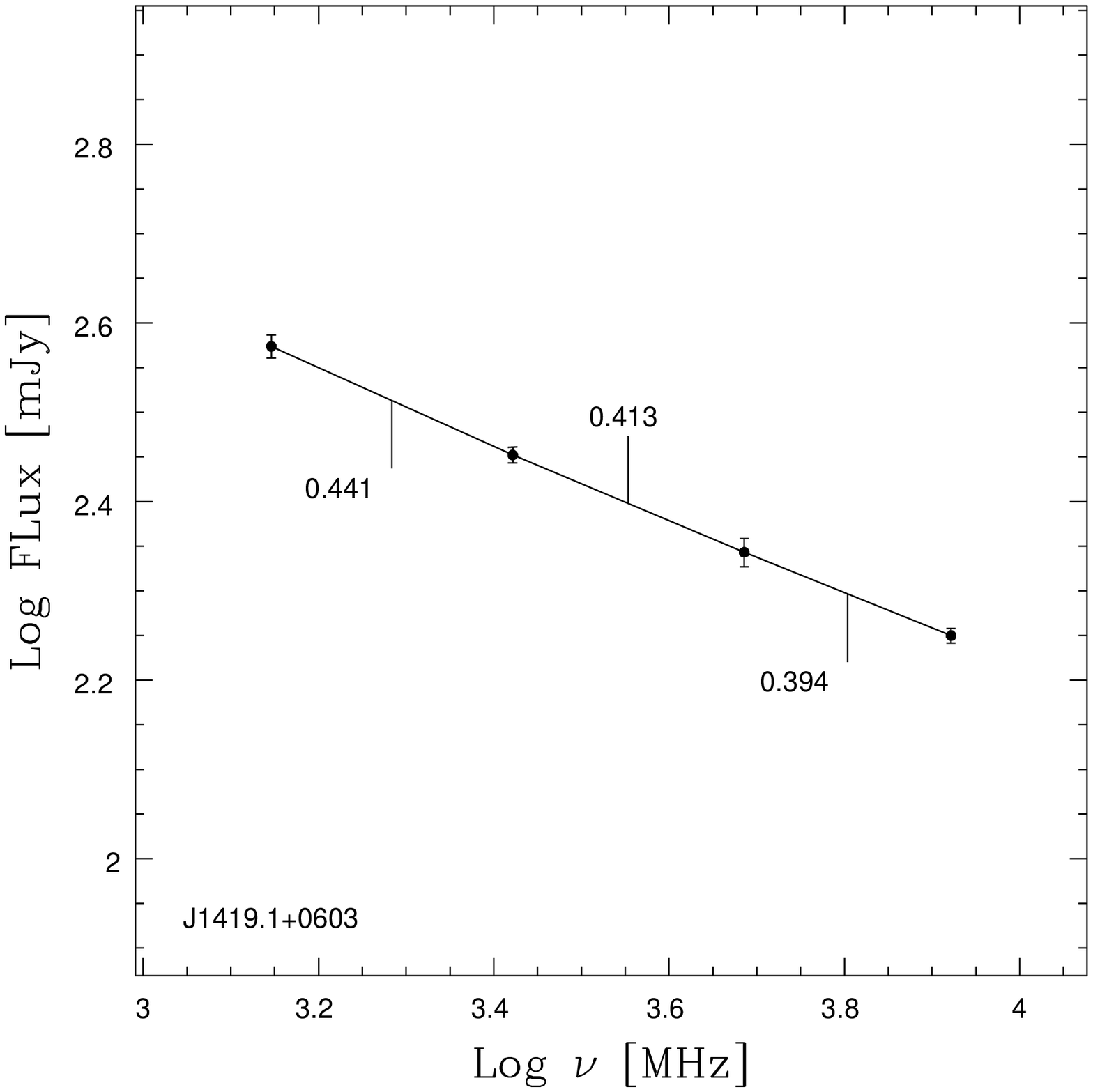}
\includegraphics[width=8cm]{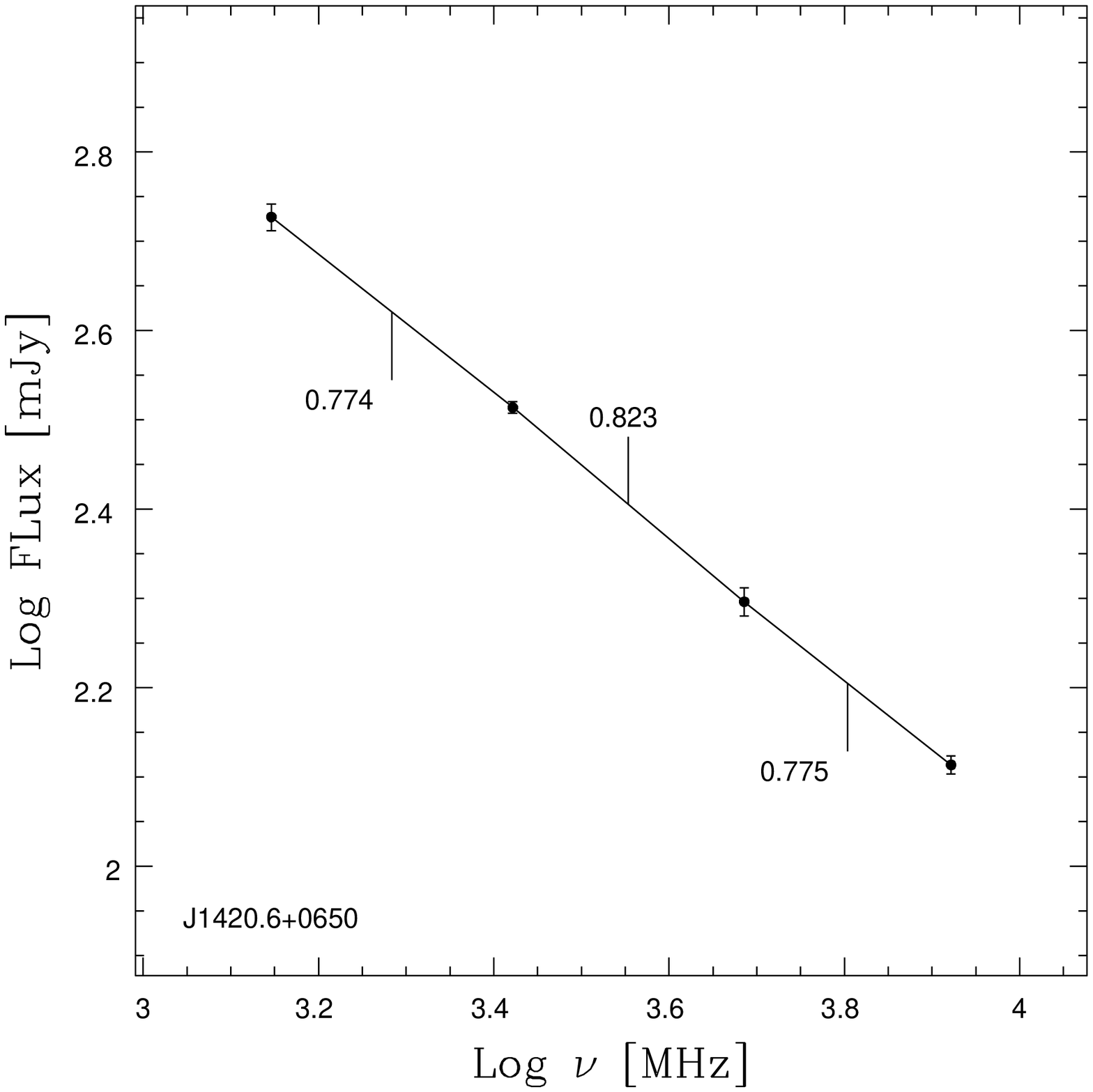}
\caption{Spectral index plots of sources in Table 3.}
\end{figure*}
\newpage
\begin{figure*}[t]
\addtocounter{figure}{+0}
\centering
\includegraphics[width=8cm]{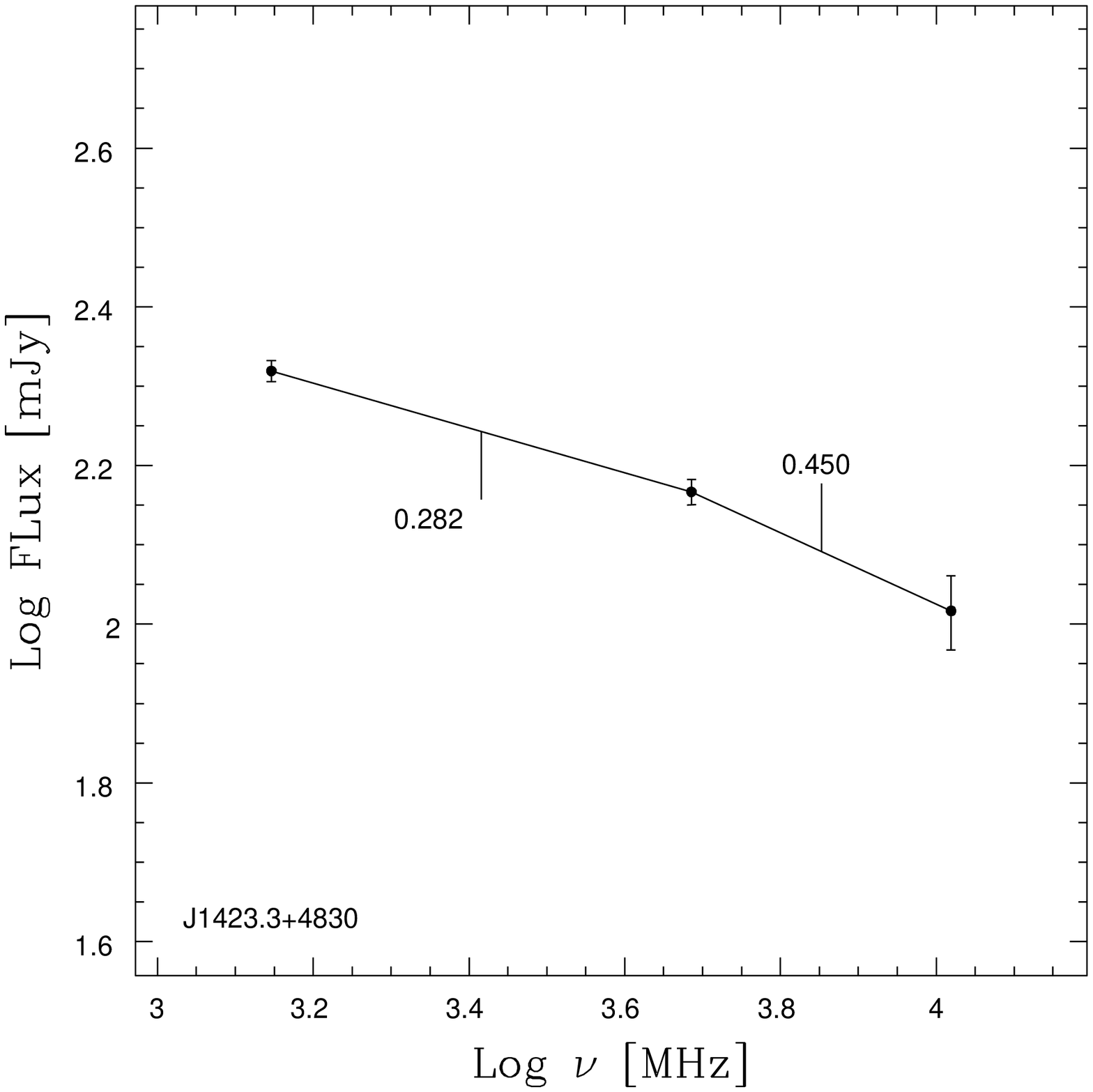}
\includegraphics[width=8cm]{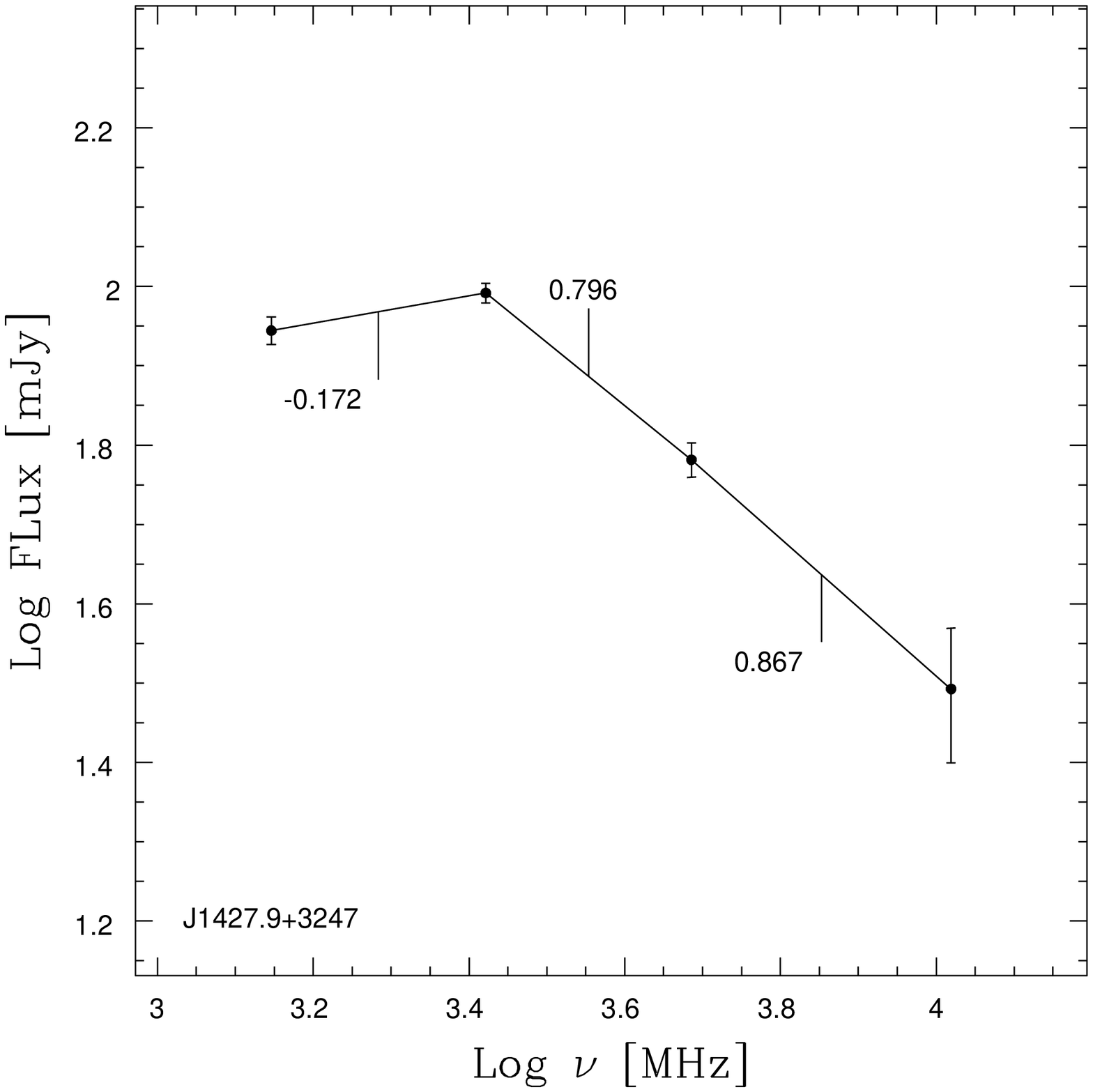}
\includegraphics[width=8cm]{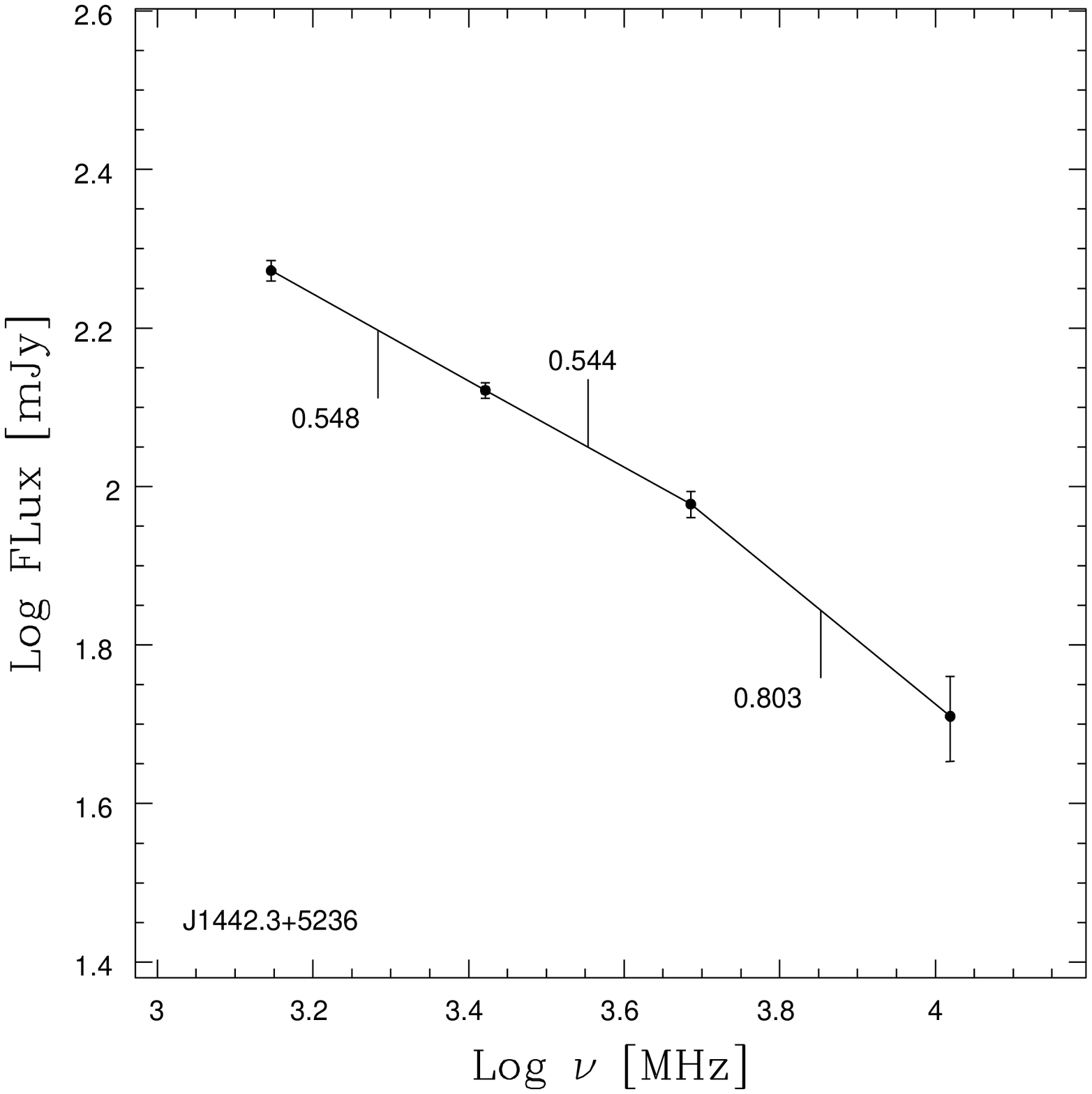}
\includegraphics[width=8cm]{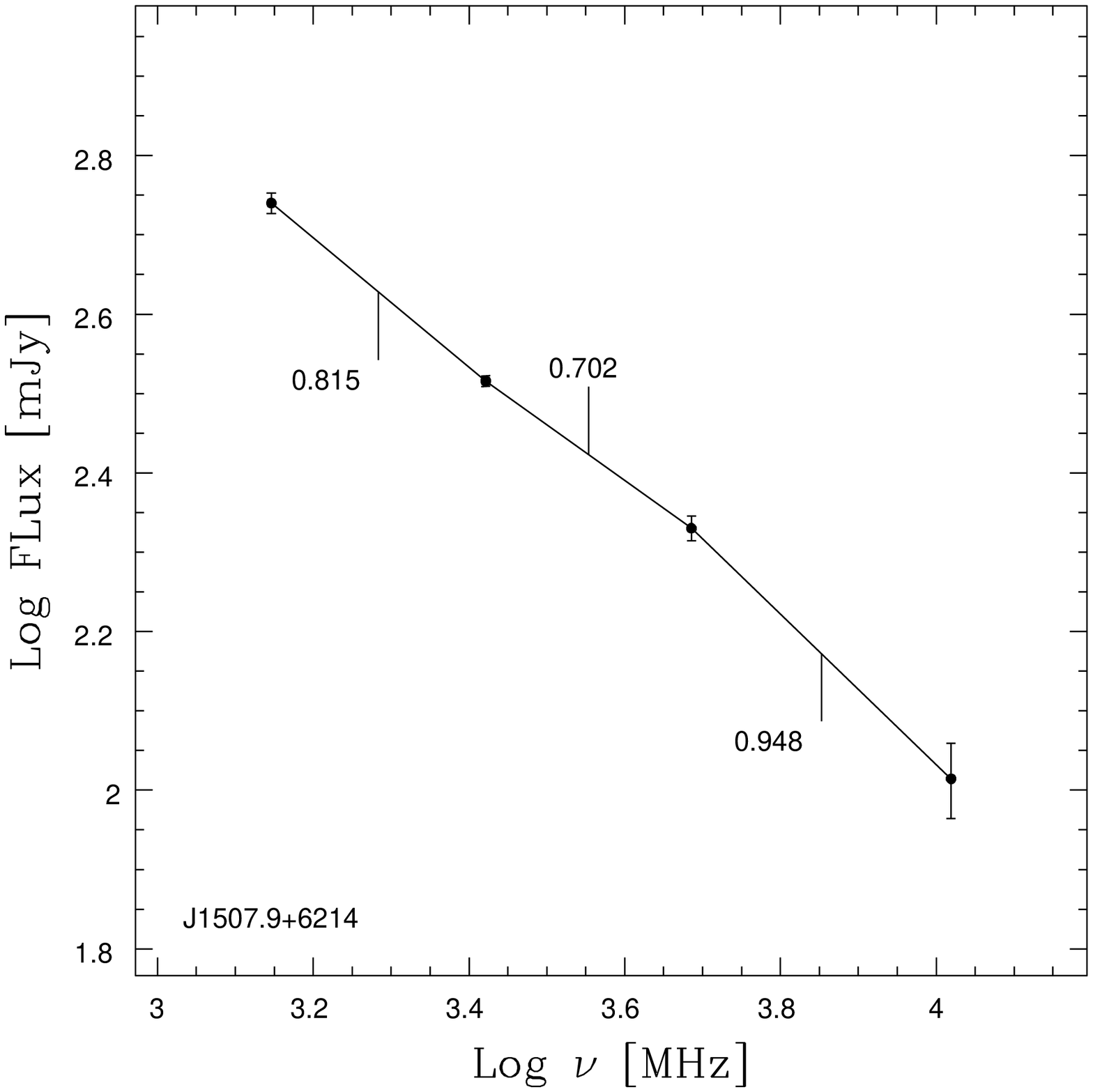}
\includegraphics[width=8cm]{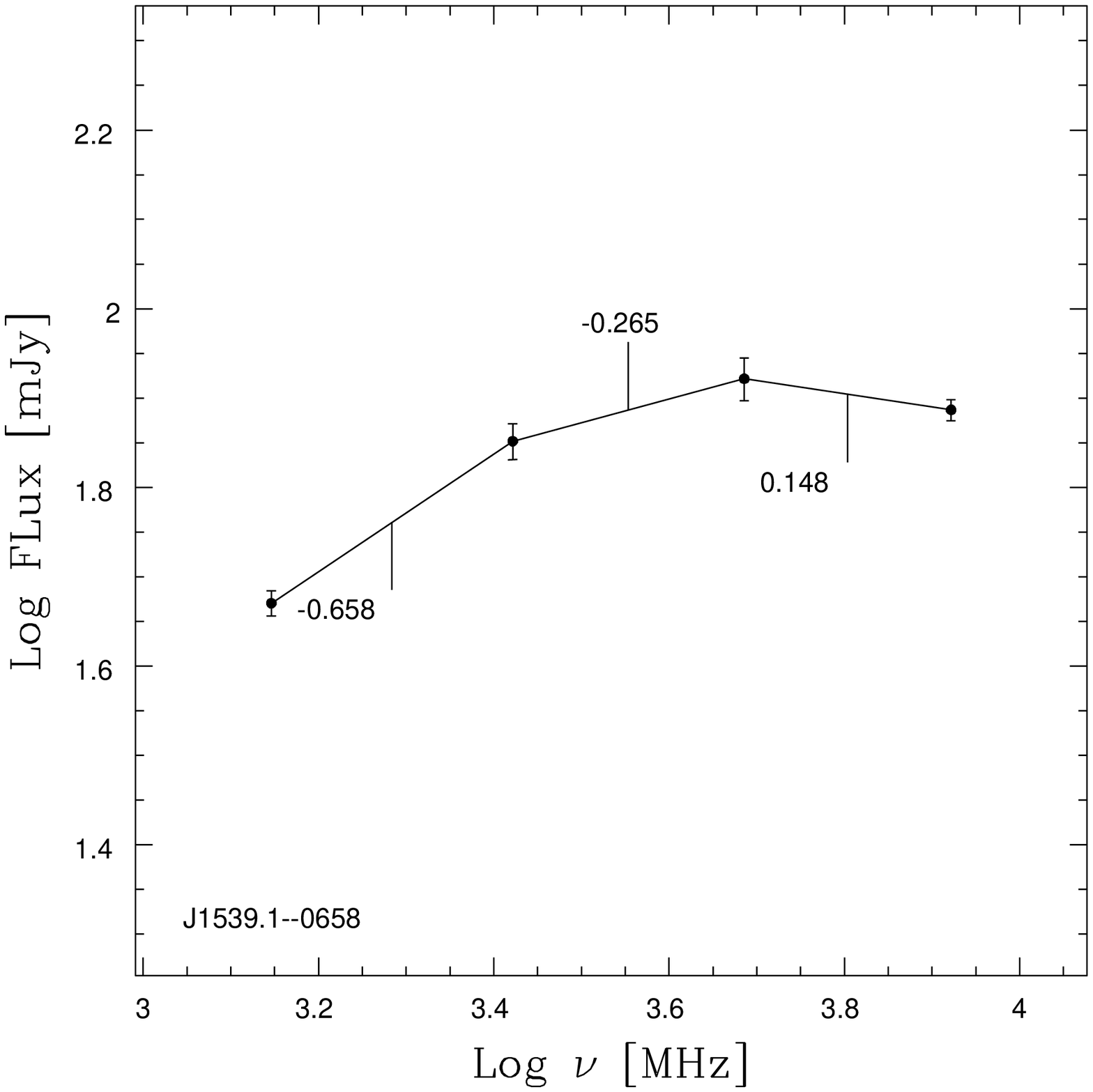}
\includegraphics[width=8cm]{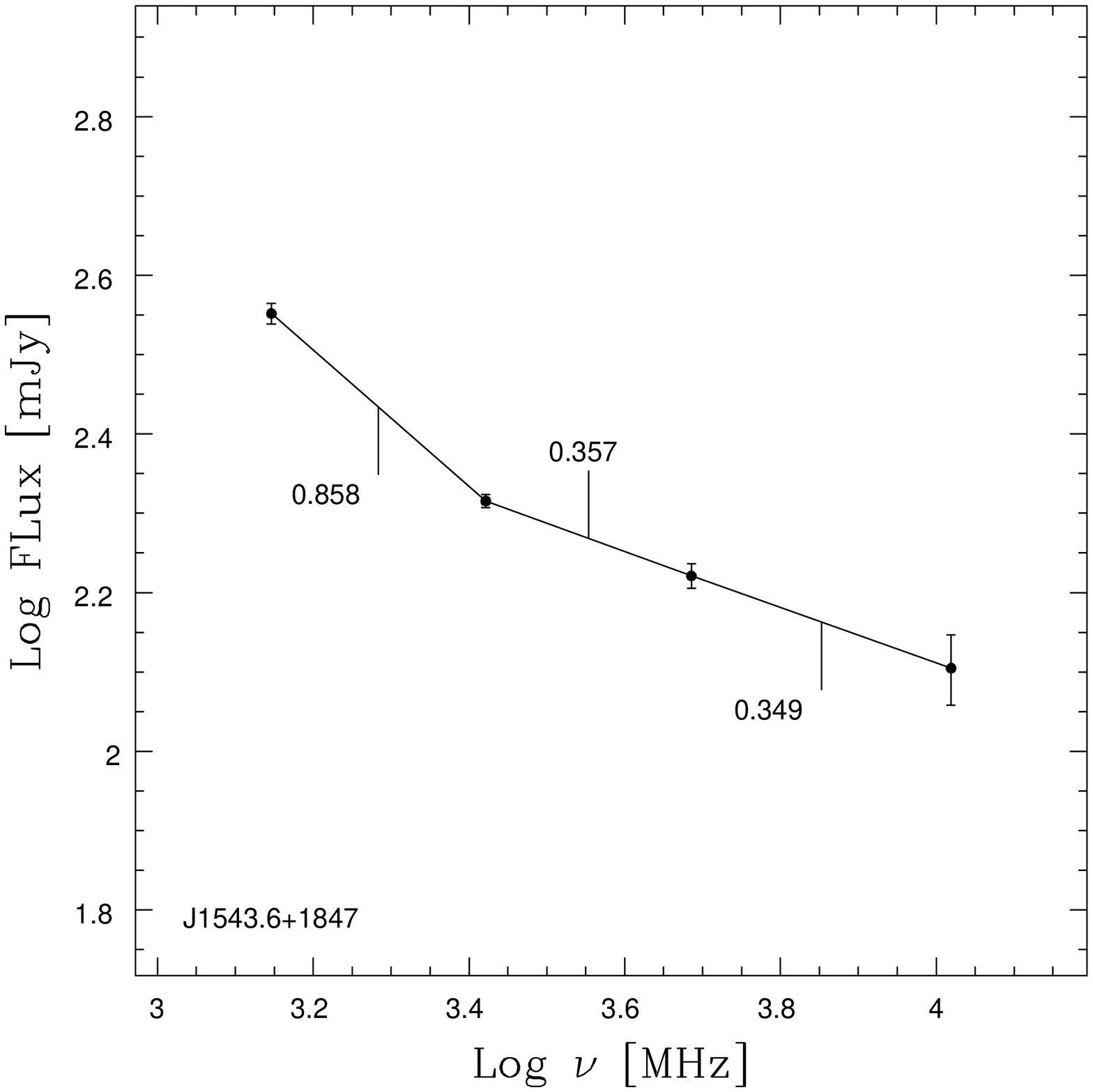}
\caption{Spectral index plots of sources in Table 3.}
\end{figure*}
\newpage
\begin{figure*}[t]
\addtocounter{figure}{+0}
\centering
\includegraphics[width=8cm]{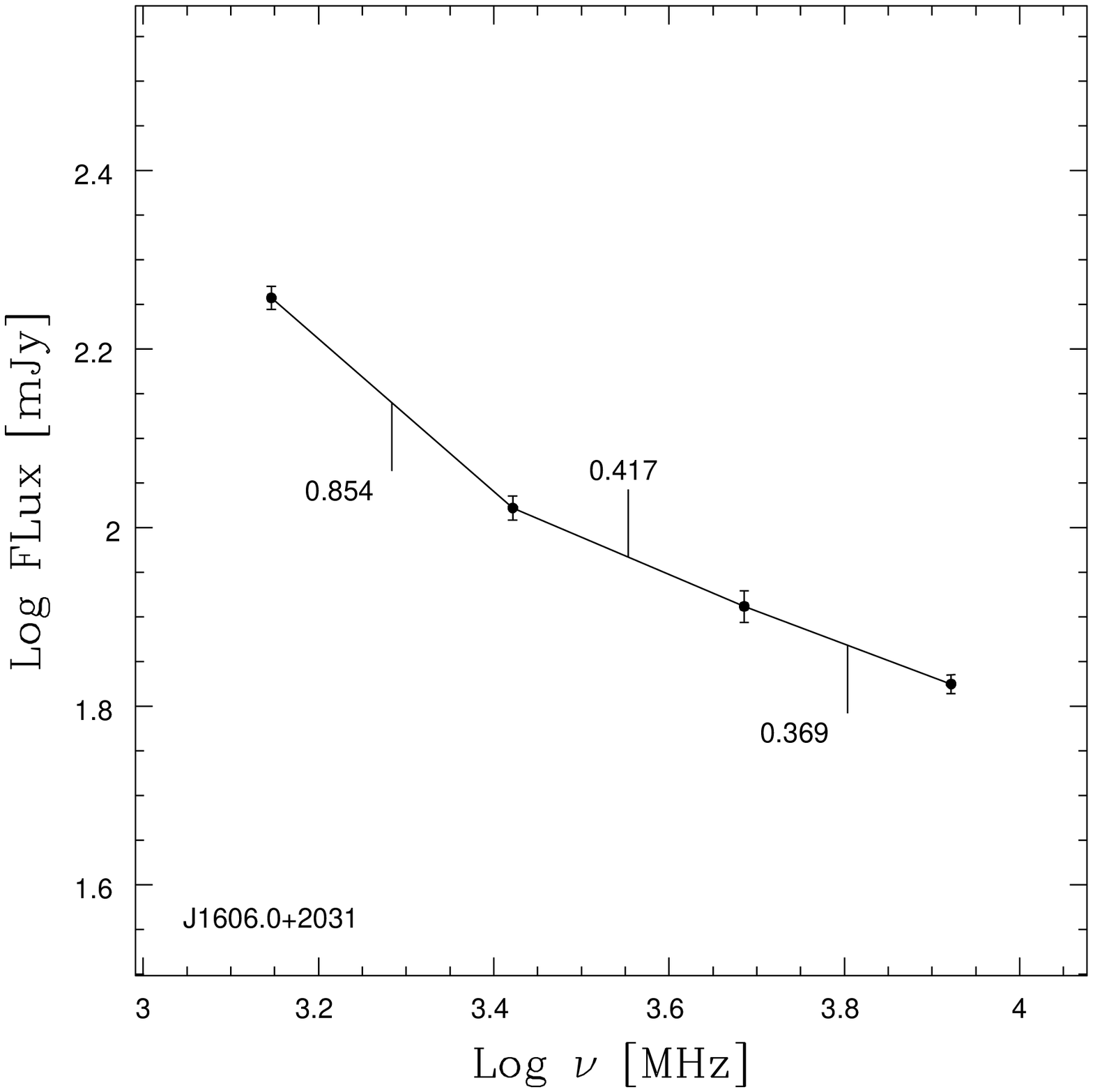}
\includegraphics[width=8cm]{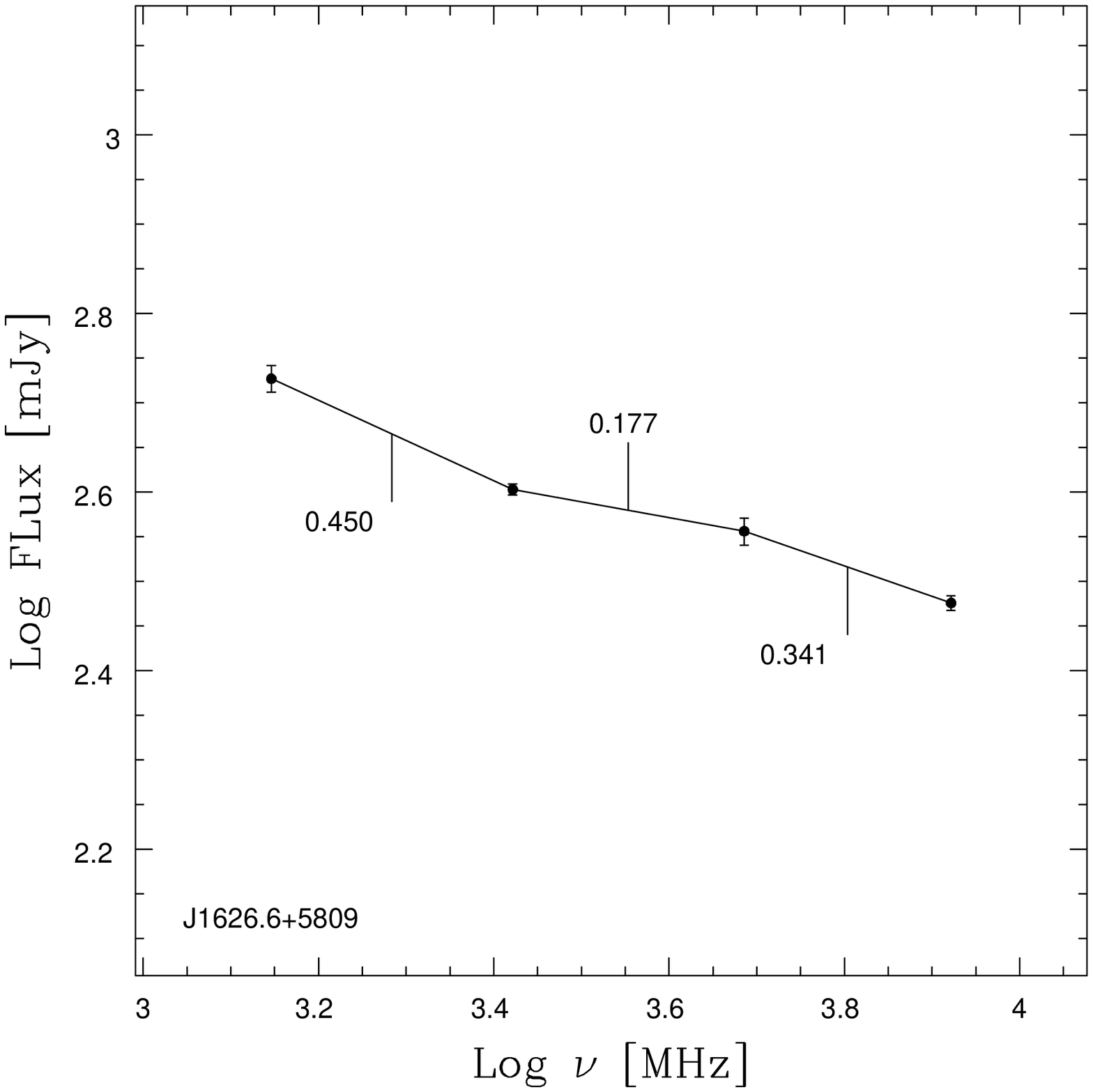}
\includegraphics[width=8cm]{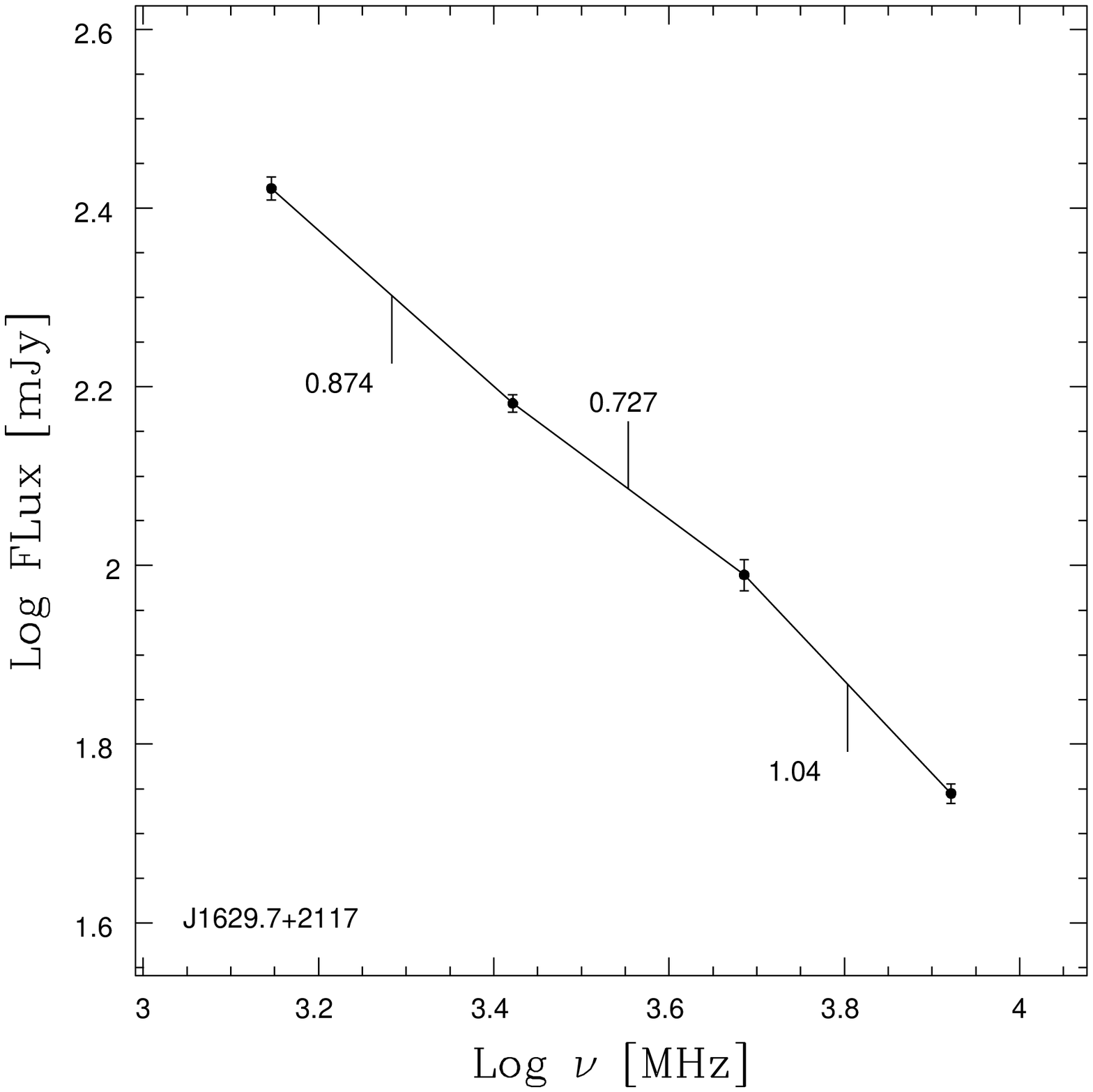}
\includegraphics[width=8cm]{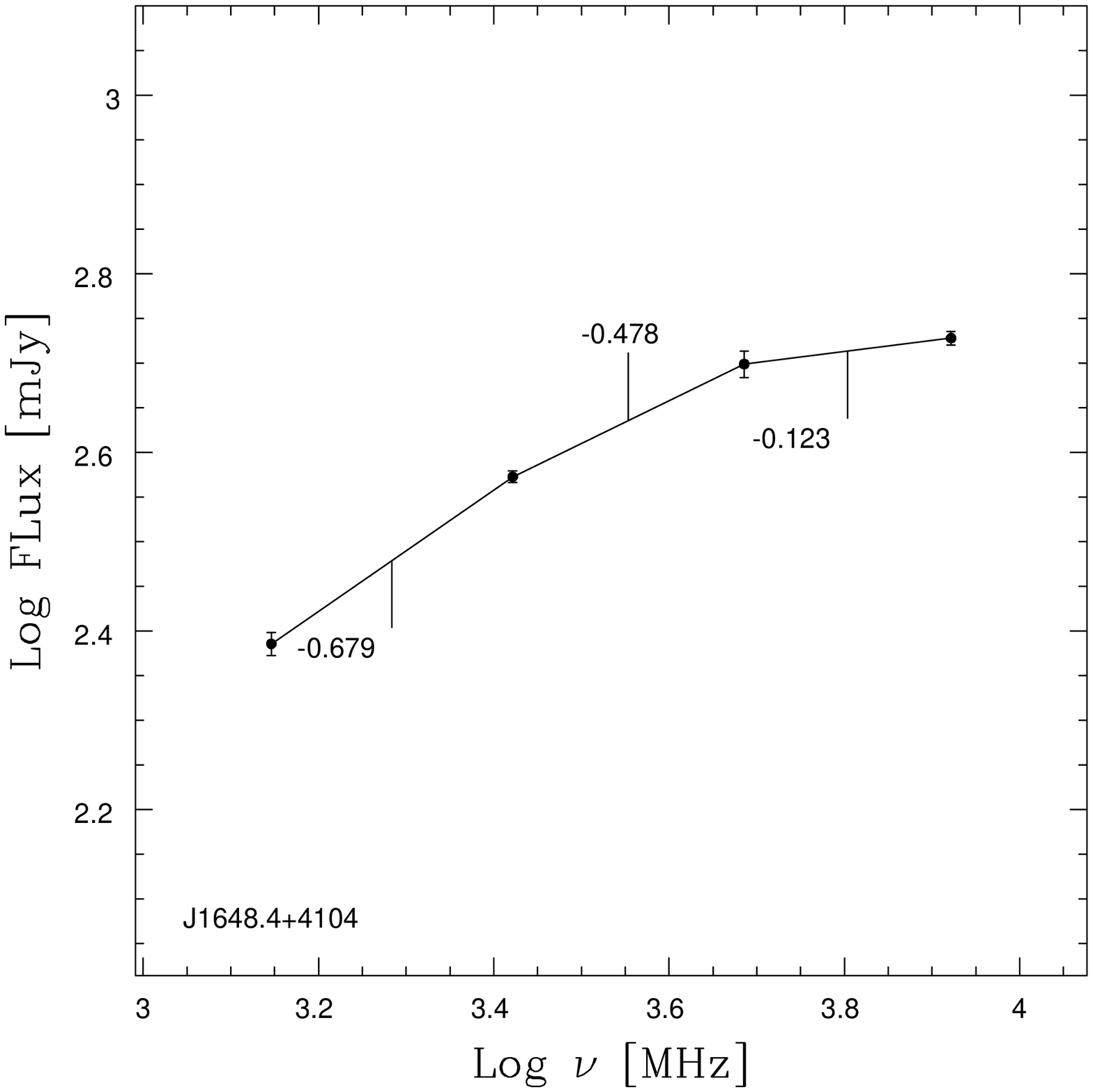}
\includegraphics[width=8cm]{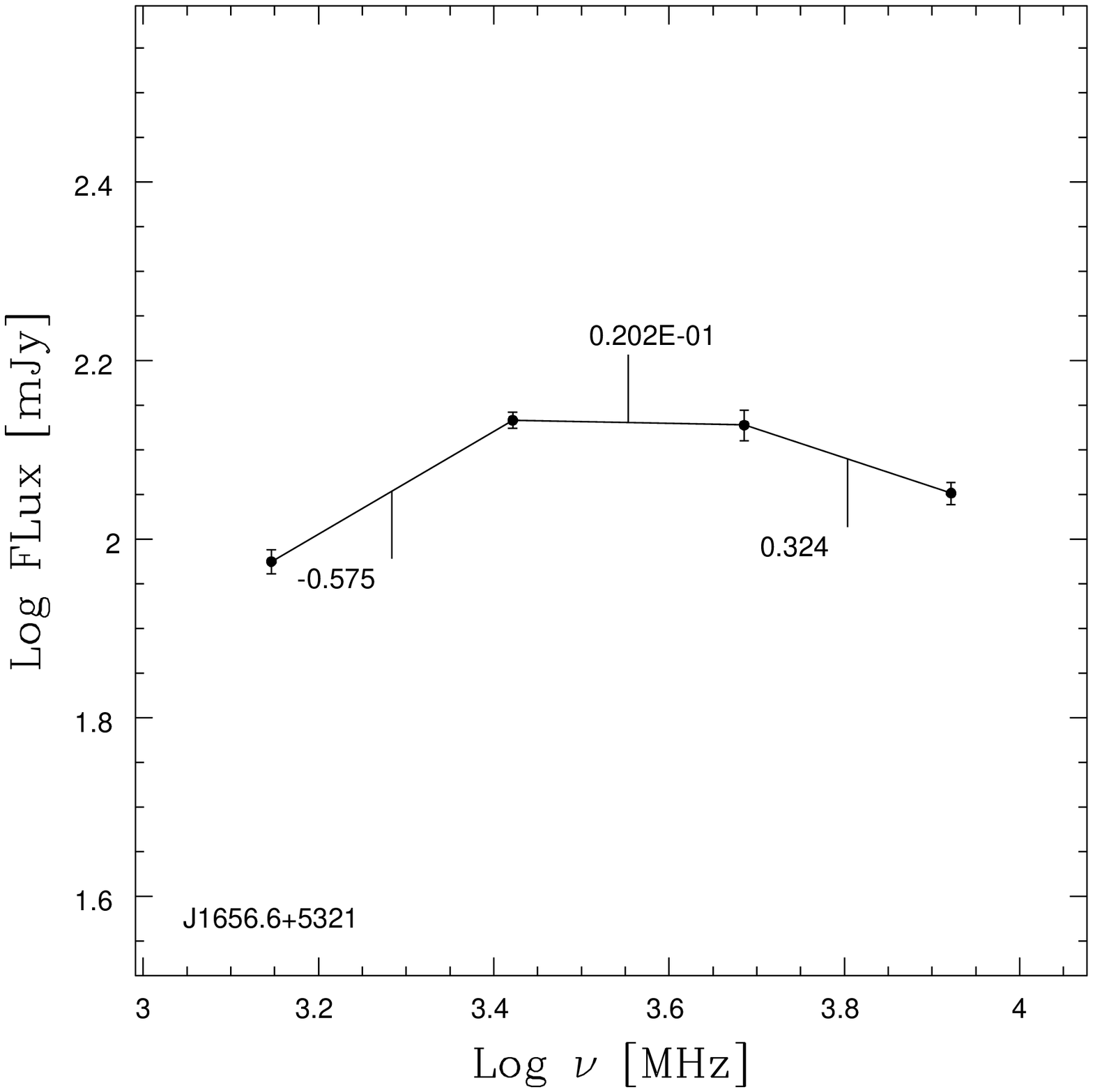}
\includegraphics[width=8cm]{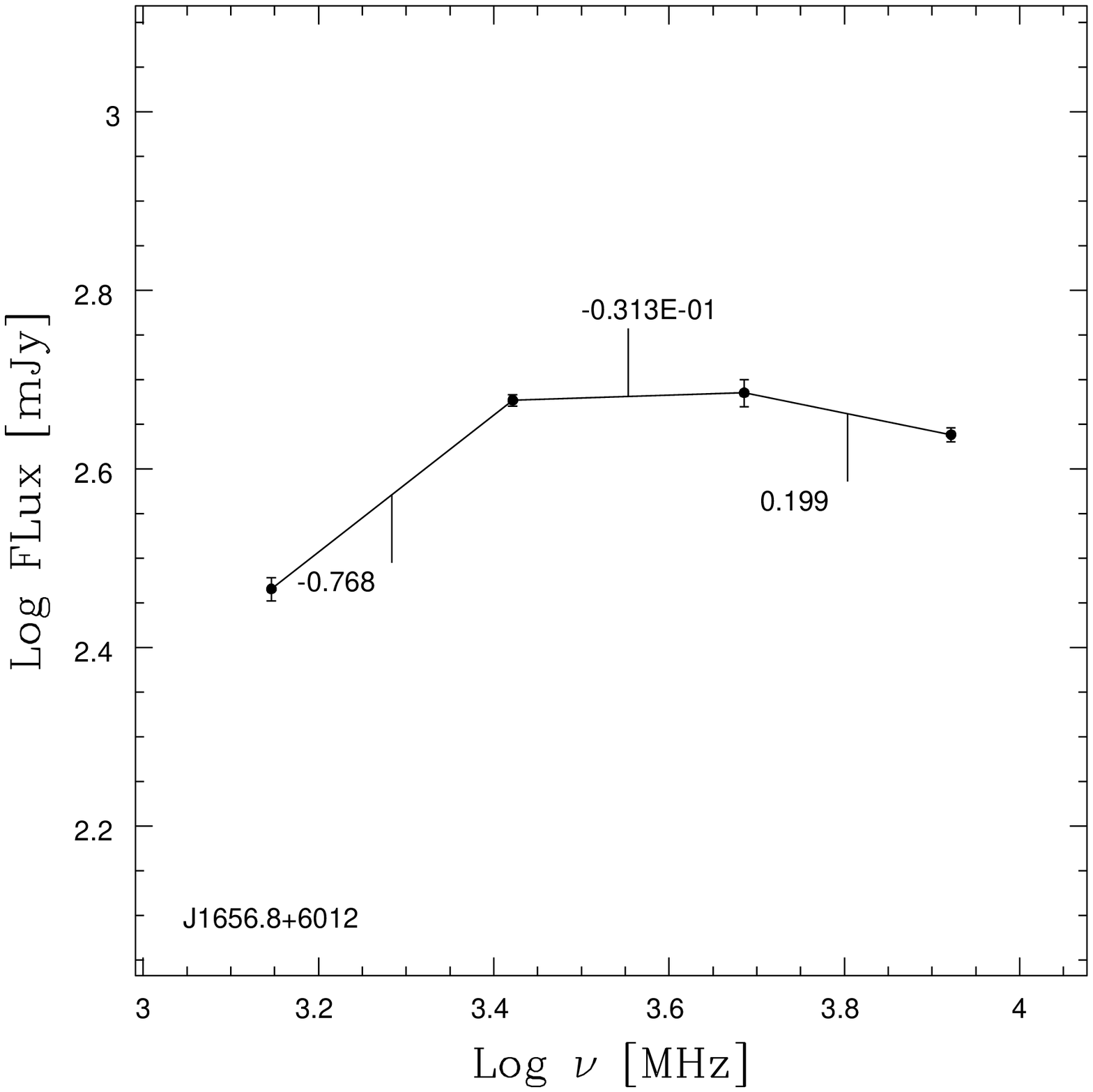}
\caption{Spectral index plots of sources in Table 3.}
\end{figure*}
\newpage
\begin{figure*}[t]
\addtocounter{figure}{+0}
\centering
\includegraphics[width=8cm]{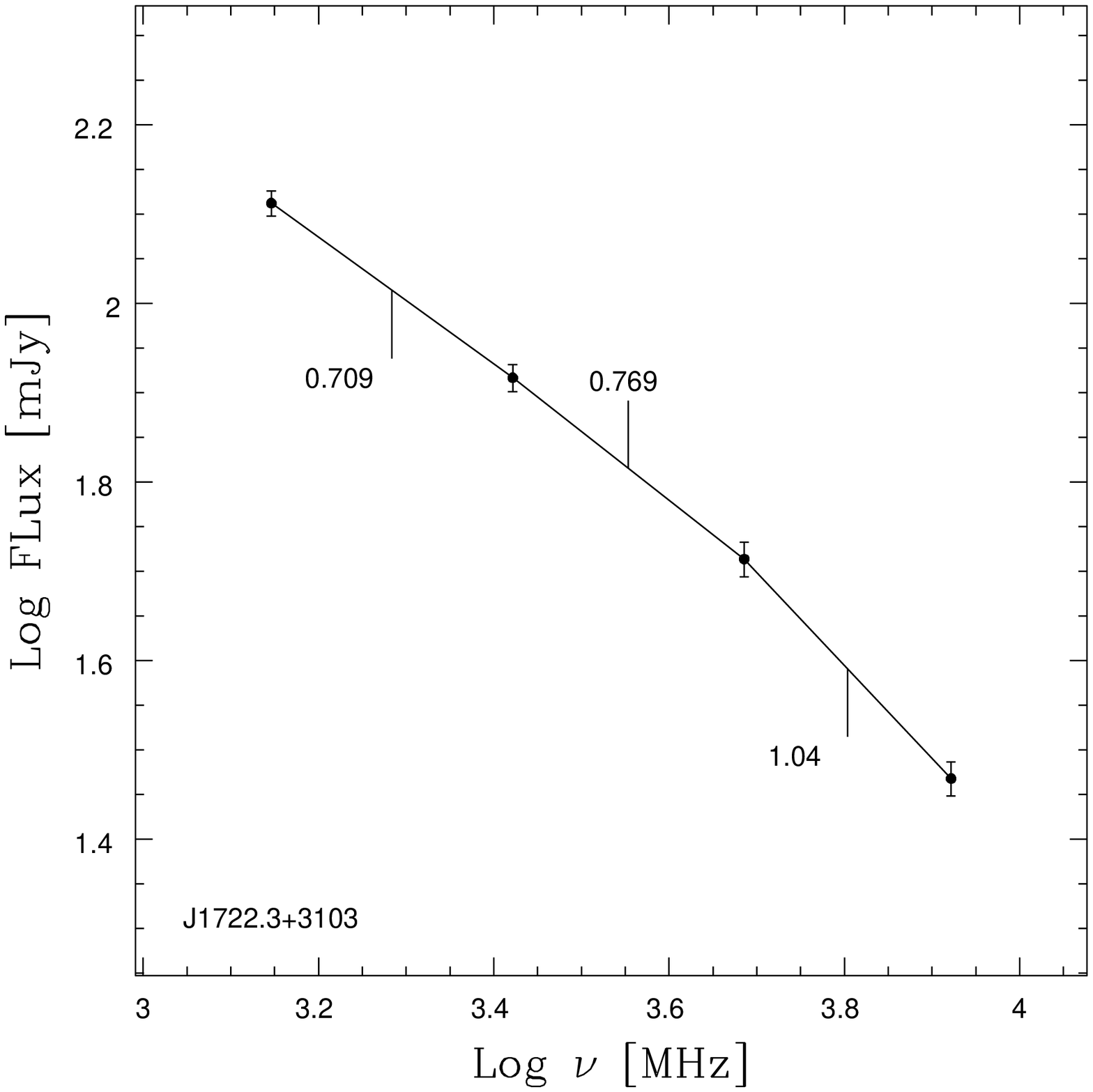}
\includegraphics[width=8cm]{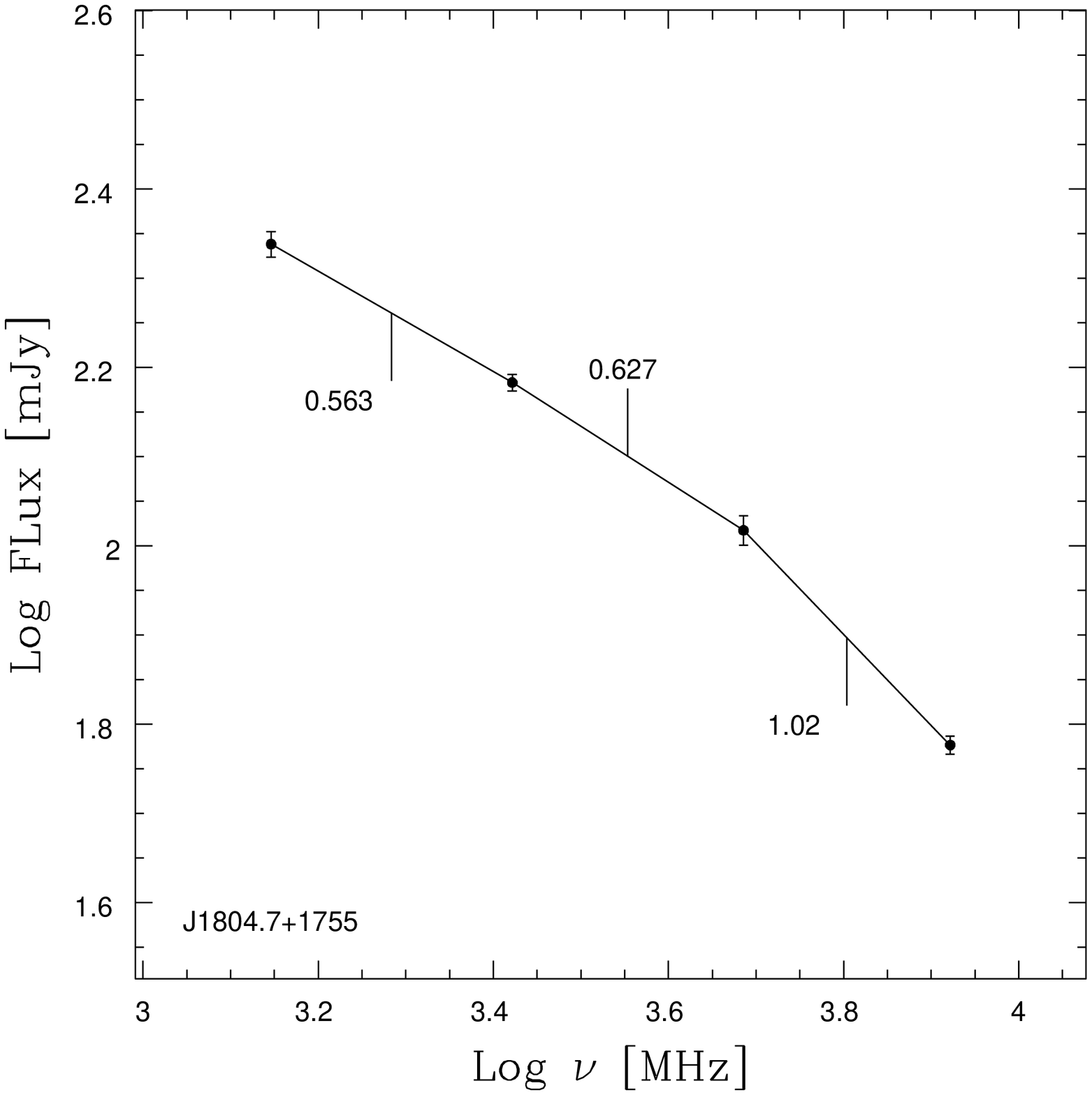}
\includegraphics[width=8cm]{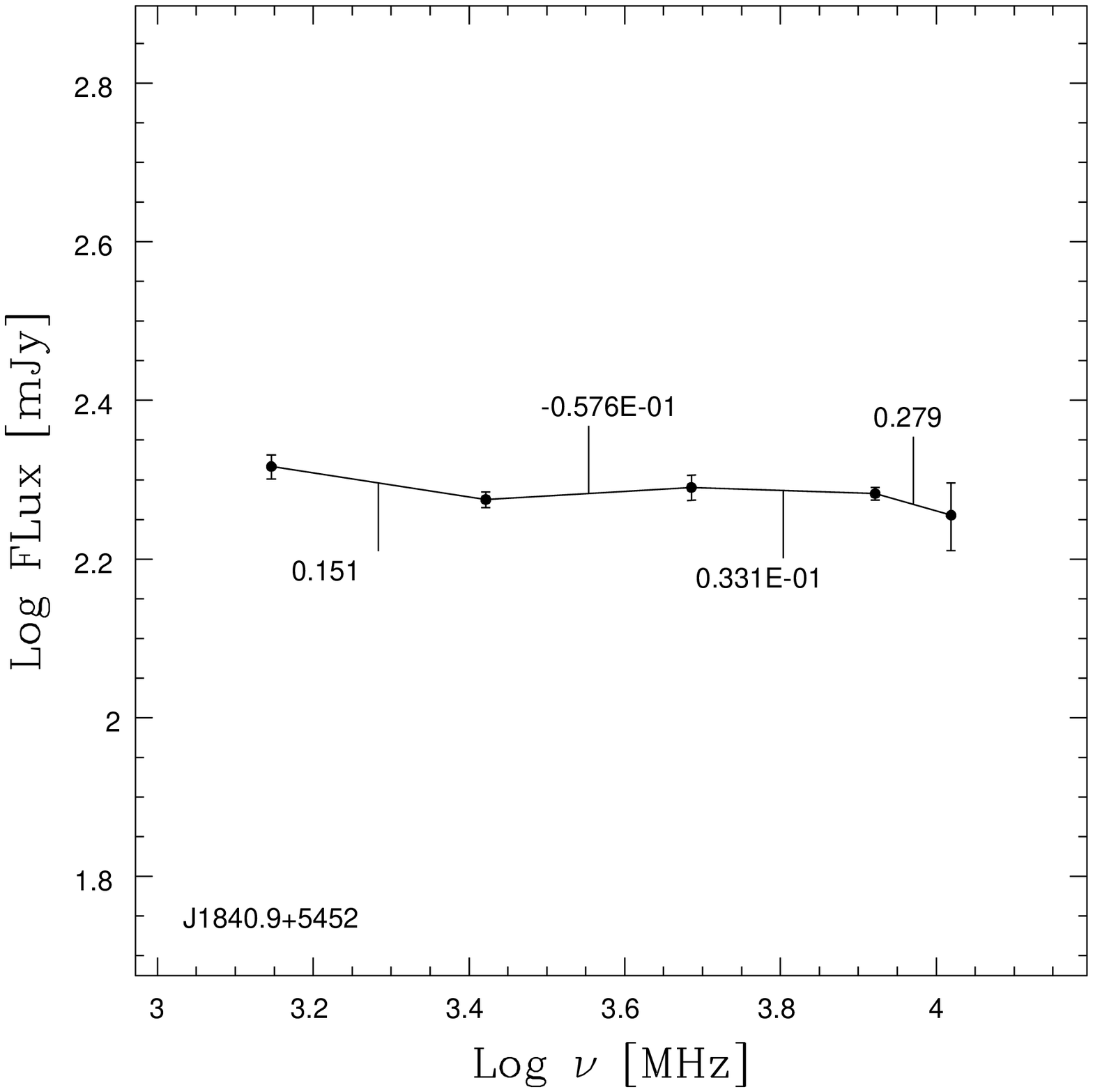}
\includegraphics[width=8cm]{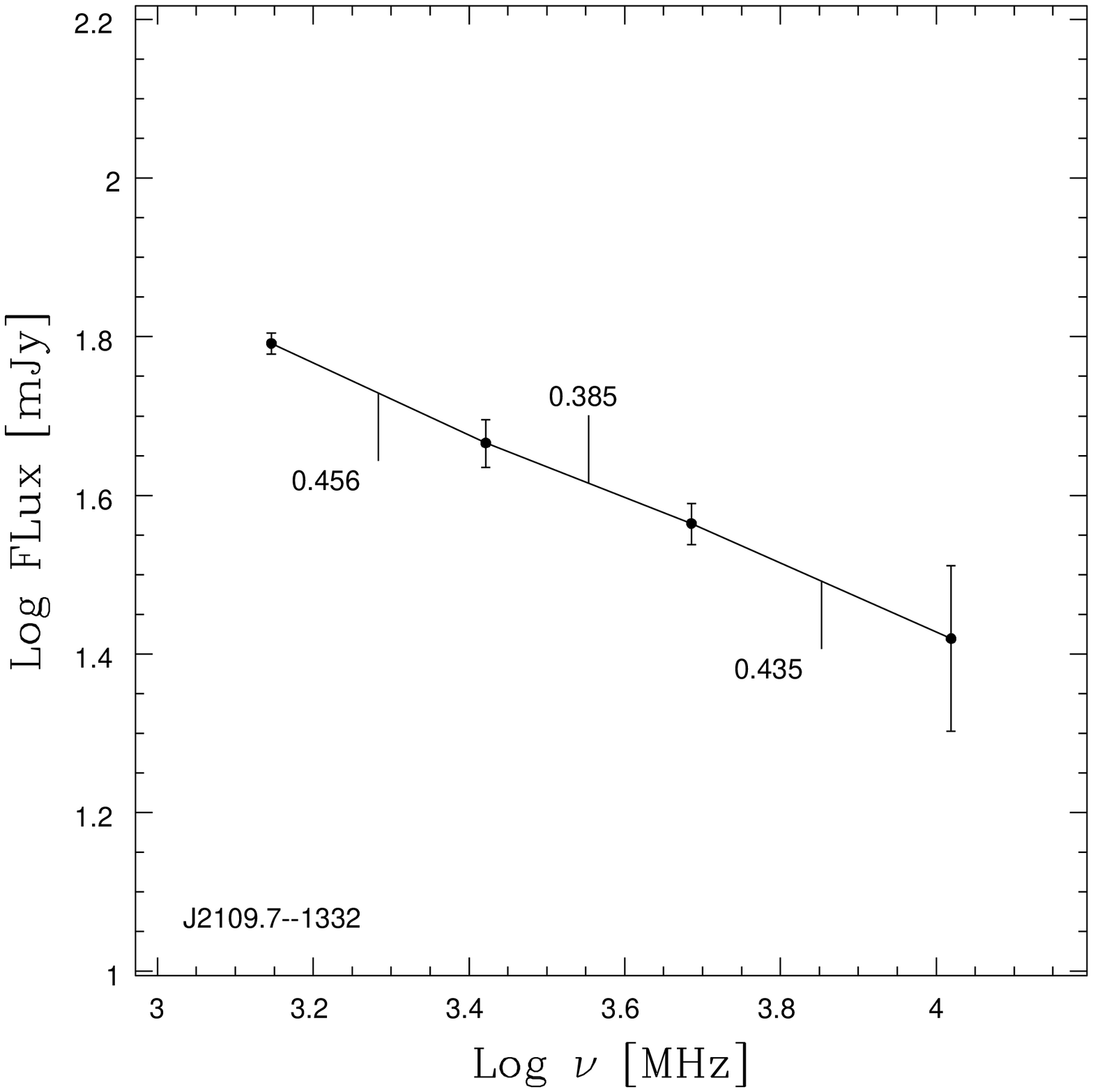}
\includegraphics[width=8cm]{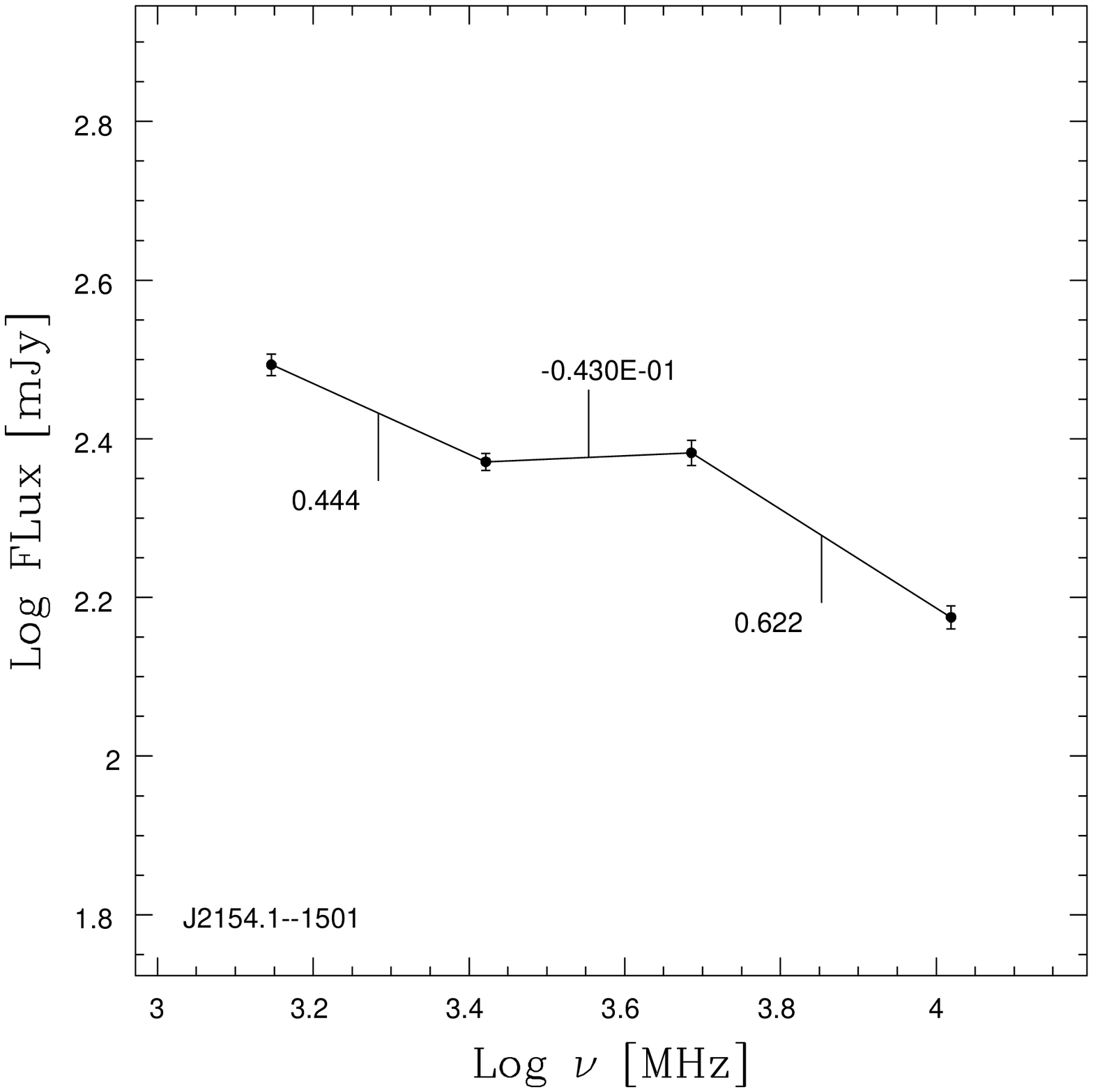}
\includegraphics[width=8cm]{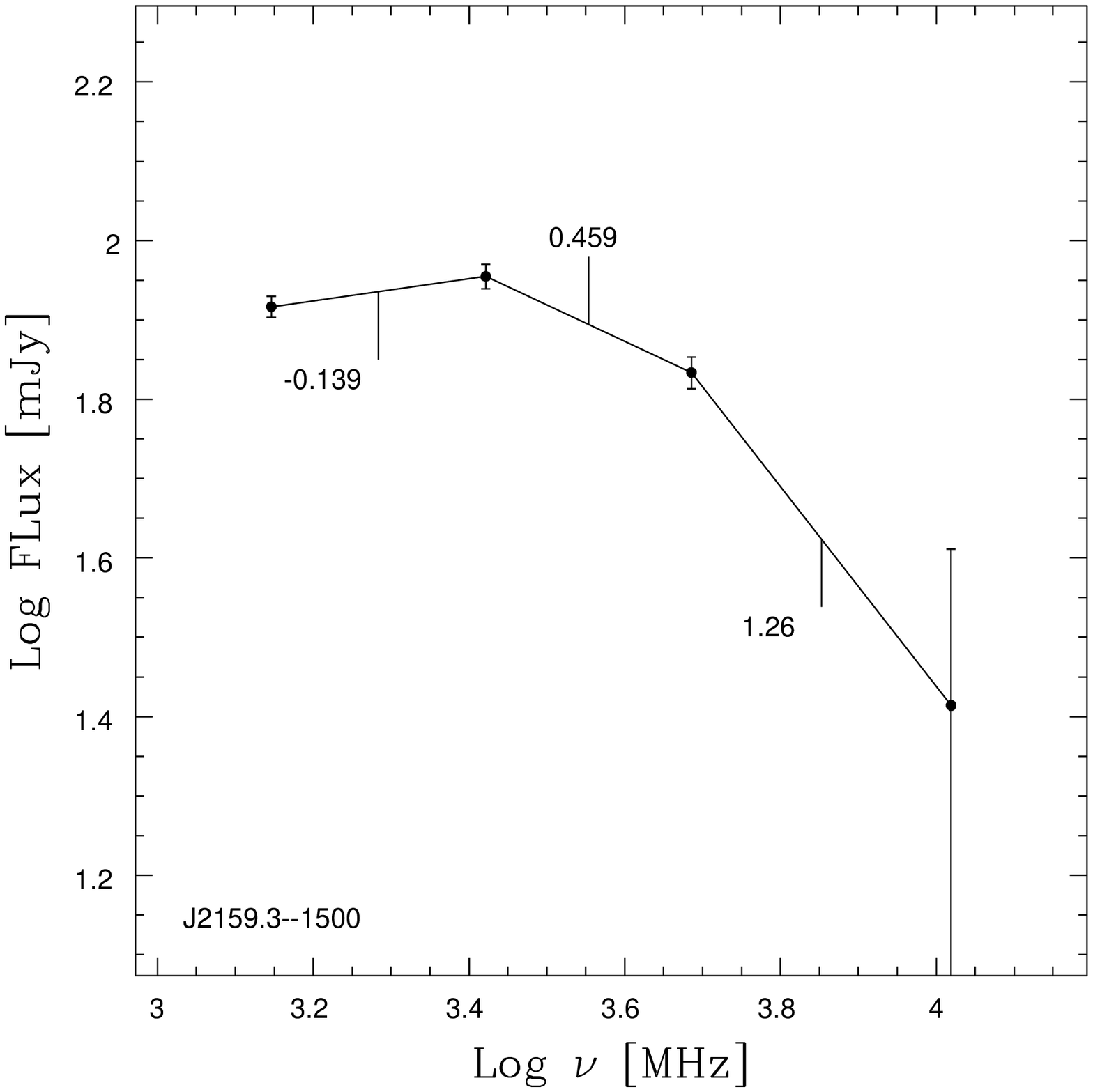}
\caption{Spectral index plots of sources in Table 3.}
\end{figure*}
\newpage
\begin{figure*}[t]
\addtocounter{figure}{+0}
\centering
\includegraphics[width=8cm]{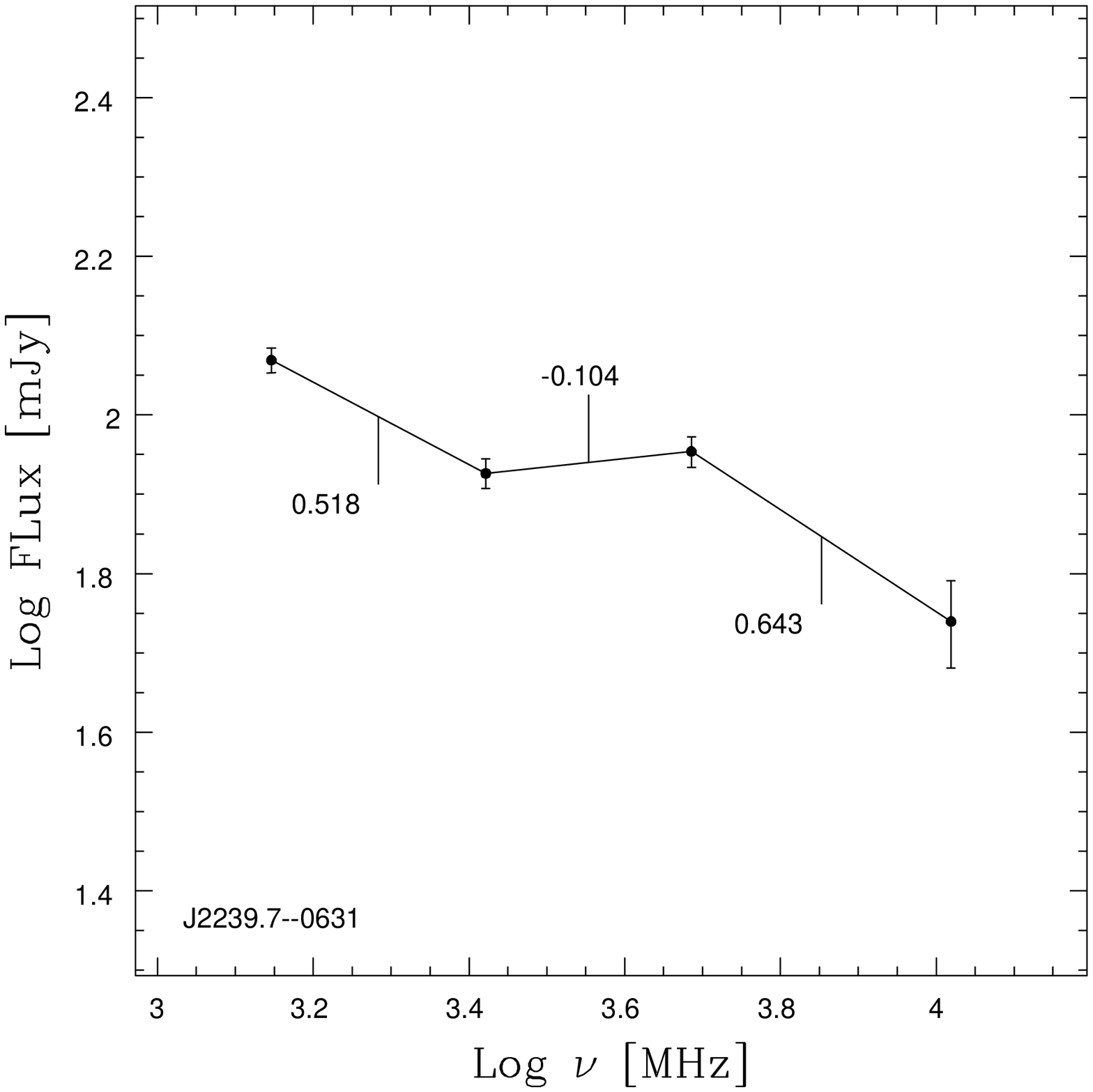}
\includegraphics[width=8cm]{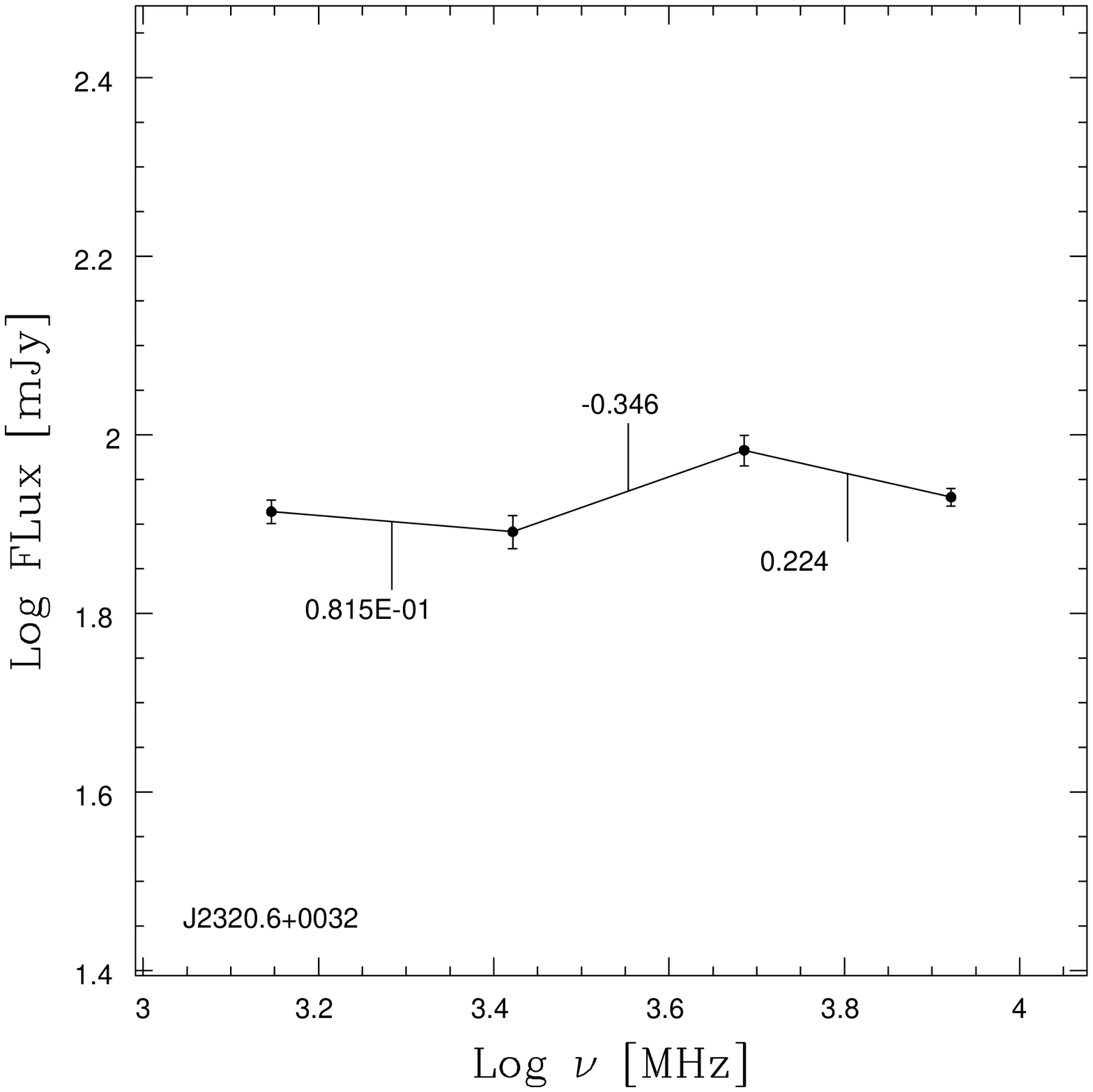}
\includegraphics[width=8cm]{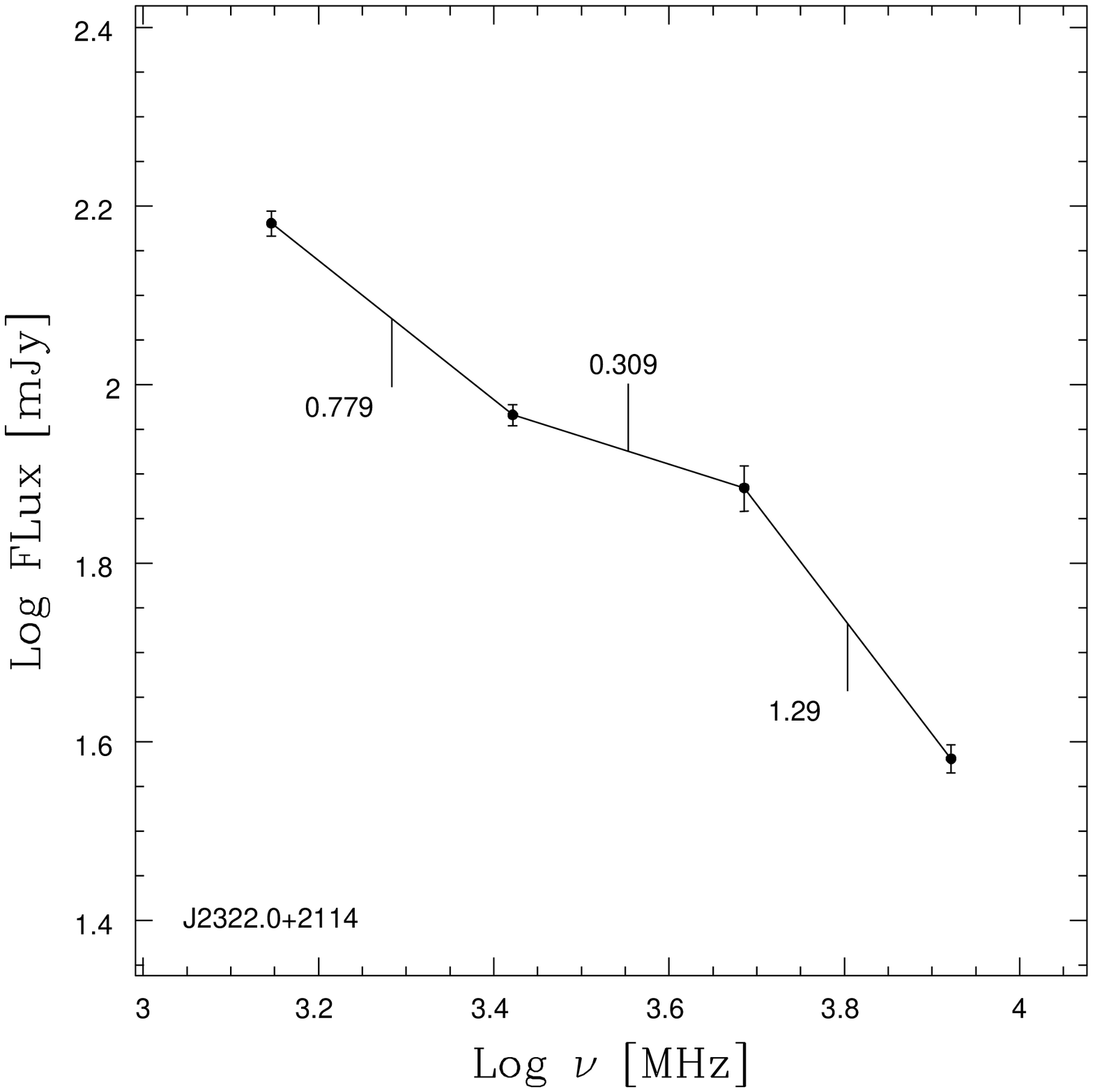}
\includegraphics[width=8cm]{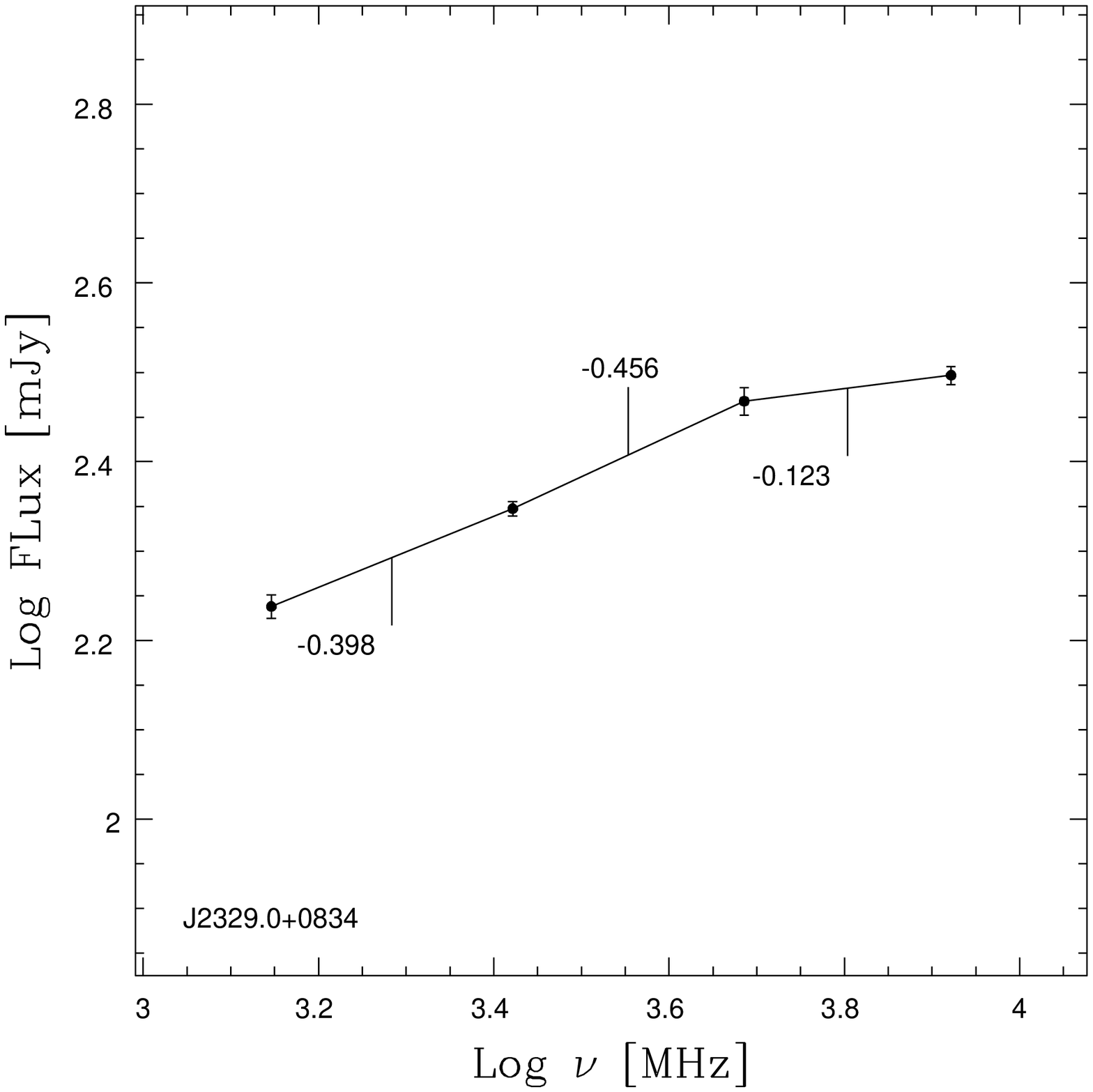}
\includegraphics[width=8cm]{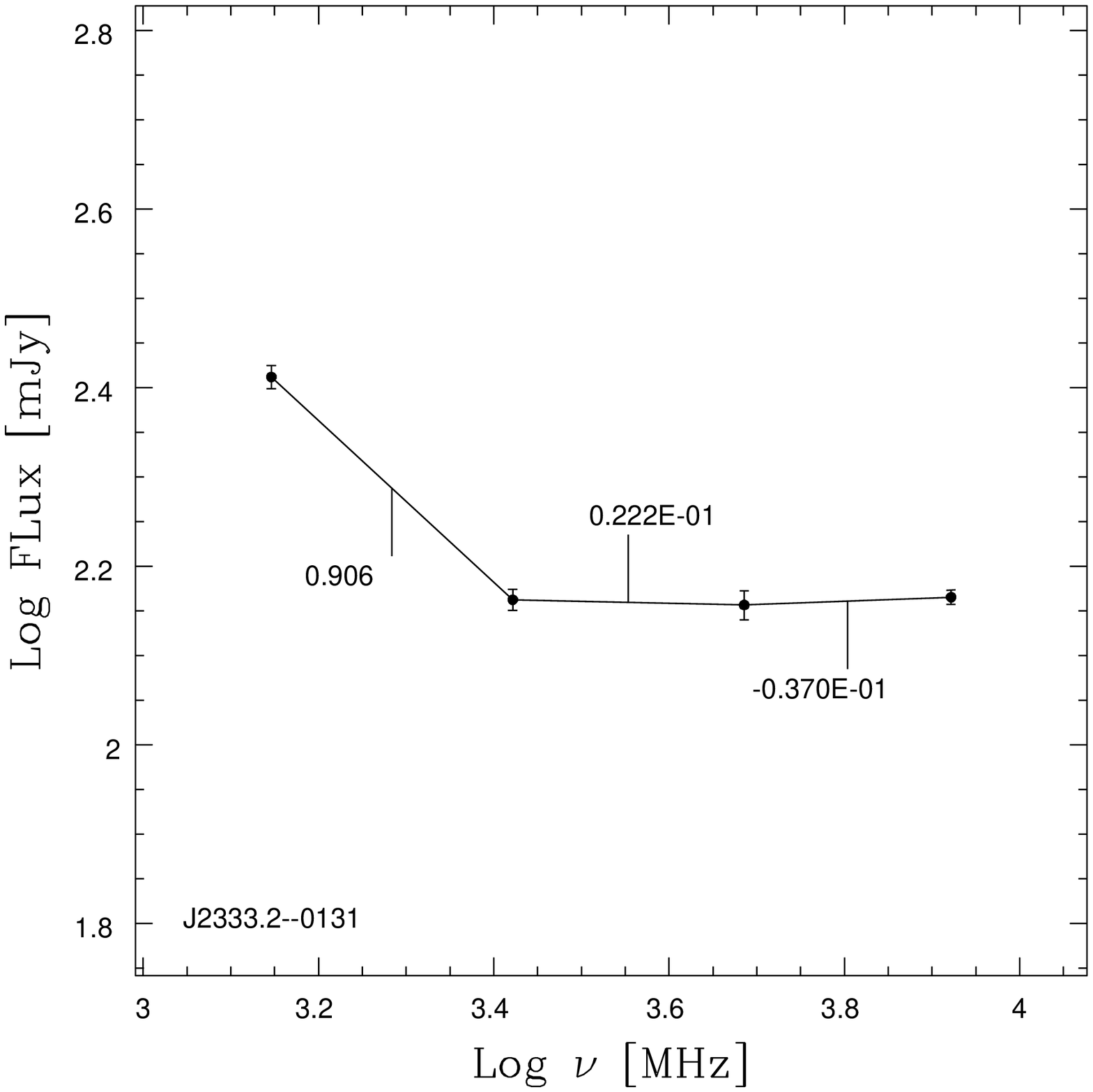}
\includegraphics[width=8cm]{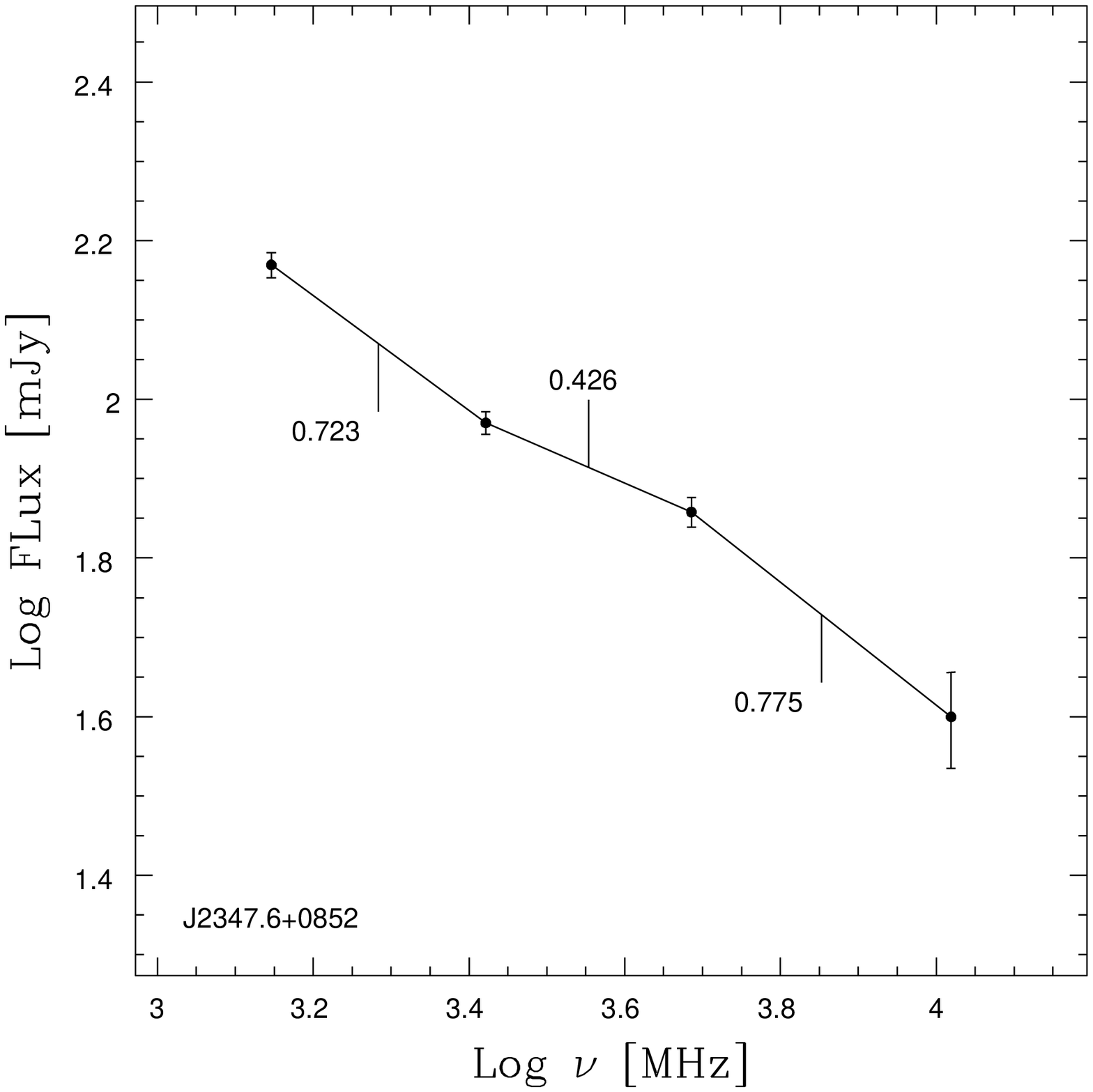}
\caption{Spectral index plots of sources in Table 3.}
\end{figure*}
\clearpage
\newpage
%
\section{RM and m plots}
\newpage
\begin{figure*}[t]
\addtocounter{figure}{+0}
\centering
\includegraphics[width=7cm]{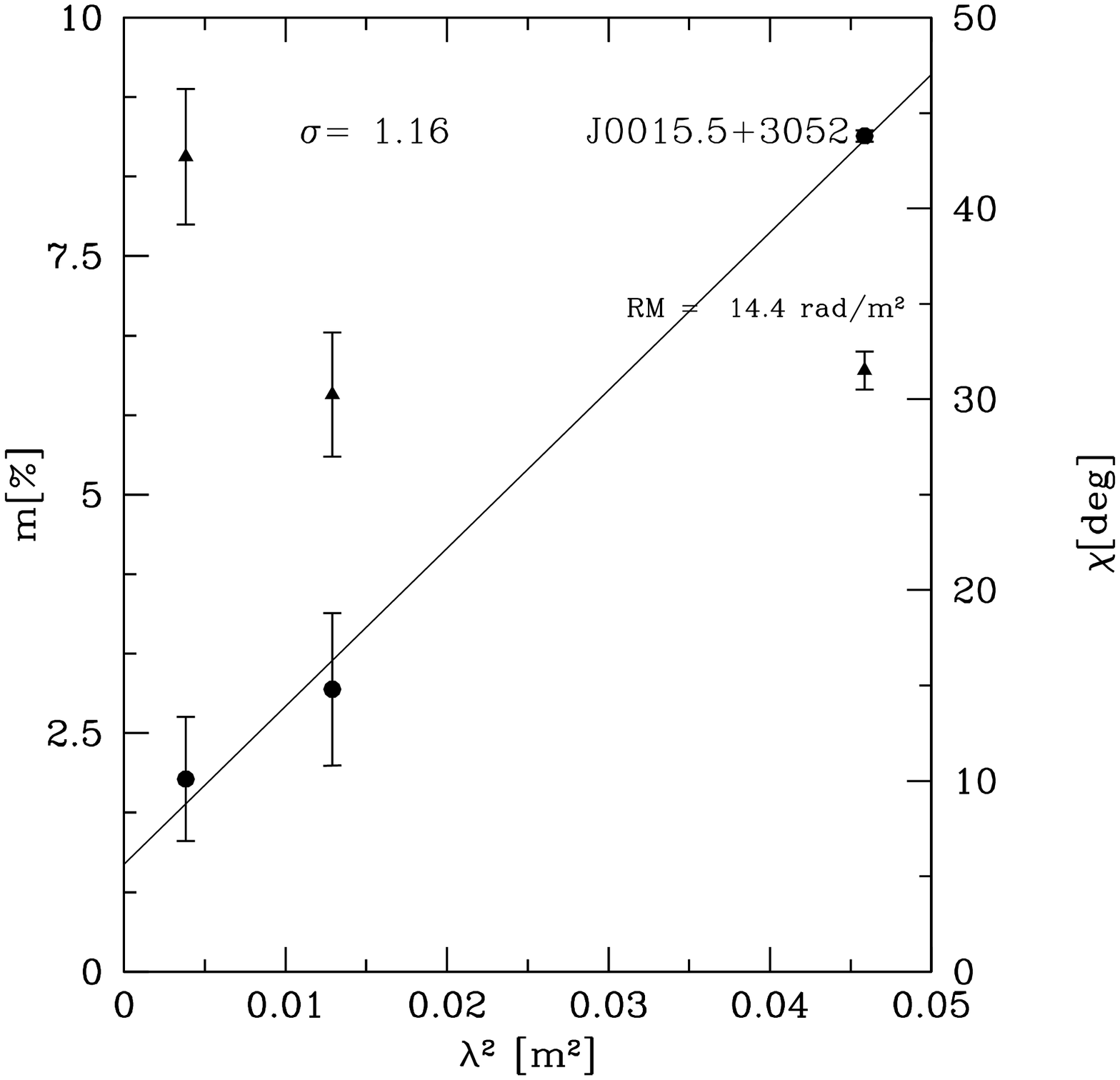}
\includegraphics[width=7cm]{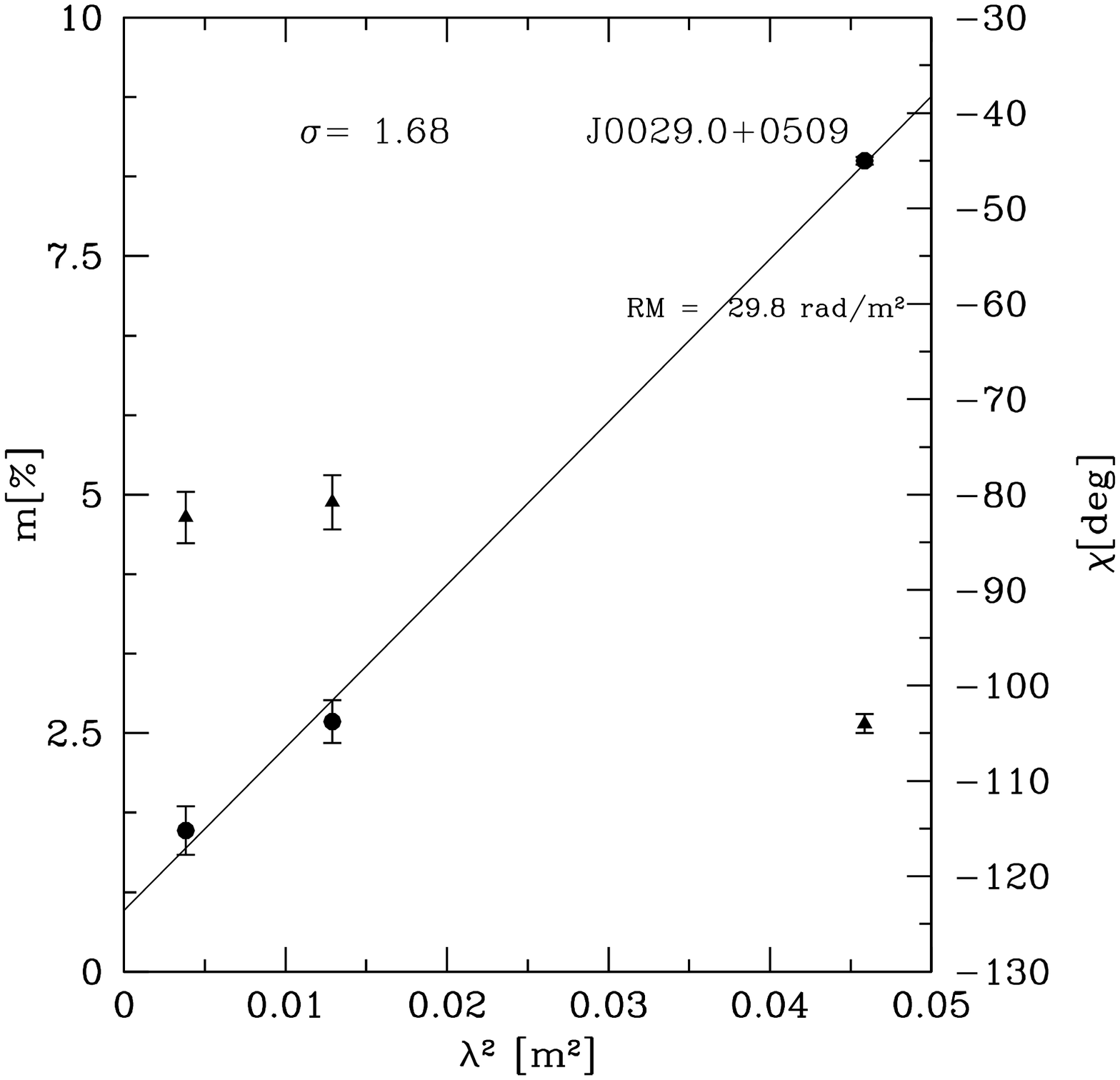}
\includegraphics[width=7cm]{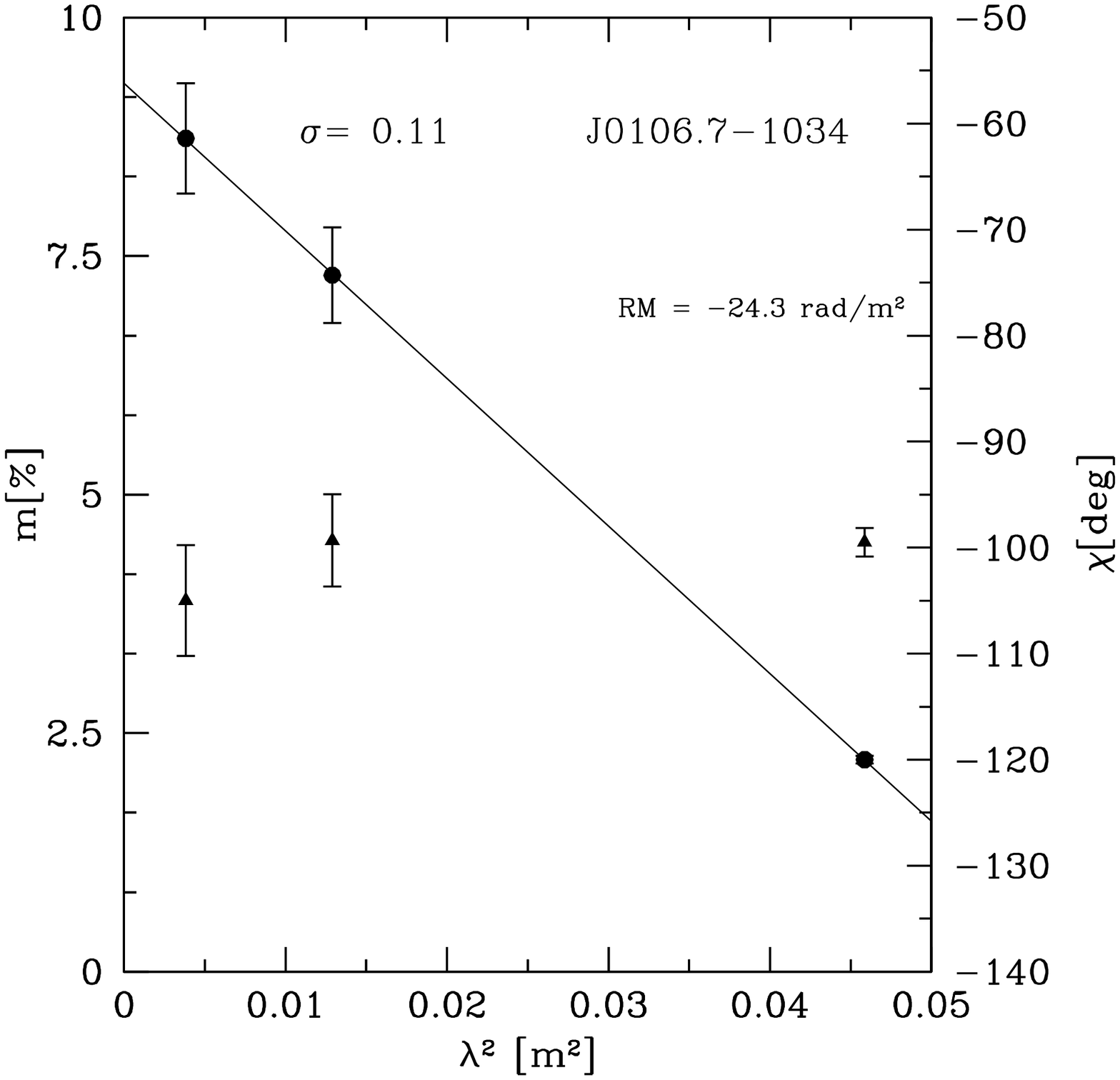}
\includegraphics[width=7cm]{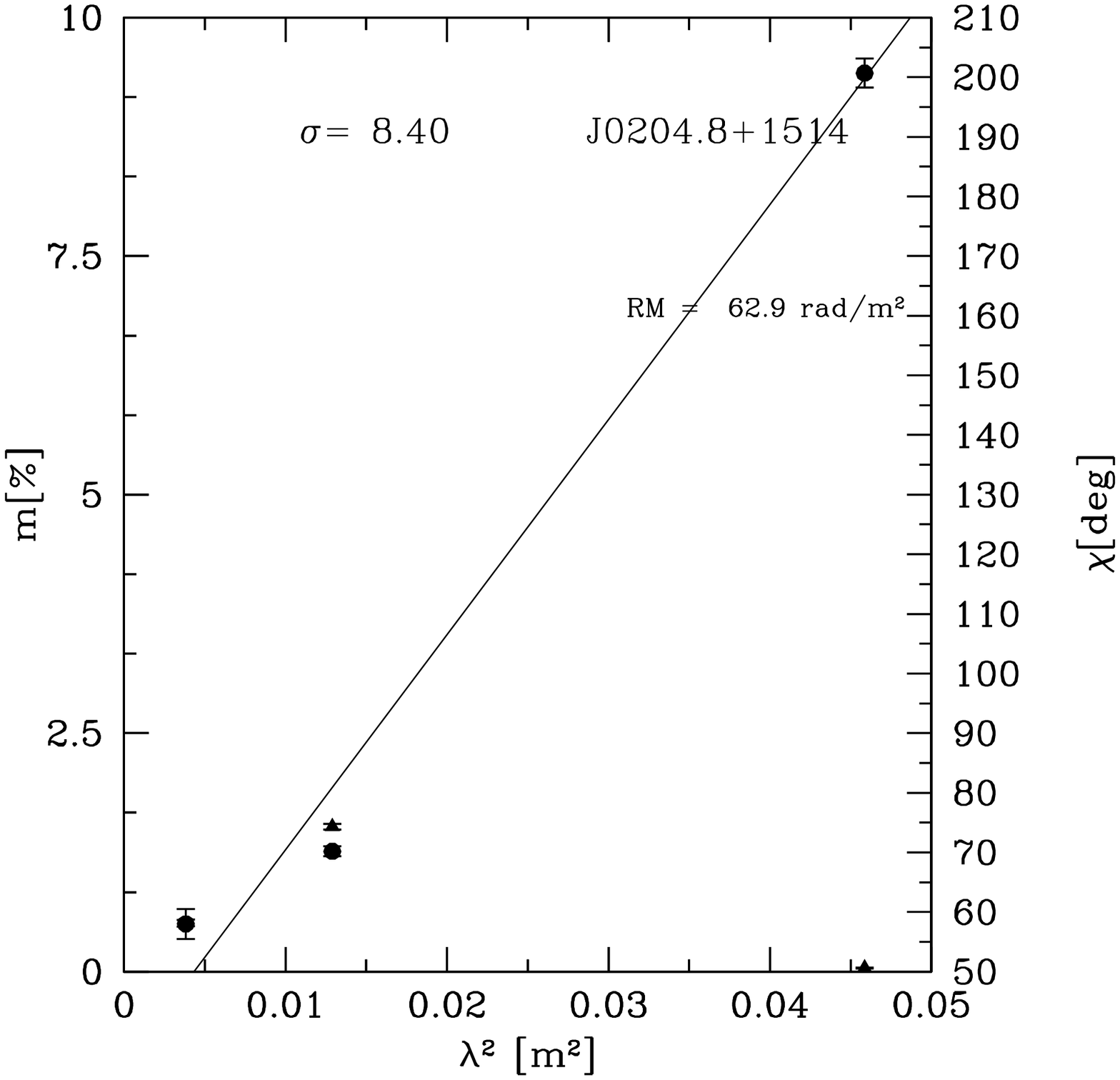}
\includegraphics[width=7cm]{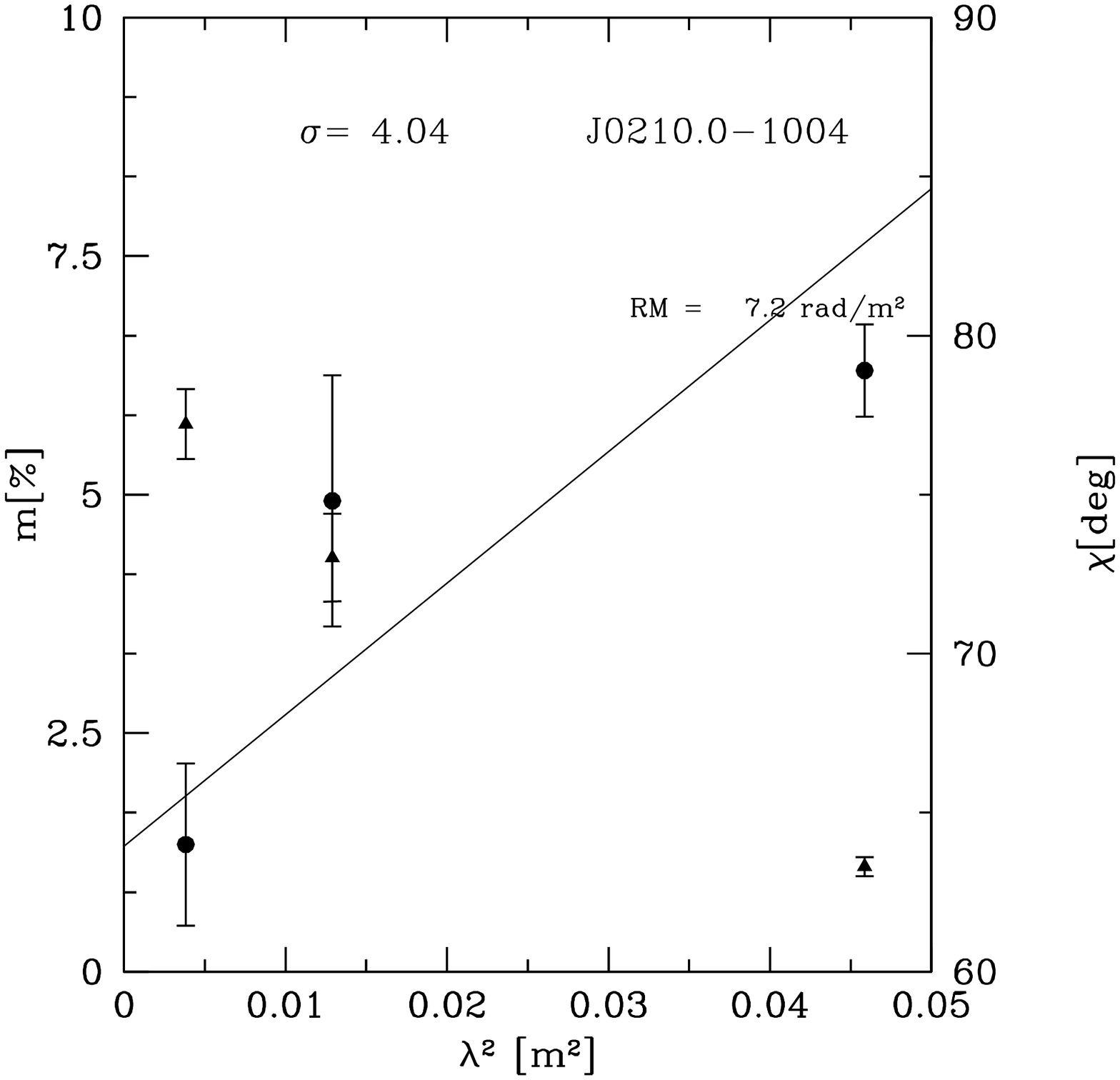}
\includegraphics[width=7cm]{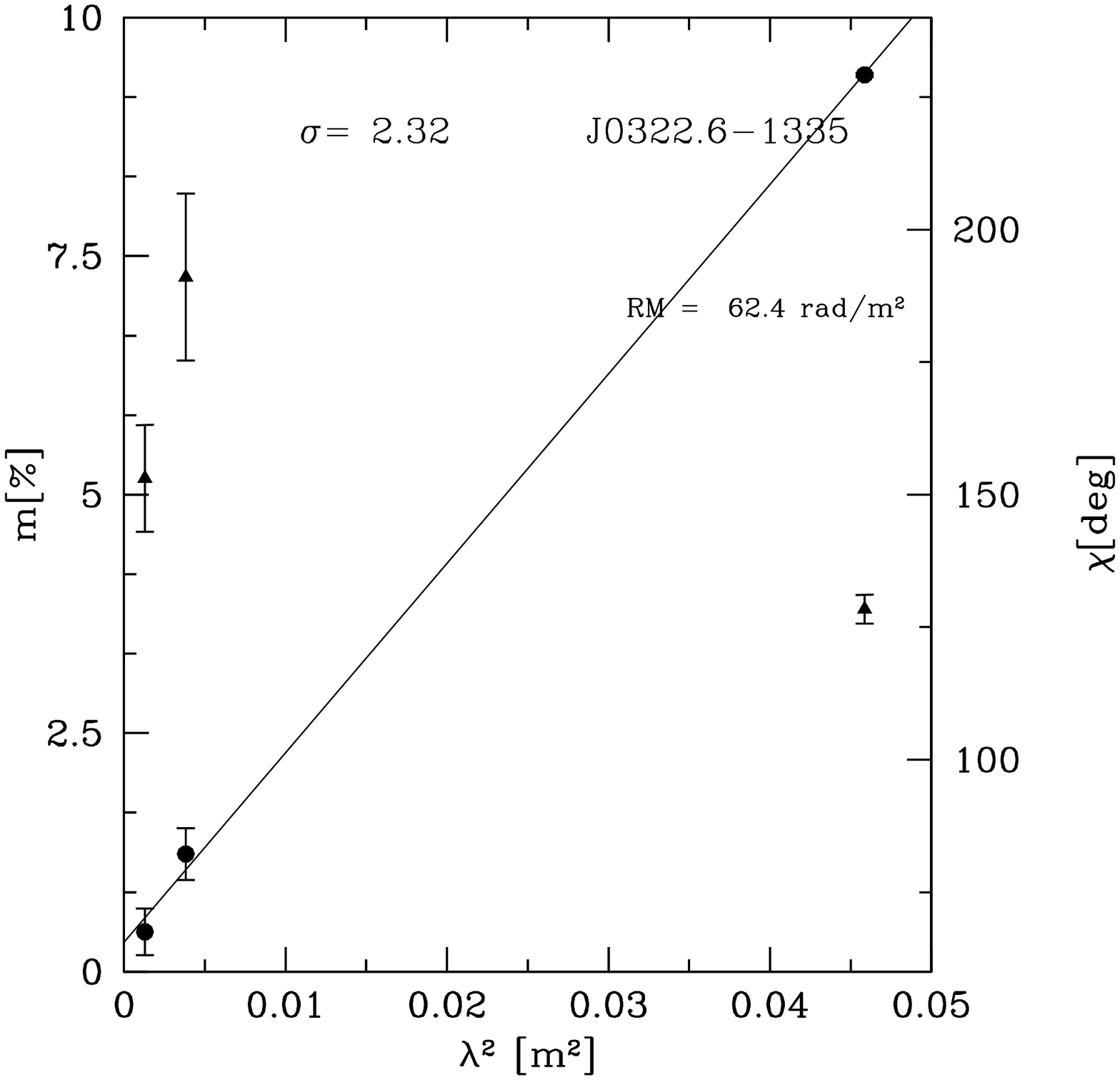}
\caption{Position angles of the electric vector $\chi$ (dots)
and fractional polarisation $m$ (triangles) versus $\lambda^2$ plots
of sources in Table 4. $\sigma$
values assess the quality of the best fit.}
\end{figure*}
\newpage
\begin{figure*}[t]
\addtocounter{figure}{+0}
\centering
\includegraphics[width=7cm]{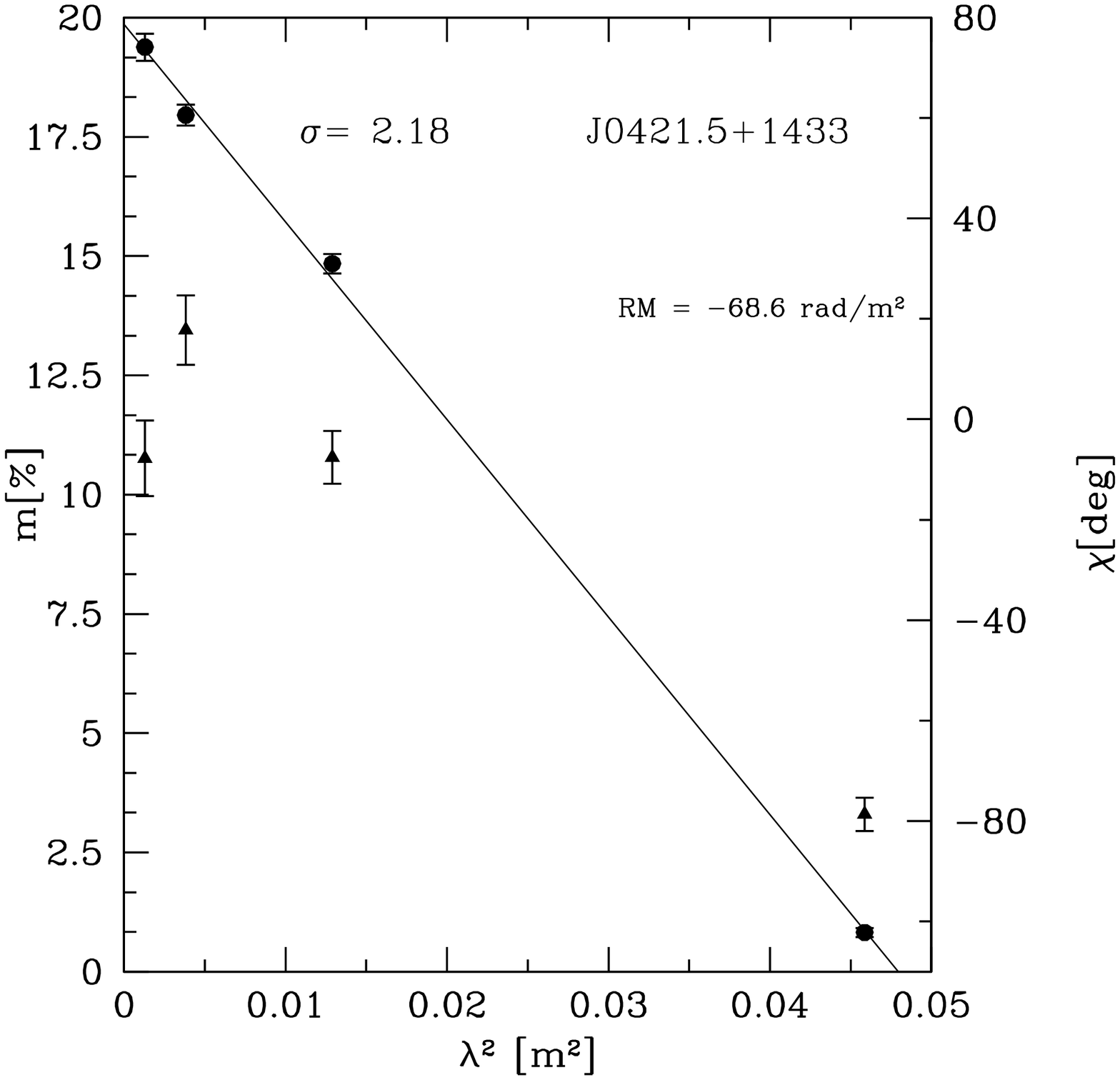}
\includegraphics[width=7cm]{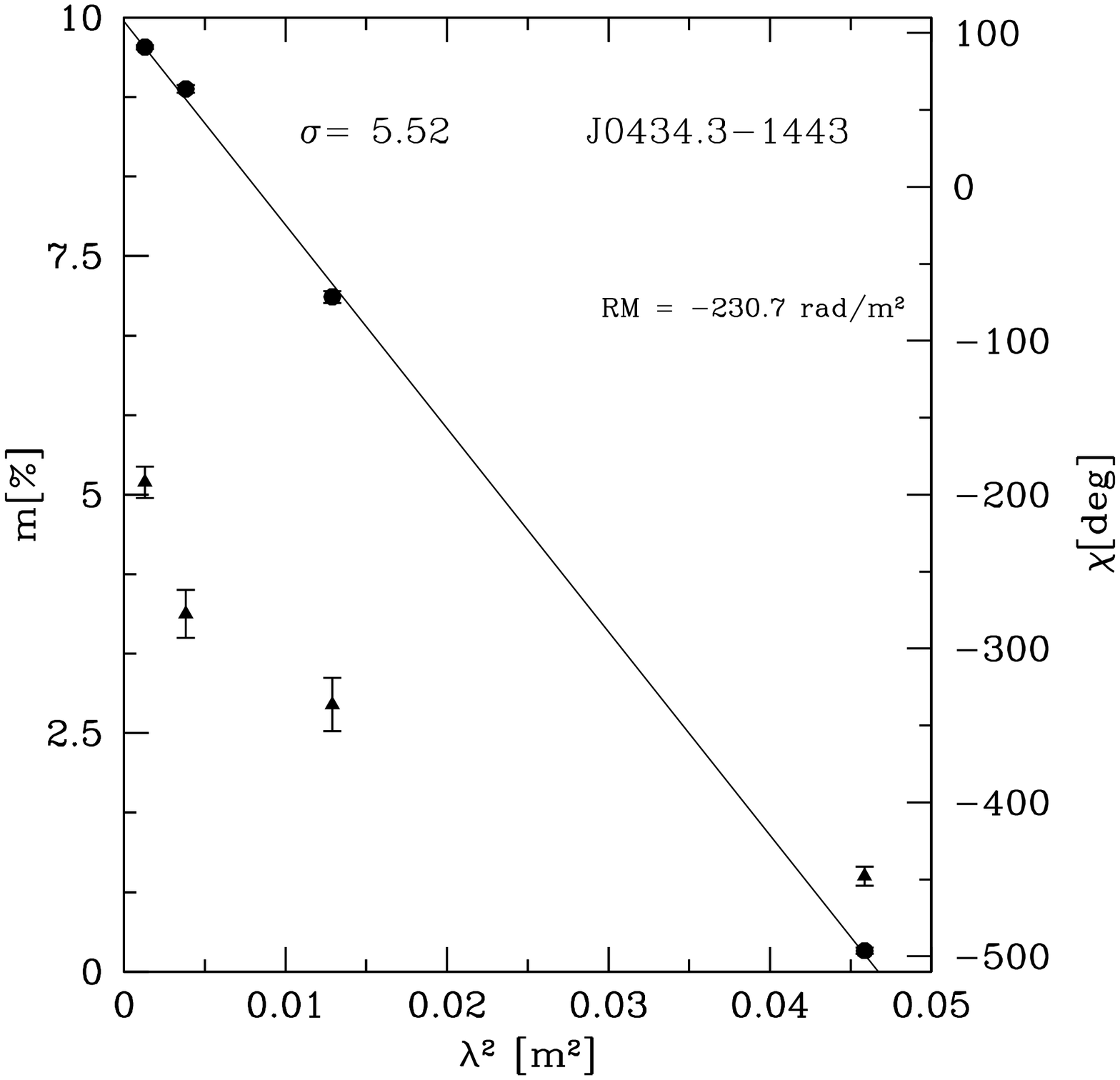}
\includegraphics[width=7cm]{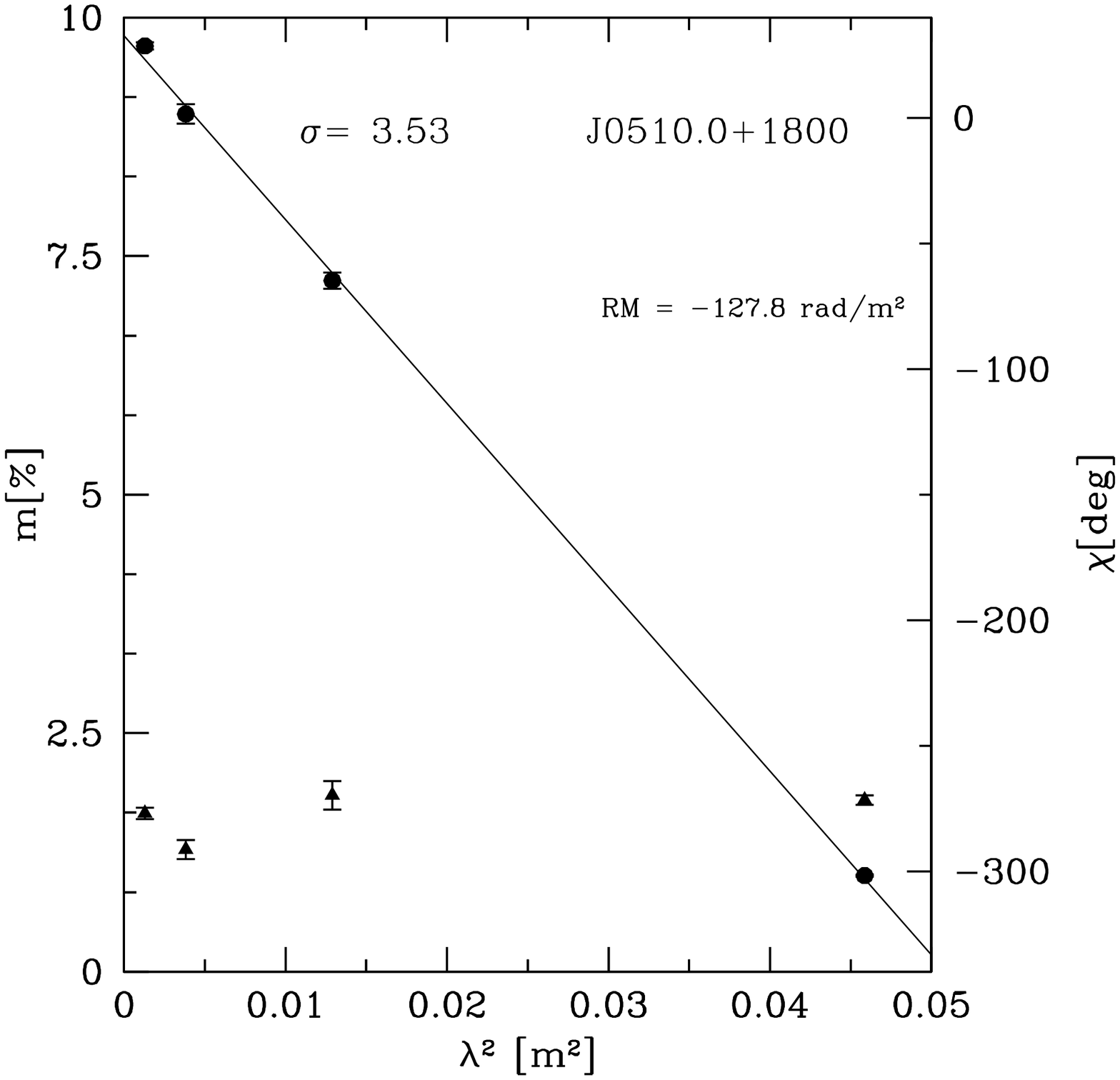}
\includegraphics[width=7cm]{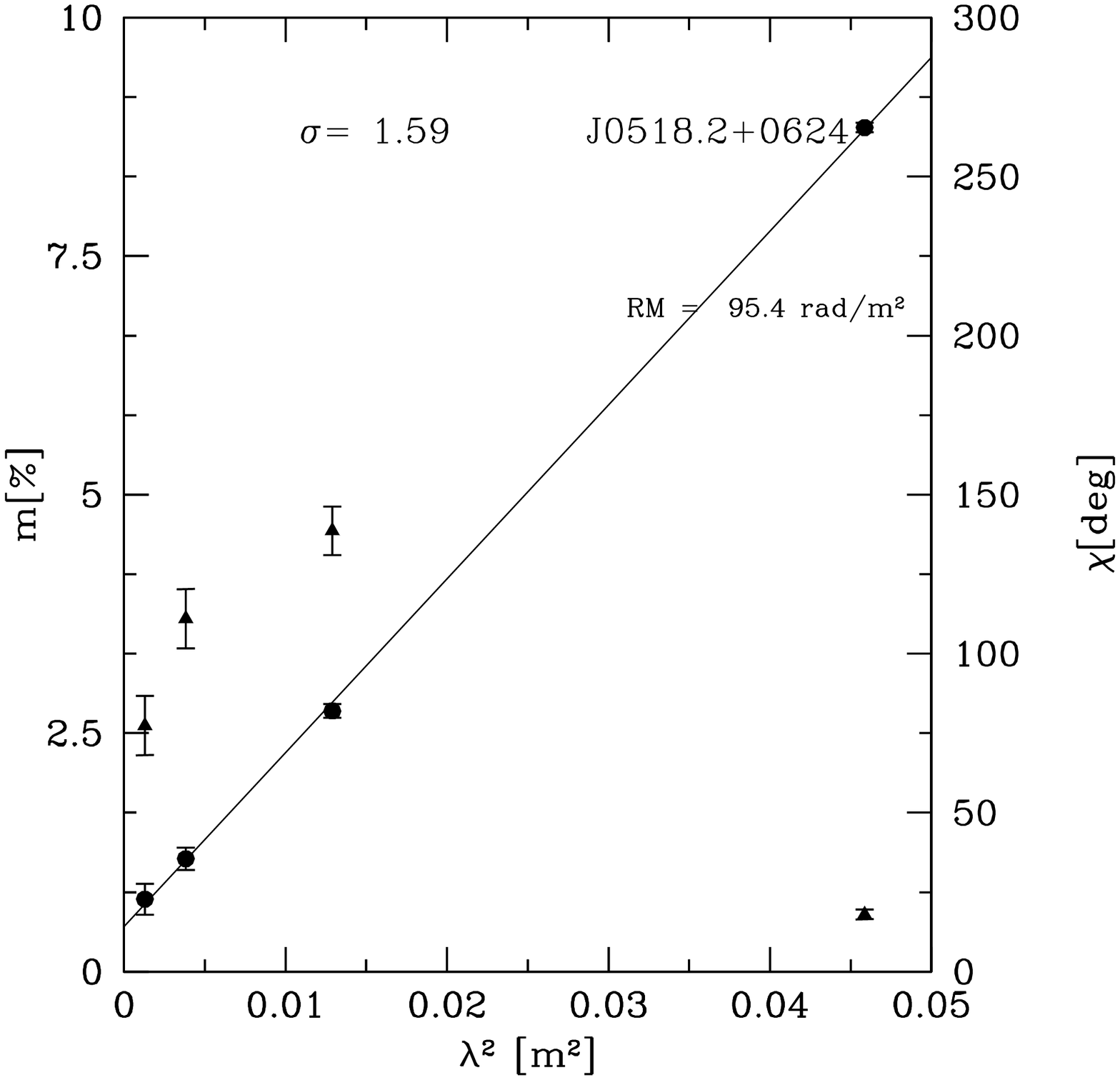}
\includegraphics[width=7cm]{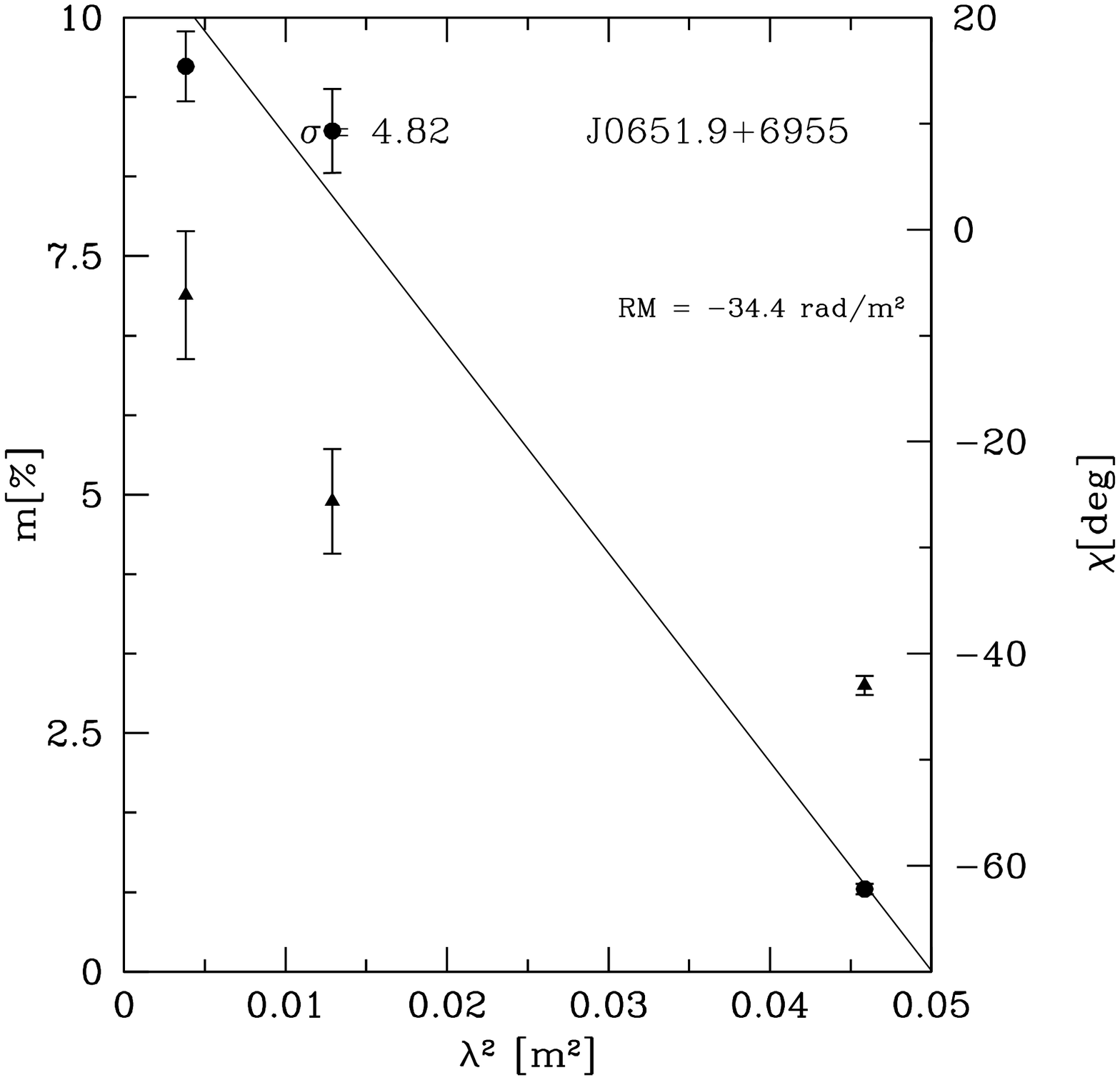}
\includegraphics[width=7cm]{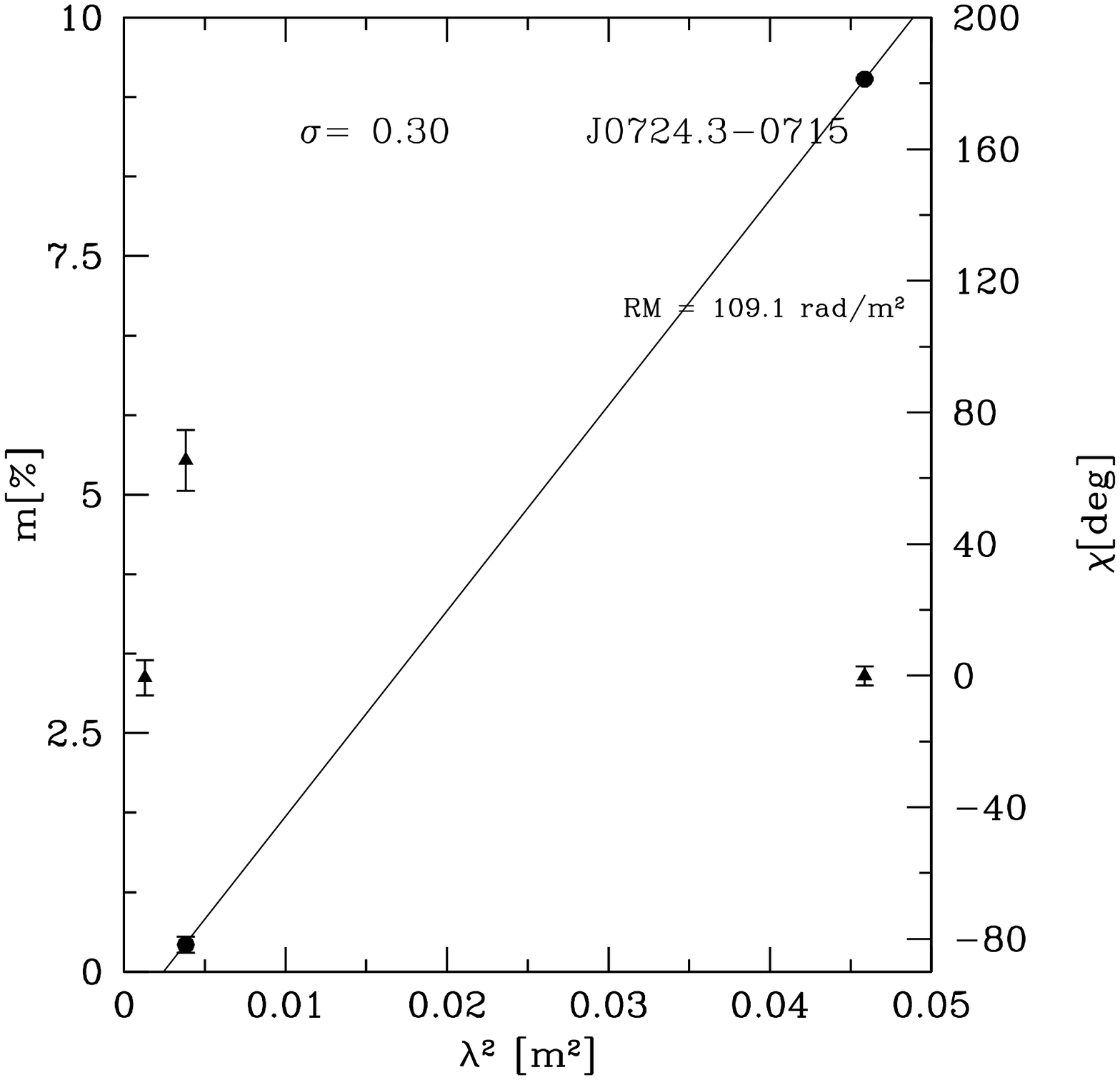}
\caption{Position angles of the electric vector $\chi$ (dots)
and fractional polarisation $m$ (triangles) versus $\lambda^2$ plots
of sources in Table 4. $\sigma$
values assess the quality of the best fit.}
\end{figure*}
\newpage
\begin{figure*}[t]
\addtocounter{figure}{+0}
\centering
\includegraphics[width=7cm]{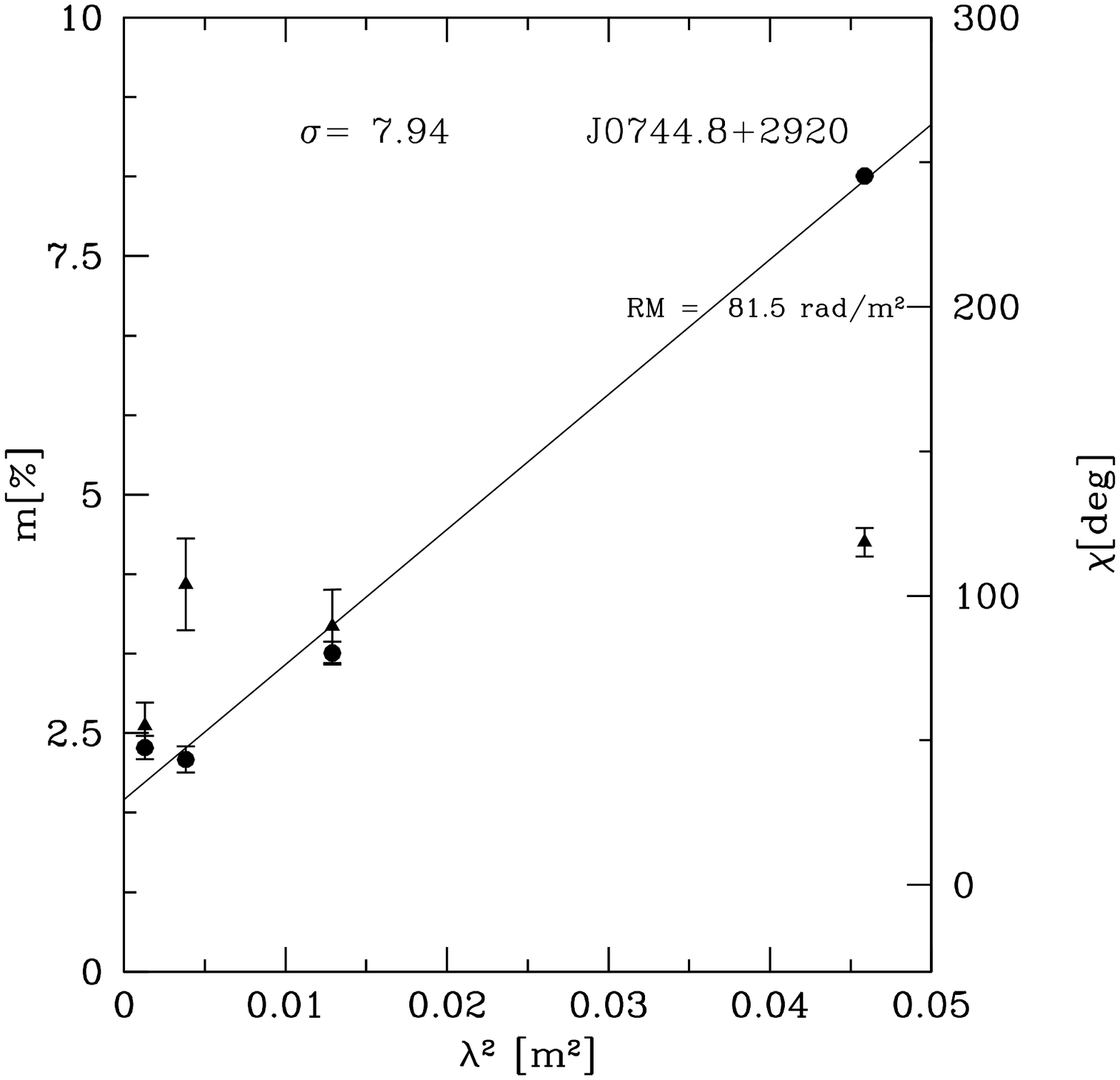}
\includegraphics[width=7cm]{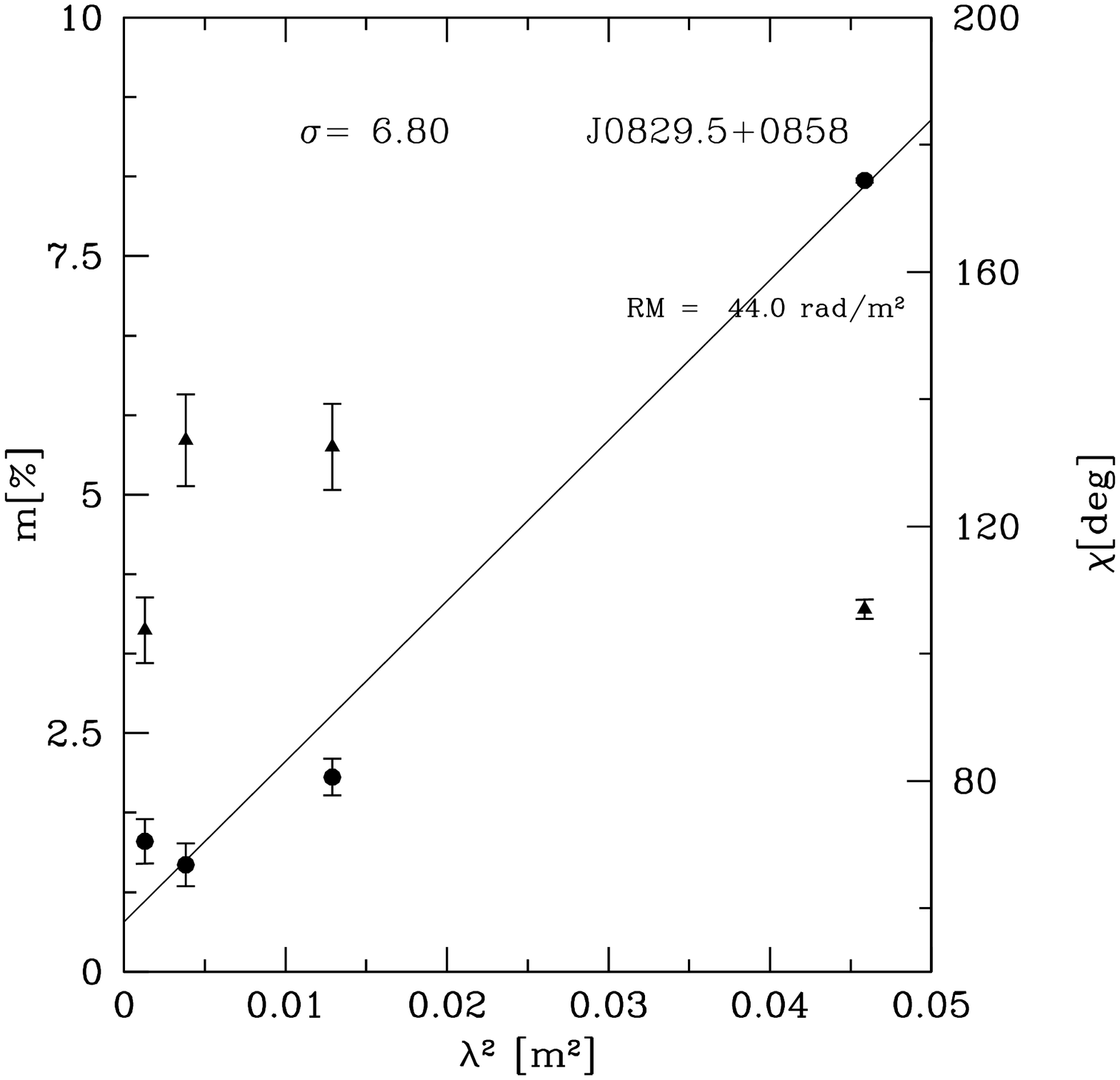}
\includegraphics[width=7cm]{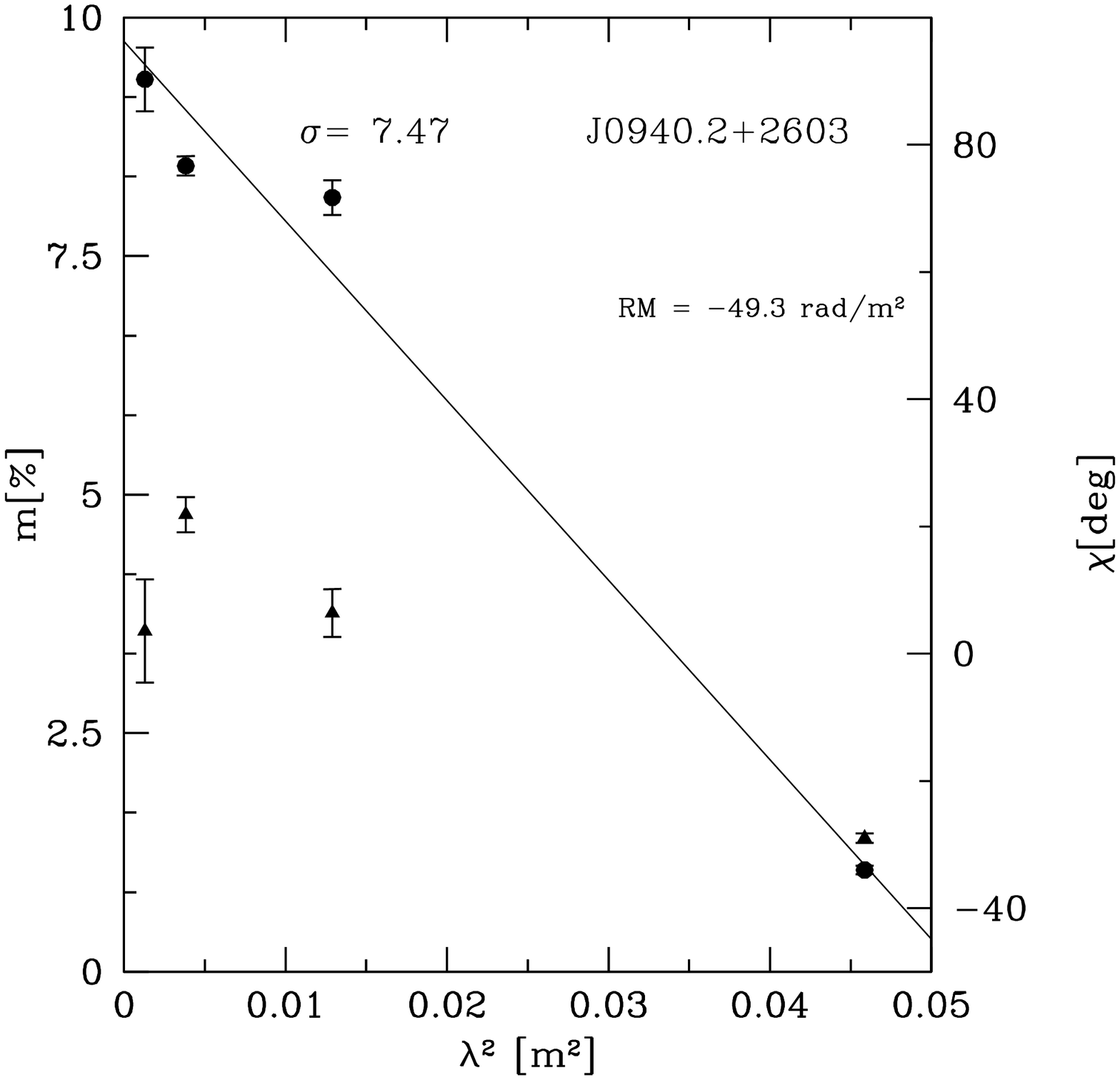}
\includegraphics[width=7cm]{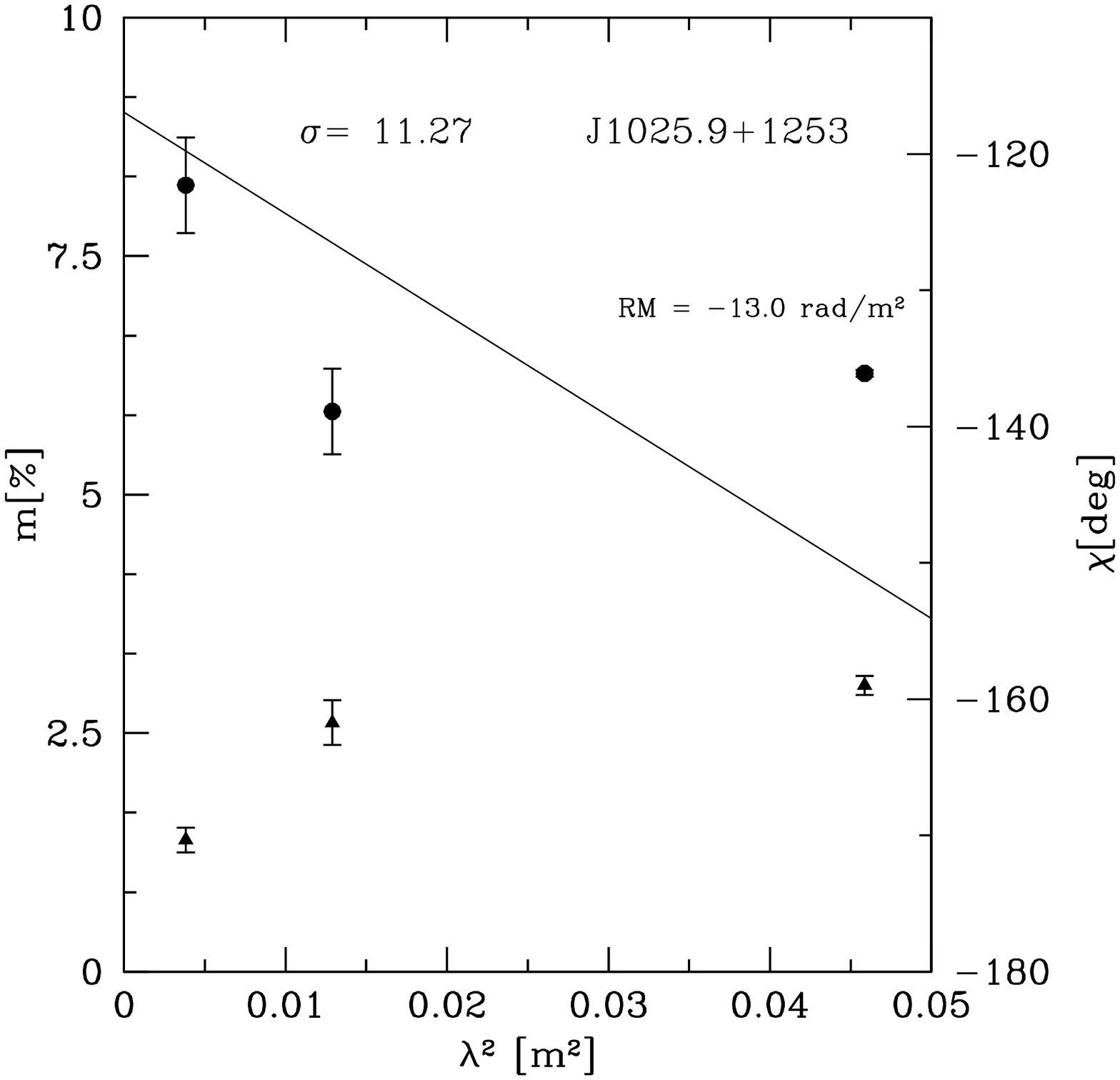}
\includegraphics[width=7cm]{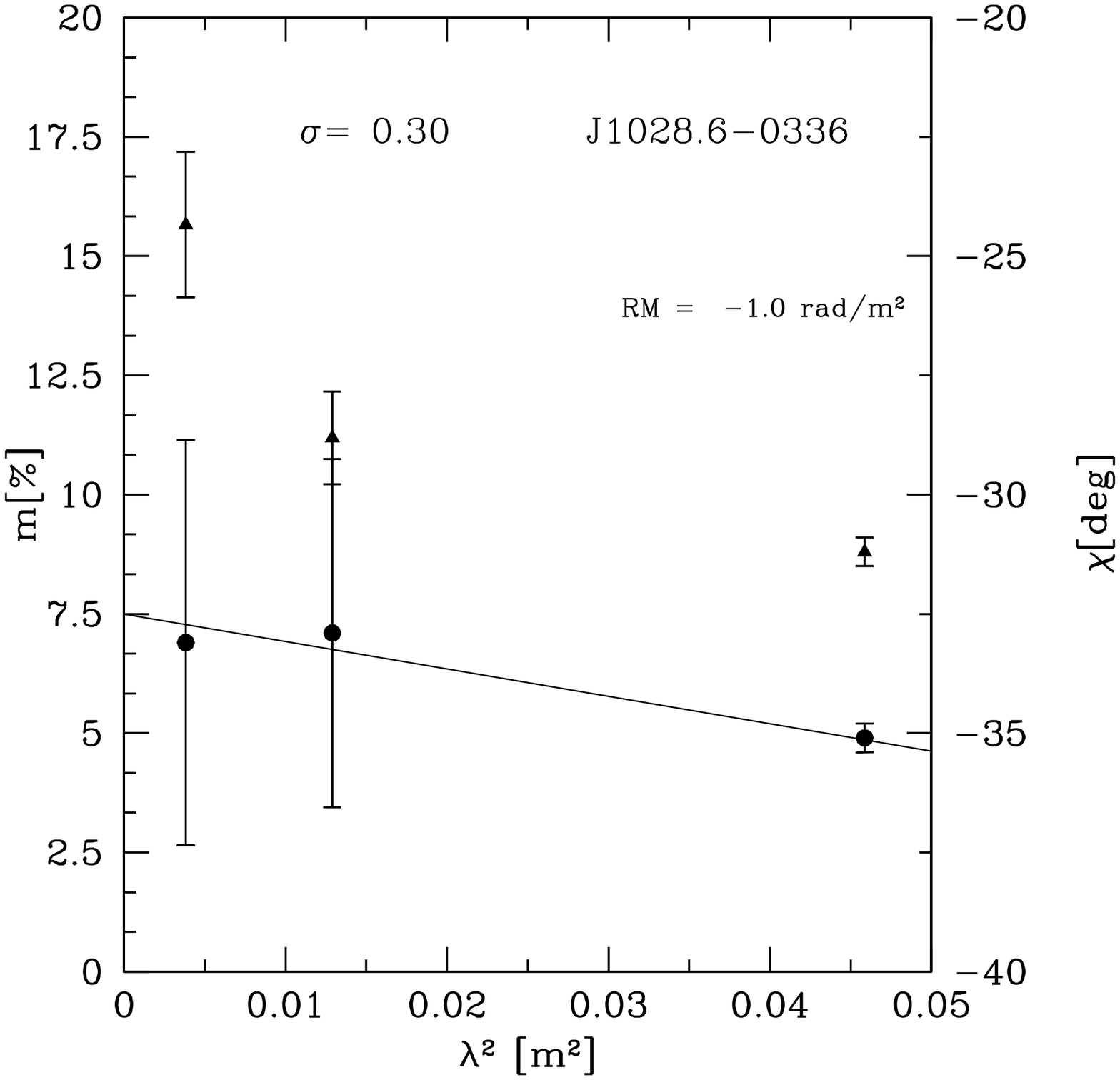}
\includegraphics[width=7cm]{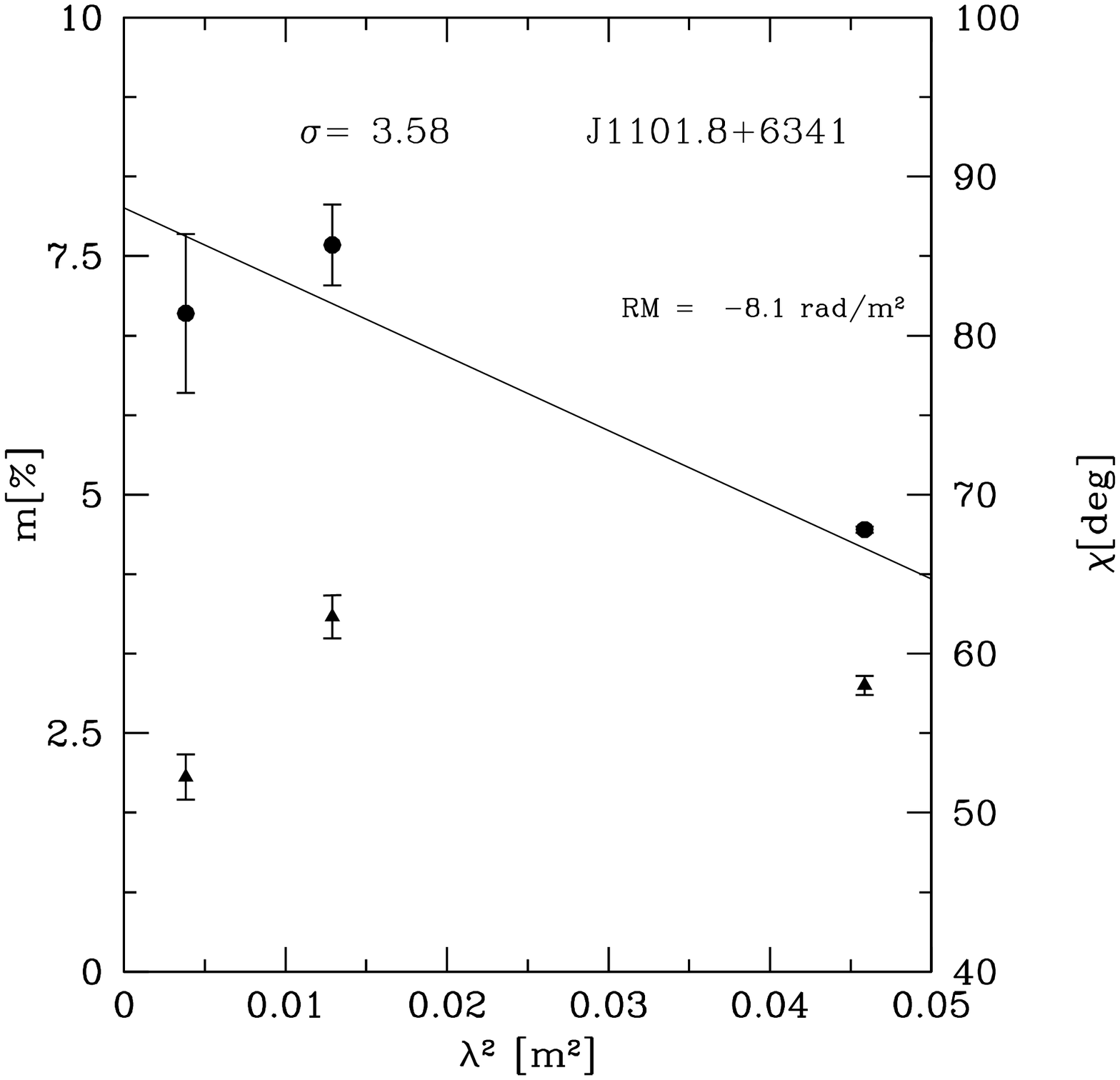}
\caption{Position angles of the electric vector $\chi$ (dots)
and fractional polarisation $m$ (triangles) versus $\lambda^2$ plots
of sources in Table 4. $\sigma$
values assess the quality of the best fit.}
\end{figure*}
\newpage
\begin{figure*}[t]
\addtocounter{figure}{+0}
\centering
\includegraphics[width=7cm]{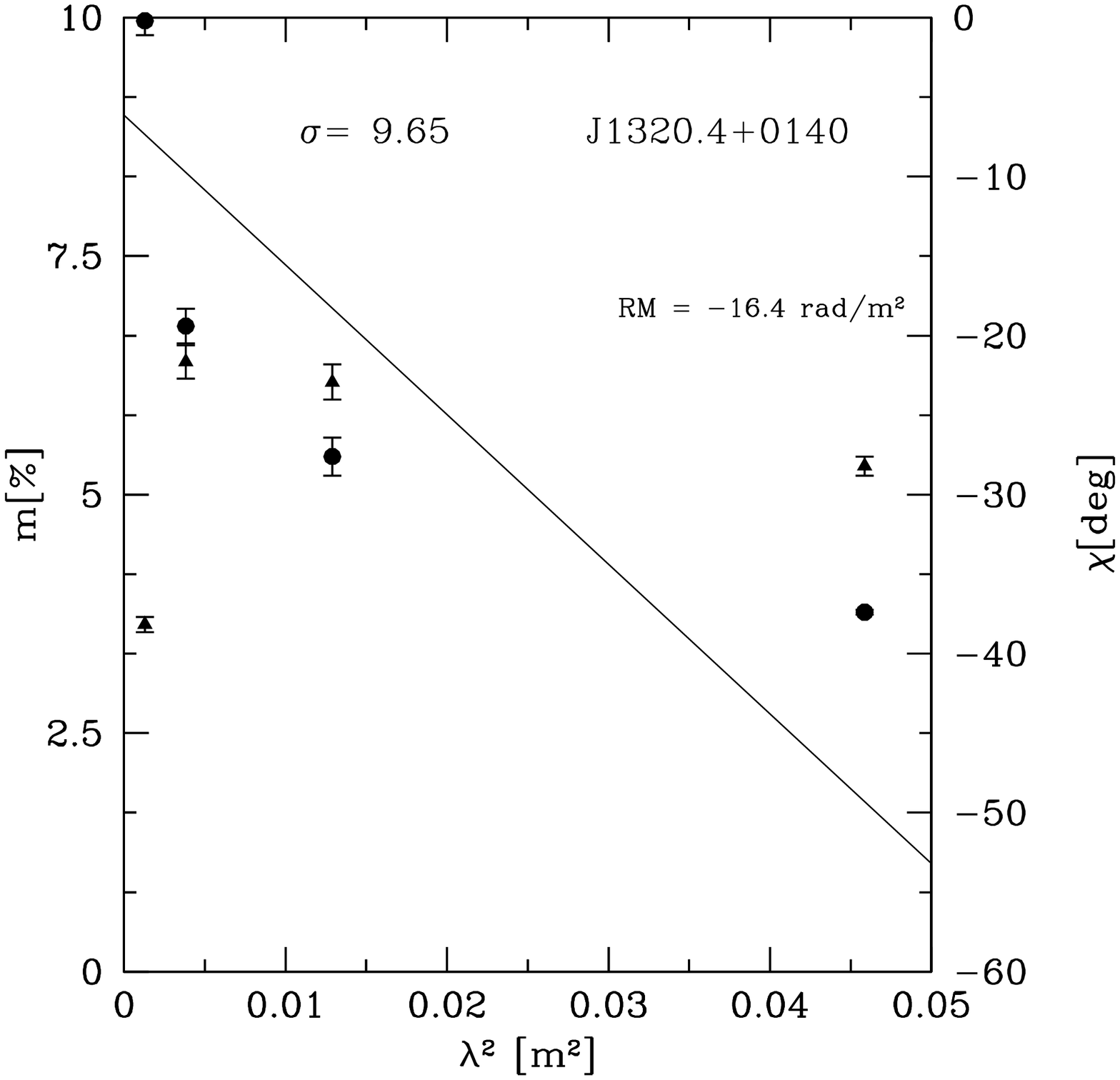}
\includegraphics[width=7cm]{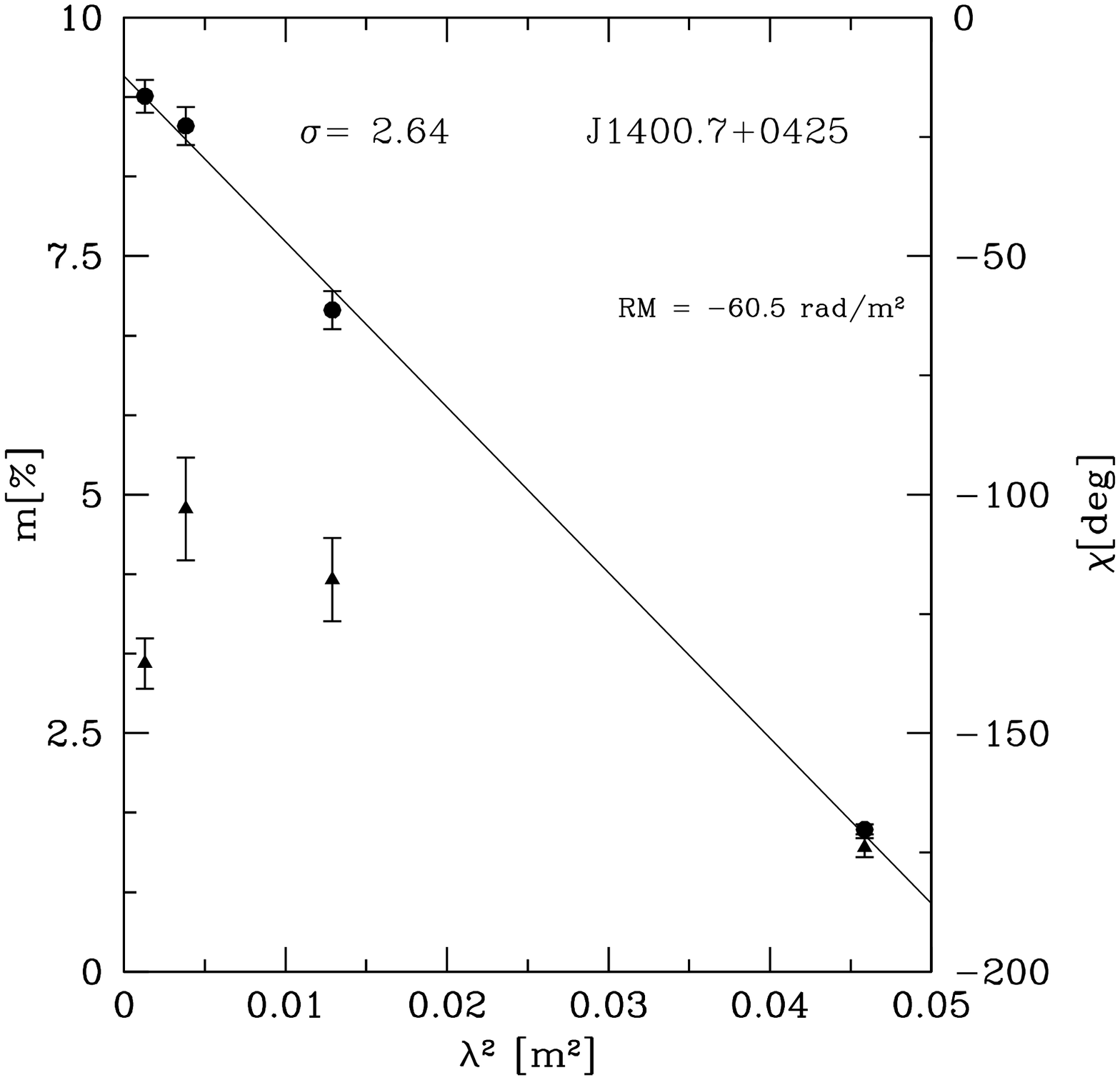}
\includegraphics[width=7cm]{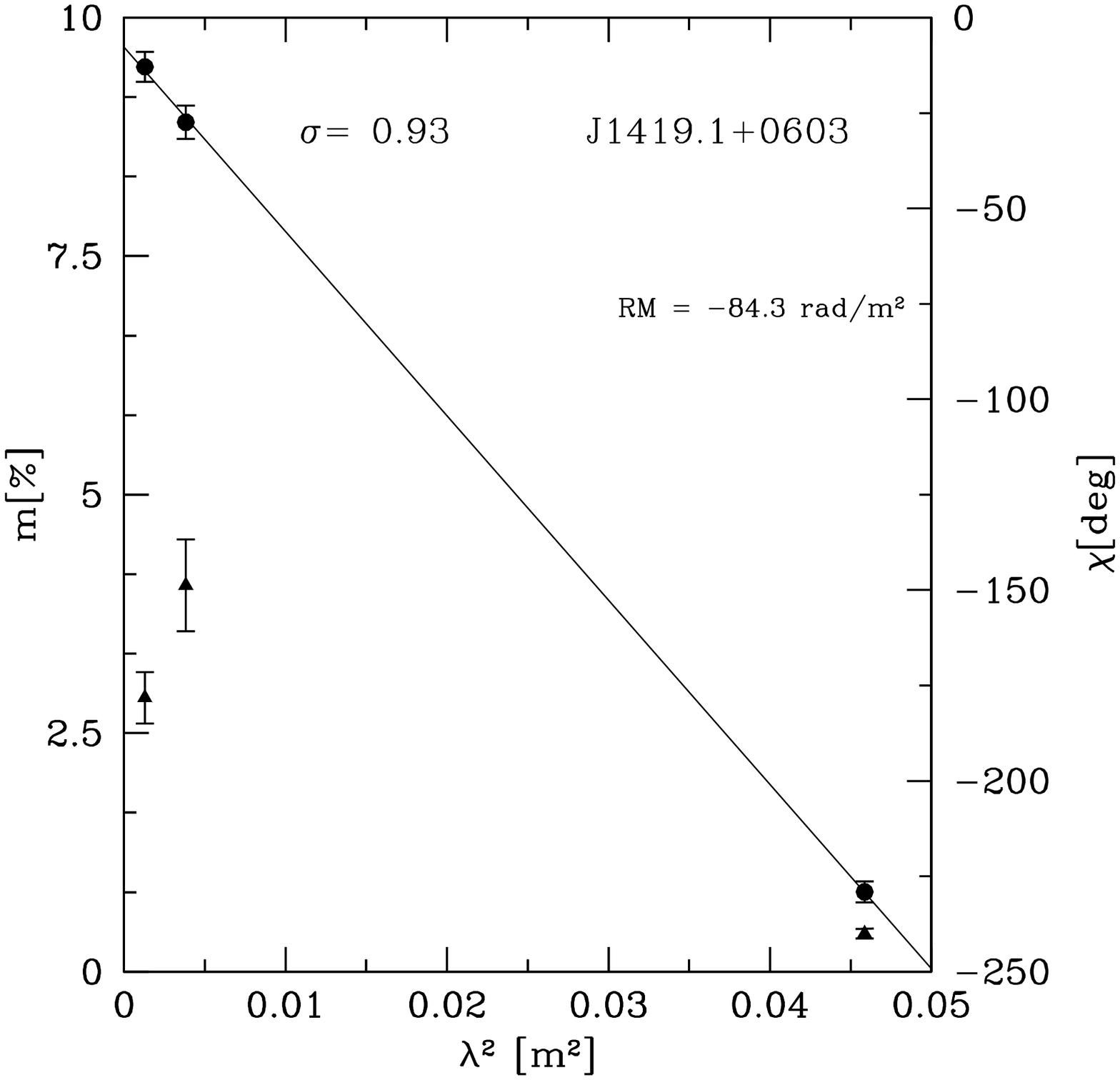}
\includegraphics[width=7cm]{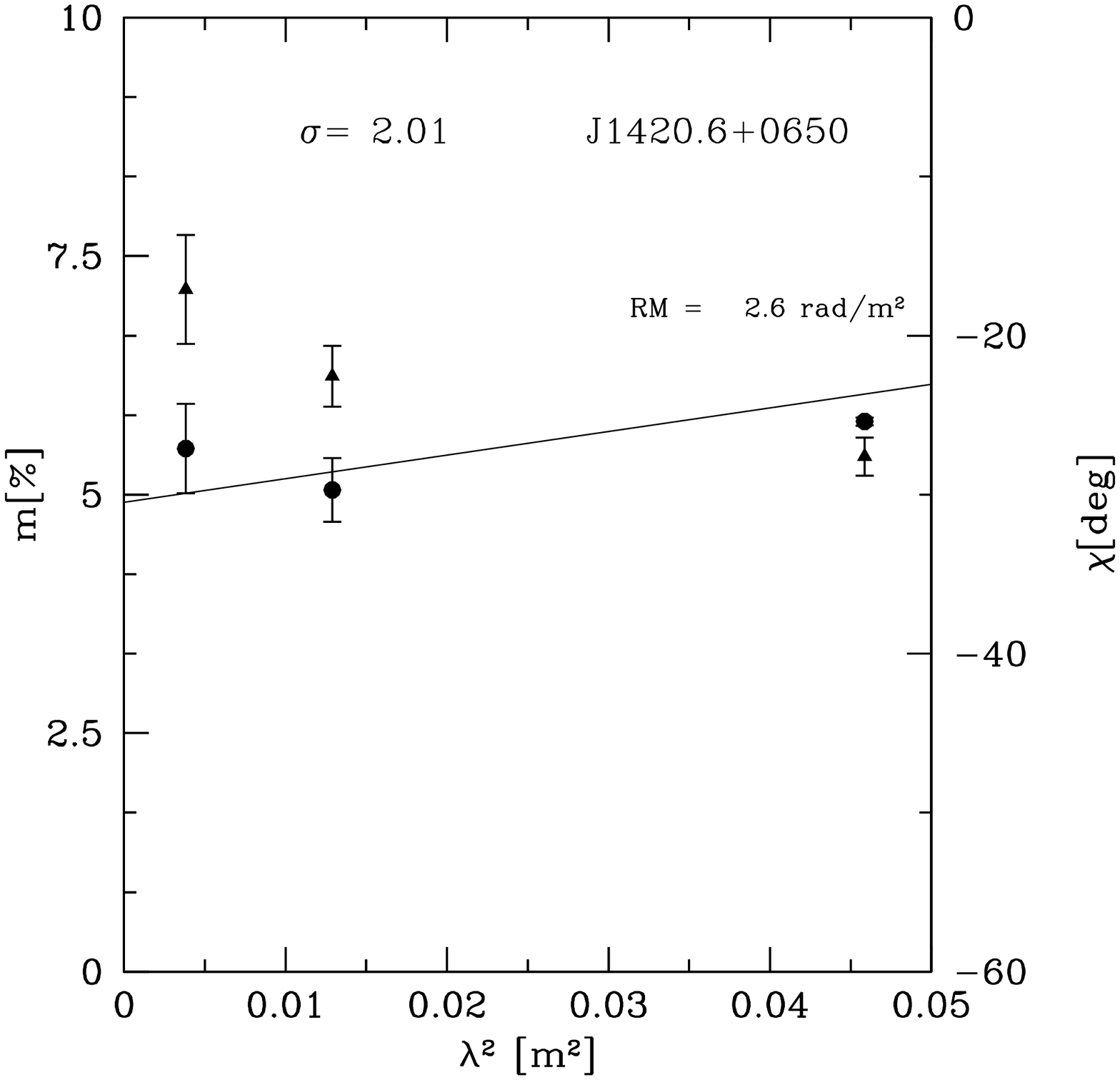}
\includegraphics[width=7cm]{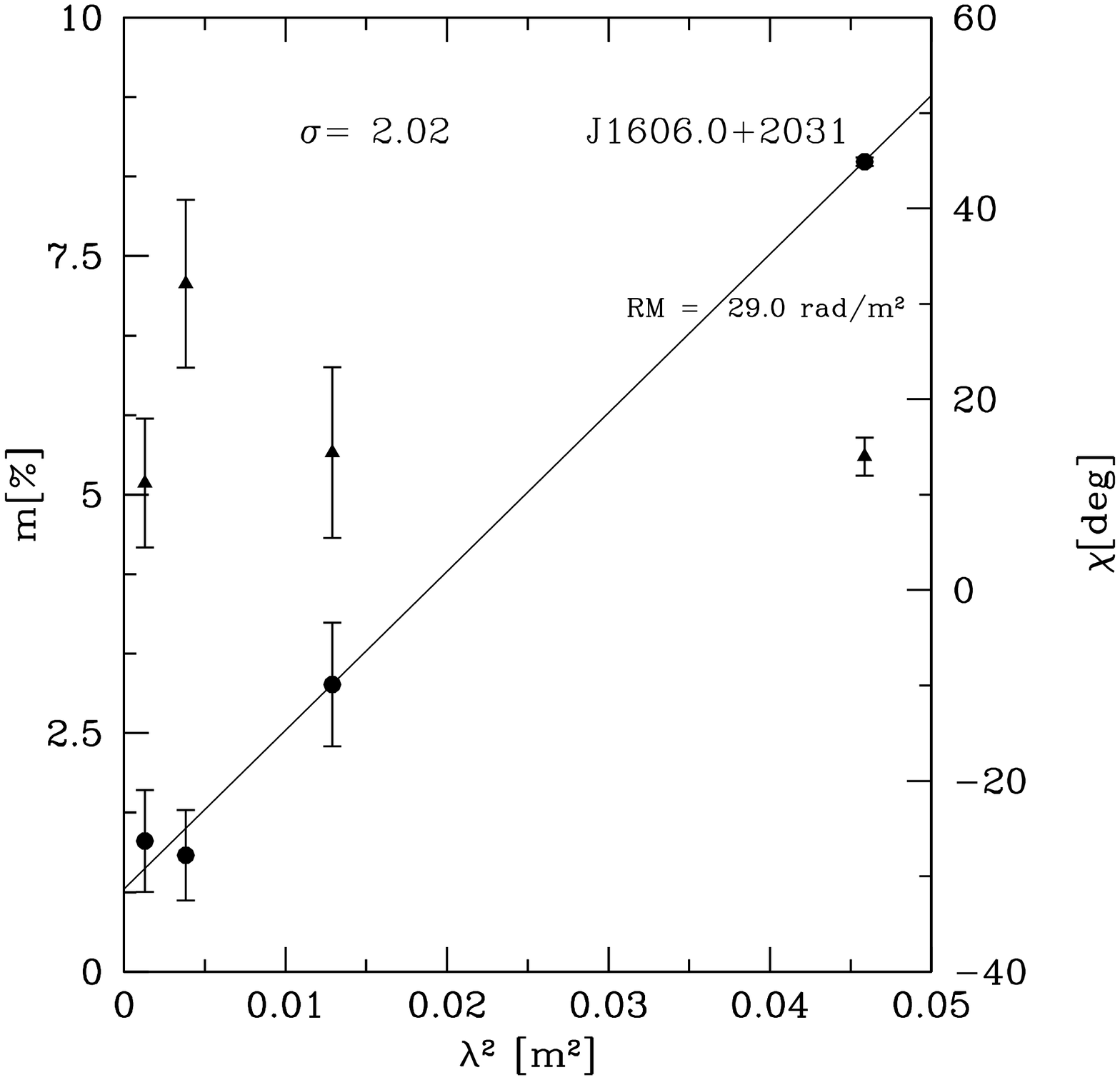}
\includegraphics[width=7cm]{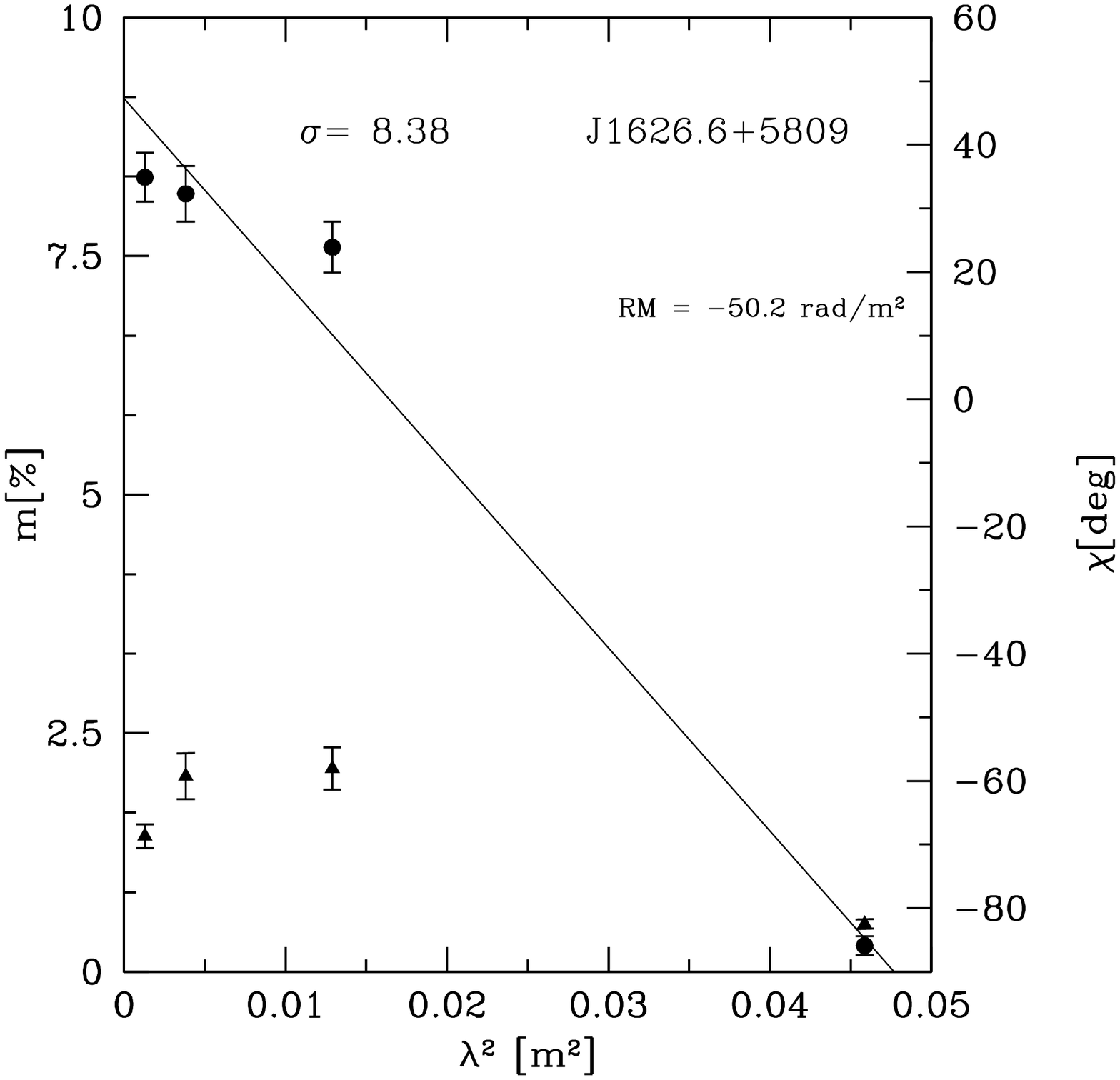}
\caption{Position angles of the electric vector $\chi$ (dots)
and fractional polarisation $m$ (triangles) versus $\lambda^2$ plots
of sources in Table 4. $\sigma$
values assess the quality of the best fit.}
\end{figure*}
\newpage
\begin{figure*}[t]
\addtocounter{figure}{+0}
\centering
\includegraphics[width=7cm]{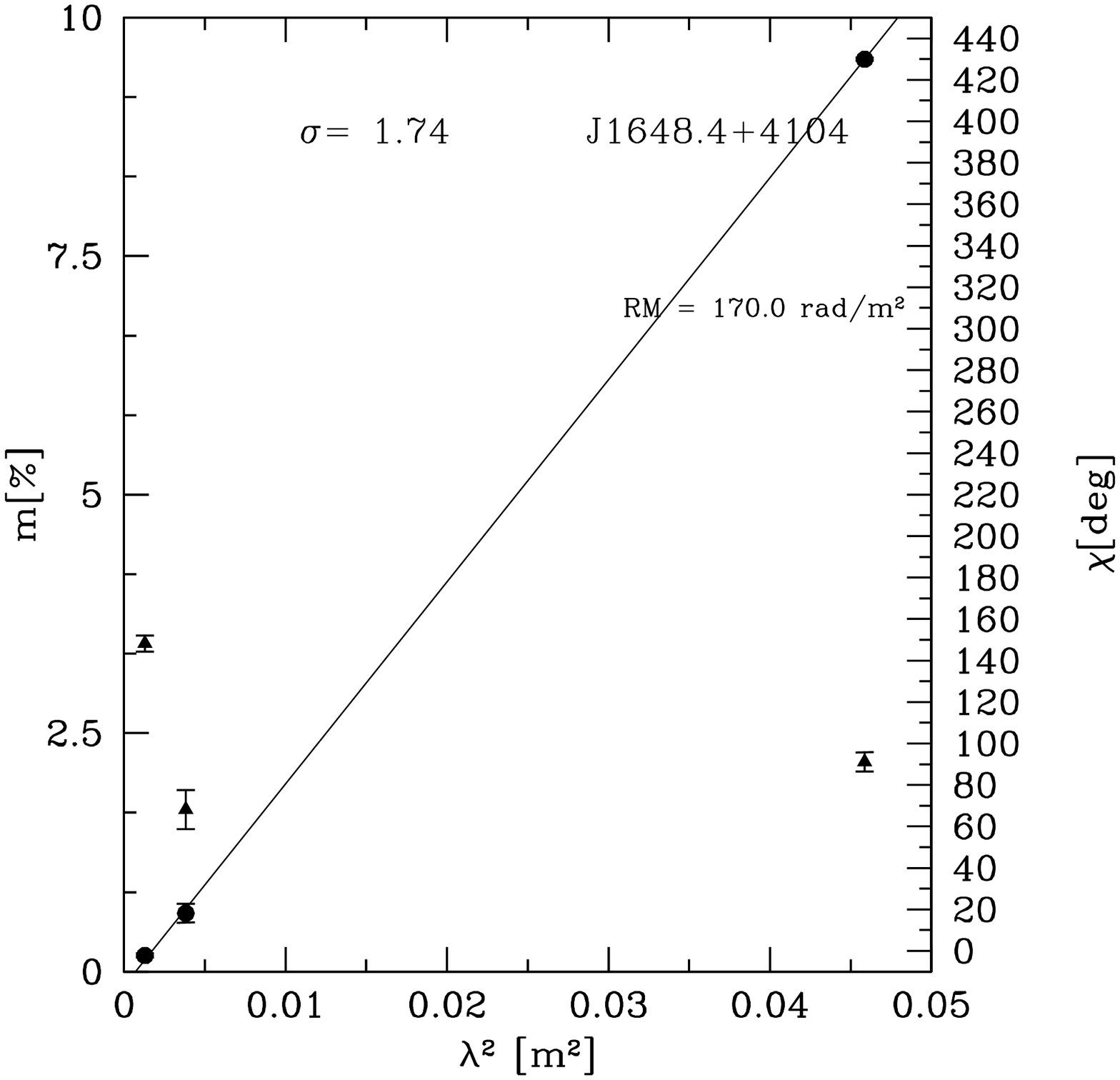}
\includegraphics[width=7cm]{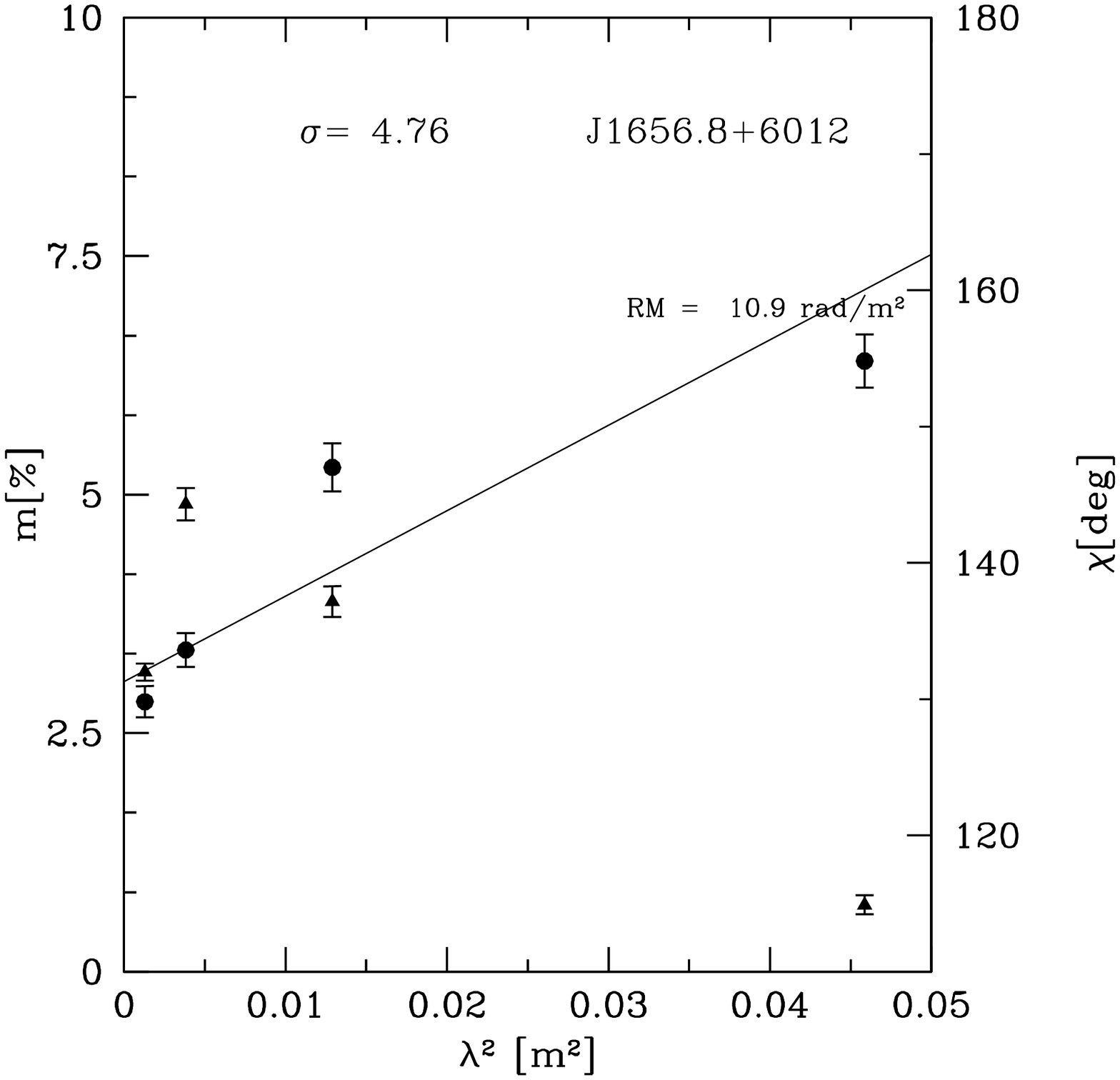}
\includegraphics[width=7cm]{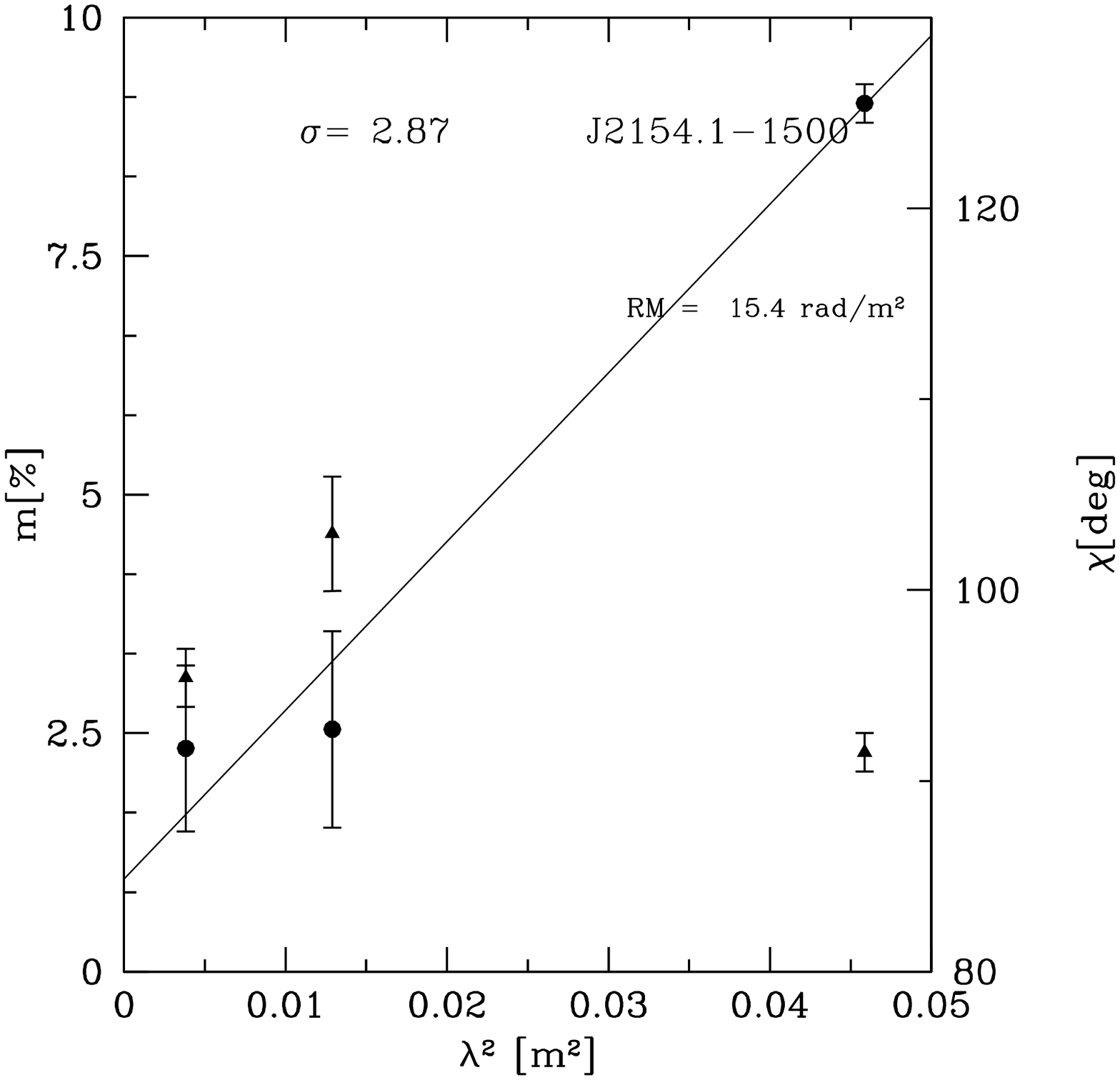}
\caption{Position angles of the electric vector $\chi$ (dots)
and fractional polarisation $m$ (triangles) versus $\lambda^2$ plots
of sources in Table 4. $\sigma$
values assess the quality of the best fit.}
\end{figure*}
\end{document}